\documentclass[12pt]{cernrep}
\usepackage{amsmath,amssymb,amsfonts,color,graphicx,cite,color,psfrag}
\evensidemargin \oddsidemargin
\graphicspath{{figs/}}

\makeatletter
\renewcommand\section{\@startsection {section}{1}{\z@}%
                           {-3.5ex \@plus -1ex \@minus -.2ex}%
                           {2.3ex \@plus.2ex}%
                           {\mathversion{bold}\normalfont\Large\bfseries}}
\renewcommand\subsection{\@startsection{subsection}{2}{\z@}%
                           {-3.25ex\@plus -1ex \@minus -.2ex}%
                           {1.5ex \@plus .2ex}%
                           {\mathversion{bold}\normalfont\large\bfseries}}
\renewcommand\subsubsection{\@startsection{subsubsection}{3}{\z@}%
                           {-3.25ex\@plus -1ex \@minus -.2ex}%
                           {1.5ex \@plus .2ex}%
                           {\mathversion{bold}\normalfont\normalsize\bfseries}}
\makeatother

\begin{document}

\thispagestyle{empty}
\setcounter{page}{0}
\def\thefootnote{\fnsymbol{footnote}}

\begin{flushright}
CERN--PH--TH/2009--166 \hfill
DCPT/09/136\\
IPPP/09/068 \hfill
SLAC-PUB-13782\\
arXiv:0909.3240 [hep-ph]\\
\end{flushright}

\vspace{1cm}

\begin{center}
{\Large {\bf From the LHC to Future Colliders}}\\

\vspace{4mm}
CERN Theory Institute Summary Report\\[1em]

A.~De~Roeck$^{a,b}$,
J.~Ellis$^{a}$, 
C.~Grojean$^{a,c}$, 
S.~Heinemeyer$^{d}$,\\
K.~Jakobs$^{e}$, 
G.~Weiglein$^{f}$,
J.~Wells$^{a}$~(organizers)\\[.3em]
G.~Azuelos$^{g}$,
S.~Dawson$^{h}$,
B.~Gripaios$^{a}$,
T.~Han$^{i}$,
J.~Hewett$^{j}$,
M.~Lancaster$^{k}$,
C.~Mariotti$^{l}$,
F.~Moortgat$^{m}$,
G.~Moortgat-Pick$^{f}$,
G.~Polesello$^{n}$,
S.~Riemann$^{p}$,
M.~Schumacher$^{e}$ (convenors)\\[.3em]
K.~Assamagan$^{h}$, 
P.~Bechtle$^{p}$, 
M.~Carena$^{q}$,
G.~Chachamis$^{r}$, 
K.F.~Chen$^{s}$,
S.~De~Curtis$^{t}$,
K.~Desch$^{u}$,
M.~Dittmar$^{a}$, 
H.~Dreiner$^{v}$, 
M.~D\"uhrssen$^{e}$, 
B.~Foster$^{au}$, 
M.T.~Frandsen$^{w,x}$,
A.~Giammanco$^{y}$,
R.~Godbole$^{z}$, 
P.~Govoni$^{aa}$,  
J.~Gunion$^{ab}$, 
W.~Hollik$^{ac}$,
W.S.~Hou$^{s}$,
G.~Isidori$^{ad}$, 
A.~Juste$^{q}$, 
J.~Kalinowski$^{ae}$,
A.~Korytov$^{af}$,
E.~Kou$^{ag}$,
S.~Kraml$^{ah}$,
M.~Krawczyk$^{ae}$,
A.~Martin$^{ai}$, 
D.~Milstead$^{aj}$,
V.~Morton-Thurtle$^{f}$,
K.~Moenig$^{o}$, 
B.~Mele$^{ak}$, 
E.~Ozcan$^{k}$, 
M.~Pieri$^{al}$, 
T.~Plehn$^{am}$, 
L.~Reina$^{an}$, 
E.~Richter-Was$^{ao,ap}$,
T.~Rizzo$^{j}$,
K.~Rolbiecki$^{f}$,
F.~Sannino$^{w}$, 
M.~Schram$^{aq}$,
J.~Smillie$^{f}$,
S.~Sultansoy$^{ar}$,
J.~Tattersall$^{f}$,
P.~Uwer$^{as}$,
B.~Webber$^{at}$,
P.~Wienemann$^{u}$.

\vspace{1em}

\end{center}
\begin{center}
{\bf Abstract}
\end{center}
Discoveries at the LHC will soon set the physics agenda for future colliders.
This report of a CERN Theory Institute includes the summaries of Working
Groups that reviewed the physics goals and prospects of LHC running with 10 to
300~fb$^{-1}$ of integrated luminosity, of the proposed sLHC luminosity upgrade,
of the ILC, of CLIC, of the LHeC and of a muon collider. The four Working Groups
considered possible scenarios for the first 10~fb$^{-1}$ of data at the
LHC in which (i) a state with properties that are compatible with a Higgs boson 
is discovered, (ii) no such state is discovered either because the Higgs
properties are such that it is difficult to detect or because 
no Higgs boson exists, (iii) a missing-energy signal beyond the
Standard Model is discovered as in some supersymmetric models, and (iv) some
other exotic signature of new physics is discovered. In the contexts
of these scenarios, the Working Groups reviewed the capabilities of the future
colliders to study in more detail whatever new physics may be discovered by the
LHC. Their reports provide the particle physics community with some tools for
reviewing the scientific priorities for future colliders after the LHC produces
its first harvest of new physics from multi-TeV collisions.

\newpage
\thispagestyle{empty}
\setcounter{page}{0}

\mbox{}\vspace{1cm}
\begin{center}

$^a$CERN, Department of Physics, Geneva, Switzerland

$^b$University of Antwerp, Wilrijk, Belgium

$^{c}$IPhT, CEA-Saclay, France

$^d$Instituto de F\'isica de Cantabria (CSIC-UC), Santander, Spain

$^e$Physikalisches Institut, Albert-Ludwigs-Universit\"at, Freiburg, Germany

$^f$IPPP, University of Durham, Durham, UK

$^g$Universite de Montr\'eal, Montr\'eal, Canada and TRIUMF, Vancouver, Canada

$^h$Physics Department, Brookhaven National Laboratory, Upton New York, USA

$^i$Department of Physics, University of Wisconsin, Madison, USA

$^j$SLAC National Accelerator Laboratory, Menlo Park, USA 

$^k$UCL, London, UK

$^l$INFN, Sezione di Torino, Italy

$^m$Department of Physics, ETH Honggerberg,  Zurich, Switzerland

$^n$INFN, Sezione di Pavia, Italy

$^o$DESY,  Zeuthen, Germany

$^p$DESY, Hamburg, Germany

$^q$Fermi National Accelerator Laboratory, Batavia, USA 

$^r$Paul Scherrer Institut, Villigen, Switzerland

$^s$Department of Physics, National Taiwan University, Taipei, Taiwan

$^t$Department of Physics, University of Florence  and INFN, Sezione di  Firenze, Italy

$^u$Universit\"at Bonn, Physikalisches Institut, Bonn, Germany 

$^v$Bethe Center for Theoretical Physics and Physikalisches Institut, Bonn
University, Germany

$^w$CP$^3$ -- Origins, University of Southern Denmark, Odense, Denmark

$^x$Rudolf Peierls Centre for Theoretical Physics, University of Oxford, UK

$^y$CP3, Universit\'e Catholique de Louvain, Louvain-la-Neuve, Belgium 

$^z$Centre for High Energy Physics, Indian Institute of Science,
Bangalore, India

$^{aa}$Universit\`a and INFN Milano-Bicocca

$^{ab}$Department of Physics, UC Davis, USA

$^{ac}$Max-Planck-Institut f\"ur Physik (Werner-Heisenberg-Institut),
M\"unchen, Germany

$^{ad}$INFN, Laboratori Nazionali di Frascati, Frascati, Italy 

$^{ae}$Physics Department, University of Warsaw, Warsaw, Poland

$^{af}$University of Florida, Gainesville, USA

$^{ag}$Laboratoire de l'Accelerateur Lineaire, Universit\'e Paris-Sud 11, Orsay, France

\end{center}

\newpage
\thispagestyle{empty}
\setcounter{page}{0}

\mbox{}\vspace{1cm}
\begin{center}

$^{ah}$LPSC, UJF Grenoble 1, CNRS/IN2P3, Grenoble, France

$^{ai}$Department of Physics, Sloane Laboratory, Yale University, New Haven,
USA 

$^{aj}$Fysikum, Stockholms Universitet, Stockholm, Sweden

$^{ak}$INFN, Sezione di Roma, and Universit\`a \ ``La Sapienza'', Roma, Italy

$^{al}$University of California San Diego, USA

$^{am}$Institute for Theoretical Physics, Heidelberg University, Heidelberg,
Germany 

$^{an}$Physics Department, Florida State University, Tallahassee, USA

$^{ao}$Institute of Physics, Jagellonian University, Krakow, Poland 

$^{ap}$Institute of Nuclear Physics IFJ-PAN, Krakow, Poland 

$^{aq}$McGill University, Montr\'eal, Canada

$^{ar}$Physics Division, TOBB University of Economics and Technology, Ankara,
Turkey 

$^{as}$Institut f\"ur Physik, Humboldt-Universit\"at zu Berlin, 
Berlin, Germany

$^{at}$Cavendish Laboratory, J.J.\ Thomson Avenue, Cambridge, UK

$^{au}$Particle Physics, University of Oxford, Keble Road, Oxford. UK

\end{center}

\def\thefootnote{\arabic{footnote}}
\setcounter{footnote}{0}



\newpage
\tableofcontents
\newpage


\addcontentsline{toc}{section}{Executive summary}
{\def\mpar#1{\marginpar{\tiny #1}}
\def\mua{\marginpar[\boldmath\hfil$\uparrow$]%
                   {\boldmath$\uparrow$\hfil}%
                    \typeout{marginpar: $\uparrow$}\ignorespaces}
\def\mda{\marginpar[\boldmath\hfil$\downarrow$]%
                   {\boldmath$\downarrow$\hfil}%
                    \typeout{marginpar: $\downarrow$}\ignorespaces}
\def\mla{\marginpar[\boldmath\hfil$\rightarrow$]%
                   {\boldmath$\leftarrow $\hfil}%
                    \typeout{marginpar: $\leftrightarrow$}\ignorespaces}

\newcommand{\lhcten}{LHC$_{10/{\rm fb}}$}
\newcommand{\lhchl}{LHC$_{300/{\rm fb}}$}

\section*{Executive summary}
\label{sec:summary}

The LHC is about to initiate the direct exploration of physics at the
TeV scale. 
Ground-breaking discoveries may be possible with the first few inverse
femtobarns of data, which would certainly have profound implications for
the future of the field of particle physics and beyond. The results
obtained at the LHC will set the agenda for the future colliders
that will be required to study any new physics in more detail. 
Once early LHC data have been analysed, the world-wide particle physics
community will need to converge on a strategy for shaping the future of
particle physics. Given the fact that the complexity and size of
possible future accelerator experiments will require a long construction
time, the decision of when and how to go ahead with a future major 
facility will have to be taken in a timely fashion. Several
projects for future colliders are being developed, and in a few years
time it will be necessary to set priorities between these
options, informed by whatever new physics the LHC may reveal.

This CERN Theory Institute brought together theorists, experimentalists
and machine physicists from around the
world to discuss --- before the actual start of data taking at the LHC
--- the physics goals, capabilities and possible results of
the LHC, and how these relate to future possible collider programmes.  
The plan of the Theory Institute was (i) to discuss recent physics
developments, (ii) to anticipate the near-term capabilities of the
Tevatron, LHC and other experiments, and (iii) to discuss the most
effective ways to be prepared to provide scientific input to plans for
the future direction of the field. 
The following points were addressed in particular:
physics progress and results prior to LHC collisions, initial goals and
prospects for the 2010 LHC physics run, and the subsequent physics
goals and prospects in the early LHC phase with 10~fb$^{-1}$ (we refer
to this as ``\lhcten'') and in the longer term with 300~fb$^{-1}$
(``\lhchl''). The programme of the Theory Institute was
structured according to the questions: 
1) What have we
learned from data accrued up to this point, and what may we expect to
know from new physics during the initial phase of LHC operation? 2) What
do we need to know from the LHC for planning future accelerators? 3)
What scientific strategies will be needed to advance from the planned
LHC running to a future collider facility? 

To answer the last two questions, participants studied in particular
what can be expected from the LHC with a specific early luminosity,
namely 10~fb$^{-1}$ (\lhcten), in different scenarios for TeV-scale physics, 
and which strategy for future colliders one would adopt in each case. In
order to address these questions, the Theory Institute efforts were
organized into four broad categories of possible signatures in the early
LHC data: (i) a Higgs candidate (and anything else), (ii) no Higgs
candidate (and anything else), (iii) missing energy, (iv) more exotic
signals of new physics. Four Working Groups studied details of each of
these different scenarios.

Key considerations for the Working Groups were the scientific benefits
of various future upgrades of the LHC compared with the feasibility and
timing of possible future colliders. For this reason, the programme also
included a series of talks overviewing future colliders, one on each
possible accelerator followed by a talk on the specific physics interest
of that collider, including the Tevatron, the (s)LHC, the ILC, the LHeC,
CLIC and a muon collider. 

Working Group 1 assessed the implications of the possible detection of a
state with properties that are compatible with a Higgs boson, whether
Standard Model (SM)-like or not. If Nature has chosen a SM-like
Higgs, then
ATLAS and CMS are well placed to discover it with 10~fb$^{-1}$ and
measure its mass. However, measuring its other characteristics (decay
width, spin, CP properties, branching ratios, couplings, \ldots) with an
accuracy better than 20 to 30\%, and correlating them with precision top
studies and electroweak precision physics, will require some future
collider. 

In the Higgs mass region below $\sim 130$~GeV the LHC experiments will
probe various production modes (gluon and weak-boson fusion) and decay
modes ($\gamma \gamma$, $\tau \tau$ and eventually ${\bar b} b$ final
states), and a 5-$\sigma$ discovery in the early phase
with 10~fb$^{-1}$ will probably
require combining these channels. In that
context, the Tevatron will add valuable complementary information, in
particular via the Higgs search in the important $W/Z H, H \to {\bar b}
b$ channels. On the other hand, the LHC with 300~fb$^{-1}$ could
measure, with mild theory assumptions,
the couplings of a low-mass Higgs boson to some fermions with an
accuracy of 15 to 30\%, and the Higgs couplings to gauge bosons to about
10\%, but will provide no access to Higgs self-couplings. The sLHC could
increase the accuracy of the measurements of the Higgs couplings and may
give some access to the Higgs self-couplings in the mass region around
160 GeV (though this requires further simulation), and would also be
sensitive to rare decays of a light Higgs boson. The ILC would allow
precise measurements of all the quantum numbers and many couplings of
the Higgs boson, in addition to its mass and width, yielding in this way
a nearly complete profile of the Higgs boson. If the Higgs boson is
relatively light, CLIC could give
access to more of its rare decay modes of the Higgs boson and improve
access to the Higgs self-couplings, and
produce any resonances weighing up to 2.5~TeV in $WW$ scattering. 

Working Group 2 considered scenarios in which no state is
detected with the first 10~fb$^{-1}$ of LHC data with properties that
are compatible with a Higgs boson. It reviewed complementary physics
channels such as gauge boson self-couplings, longitudinal vector-boson
scattering, exotic Higgs scenarios, and scenarios with invisible Higgs
decays. 
If no clear Higgs-like signal has been
established with the first 10~fb$^{-1}$ of LHC data, one needs to
consider two generic classes of scenarios: those
in which a Higgs exists but is difficult to see, and those in which no
Higgs exists at all. 

Three specific examples of the former scenarios were studied: models
with complex parameters such as the supersymmetric CPX scenario, models
with unexpected visible Higgs decays, and models with invisible Higgs
decays. Also studied were four ``Higgsless'' scenarios: walking
technicolour models, scenarios with extra dimensions, models in which
extra dimensions are deconstructed (i.e., replaced by a set of discrete
points) and models with strong $WW$ scattering. 

In many of these scenarios, with higher LHC luminosity (e.g., with the
sLHC) it should be possible to determine whether or not a Higgs boson
exists, e.g., by improving the sensitivity to the production and decays
of Higgs-like particles or vector resonances, or by measuring $WW$
scattering. The ILC would enable precision measurements of even the most
difficult-to-see Higgs bosons, as would CLIC. The latter would be also
good for producing heavy resonances. Which future collider option is to
be preferred may well depend on other early LHC physics results, e.g.,
whether the LHC discovers other new physics such as supersymmetry or
extra dimensions, or whether there is other evidence from the LHC or
elsewhere for CP-violating effects beyond the SM. In particular, if
other new physics is detected that seems to hint at the realisation of 
(at least one) fundamental Higgs state in nature, e.g., 
supersymmetric particles are produced and / or the gauge sector does not
show indications of strong electroweak symmetry breaking dynamics,
then this could be a strong case for an $e^+e^-$ linear collider to explore 
the expected mass range for the Higgs and to determine precisely the nature
of the other observed new physics.

In considering missing-energy signatures at the LHC, Working Group 3
used supersymmetry as a representative model. The signals studied
included events with leptons and jets, with a view to measuring the
masses, spins and quantum numbers of any new particles produced. 

Studies of the LHC capabilities at $\sqrt{s} = 14$~TeV show that
100~pb$^{-1}$ of luminosity would enable the overall energy scale of
the missing-energy physics to be determined with an accuracy of 8\%, if
the mass scale of supersymmetry (or other missing-energy physics)
is near the lower limit of the range
still allowed by lower-energy experiments. With 1~fb$^{-1}$ of LHC
luminosity, signals of missing energy with one or more additional
leptons would give sensitivity to a large range of supersymmetric mass scales.  
Several ways to measure the masses of individual sparticle
masses were discussed, aimed at dealing with the difficulties presented
by the missing energy-momentum vector. This also creates difficulties
for spin measurements, which would benefit from more information about
the reference frame of the decaying particle. 

In all the missing-energy scenarios studied, early LHC data would
provide important input for the technical and theoretical requirements
for future linear collider physics, such as the detector capabilities
(e.g., resolving mass degeneracies could require exceptionally
good jet energy
resolution), running scenarios, required
threshold scans and upgrade options (e.g., for a $\gamma
\gamma$ collider and/or GigaZ). 

In many scenarios, the missing energy is carried away by dark-matter
particles, generating a very important connection with
cosmology. Characterizing the nature of the missing-energy scenario,
e.g., so as to be able to use data to calculate the dark-matter density,
will (for many supersymmetric scenarios) be difficult with LHC data
alone, and a future linear collider would help greatly in this
analysis. 

Working Group 4 studied examples of phenomena that do not involve a
missing-energy signature, such as the production of a new $Z^\prime$
boson, other leptonic resonances, a fourth generation of fermions and
exotic quarks, lepton-number-violating signals, the impact of new
physics on observables in the flavour sector, TeV-scale gravity
signatures, heavy stable charged particles and other exotic signatures
of new physics. In general, determining the properties of such new
physics phenomena, for instance by measuring couplings and angular
distributions, will require a much larger number of events and more
precise measurements than what is needed for the discovery of such
phenomena. 

The sLHC luminosity upgrade has the capability to add crucial
information on the properties of any new physics discovered during early
LHC running, as well as increasing the search sensitivity. On the other
hand, a future linear collider, with its clean environment, known
initial state and polarised beams, is unparalleled in terms of its
abilities to conduct ultra-precise measurements of new and SM phenomena,
as long as the new physics scale is within reach of the machine. For
example, in the case of a $Z^\prime$, high-precision measurements at a
future linear collider would provide a mass reach that is about ten
times higher than the centre-of-mass energy of the linear collider
itself. 

Generally speaking, the physics capabilities of the sLHC, the ILC and
CLIC are relatively well understood, but will need refinement in light of
initial LHC running. In cases where the exploration of new physics might be 
challenging at the early LHC, synergy with a linear collider could be very
beneficial. In particular a staged approach to linear-collider energies
seems very promising from the physics point of view, 
and should be further investigated. On the other hand, the physics cases for
the LHeC 
or a muon collider have yet to be established. The prospects of the LHeC
for enlarging the coverage of the LHC will depend crucially on the
specific scenario of TeV-sale physics realised in Nature. In the case of
the muon collider, a background-saturated environment, the challenge of making
vertex measurements and the lack of polarised beams, as well as the
significant loss of 
forward coverage due to shielding, will make precision measurements more
challenging than at a linear collider. The exploration of some new
physics phenomena may ultimately call for very high energies. For
instance, in a scenario where the fundamental Planck scale is in the TeV
range, a complete mapping of the energy regime five to ten times above
the Planck scale would require an energy upgrade of the LHC (DLHC) or
even a Very Large Hadron Collider (VLHC). 

As already emphasized, the physics cases for all these projects will
need to be reviewed after data from the initial LHC running are
analyzed. Some physics scenarios envisage new physics at a relatively
low energy scale, such as a light Higgs boson or some low-mass
supersymmetric particles. On the other hand, some scenarios such as
those with high-mass vector resonances or strong $WW$ scattering suggest
that new physics may appear only at high energies. Even in scenarios
with new low-mass physics, their full exploration may also require
higher energies, e.g., to measure the properties of strongly-interacting
sparticles or higher-lying Kaluza--Klein excitations in scenarios with
extra dimensions. 

The purpose of this Theory Institute was not to arbitrate between these
scenarios and the corresponding priorities for future colliders, but
rather to provide the particle physics community with some tools for
such considerations when the appropriate time comes. 
Novel results from the early LHC data may open the way towards
an exciting future for particle physics
made possible by a new major facility. In order to
seize this opportunity, the particle physics community will 
need to agree on convincing and
scientifically solid motivations for such a facility.
The intention of this Theory Institute was to provide a framework for
discussing now how this could be achieved,
before actual LHC results start to come in. 
We now look forward to the first multi-TeV collisions in the LHC, and to the
harvest of new physics they will provide.


 }
\newpage
{
\subsection*{Acknowledgements}

The CERN TH Institute has
been funded by the CERN TH unit and by 
`MassTeV' ERC advanced grant 226371.
This work has been supported
in part by the European Community's Marie-Curie Research
Training Network under contract MRTN-CT-2006-035505
`Tools and Precision Calculations for Physics Discoveries at Colliders'
(HEPTOOLS) and MRTN-CT-2006-035657
`Understanding the Electroweak Symmetry
Breaking and the Origin of Mass using the First Data of ATLAS'
(ARTEMIS).
This work was supported in part by the DOE under Task TeV of contract
DE-FGO2-96-ER40956.
J.H.\ and T.R.\ were supported by the
U.S. Department of Energy under Contract DE-AC02-76SF00515.
M.L. was supported by the Science and Technology Facilities Council, UK.
 }
\newpage
{\setcounter{equation}{0}
\setcounter{figure}{0}
\setcounter{table}{0}

\input paperdef

\section{WG1: Higgs}
\label{wg1}
{\it 
S.~Dawson, 
S.~Heinemeyer, 
C.~Mariotti,
M.~Schumacher (convenors)\\ 
K.~Assamagan, 
P.~Bechtle, 
M.~Carena,
G.~Chachamis, 
K.~Desch,
M.~Dittmar, 
H.~Dreiner, 
M.~D\"uhrssen, 
R.~Godbole, 
S.~Gopalakrishna,
W.~Hollik,
A.~Juste, 
A.~Korytov,
S.~Kraml,
M.~Krawczyk,
K.~Moenig, 
B.~Mele, 
M.~Pieri, 
T.~Plehn, 
L.~Reina, 
E.~Richter-Was,
P.~Uwer,
G.~Weiglein
}

\bigskip
The prospects for a Higgs boson discovery with 10 \ifb\ at the LHC are
summarized and the implications of such a discovery for future
colliders such as the sLHC, the ILC, and CLIC are discussed in this section.


\newcommand{\lhcten}{LHC$_{10/{\rm fb}}$}
\newcommand{\lhchl}{LHC$_{300/{\rm fb}}$}

\subsection{Introduction and scenarios}

Identifying the mechanism of electroweak symmetry
breaking will be one of the main goals of the LHC and other future
high-energy physics experiments.
Many possibilities have been studied in the literature, of which 
the most popular ones are the Higgs mechanism in the Standard Model
(SM) and the Minimal Supersymmetric Standard Model (MSSM).
Assuming that a new state which is
 a possible candidate for a Higgs boson 
has been observed, the full identification of the mechanism of
electroweak symmetry breaking will require the measurement of all its
characteristics. This comprises an accurate mass determination, 
a (model-independent) measurement of its individual couplings to other
particles (i.e.\ not only the ratio of couplings), 
a determination of the self-couplings to confirm the ``shape'' of the
Higgs potential, as well as measurements of its spin and $\cp$-quantum
numbers, etc. 
These measurements will most probably only be partially  possible at
the LHC, even running at high luminosity. It will be up to future
colliders to complete the Higgs profile.

We first review what we might know about the Higgs sector 
once the LHC has collected 10\,\ifb\ at a center-of-mass energy  
of 14 TeV (called \lhcten\ in the following) and has observed 
an object compatible with a Higgs boson.
Secondly, we investigate the capabilities
of  future colliders to  
further unravel the mechanism responsible for 
electroweak symmetry breaking and to confirm that a Higgs boson 
has indeed been observed.
The discussion of the second step will be split into three scenarios:
\begin{itemize}
\item {A:} Observation of a SM-like Higgs boson with a mass
$130 \gev \lsim \MH \lsim 180 \gev$.  This mass range
theoretically
allows the SM to be valid until the Planck scale. 
SM-like means that no statistically significant deviations of the
 properties of the Higgs boson from the expectations of the SM can be
observed at the \lhcten.
It should be kept in mind that a SM-like Higgs boson in the mass range
of $160 \gev \le \MHSM \le 170 \gev$ has recently been excluded at the
95\%~C.L.\ by the Tevatron~\cite{Phenomena:2009pt}.
\item {B}: Observation of a SM-like Higgs boson outside the above 
mass range of $130 \gev$ to $180 \gev$.
\item {C}: Observation of a non-SM-like Higgs boson (e.g.\
signal rates or coupling structures  deviate from  SM expectations). 
See Section~\ref{sec:WG2InvHiggs}  for a discussion of various scenarios with unusual
couplings.
\end{itemize}
Scenario C, or Scenario B with $\MH\gsim 180 \gev$,
typically imply additional signs of new physics besides a single
Higgs boson.


\subsection{Observations at the \boldmath{\lhcten} for a SM-like Higgs boson}

Most quantitative analyses at ATLAS and CMS have been performed for a
SM-like Higgs boson. Consequently, we 
summarize and comment on the potential of \lhcten\ to observe
a Higgs boson assuming SM-like couplings. 

We will not try to disentangle and explain
differences in the discovery potentials between ATLAS and 
CMS. Details can be found in the original publications of the
ATLAS~\cite{Aad:2009wy} and CMS~\cite{Ball:2007zza}
collaborations, which contain information on
how the discovery potentials have been evaluated.
 However, we will briefly mention existing
differences between
the experimental results
 that might be relevant for the subsequent discussion.

\begin{table}[htb!]
\renewcommand{\arraystretch}{1.2}
\begin{center}
\begin{tabular}{|l||c|c|c|c|c|} \hline
channel / $\MH$ [GeV]                        &110&115&120&125&130\\ 
                                                              \hline\hline
ATLAS $H \to \ga\ga$ cuts             & --&2.0&2.4&-- &2.7\\ \hline
ATLAS $H \to \ga\ga$ opt.             & --&-- &3.6&-- &4.3\\ \hline
ATLAS $qq\to qqH$,$H \to \tau \tau$   &2.4& --&2.9&-- &2.5\\ \hline
ATLAS $H \to WW\to e\mu\nu\nu$+ 0 Jets& --& --&-- &-- &3.4\\ \hline
ATLAS $H \to WW\to e\mu\nu\nu$+ 2 Jets& --& --&-- &-- &2.0\\ \hline
ATLAS $H \to ZZ\to$ 4 leptons         & --& --&1.5&-- &3.5\\ \hline
CMS   $H \to \ga\ga$ cuts             & --&3.1&3.3&-- &3.5\\ \hline
CMS   $H \to \ga\ga$ opt.             & --&5.3&5.7&-- &4.7\\ \hline
CMS   $qq\to qqH$, $H \to \tau \tau\to l$\,had
                                      & --&2.2&--&2.0&--\\ \hline
CMS   $H \to WW\to ll\nu\nu$          &-- &-- &0.4&-- &0.9\\ \hline
CMS   $H \to ZZ\to$ 4 leptons         &-- &2.6&2.3&-- &5.3\\ \hline
\end{tabular}
\end{center}
\vspace{-0.5em}
\caption{\label{tab_obspot1}
Summary of the significances for observation of a SM-like Higgs boson in
various search channels for masses below $130 \gev$
in the ATLAS and CMS experiments after collecting 10 fb$^{-1}$.}
\renewcommand{\arraystretch}{1.0}
\end{table}

\begin{table}[htb!]
\renewcommand{\arraystretch}{1.2}
\begin{center}
\begin{tabular}{|l||c|c|c|c|c|c|c|} \hline
channel / $\MH$ [GeV]                &135&140&145&150&160& 170 &180\\ 
                                                              \hline\hline
ATLAS $H \to \ga\ga$ cuts             & --&2.2&-- &-- &-- & --  &-- \\ \hline
ATLAS $H \to \ga\ga$ opt.             & --&4.0&-- &-- &-- & --  &-- \\ \hline
ATLAS $qq\to qqH$,$H \to \tau \tau$   &-- &1.9&-- &-- &-- & --  &-- \\ \hline
ATLAS $H \to WW\to e\mu\nu\nu$+ 0 Jets& --&5.8&--&8.4 &10.6&10.2&7.1\\ \hline
ATLAS $H \to WW\to e\mu\nu\nu$+ 2 Jets& --&3.0&--&4.1 &5.1 &5.1&4.2 \\ \hline
ATLAS $H \to ZZ\to$ 4 leptons         &-- &6.3&--&7.3 &4.1 &-- &2.9 \\ \hline
CMS   $H \to \ga\ga$ cuts             & --&3.2&-- &2.3&-- &-- &-- \\ \hline
CMS   $H \to \ga\ga$ opt.             & --&3.9&-- &-- &-- &-- &-- \\ \hline
CMS   $qq\to qqH$, $H \to \tau \tau\to l$\,had
                                     &2.1 &-- &0.8&-- &-- &-- &--  \\ \hline
CMS   $H \to WW\to ll\nu\nu$          &-- &1.3&-- &2.9&6.3&6.3&4.8 \\ \hline
CMS   $H \to ZZ\to$ 4 leptons         &-- &7.8&-- &9.0&5.4&2.6&4.5 \\ \hline
\end{tabular}
\end{center}
\vspace{-0.5em}
\caption{\label{tab_obspot2}
Summary of the significances for observation of a SM-like Higgs boson in
various search channels for $130 < \MH \le 180 \gev$
in the ATLAS and CMS experiment after collecting 10 fb$^{-1}$.}
\renewcommand{\arraystretch}{1.0}
\end{table}

\begin{table}[htb!]
\renewcommand{\arraystretch}{1.2}
\begin{center}
\begin{tabular}{|l||c  |c  |c  |c  |c  |c  |c  |c  |c  |c|}\hline
channel / $\MH$ [GeV]         &190&200&250&300&350&400&450&500&550&600\\ 
                                                              \hline\hline
ATLAS $H \to ZZ\to$ 4 leptons &-- &8.3&-- &7.2&-- &6.0&-- &2.9& --&1.8\\ \hline
CMS   $H \to WW\to ll\nu\nu$  &2.2&1.3&-- &-- &-- &-- &-- &-- & --&-- \\ \hline
CMS   $H \to ZZ\to$ 4 leptons &9.1&9.2&7.7&8.0&8.1&7.8&6.6&5.2&4.1&3.2\\ \hline
\end{tabular}
\end{center}
\vspace{-0.5em}
\caption{\label{tab_obspot3}
Summary of the significances for observation of a SM-like Higgs boson in
various search channels for masses above $180 \gev$
in the ATLAS and CMS experiment after collecting 10 fb$^{-1}$.}
\renewcommand{\arraystretch}{1.0}
\end{table}

The discovery potentials  for a SM-like Higgs boson  in 
three different mass ranges of the Higgs boson
using the ATLAS and CMS
detectors
 are  shown in 
\reftas{tab_obspot1} - \ref{tab_obspot3}.
As mentioned above, some differences between ATLAS and CMS can be
observed. One difference can be seen in the channel 
$pp \to H \to \ga\ga$, where the CMS results look more optimistic,
especially in the ``optimized'' analysis~\cite{Ball:2007zza}. 
ATLAS has also performed an optimized analysis for the 
$H \to \ga \ga$ decay mode and it is expected that the sensitivity can
be increased by 50\% relative to the results shown in
\refta{tab_obspot1}. 
The $H \to W^+W^-$ decay mode looks more promising in the ATLAS analysis
than in the CMS studies.
In the ATLAS study, only the final state with one
electron and  one muon has been analyzed. Taking into account also the
di-electron and di-muon final states,
 it is expected that the significance of an
observation will increase by a factor of up to about $\sqrt{2}$\,.
For the $H \to \tau^+\tau^-$ decay mode, ATLAS has investigated the 
$\tau^+\tau^- \to l^+ l^-X$ and  $\tau^+\tau^- \to l^\pm \textrm{had}^\mp X$ final states,
whereas CMS has only considered the latter one ($\tau\tau \to  l^\pm \textrm{had}^\mp$
refers to one $\tau$ decaying leptonically and the other one hadronically).

Most analyses up to now have been performed for 30 \ifb. In order to
arrive at the data shown in \reftas{tab_obspot1} - \ref{tab_obspot3}
the following rescaling and extrapolation methods were applied:
The ATLAS numbers are taken from Ref.~\cite{Aad:2009wy} and have been
obtained by taking the square root of the $-2 \ln \la$ values quoted
there (where $\la$ denotes the profile likehood ratio of the
background over the signal plus background hypothesis).
The CMS numbers are based on Ref.~\cite{Ball:2007zza}, where
numbers for 30~\ifb\ are reported. The numbers above are obtained by
scaling the number of signal and background events by a factor of $1/3$,
but using the relative uncertainties from the original analysis (which
might be optimistic). 

\smallskip
Both experiments  currently do not consider the associated production
with a pair of top-quarks and subsequent decay to a pair of b-quarks 
($t \bar t H, H \to b \bar b$) as a discovery mode for initial
data taking. The latest sensitivity studies quote statistical
significances at a mass of $\MH = 120 \gev$ corresponding to 1.8 to 2.2 with
30~\ifb\ in the ATLAS experiment~\cite{Aad:2009wy} using the
semileptonic decay mode only. 
In the CMS experiment, combining all possible final states 
a significance of 1.6 to 2.4 with 60~\ifb\ is reached~\cite{Ball:2007zza}.
Including current estimates of systematic background uncertainties, the
significance is below 0.5 for the integrated luminosities
assumed~\cite{Aad:2009wy,Ball:2007zza}. 
According to recent NLO calculations  of the 
background \cite{Bredenstein:2008zb,Bredenstein:2009aj}, yielding a
relatively large $K$-factor of~$\sim 1.8$, the prospects for this
channel are even more doubtful.
Alternatives to possibly recover some sensitivity for the 
$H \to b \bar b$
channel are discussed in the next section.

In \reffi{results10fb}, the expected
performances of the  ATLAS and CMS detectors are shown
as a function of $\MH$ (assuming SM rates).
CMS shows the luminosity needed for a 5$\sigma$ discovery, while ATLAS
shows the expected significance after 10~\ifb.

\begin{figure}
  \centering
   \includegraphics[width=7 cm,height=7 cm]{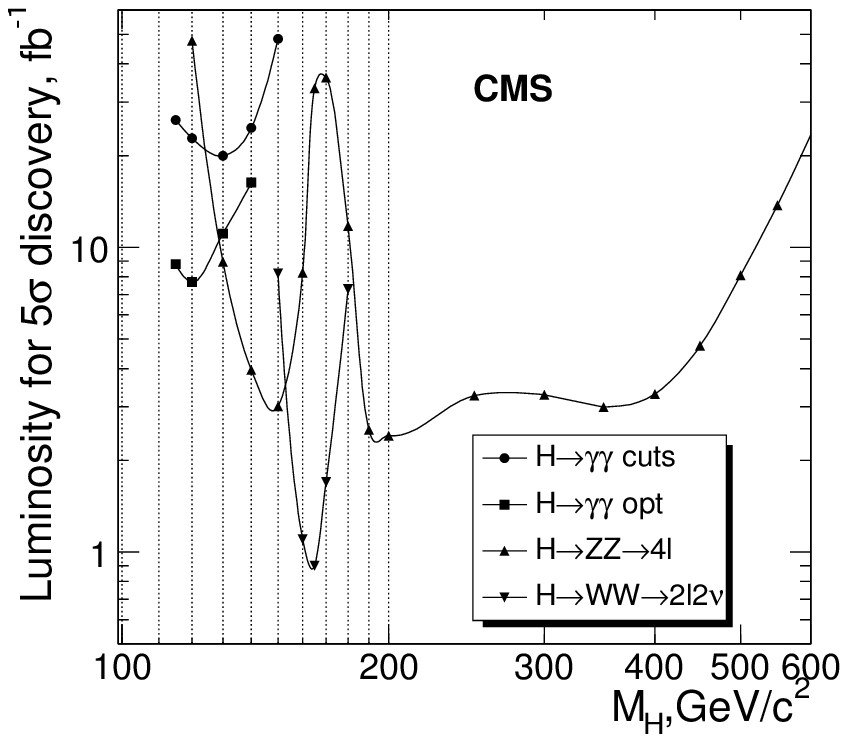}
   \includegraphics[width=7 cm,height=7 cm]{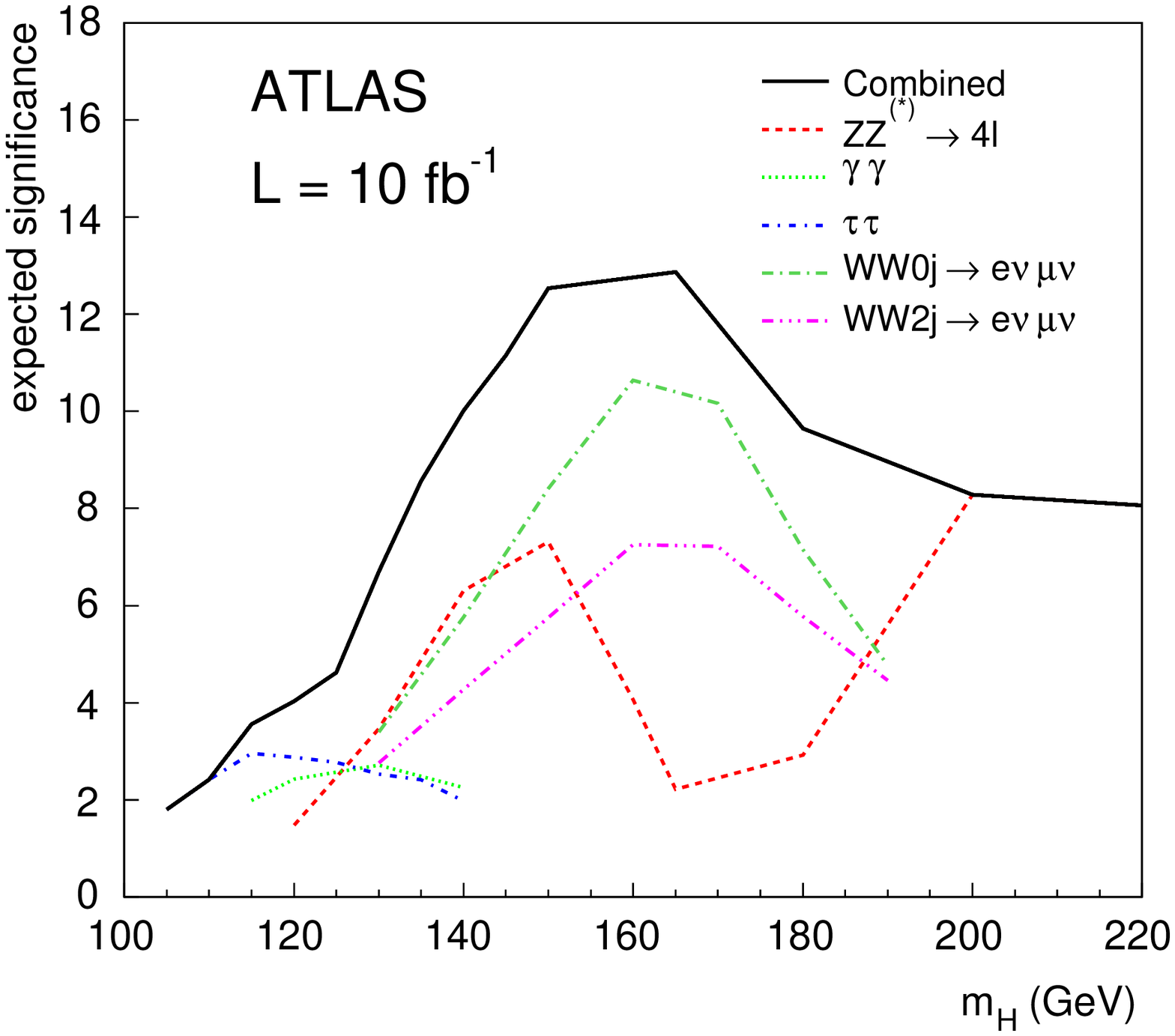}
\caption{
Left: Luminosity needed for a $5\,\si$ discovery at CMS~\cite{Ball:2007zza}.
Right: Expected significance at ATLAS with 10~\ifb~\cite{Aad:2009wy}.
}
\label{results10fb}
\end{figure}


\subsubsection*{LHC at \boldmath{$\sqrt{s} = 10 \tev$}}

The LHC is expected to initially run at $\sqrt{s} = 10 \tev$.  At this
energy, production rates are typically reduced by about a factor of
two from those at $\sqrt{s} = 14 \tev$. Both CMS and ATLAS have performed
{\em preliminary} studies of the expected Higgs sensitivities at 
$\sqrt{s} = 10 \tev$~\cite{ATLAS10tev,Baffioni:10TeV}, see also
\cite{Assamagantalk}. 
In the combined $H \to ZZ+W^+W^-$  channel, CMS estimates that the
required luminosity for a 95\%~C.L.\ exclusion limit is roughly doubled at
$\sqrt{s} = 10 \tev$ for $\MH$ between $120$ and $200 \gev$ from that at
$\sqrt{s}= 14 \tev$.  In the mass range $\MH = 160 - 170 \gev$, 
a 95\%~C.L.\ exclusion limit is obtained
in this channel with $\sim 0.2$~\ifb\ (as compared with $0.1$~\ifb\ 
at $\sqrt{s} = 14 \tev$).
Similarly, ATLAS has examined the combined $H \to W^+W^- \to 2l$ with 0 and
2~jets channel and finds that a 5$\sigma$ discovery is possible with
$\sim 1$~\ifb\ for 
$\MH\sim 160 - 170 \gev$ at $\sqrt{s}= 10 \tev$.

The preliminary findings above are partially obtained using a fast
simulation
of the detectors, without optimisation of the selections
for running at $10 \tev$ and using the relative systematic uncertainties
from earlier studies assuming sometimes larger integrated luminosities.
Hence the results have not the same level of maturity as those
in Refs.~\cite{Aad:2009wy,Ball:2007zza}, but yield an indicative estimate of
the sensitivity during early data taking.


\subsubsection*{Higgs searches at the Tevatron}

By the time the ATLAS and CMS collaborations have analyzed 10~\ifb, 
the Tevatron Run~II will have been completed~\cite{:2009pt}. If the Tevatron
runs 
in 2011, a total of 10~\ifb\ analyzed per experiment (CDF and D\O) is
expected. The latest projections by the Tevatron
experiments~\cite{talkBernardi} suggest
that with this luminosity, a 95\%~C.L.\ exclusion of the SM Higgs in the
mass range $114 - 185 \gev$ could be achieved (where $114.4 \gev$ is the
limit obtained for a SM-like Higgs boson at 
LEP~\cite{Barate:2003sz,Schael:2006cr}). 
In addition, a $3\si$ sensitivity is expected for $\MH < 115 \gev$ and 
$150 \gev < \MH < 180 \gev$. This means
that the significance of a SM Higgs signal would be $\sim 2-3 \si$ for 
$\MH < 150 \gev$. While this is the overall sensitivity from the combination
of all search channels, in particular, for 
$\MH < 130 \gev$, most of the sensitivity comes from $VH$ ($V=W,Z$), 
with $H\to b\bar b$. This is complementary with
the LHC, which in this mass range mainly probes 
$H\to \ga \ga$ and $H\to \tau^+\tau^-$,
demonstrating the complementarity between both machines. Therefore, the 
Tevatron could potentially yield interesting information on 
$\si(p \bar p \to V^* \to VH) \times \br(H\to b\bar b)$ 
for $\MH < 130 \gev$, which could be used
in a global analysis of Higgs couplings at the LHC, (see below).


\subsection{Investigation of the Higgs sector at \boldmath{\lhcten}}
\label{sec:HiggsAna}

After the observation of a new resonance at the \lhcten\ the first goal
will be to measure its characteristics (mass, width, branching ratios,
couplings, \ldots).
Only if the profile agrees completely (within sufficiently small
experimental errors) with that predicted for 
a SM Higgs boson, one could be convinced that the SM Higgs mechanism is
realized in nature.

The accuracy of the determination of a Higgs boson mass will crucially depend
on the decay modes observable. Assuming SM properties, the precision
is dominated by the decay $H \to \ga\ga$ at low masses and 
by $H \to ZZ^{(*)}$ at higher masses. 
From the $H \to \ga\ga$ channel a precision better than $\sim 1\%$ can
be expected at the \lhcten\ (rescaling the numbers from 
\citere{Ball:2007zza}). For higher masses CMS has shown that a statistical
error of $\lsim 0.4\%$ on the mass measurement
can be reached assuming 30~\ifb\ in 
the $H \to ZZ^*\to 4$ lepton channel
for Higgs boson masses below 
$180 \gev$~\cite{Ball:2007zza}.  Even for a Higgs boson mass of $600 \gev$,
the expected precision is $2.4\%$.

In order to verify that the resonance observed is indeed a Higgs boson,
it will be crucial  to measure its couplings to all particle
species. 
A study was performed in 2004 assuming at least an integrated luminosity 
of $2 \times 30$~\ifb~\cite{Duhrssen:2004cv} (``combining'' ATLAS and CMS). 
This analysis, however, 
used now outdated results from ATLAS and CMS.
The analysis assumed SM production and decay rates. Another assumption
employed was that the coupling to SM gauge bosons is bounded from above by
$g_{HVV}^2 < (g_{HVV}^{\rm SM})^2 \times 1.05$ (which is realized in all
models with Higgs singlets and doublets only, including the MSSM). 
For Higgs boson masses below $150 \gev$ the results depend strongly on the 
observability  of the $H \to b \bar b$ decay mode since it dominates the
total decay width. 
The corresponding results for the \lhcten\ are obtained from
\citere{Duhrssen:2004cv} by rescaling.
Table~\ref{tab:LHC10coup} summarizes estimated precisions 
on the absolute couplings as well as the total and invisible (or
undetectable) Higgs 
width. It has been found that new negative contributions to the $ggH$
and $\ga\ga H$ (loop 
induced) couplings could be detected at the $-50\%$ level. 
However, it should be kept in mind that these analyses assume a
measurement of the $t\bar t H, H\to b \bar b$ and $H \to WW^{(*)}$
channel and are thus to be taken with care.

Given the new findings for the
$t \bar t H, H \to b \bar b$ channel~\cite{Aad:2009wy,Ball:2007zza}, the
decay to $b \bar b$ will hardly be observable, thus missing a large
contribution to the total width, and consequently no
coupling determination 
seems to be possible for Higgs boson masses below $150 \gev$ at the \lhcten.
New methods to recover the observability of $H \to b \bar b$ need to be
studied experimentally in order to regain at least some sensitivity
in the low mass region. 
Several methods have been suggested; e.g., 
$W H, H \to b \bar b$ with a large boost of the 
Higgs bosons~\cite{Butterworth:2008iy},
or Higgs production in vector boson fusion in  association with either
a central photon in $pp\to q q H \ga \to q q b\bar b \ga$, where the
requirement of an extra high-$p_T$ photon in the  $q q H \to q q b \bar b$
final state dramatically enhances the $S/B$ ratio~\cite{Gabrielli:2007wf},
or an additional $W$ boson in
$pp\to q q H  W \to q q b \bar b \ell \nu$, with the final high-$p_T$ lepton
improving the trigger efficiency~\cite{Rainwater:2000fm,Ballestrero:2008kv}.
These strategies are currently being investigated by the ATLAS and 
CMS collaborations.
An updated study of Higgs coupling measurements has been presented
in Ref.~\cite{Lafaye:2009vr} for $\MH = 120 \gev$ (based on the
parton-level study in 
Ref.~\cite{Butterworth:2008iy} to recover the decay $H \to b \bar b$) 
with the conclusion that coupling
constant measurements with accuracies in the 20-40$\%$ region
should be possible with 30 \ifb.

Without assumptions about the Higgs model (for instance an upper bound on 
$g_{HVV}$, see above), one would be left with 
measurements of ratios of Higgs boson decay widths. 
The accessible ratios directly correspond
to the visible production and decay channels at a given value of $\MH$.  
A rough summary of the
estimated precision on the ratios is given in \refta{tab:ratios}. 
It is found from \citere{ATL-PHYS-2003-030} by rescaling to
lower luminosity at the \lhcten.

\begin{table}[htb!]
\renewcommand{\arraystretch}{1.2}
\begin{center}
\begin{tabular}{|l||c  |c  |c  |c  |c  |c  |c  |c  |}\hline
channel / $\MH$ [GeV] & 120  & 130  & 140  & 150 & 
                        160  & 170  & 180  & 190  \\ \hline\hline
$g_{HWW}$             & 29\% & 25\% & 20\% & 14\% & 
                         9\% &  8\% &  8\% &  9\% \\ \hline
$g_{HZZ}$             & 30\% & 27\% & 21\% & 16\% & 
                        15\% & 19\% & 14\% & 11\% \\ \hline
$g_{H\tau\tau}$       & 63\% & 39\% & 38\% & 50\% & 
                             &      &      &      \\ \hline 
$g_{Hbb}$             & 72\% & 54\% & 56\% & 73\% &  
                             &      &      &      \\ \hline
$g_{Htt}$             & 87\% & 62\% & 45\% & 36\% &  
                        31\% & 32\% & 36\% & 45\% \\ \hline\hline
$\Ga_H$               &      & 77\% & 60\% & 42\% &
                        27\% & 25\% & 26\% & 29\% \\ \hline
$\Ga_{\rm inv}/\Ga_H$  & 81\% & 72\% & 56\% & 39\% &
                        23\% & 20\% & 22\% & 24\% \\ \hline
\end{tabular}
\end{center}
\vspace{-0.5em}
\caption{\label{tab:LHC10coup}
Summary of the precisions at the \lhcten, assuming 
$g^2_{HWW} < (g_{HWW}^{\rm SM})^2 \times 1.05$~\cite{ATL-PHYS-2003-030,Duhrssen:2004cv}. 
Upper part: $\de g_{Hxx}/g_{Hxx}$; lower part:
$\Ga_H$ is the total Higgs width, $\Ga_{\rm inv}/\Ga_H$ denotes the
sensitivity to an invisible or undetectable width with respect to the
total width. 
``Precisions'' larger than~100\% are omitted. 
}
\renewcommand{\arraystretch}{1.0}
\end{table}

\begin{table}[htb!]
\renewcommand{\arraystretch}{1.2}
\begin{center}
\begin{tabular}{|l||c  |c  |c  |c  |c  |c  |c  |c  |}\hline
channel / $\MH$ [GeV] & 120  & 130  & 140  & 150 & 
                        160  & 170  & 180  & 190  \\ \hline\hline
$\Ga_{HZZ}/\Ga_{HWW}$   &      & 55\% & 36\% & 32\% & 
                        47\% & 78\% & 46\% & 27\% \\ \hline
$\Ga_{H\tau\tau}/\Ga_{HWW}$ &   & 58\% & 62\% & 85\% & 
                              &      &      &      \\ \hline 
$\Ga_{H\ga\ga}/\Ga_{HWW}$ & 79\% & 53\% & 53\% & 68\% &
                               &      &      &      \\ \hline
\end{tabular}
\end{center}
\vspace{-0.5em}
\caption{\label{tab:ratios}
Summary of the estimated
precision, 
$\de(\Ga(H \to XX)/\Ga(H \to WW^{(*)}))/(\Ga(H \to XX)/\Ga(H \to WW^{(*)}))$
 on ratios of couplings at the \lhcten\
(see text)~\cite{ATL-PHYS-2003-030}. ``Precisions'' larger than~100\% are
omitted. 
}
\renewcommand{\arraystretch}{1.0}
\end{table}

The Higgs tri-linear self-coupling, $g_{HHH}$, is a key parameter in the
Higgs sector since it describes the ``form'' of the Higgs potential.
The measurement of $g_{HHH}$ allows  a stringent
test of the SM potential 
and some discrimination between different models (2HDM,
MSSM, baryogenesis, Higgs-Radion mixing, \ldots) where the coupling 
may be significantly enhanced. 
Unfortunately, at the LHC, even with $\cL = 300$~\ifb, no measurement of
a SM-like Higgs self-coupling can be expected.

Another measurement that can be made at the \lhcten\ concerns 
the structure of the tensor coupling of the 
putative Higgs 
resonance to weak gauge
bosons. This can  be studied at \lhcten\ with good precision for
some values of $\MH$. A study exploiting the difference in the
azimuthal angles of the two tagging jets in weak vector boson 
fusion has shown that for $\MH = 160 \gev$ 
the decay mode into a pair of $W$-bosons (which is maximal at 
$\MH = 160 \gev$) allows the discrimination between
the SM tensor structure and purely anomalous $\cp$-even and -odd coupling
structures at a level of 4.5 to  5.3\,$\si$ assuming the production rate
is that of the SM\cite{Ruwiedel:2007zz,Hankele:2006ma,Plehn:2001nj}. 
A discriminating power of two standard deviations for the distinction of
the $\cp$-even and -odd tensor structure at a mass of $120 \gev$
in the tau lepton decay mode requires an integrated luminosity of 30~\ifb.


\subsection{From the \boldmath{\lhcten} to future colliders}

The LHC running at high(er) luminosity (subsequently called \lhchl, 
assuming the collection of $\sim 300$~\ifb\ per detector)
will follow the
\lhcten, expanding the knowledge about the Higgs sector. 
In this section we will analyze what can be gained from future
colliders in the various scenarios beyond what is anticipated from the
\lhchl. As future colliders, we consider the 
sLHC~\cite{Gianotti:2002xx}, 
the ILC~\cite{Djouadi:2007ik} and CLIC~\cite{Braun:2008zzb}. 

Other options could be an LHC with double energy
(DLHC), see \citere{Scandale:2008zzc} and references therein, 
and a VLHC (Very Large Hadron Collider), with an energy of 
$\sqrt{s}=40-200 \tev$~\cite{Ambrosio:2001ej,Scandale:2008zzc}. 
More information can also be found in Refs.~\cite{CAREconf,HHH}. 
The physics case for a DLHC or VLHC
will only emerge after discoveries at the LHC, e.g.\ concerning the
potential measurement of  the Higgs tri-linear coupling, $g_{HHH}$.
Another option could be a 
$\mu^+\mu^-$~collider~\cite{Autin:1999ci,Raja:2001be},
with an energy of $\sqrt{s} \sim \MH$. 
At a $\mu^+\mu^-$ collider, with an integrated
luminosity of $\cL^{\rm int} \lsim 10\,{\rm pb}^{-1}$ 
an ultra-precise measurement of a Higgs
boson mass and width would be possible\cite{Blochinger:2002hj}
and coupling measurements up to the same level as at
the ILC could be performed. The $\mu^+\mu^-$~collider could thus
help to determine the Higgs profile.
In the following, however, we will not discuss the physics
capabilities of a DLHC, VLHC or a  $\mu^+\mu^-$~collider, as
the technical feasibility studies are in very preliminary stages.

We start by briefly summarizing the existing analyses 
in the Higgs sector for the \lhchl, sLHC, ILC and CLIC.

\noindent\ul{\lhchl:}\\
Going to the \lhchl\ will allow the
observation of a Higgs boson candidate in more production and decay
modes compared to the \lhcten. This will yield a better determination of
ratios of partial widths as well as absolute couplings, provided the 
$H \to b\bar b$ channel
is accessible and assuming that the coupling to weak gauge
bosons $g_{HVV}$ is bounded from above by 
$g_{HVV}^2 < (g_{HVV}^{\rm SM})^2 \times 1.05$~\cite{Duhrssen:2004cv}. 
In this study, for $\MH \lsim 150 \gev$, couplings to fermions could
be determined between $\sim 13\%$ and $\sim 30\%$, whereas Higgs couplings
to gauge bosons could be measured to $10 - 15\%$ $(5-10\%)$ for 
$\MH \lsim (\gsim) 150 \gev$ (see also Ref.~\cite{Lafaye:2009vr}).

Several studies for the measurement of the tri-linear Higgs coupling,
$g_{HHH}$, have been performed, assuming $\MH \gsim 140 \gev$ with 
$H \to WW^{(*)}$ as the dominant decay
mode~\cite{Gianotti:2002xx,Baur:2003gp,ATL-PHYS-2002-029}. The studies
conclude that at the \lhchl\ a determination of $g_{HHH}$ will not be
possible. 

With a larger data sample the spin and $\cp$~quantum numbers can 
be inferred from the angular distributions of the leptons in the
$H\to ZZ \to 4 \ell$ decay mode (see
\citeres{Choi:2002jk,Godbole:2007cn,BhupalDev:2007is} and references
therein for theory studies). 
The CMS collaboration considered the case that the observed
scalar boson $\phi$ is a mixture of a $\cp$-even ($H$)~and $\cp$-odd~($A$)
boson according to $\Phi= H+\eta A$. Assuming the SM production rate
the parameter $\zeta = \arctan\eta$ can be determined to 
10-20\% for $\MH = 200 - 400 \gev$ with an integrated luminosity
$\cL^{\rm int} = 60$~\ifb~\cite{Ball:2007zza}. 
Using the same observables the ATLAS collaboration found that the
hypothesis of non-SM $\cp$ and spin combinations can be distinguished from 
the SM value at the 95\%~C.L.\ for $\MH \gsim 250 \gev$ and 
$\cL^{\rm int} = 100$~\ifb~\cite{Buszello:2002uu}.

\noindent\ul{sLHC:}\\
The sLHC is a luminosity upgrade of the LHC which aims for an
ultimate luminosity of\\ 
1000~\ifb/year sometime after 2018.
Assuming that the detector capabilities remain roughly the 
same as those anticipated for the LHC, the sLHC\cite{Gianotti:2002xx} 
will increase the
discovery potential for high mass objects by $25-40\%$.

By the time the sLHC is realized, the Higgs discovery phase at the
LHC will be largely completed.  For processes which are limited
by statistics at the LHC, the sLHC may be useful.
The increased luminosity of the sLHC could enable the observation
of rare Higgs decays. The sLHC
could also potentially increase the accuracy of the measurements
of Higgs couplings. 
There might be some sensitivity on the tri-linear Higgs
self-coupling~\cite{ATL-PHYS-2002-029}; however, some background
contributions might have been underestimated. Further studies to clarify
these issues are currently in progress, see \citere{Jakobs:2009zza} for
a discussion.
A key concern is maintaining detector performance, since the increased
luminosity will result in significantly more pileup per
beam crossing, increasing occupancy rates in the tracking systems.

\noindent
\ul{ILC:}\\
The following details are based on the Technical Design Report (TDR)
that appeard in 2001 (for the TESLA
design~\cite{AguilarSaavedra:2001rg}), the Reference Design Report
(RDR)~\cite{Brau:2007zza}  and subsequent documents (see also
\citere{Heinemeyer:2005gs}). 
The initial stage of the ILC is expected to have 
an energy of $\sqrt{s}=500 \gev$
with a luminosity of $2\times 10^{34}$/cm$^2$/s, along with $90\%$
polarization of the $e^-$ beam and $30-45\%$ polarization of the $e^+$
beam.  A future upgrade to $\sqrt{s} = 1 \tev$ (with an even higher
luminosity) is envisioned. An advantage of the
machine is that it is designed to have low beamstrahlung and a 
precise knowledge of the luminosity ($\de \cL/\cL < 10^{-3}$) and 
energy ($(\delta\sqrt{s})/\sqrt{s} < 200$~ppm), along with excellent 
detector resolution.   The tunable energy scale allows for a
scan of particle production thresholds.

The ILC offers a clean environment for the precision measurement of
all quantum numbers and couplings of the Higgs boson, in
addition to precision measurements of its mass and width. While the
mass range of a  SM-like Higgs boson can be covered completely
by an ILC with $\sqrt{s}=500 \gev$ up to $\MH\lsim 400 \gev$, the
achievable precision on the measurements of couplings and other
properties is strongly dependent on the Higgs mass and
differs for the various decay modes. A set of studies of 
properties of the Higgs boson has been collected
in~\cite{AguilarSaavedra:2001rg}, many of which are being  updated using
the most recent designs for the accelerator and the detectors and fully
simulated Monte Carlo events in~\cite{ILDLetterOfIntent} (see 
also \citeres{Heinemeyer:2005gs,Ackermann:2004ag}).
A summary of the current analyses is given in \refta{tab:ILCprecisions}
for $\MH \approx 120 \gev$ and $\MH \approx 200 \gev$.

\begin{table}
\renewcommand{\arraystretch}{1.2}
  \begin{center}
    \begin{tabular}{| l | c     | c |}
      \hline
      Observable                  & Expected precision & Reference\\
      \hline\hline
      \multicolumn{3}{|c|}{SM-like Higgs with $\MH\approx120 \gev$}\\
      \hline
 $\MH$ [GeV]        & 0.04\,\%      &   \cite{AguilarSaavedra:2001rg}\\ 
 $\Ga_H$ [GeV]      & 0.056\,\%     &   \cite{AguilarSaavedra:2001rg}\\ 
 $g_{HWW}      $     & 1.2\,\%       &   \cite{AguilarSaavedra:2001rg}\\ 
 $g_{HZZ}      $     & 1.2\,\%       &   \cite{AguilarSaavedra:2001rg}\\ 
 $g_{Htt}      $     & 3.0\,\%       &   \cite{AguilarSaavedra:2001rg}\\ 
 $g_{Hbb}      $     & 2.2\,\%       &   \cite{AguilarSaavedra:2001rg}\\ 
 $g_{Hcc}      $     & 3.7\,\%       &   \cite{AguilarSaavedra:2001rg}\\ 
 $g_{H\tau\tau}$     & 3.3\,\%       &   \cite{AguilarSaavedra:2001rg}\\ 
 $g_{Htt}      $     & 7\,\%         &  \cite{Gay:2006vs} \\
 $g_{HHH}      $     & 22\,\%        &   \cite{AguilarSaavedra:2001rg}\\ 
 $\br(H \to\ga\ga)$  & 23\,\%        &   \cite{AguilarSaavedra:2001rg}\\ 
 $\cp_H$      & 4.7$\sigma$ diff. between even and odd & \cite{Andreas}\\
 \hline
 GigaZ Indirect $\MH$ [GeV]   & 7\,\% &  \cite{Erler:2000jg,Flacher:2008zq} \\
 \hline
 \multicolumn{3}{|c|}{Heavy SM-like Higgs with $\MH\approx200 \gev$}\\
 \hline
 $\MH$ [GeV]          & 0.11\,\%       & \cite{Meyer:2004ha} \\ 
 direct $\Ga_H$~[GeV] & 34\,\%       & \cite{Meyer:2004ha} \\
 $\br(H \to WW)$      & 3.5\,\%        & \cite{Meyer:2004ha} \\
 $\br(H \to ZZ)$      & 9.9\,\%        & \cite{Meyer:2004ha} \\
 $\br(H \to b\bar b)$ & 17\,\%         & \cite{Battaglia:2002av} \\
 $g_{Htt}$            &  14\,\%         & \cite{Gay:2006vs} \\
 \hline
 \multicolumn{3}{|c|}{Additional Measurements for Non-SM Higgs with $\MH \approx120 \gev$}\\      
 \hline
 $\br(H \to \mathrm{invisible})$ & $<$ 20\,\% for BR$>0.05$     &   
                                        \cite{AguilarSaavedra:2001rg}\\ 
 \hline
 \end{tabular}
\end{center}
\caption{
Examples of the precision of SM-like Higgs observables
  at a $\sqrt{s}=500 \gev$ ILC. For the direct measurements, an
  integrated luminosity of $\cL^{\rm int} = 500~\mathrm{fb}^{-1}$ is
  assumed  (except for the $b\bar b$ channel at $\MH \approx 200 \gev$ 
  and the $t \bar t$ channel, which assume $\sim 1$~\iab\ 
  at $\sqrt{s} = 800 \gev$). 
  For the indirect measurements at GigaZ, a running time of
  approximately one year is assumed, corresponding to 
  $\cL = $~\order{10~\mathrm{fb}^{-1}}.}
\label{tab:ILCprecisions} 
\renewcommand{\arraystretch}{1.0}
\end{table}

In addition, the options of GigaZ ($10^9$ $Z$'s at 
$\sqrt{s} \approx \MZ$)~\cite{Hawkings:1999ac,Erler:2000jg}, 
and MegaW ($\sqrt{s} \approx 2 \MW$)~\cite{mwgigaz} allow  precision
tests of the SM with uncertainties reduced approximately by one order
of magnitude from the predictions of current ILC studies. 
This would allow the mass of the SM Higgs
boson to be constrained quite strongly by indirect methods
and could potentially exclude a SM-like Higgs boson
with $\MH \gsim 130 \gev$~\cite{Erler:2000jg,Flacher:2008zq} (see 
\citere{Buchmueller:2007zk} for a corresponding MSSM analysis).

Another ILC option are $e\ga$ or $\ga\ga$ collisions
(the Photon Linear Collider, PLC)~\cite{Badelek:2001xb}, 
with $\ga$~beams obtained from the
backscattering on laser beams. The energy of the photons would be 
$\sim 80\%$ of the electron beam, maintaining a high degree of
polarization and a luminosity of $\sim 1/3$ of the ILC in the high
energy peak. The PLC could potentially
perform precision measurements of resonantly produced Higgs bosons.
Combining ILC and PLC measurements, the $H\ga\ga$ coupling could be
determined at the level of $\sim 3\%$.

\noindent
\ul{CLIC:}\\
CLIC is proposed as a multi-TeV $e^+e^-$ collider with an energy
of $\sqrt{s} \sim 1-3 \tev$ and a luminosity of $\cL \sim 10^{34}$/cm$^2$/s.
The goal of the current studies is to demonstrate technical
feasibility and to have
a Conceptual Design Report by 2010 and a Technical Design Report by
2015. Examples of anticipated precisions for Higgs boson couplings are
given in \refta{tab:CLICprecisions}. Analyses mostly focused on channels
that are challenging at the ILC, see \citere{Accomando:2004sz} and
references therein.

\begin{table}
\renewcommand{\arraystretch}{1.2}
\begin{center}
\begin{tabular}{|l||c  |c  |c  |c  |}\hline
coupling / $\MH$ [GeV]    & 120 & 150 & 180 & 220 \\ \hline\hline
$g_{Hbb}$     &       &        &  1.6\% & 3.4\% \\
$g_{H\mu\mu}$ & 4.2\% & 11.0\% &        &       \\
$g_{HHH}$     & 9.3\% &        & 11.5\% &       \\
\hline
\end{tabular}
\end{center}
\caption{Examples of the precision, $\de g_{Hxx}/g_{Hxx}$,
 for measurements of  Higgs couplings
  at a $\sqrt{s}=3 \tev$ CLIC with 3~ab$^{-1}$\cite{Accomando:2004sz}.}
\label{tab:CLICprecisions} 
\renewcommand{\arraystretch}{1.0}
\end{table}

\subsubsection{Scenario A: SM-like Higgs with 
            \boldmath{$130 \gev \le \MH \le 180 \gev$ } }

In the region of masses $130 - 180 \gev$, the dominant decay modes considered
are to two vector bosons, $ZZ^{(*)}$ and $WW^{(*)}$, yielding a
discovery at the \lhcten.
New studies~\cite{Baffioni:10TeV} indicate that even 200~\ipb\ could be
sufficient to 
probe the region of $160 \gev \le \MH \le 170 \gev$, which is currently
excluded at $95\%$~C.L.\ at the Tevatron~\cite{Phenomena:2009pt}. 

For $\MH \lsim 200 \gev$, the total Higgs width cannot be 
measured at the LHC and there is expected to be only an upper
limit of \order{1 \gev}.  Hence in this mass region, only
ratios of Higgs couplings can be measured in a model independent
fashion, see \refta{tab:ratios} for the \lhcten\ expectations.
In the lower part of the mass range, $130 \gev \le \MH \le 150 \gev$,
the $\ga\ga$ and $\tau^+\tau^-$ final states are accessible.
As discussed above, the final state with $b$ quarks ($H \to b \bar{b}$)
seems not to be accessible during the first years of the
LHC, because of the very difficult background environment. 

In this scenario the large mass reach of the sLHC could be helpful to
detect new scales beyond the SM. Furthermore, 
the sLHC can in principle improve the accuracy of Higgs coupling
constant measurements in this regime\cite{Gianotti:2002xx}.  
For $\MH \gsim 150 \gev$, the
decays $H \to Z Z^*\to 4l$  
and $H \to W^+ W^- \to l \nu l \nu$ provide a direct
measurement of the ratio of the partial Higgs widths, $\Ga_{HZZ}/\Ga_{HWW}$.  
For $\MH = 170 \gev$, 3000 \ifb/experiment
could improve the measurement of  
$\delta(\Ga_{HZZ}/\Ga_{HWW}) /(\Ga_{HZZ}/\Ga_{HWW})$ 
from the \lhchl measurement by about a factor 
of~$1.5$.  The improvement at the sLHC
for most other masses values in the intervall $\MH = 150 - 180 \gev$  
is quite small, however.  In this mass region, the decays 
$H \to \tau^+\tau^-$ and $H \to W^+W^-\to l\nu l\nu$ 
provide a direct measurement of $\Ga_{H\tau\tau}/\Ga_{HWW}$.
The improvement in this channel at the sLHC 
over the \lhchl\ result is not known, since it
depends crucially on $\tau$ identification, missing $E_T$ capabilities, and
identification of forward jets at $\cL = 10^{35}$/cm$^2$/s. 

The sLHC is sensitive to rare Higgs decays for light Higgs
bosons~\cite{Gianotti:2002xx}.  The 
decay $H \to \mu^+ \mu^-$ has a branching ratio $\sim 10^{-4}$
and almost certainly cannot be observed at the \lhchl.
For $\MH = 140 \gev$, the sLHC can obtain a $5.1\sigma$ observation 
and an accuracy of $\delta (\sigma \times \br(H \to \mu^+\mu^-))/
 (\sigma \times \br(H \to \mu^+\mu^-)) = 0.2$
with 3000 \ifb/experiment.  
The accuracy rapidly decreases with increasing Higgs
mass and for 
$\MH = 150 \gev$ the significance is $2.8\sigma$ with 
an accuracy of $\delta (\sigma \times \br(H \to \mu^+\mu^-))/
 (\sigma \times \br(H \to \mu^+\mu^-)) = 0.36$.
Similarly, the decay $H \to Z\ga$ can be observed
at $11\sigma$ for $100 \gev \le \MH\le 160 \gev$. This is to be
compared with $3.5\sigma$ at \lhchl.

Since in this scenario the coupling of the Higgs to $ZZ$ and/or $W^+W^-$
will have been observed at the \lhcten, the production of this new state
via $e^+e^- \to Z^* \to ZH$ or 
$e^+e^- \to \nu\bar\nu W^+W^- \to \nu\bar\nu H$ at lepton colliders is
guaranteed. Lepton colliders can potentially improve the precision
measurements of the Higgs couplings, self-coupling, width and
spin in this Higgs mass region. At the ILC, the total cross section can
be measured in a decay mode independent analysis via $Z$ recoil in the channel $e^+e^- \to ZH$,
 from which, in
conjunction with the branching fractions, the absolute values of the
couplings can be derived. 
For $\MH<150 \gev$, a precision measurement of the absolute values of the
Higgs boson couplings to $W$, $b$, $\tau$, $c$, $t$ and $g$, $\ga$ 
(through loops, possibly combining with the PLC option) can be
achieved~\cite{AguilarSaavedra:2001rg}, see 
\refta{tab:ILCprecisions}. The mass can be measured with a precision of
around $\Delta \MH\approx 50$~MeV. In addition, the $\cp$ quantum numbers
can be measured in $\tau$ decays~\cite{Andreas} and the spin can be
determined both in production and in decay. For $\sqrt{s}=800-1000 \gev$,
the $t\bar t H$ associated production 
allows the measurement of the coupling to the top quark~\cite{Gay:2006vs}, 
and for 
very high luminosities also the Higgs self-coupling can be measured in 
$ZHH$ final states~\cite{Battaglia:2001nn}.
Thus, a nearly complete precise Higgs boson profile could be determined,
and possible signals of scales beyond the SM could be detected.

For $\MH=160-180 \gev$ (currently probed by the Tevatron Higgs
searches), the Higgs decays are dominated by 
$H \to W^+W^-$, suppressing 
the branching fractions of the Higgs into most of the
particles mentioned above below the per-mille range, making precision
measurements of those couplings impossible with $\sqrt{s} = 500 \gev$.
Besides the decay to vector bosons, the $b \bar b$ channel remains a
relatively precise observable at the ILC. 
Consequently an important part of the Higgs
profile could be determined in a model-independent way (including as
well possible exotic or invisible decay channels). 
Also at CLIC the detectable channels remain 
the same, see \refta{tab:CLICprecisions}.
In this way these colliders could be sensitive to new scales beyond the
SM.

In this Higgs mass range, on the other hand, 
the ILC with the GigaZ ($\sqrt{s} \approx \MZ$)
and MegaW ($\sqrt{s} \approx 2 \MW$) options could
indirectly exclude a SM Higgs,
based on precision observables~\cite{Erler:2000jg,Flacher:2008zq},
at more than $3\sigma$ throughout (if the order of magnitude of the
current measurements of the precision observables is in the right
range). Given the possible precision, $\MH \gsim 130 \gev$ could be excluded
in the SM at the $3\,\sigma$ level if the true SM Higgs mass stays at
its current best fit point of around $\MH\approx 90 \gev$. The
combination of Higgs observables and more precise SM observables would
offer new realms for the precision tests of New Physics theories
explaining the apparent difference between precision observables and
an observed Higgs mass above $\sim 160 \gev$.


\subsubsection{Scenario B-I: SM-like Higgs with \boldmath{$\MH \le 130 \gev$}}

In the region of low masses, $114 \gev \le \MH \le 130 \gev$, a
channel where the LHC has the
potential to  discover the Higgs at the $5\,\si$ level is
the $H \to \ga\ga$ final state. 
The CMS optimized analysis shows a discovery potential in this
region with $\sim 10$ \ifb.
Close to $\MH = 130 \gev$  the channel $H \to ZZ^* \to$~4~leptons could
also reach the $5\,\si$ level.
On the other hand, in the mass region $\MH \lsim 130 \gev$ 
effects smaller than $5\,\si$ could be studied separately for the vector
boson fusion and $gg$ production channels. 
The $qq \to qqH, H \to \tau^+ \tau^-$ channel can reach only the 2 to
3$\si$ level.

Rare decays for a light Higgs can be studied at the sLHC. For
example, for $\MH = 120 \gev$, the sLHC can obtain a $7.9~\sigma$
observation of the $\mu\mu$ channel 
and an accuracy of $\delta (\sigma \times \br(H \to \mu^+\mu^-))/
 (\sigma \times \br(H \to \mu^+\mu^-)) = 0.13$
with 3000 \ifb/experiment. 

For the ILC, even running at $\sqrt{s}$ substantially below $500 \gev$, 
the range
$\MH \lsim 130 \gev$ should be particularly
``easy''. The ILC will be able to measure many Higgs properties in the light
Higgs mass range: the mass, couplings (in a model-independent way) to
nearly all fermions of the third family, to the SM gauge bosons and
(running at $\sqrt{s} = 800 \gev$) the tri-linear Higgs
self-couplings. Also a determination of the Higgs boson spin and quantum
numbers should be easily feasible. The anticipated precisions are
summarized in \refta{tab:ILCprecisions}. 
Thus, a nearly complete Higgs boson profile could be determined, 
and possible signals of scales beyond the SM could be detected.

CLIC operating at $\sqrt{s} = 3 \tev$ can observe many Higgs boson
decays. In addition to what could be measured at the ILC, particularly
interesting would be the Higgs decay  
$H \to \mu^+ \mu^-$ through the vector boson fusion process,
$e^+e^-\to H\nu{\overline \nu}$\cite{Battaglia:2008aa}.
With  5 ab$^{-1}$,
CLIC can obtain a precision on the coupling constant of $\delta g_{H\mu\mu}/
g_{H\mu\mu} = 0.04$ for $\MH = 120 \gev$. CLIC can also obtain a 10$\%$
measurement of $g_{HHH}$ for $\MH = 120 \gev$.   


\subsubsection{Scenario B-II: SM-like Higgs with \boldmath{$\MH \ge 180 \gev$}}

This region is severely constrained by the
electroweak precision fits~\cite{Flacher:2008zq,ewpoMoriond2009,lepewwg}.
Excluding the direct search results, the $3\sigma$ allowed region
 is $\MH \le 209 \gev$.
When the direct search results from LEP2 and the Tevatron are included,
the $3\sigma$ allowed region  is $\MH \le 168 \gev$ or 
$180 \gev \le \MH \le 225 \gev$.
Consequently, the discovery of a Higgs 
boson above this mass range, even in the absence of any
other signal, would point to new physics beyond the SM.

The main channel for the Higgs discovery at the LHC in this
mass region is  $H \to ZZ \to 4l$
and with only 2~\ifb\ the mass range $M_H\sim 190 - 500 \gev$ can be covered.
Starting at a luminosity around 30~\ifb, the vector
boson fusion Higgs production channel can be
studied and Higgs masses up to $1 \tev$ can be explored using
$H \to W^+W^- \to llqq$.

With
300~\ifb, spin-parity quantum numbers $0^{-+}$ and $1^{--}$ can be excluded
for $\MH = 230 \gev$~\cite{Buszello:2002uu}.  The total width  can be
measured to a precision of 
35$\%$ with 30~\ifb\ from $H \to Z Z^*\to 4$ leptons above
$\MH\sim 200 \gev$~\cite{Ball:2007zza} and ultimately to $5-8\%$ with
300~\ifb~\cite{ATLASold}.  

In the mass range $170 \gev \le \MH\le 200 \gev$, the sLHC may be able
to observe Higgs boson pair production through the process 
$gg\to HH \to l^+l^{\prime +} 4j$, thus getting sensitivity for the
tri-linear Higgs self-coupling. A critical feature of the analysis
is the assumption that detector capabilities at the sLHC are roughly
the same as at the LHC. Further studies to clarify
these issues are currently in progress, as discussed in 
\citere{Jakobs:2009zza}.

As above, the precision SM observables of the ILC GigaZ and MegaW
options could indirectly rule out this mass range in the SM. 
On the other hand, the 
precision measurement of the Higgs mass, the couplings to the top, $W$
and $Z$ and the direct measurement of the Higgs width~\cite{Meyer:2004ha}
from the lineshape would be possible at the ILC if the Higgs boson is
kinematically 
accessible. These measurements will yield stringent constraints on
potential New Physics models explaining the high Higgs boson mass and
point to new scales beyond the SM.


\subsubsection{Scenario C: A non-SM-like Higgs}

Another possibility at the \lhcten\ would be to observe a state
compatible with a Higgs boson that appears to be in disagreement with
the SM predictions. This could be due to an $\MH$ value outside the
region allowed by the precision data. A mass above $\sim 170 \gev$ would
indicate a disagreement with today's precision data 
(not taking into account the direct searches, see above)~\cite{lepewwg}.
We will not pursue this option further and assume for the rest of the
discussion a mass in the range $110 \gev \lsim \MH \lsim 160 \gev$.

Another possibility for a non-SM-like Higgs boson would be
production and/or decay rates different from the SM prediction. 
While a suppression of a decay could only be observed in the sensitive
channels, a strong enhancement could appear in any of the search channels
(and is consequently more arbitrary).  It is possible to suppress all
of the Higgs couplings by the simple mechanism of adding extra
Higgs singlets, which can make the Higgs search at the LHC quite
challenging\cite{Barger:2007im,Dawson:2009yx}.

Of particular interest are the loop-mediated Higgs couplings, 
$H\to \ga\ga$ and $H\to gg$, which can receive sizable contributions 
from New Physics (NP). In many NP models 
the Higgs couplings to $W$ and $Z$ remain essentially unaffected. 
Several channels at the LHC are sensitive to different combinations of
these loop-mediated Higgs couplings,
which allows a quasi-model independent
analysis~\cite{Cacciapaglia:2009ky}, potentially shedding  
light on the nature of new states discovered at the LHC and on the
underlying model of electroweak symmetry breaking. The simultaneous
measurement of the inclusive $H\to \ga\ga$ and 
$H\to W^+W^-$ cross sections, as well as the vector
boson fusion process, $qq\to qqH, H\to \ga\ga$ cross section, 
allows the placement of constraints in the 
two-dimensional plane of the $H\ga\ga$ and $Hgg$ couplings. 
From a survey of NP models performed in Ref.~\cite{Cacciapaglia:2009ky}
contributions to some of these cross sections as
large as 50\% were found (see also \citere{Ginzburg:2001ss} for a 2HDM
analysis). Therefore, measurements of these cross 
sections at the LHC with 10-20\% accuracy should allow some discrimination 
of NP models. At the ILC, the percent-level measurement of the $H\to \ga\ga$ 
and $H\to WW$ branching ratios will allow a much more sensitive probe of
NP models. 

\bigskip
For $\MH \lsim 130 \gev$ a channel suppression at the \lhcten\ would
only be possible in the decay $H \to \ga\ga$, which could be due to
either a suppressed $HWW$ coupling (or a large new loop correction
interfering negatively with the $W$~loop contribution) or an enhancement
of a Higgs branching ratio to a channel invisible at the \lhcten. 
Invisible could mean the Higgs boson decays
either to known particles that are difficult to
detect at the LHC, such as decays to light quarks, or to truly invisible
particles such as neutrinos, the lightest SUSY particle (assuming
$R$-parity conservation), the lightest Kaluza-Klein mode etc.
A sensitivity at the \lhcten\ to invisible decays (assuming SM
production rates) would only be possible if the BR into the invisible
channel is close to 100\%~\cite{Hinvtalk} (see also WG2 report).

For $130 \gev \lsim \MH \lsim 170 \gev$ a suppression of the decays to 
$WW^{(*)}$ and/or $ZZ^{(*)}$ could be observable. As for the light Higgs
case this could be due to
either a suppressed $HWW$ and/or $HZZ$ coupling or an enhancement of a
Higgs branching ratio to a channel invisible at the \lhcten. 

A measurement of decays that are suppressed due to the (enhancement of
the) decay to difficult or invisible channels at the \lhcten\ 
would improve with higher luminosity. 
The sensitivity could (assuming SM Higgs production rates)
go down to a BR of $\sim 15\%$~\cite{Hinvtalk}. Consequently, the
observation of a suppression could be backed-up by the observation of
``invisible'' decays. 

At the ILC, see \refta{tab:ILCprecisions}, any 
kinematically accessible decay channel with a substantial decay rate
will be detectable, including the decay to truly invisible
particles. Therefore at the ILC a nearly complete Higgs
boson profile could be determined. The ILC would be ideal to shed light
on this case.
Similar results could be expected for CLIC.

A suppressed coupling of the Higgs to vector bosons would strongly hint
towards an extended Higgs sector where several Higgs bosons share the
couplings to the $W^+W^-$ and $ZZ$. 
A maximum coverage of the mass range would be needed to discover these
additional Higgs bosons (or to reject this solution). 
The LHC could cover masses up to $\lsim 1 \tev$ assuming SM-like
decays. The situation improves slightly at the sLHC. The ILC (especially
with the $\ga\ga$ option~\cite{Badelek:2001xb}) would have a good chance
to discover heavier Higgs bosons (with
$M_H\lsim 400 \gev$) also with non-SM-like decay
rates. CLIC, with the high $\sqrt{s} \approx 3 \tev$ would cover an even
larger Higgs mass range.

This search could be supplemented by the analysis of $WW$
scattering at very high energies to investigate whether other forces
than the Higgs mechanism might be at work. 
In this way the measurement of the cross section of $WW$ scattering as a
function of the invariant mass of the di-boson is a key ingredient to the
understanding of the symmetry breaking mechanism. 
Within the SM the Higgs particle is essential to the renormalization of the
theory and ensures that the unitarity bounds are not violated in high energy
interactions, i.e.\ $\si(WW \to WW)$ does not rise with $M(WW)$ for 
$M(WW) \gsim 1-2 \tev$, and a resonance at $M(WW) = \MH$ should be
observable. 
If another mechanism of electroweak symmetry breaking is realized in
nature, see e.g.~\cite{Giudice:2007fh} and references therein, 
the behavior of $\si(WW \to WW)$ will deviate from the SM
expectations. Corresponding LHC studies can be found in \cite{Aad:2009wy}
(p.\,1695) and \cite{WW_CMS,Govoni:2009zz}. 
Both ATLAS and CMS  expect to be able to discover strongly interacting
resonances with $M(WW) > 1 \tev$ only with 100~\ifb\ or more.
Integrated luminosities of 300 - 500~\ifb\ will be necessary to
understand the shape of the $\si(WW \to WW)$ vs.\ $M(WW)$ distribution
at high energies in order to investigate if a light Higgs is present or
a different mechanism of electroweak symmetry breaking is realized. 
In order to explore this region fully, the sLHC will be necessary.
At the ILC new resonance scales up to $\sim 30 \tev$ could be
probed indirectly~\cite{Giudice:2007fh,Ackermann:2004ag}, while
at CLIC direct resonances up to $\sim 2.5 \tev$ could be accessed (for
$\sqrt{s} = 3 \tev$)~\cite{Accomando:2004sz}.
More details can be found in \refse{wg2}.

There are of course, many possibilities for non-SM Higgs bosons, such
as scenarios with multiple Higgs bosons or with very light Higgs bosons
which evade the LEP Higgs searches. Some of these options are discussed
in the \refse{wg2}.


\subsection{Summary and conclusions of WG1}

The LHC will explore  the mechanism responsible for
electroweak symmetry breaking.
Assuming that a new state as a possible candidate for a Higgs boson will
have been observed at the \lhcten, the full identification of the
mechanism of electroweak symmetry breaking will require to measure all its
characteristics. This comprises an accurate mass determination, 
a (model-independent) measurement of its individual couplings to other
particles, a determination of the self-couplings to confirm the
``shape'' of the Higgs potential, as well as measurements of its spin
and $\cp$-quantum numbers . 
At the LHC, even running at high luminosity, these measurements will 
only partially be possible.

We reviewed what we might know about the Higgs sector 
once the LHC has collected 10\,\ifb\ (of understood data) 
at a center-of-mass energy  of $14 \tev$. 
Based on the anticipated future knowledge of the Higgs
sector, we investigated the capabilities
of future colliders to further unravel the Higgs mechanism.

While the sLHC will be able to extend the reach and precision of the
\lhchl\, it seems clear that a full exploration of the Higgs sector
will require either the ILC or CLIC.
For a SM-like Higgs with $\MH \lsim 150 \gev$ at the ILC 
a nearly complete Higgs boson profile could be determined. For larger
masses (currently probed at the Tevatron) the decay to SM gauge
bosons becomes dominant, suppressing other decay modes and making them
more difficult to measure. 
In the case of a non-SM-like Higgs nearly all channels visible at the ILC
can be determined with high accuracy. 
The corresponding CLIC analyses have focused mostly on measurements that
are challenging at 
the ILC. Due to its high luminosity and high center-of-mass energy up to
$\sqrt{s} \approx 3 \tev$ very heavy Higgs bosons, for instance from
extended Higgs sectors, could be probed.
The precision measurements obtainable 
at the ILC and CLIC could point to  New Physics beyond the SM, 
opening the window to energy scales beyond the LHC.


 }
\newpage
{\setcounter{equation}{0}
\setcounter{figure}{0}
\setcounter{table}{0}

\section{WG2: No Higgs boson}
\label{wg2}
{\it 
G.~Azuelos,
C.~Grojean,
M.~Lancaster,
G.~Weiglein
(convenors)\\ 
S.~Dawson,
S.~De~Curtis,
M.T.~Frandsen,
R.~Godbole,
P.~Govoni,
J.~Gunion,
T.~Han,
S.~Heinemeyer,
G.~Isidori,
A.~Martin,
E.~Ozcan,
T.~Plehn,
F.~Sannino, 
M.~Schram
}

\bigskip
If no Higgs candidate is found in the first 10~fb$^{-1}$ at the LHC,
two options will have to be considered: (i) A Higgs boson exists (or more
than one) but it has non-standard properties that make it difficult to
detect because of suppressed couplings to gauge bosons and/or
fermions or because of unusual decays; or (ii) There really is no fundamental
Higgs boson and new degrees of freedom or new dynamics beyond the
Standard Model are needed to maintain unitarity at high energy. The
implications of these two scenarios for future colliders are
discussed in this section.

\newcommand{\lhcten}{LHC$_{10/{\rm fb}}$}
\newcommand{\lhchl}{LHC$_{300/{\rm fb}}$}
\newcommand{\lsim}
{\;\raisebox{-.3em}{$\stackrel{\displaystyle <}{\sim}$}\;}
\newcommand{\gsim}
{\;\raisebox{-.3em}{$\stackrel{\displaystyle >}{\sim}$}\;}

\subsection{Introduction}

The exploration of the Terascale at the LHC will probe directly the
dynamics responsible for electroweak symmetry breaking (EWSB). While
the evidence in favour of spontaneous breaking of the electroweak
symmetry is very strong, the
fact that this breaking occurs via a single fundamental Higgs field,
with a non-trivial vacuum expectation value, is far from being clearly
established.  The Higgs mechanism is certainly the most
economical way of explaining this spontaneous breaking, and a light Higgs
mass ($m_{h}\approx $~100~GeV) is also an efficient way to account for
all the existing electroweak precision tests (EWPTs). However, the
strong sensitivity of $m_{h}$ to short-distance scales poses a serious
naturalness problem at least for the Standard Model (SM) 
and motivates the search for alternative
symmetry-breaking mechanisms.  The SM Higgs boson plays
the role of moderator of the strength of longitudinal $W$ interactions
and allows the model to be extrapolated to very short distances. Thus
in the absence of a Higgs boson, the dynamics behind EWSB is expected 
to become strong around a TeV and to deviate significantly from the SM.

In this report we consider the situation where a Higgs-like signal is
absent in the early LHC data and we investigate in how far the possible
physics scenarios can be constrained in such a case.
The non-observation of a
Higgs candidate in the first 10~fb$^{-1}$ at the LHC could evidently
point to one of the two following options:
\begin{itemize}
\item There exists a Higgs boson (or more than one) but it has
  non-standard properties that make it difficult to detect, either
  because of suppressed couplings to gauge bosons and/or fermions, or
  because of unusual decays.
\item There really is no fundamental Higgs boson. Then new degrees of
  freedom or new dynamics beyond the Standard Model are required 
  to maintain unitarity at high energy.
\end{itemize}
In the case of the absence of a Higgs-like signal in the early LHC data
it is of particular importance to investigate the behavior of the
$W_LW_L$ scattering amplitudes, which will be directly affected by the
dynamics responsible for restoring unitarity.

\subsection{Unobserved Higgs boson scenarios}
\label{sec:WG2InvHiggs}

With 10~fb$^{-1}$, a truly SM-like Higgs boson will be discovered by ATLAS and
CMS, since the LHC covers the full kinematic range expected for a SM
Higgs boson from unitarity arguments and EWPTs (see WG1 summary). However, 
for a Higgs boson with non-standard properties it could be more
difficult to extract a Higgs signal from the data in the experimental
environment at the LHC. Thus, if the Higgs 
mechanism is realized in nature, the absence of a conclusive sign of a Higgs
boson at the \lhcten\ would point towards a Higgs sector with a 
more involved structure than in the SM.

In models with an extended Higgs sector and/or an enlarged particle
content Higgs phenomenology can drastically differ
from the SM case. On the one hand, it should be recalled that the SM
exclusion bound obtained at LEP of
$m_h^{\rm SM} > 114.4$~GeV at 95\%~C.L.~\cite{Barate:2003sz} is not
applicable for a more complicated Higgs sector giving rise to a
suppression of the coupling of a light Higgs to gauge bosons and
possibly also yielding 
unusual decay properties. Thus, the possibility that a Higgs boson
much lighter than the SM exclusion bound has escaped detection in the
searches carried out up to now cannot be excluded. Such a light Higgs boson
could be very difficult to detect also at the LHC. A well-known
realisation of such a scenario is the Minimal Supersymmetric Standard
Model (MSSM) with non-vanishing complex phases in the CPX benchmark
scenario~\cite{Carena:2000ks}
giving rise to a Higgs boson as
light as about 45~GeV that is unexcluded by the LEP 
searches~\cite{Schael:2006cr}. The Higgs bosons in this scenario would
also be difficult to detect with the standard search channels at the
LHC~\cite{Buescher:2005re,Schumacher:2004da,Accomando:2006ga}.

On the other hand, the lightest Higgs boson could also be much heavier
than the mass range preferred for a SM-type Higgs by the EWPTs. In this
case new physics contributions to electroweak precision observables
would be necessary to 
compensate the effects of a heavy Higgs boson, mimicking in
this way the contribution from a light SM-like Higgs boson.

Even if there exists a light Higgs boson not far above the exclusion bound
for a SM-type Higgs, its properties could still be very different from a
SM Higgs. In the SM scenario, a light Higgs boson with a mass below the $WW$
threshold has a rather narrow decay width as the heaviest SM
particles it can decay to is a $b$ quark pair. The $Hbb$ 
coupling is quite small, only about 1/40. 
As a consequence, any new particle with less than half the
Higgs mass which interacts with the Higgs boson could 
modify the decay branching fractions very substantially. If these new particles carry no colour or electroweak charge then they
will be difficult to produce directly at the LHC.  However, they are
likely to alter Higgs decays, leading to final states that
could be either visible but complicated or invisible. An example of the latter
is Higgs decays to a pair of LSPs in supersymmetric models. In this
case, the LEP lower bound on the mass of a Higgs boson with SM-like
$WW,ZZ$ couplings is $114$~GeV.  An example of the former is Higgs
decays to a pair of light CP-odd scalars that are primarily SM
singlets, as strongly motivated in the Next to Minimal Supersymmetric
Model (NMSSM) \cite{Dermisek:2005ar,Dermisek:2005gg}. In this case, the LEP
lower bound on the Higgs mass could be as small as $82$~GeV.

Couplings in the Higgs sector can differ from the SM case both because
of a different tree-level structure and because of higher-order
contributions that can often be very large. In the MSSM, for instance,
the couplings of neutral Higgs bosons to a pair of down-type fermions can
be very strongly enhanced for large values of the parameter $\tan\beta$,
giving rise to a simultaneous suppression of the branching ratios 
into $\gamma\gamma$, $WW^{(*)}$, $ZZ^{(*)}$. Higgs production in gluon
fusion can be suppressed if there is a destructive interference of
the top-quark loop with the contributions from other new particles, like the
superpartners of the top quark in the case of the MSSM (see
e.g.\ Refs.~\cite{Djouadi:1998az,Manohar:2006gz, Giudice:2007fh}).

Higgs properties can also be modified very significantly if a Higgs
boson mixes with other states of new physics. An example for such a
scenario is the mixing of a Higgs boson with a radion, a state that is
predicted in models with 3--branes in extra dimensions. As the 
radion has the same quantum numbers as the Higgs boson the two states
will mix with each other in general. Since the radion has couplings that
are different from those of the SM Higgs boson, the two physical
eigenstates will have unusual properties that differ substantially from
the ones of the pure Higgs state~\cite{Giudice:2000av}.

The impact of new physics contributions in the Higgs sector 
can, on the one hand, be to enhance the prospects for Higgs searches, 
for instance by opening up new discovery channels, while, on the other
hand, unusual Higgs properties can also make it much more difficult to
extract a Higgs signal from the data than is the case for a SM Higgs.
Examples that could lead to a situation where no Higgs signal can 
be established at least with the first 10~fb$^{-1}$ at the LHC
are the case of a Higgs boson that decays primarily into
hadronic jets, possibly without definite flavour content, or the
possibility that a Higgs boson with SM-type couplings could primarily decay
into a pair of very light pseudoscalar Higgs bosons. Another very
difficult scenario for Higgs boson detection would be the
case of a ``continuum'' Higgs model, i.e.\ a large number of doublet
and/or singlet fields with complicated self interactions (see, e.g., Refs.~\cite{Binoth:1996au, Espinosa:1998xj, Dawson:2009yx}). This could result in
a very significant diminution of all the standard LHC signals.

A review of phenomenological consequences of non-standard Higgs boson
decays can be found in Ref.~\cite{Chang:2008cw}.
Table~\ref{table:non-standardHdecays} gives the 95\% C.L.\ lower LEP limits
on $m_h$ for a Higgs boson assuming that it has SM-like couplings to gauge
bosons, obtained from searches in specific channels.

\begin{table}[htb!]
\begin{center}
\begin{tabular}{|c|c|c|c|c|c|c|c|c|c|c|c|c|c|}
\hline
Decay mode & SM & 2$\tau$, 2$b$ & 2j & $WW^\star$ & $\gamma\gamma$ & E\hspace{-.35cm} \big/ & $4e, 4\mu, 4\gamma$ & $4b$ & $4 \tau$ & anything \\
\hline
\hline
$m_h$ bound [GeV] & 114.4 & 115 & 113 & 109.7 & 117 & 114 & 114 & 110 & 86 & 82\\
\hline
\end{tabular}
\end{center}
\caption{\label{table:non-standardHdecays}
LEP bounds on the Higgs mass for a Higgs boson with SM-like couplings to gauge
bosons, as obtained  from searches in specific decay modes (for a discussion of the assumptions used for obtaining these limits, see Ref.~\cite{Barate:2003sz,Schael:2006cr,Chang:2008cw} and references therein).}
\end{table}

The main discovery channels for a light SM-like Higgs boson at the 
\lhcten~ are~\cite{Aad:2009wy,Ball:2007zza}, 
\begin{align}
\label{lowmass}
m_h \lsim 140 \mbox{ GeV} &: q \bar q H, H \to \tau^+\tau^- ~, ~ pp \to H, H \to \gamma\gamma ~, \\
\label{highmass}
130 \mbox{ GeV} \lsim m_h &: q \bar q H, H \to WW^{(*)}, ZZ^{(*)} ~,  \\
\label{ggHZZ}
140 \mbox{ GeV} \lsim m_h &: pp \to H + X, H \to ZZ^{(*)} ~,\\
\label{ggHWW}
150 \mbox{ GeV} \lsim m_h &: pp \to H + X, H \to WW^{(*)} ~.
\end{align}

The ATLAS and CMS sensitivities in the search of the Higgs boson can be
significantly affected in the low mass region for the following two
main reasons: (i) the SM Higgs boson is dominantly produced by gluon fusion,
a one-loop process that can receive large negative corrections from new
physics, such as from 
squarks~\cite{Djouadi:1998az,Manohar:2006gz, Giudice:2007fh} and (ii)
the narrow width of the Higgs boson makes it more vulnerable to any new
decay channels. It should be noted that if any non-zero coupling of the Higgs boson to 
gauge bosons can be observed at \lhcten, this would guarantee the Higgs 
production at the ILC via Higgs-strahlung or weak boson fusion. In this case
a determination of the Higgs boson properties can be performed (see
WG1 summary). Even if there are multiple mixed Higgs bosons which
  overlap within the mass resolution and which decay in
  multiple ways in such a way that no LHC signal is seen, the ILC
  would be guaranteed to see an enhancement in the $M_X$ distribution
  in $e^+ e^- \to ZX$ and be able to study the decays of the individual
  Higgs bosons \cite{Espinosa:1998xj}.

In the following we briefly discuss three examples of scenarios that could
give rise to the absence of a clear Higgs signal in the first
10~fb$^{-1}$ at the LHC: (i) a
light Higgs boson with reduced gauge couplings in the
MSSM with non-vanishing complex phases; (ii) a light Higgs boson with
full-strength gauge couplings but unusual decays; and (iii) a light Higgs boson
with invisible  decays.

\subsubsection{A light gaugephobic Higgs boson in the MSSM with complex parameters}

The Higgs sector of the MSSM appears to be a ``minimal'' extension of
the SM Higgs sector in the sense that is characterized by two free
parameters at lowest order (conventionally chosen as $\tan\beta$, the
ratio of the two vacuum expectation values, and either the mass of the 
pseudoscalar Higgs boson, $m_A$, or the mass of the charged Higgs bosons,
$m_{H^\pm}$) instead of the single free parameter in the case of the SM
(the mass of the Higgs boson). Nevertheless the different tree-level
structure can change Higgs phenomenology very substantially.
Furthermore, large higher-order effects also play an important role.

While the MSSM Higgs sector is CP-conserving at lowest order,
CP-violating effects can be induced by non-vanishing phases entering via
potentially large higher-order contributions. As a consequence, all
three neutral Higgs bosons $h$, $H$ and $A$ mix with each other, leading
to the mass eigenstates $h_1$, $h_2$, $h_3$. The three mass eigenstates 
share their couplings to gauge bosons, i.e.\ the sum of the squares of
the couplings of $h_1$, $h_2$, $h_3$ to gauge bosons is equal in good
approximation to the squared coupling of the SM Higgs boson to gauge bosons.
Depending on the mixing between the three Higgs bosons, the coupling of
at least one of the three Higgs states to gauge bosons can be heavily 
suppressed compared to the SM case. Such a situation occurs in the CPX
benchmark scenario of the MSSM~\cite{Carena:2000ks}, where over a large
part of the parameter space, the lightest Higgs state, $h_1$, decouples from
gauge bosons, while the second-lightest Higgs boson can have a large
branching ratio into a pair of the lightest Higgs bosons, 
$h_2 \to h_1h_1$. 

In the LEP Higgs searches~\cite{Schael:2006cr} carried out in the CPX
scenario, an unexcluded parameter region remained for a mass of the
lightest Higgs boson of about 45~GeV and moderate $\tan\beta$. This is
illustrated in Fig.~\ref{fig:cpx}, which shows the parameter
regions excluded at 95\% C.L.\ by the topological cross section limits
obtained at LEP~\cite{Barate:2003sz,Schael:2006cr} (as implemented in the
program {\tt HiggsBounds}~\cite{Bechtle:2008jh}), making use of the
currently most advanced theory 
prediction for Higgs boson cascade decays obtained in Ref.~\cite{Williams:2007dc}.

\begin{figure}[htbp]
\centering
\includegraphics[width=0.35\textwidth]{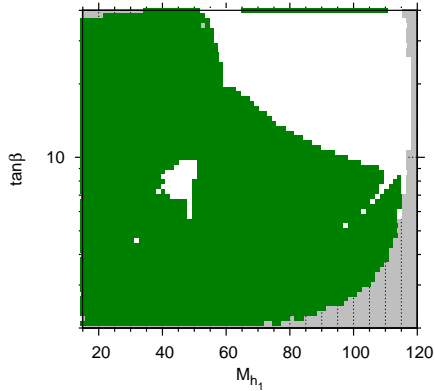}
 \caption
{Coverage of the LEP Higgs searches 
in the $(M_{h_1}, \tan\beta)$ plane of the 
CPX benchmark scenario of the MSSM. The plot (from Ref.~\cite{Bechtle:2008jh})
shows the parameter
regions excluded at 95\% C.L.\ (green) by the topological cross section limits
obtained at LEP~\cite{Barate:2003sz,Schael:2006cr}, using the theory
prediction for Higgs cascade decays from Ref.~\cite{Williams:2007dc}.}
\label{fig:cpx}
\end{figure}

The case of a low Higgs mass in this scenario has only partially been explored at the
LHC so far, and it seems difficult to close the ``hole'' shown in the
LEP coverage of Fig.~\ref{fig:cpx} with the standard search channels,
see Refs.~\cite{Buescher:2005re,Schumacher:2004da,Accomando:2006ga} for
experimental studies of the ATLAS collaboration and
Refs.~\cite{Ghosh:2004cc,Bandyopadhyay:2007cp} for a discussion 
of other possible search channels for covering this parameter region.

Thus, the described scenario of the MSSM with non-vanishing phases of
the parameters could be a case where no Higgs signal appears
in the first 10~fb$^{-1}$ at the LHC. On the other hand, in this
scenario supersymmetric particles would be detected at the LHC with the 
first 10~fb$^{-1}$. One would therefore be in a situation where on the
one hand there would be clear evidence for new physics, compatible with
the discovery of supersymmetry. On the other hand, the experimental
evidence for a light Higgs boson, which is required in this framework, would
be lacking. This could be a strong case for an $e^+e^-$ linear collider
(LC) with a center-of-mass energy of at least 250~GeV to explore whether or not the
Higgs mechanism is realized in nature.

\subsubsection{A Higgs boson with SM-like $WW/ZZ$ couplings but unusual visible
  decays}

It is even possible that \lhcten~ would not see a signal for a
(possibly quite light, $m_h\sim 100$~GeV) Higgs boson with SM-like
$WW/ZZ$ couplings due to the fact that the Higgs decays to a pair of
particles each of which subsequently pair-decays.
Possibilities for the latter pairs include: $gg; c\bar{c}$; 
$\chi_2\chi_1$ where $\chi_1$ is invisible and $\chi_2$ decays to $f \bar{f}
\chi_1$; and to a less extent, $\tau^+\tau^-$. The first two cases arise in the NMSSM scenario explored 
in Refs.~\cite{Dermisek:2005ar,Dermisek:2005gg}. In the NMSSM, even in the
absence of CP violation there are three CP-even Higgs bosons,
$h_{1,2,3}$, two CP-odd Higgs bosons, $a_{1,2}$ and a charged Higgs boson
pair, $h^\pm$. At moderate to high $\tan \beta$ ($\tan\beta \gsim 3$) the most
attractive scenarios are such that the $h_1$ has quite SM-like
couplings to $WW,ZZ, f\bar{f}$ but decays predominantly via $h_1\to
a_1a_1$ which dominates over $h_1\to\ b\bar{b}$. In these scenarios,
$a_1$ has a mass smaller than $2m_b$ so that $a_1$ decays to $\tau^+ \tau^-$ or to a mixture of
$gg,cc,\mu^+\mu^-$, depending on the precise value of $m_{a_1}$.  Since such
a scenario allows $m_{h_1}$ to be as low as $86$~GeV, precision electroweak
constraints are robustly satisfied and no fine-tuning is required to
get the observed value of $M_Z$ since supersymmetric particle masses
can be low if the Higgs boson with SM $WW,ZZ$ couplings has $m_{h_1}\lsim 105$~GeV.

As in the previous MSSM case, this kind of scenario predicts
robust signals for supersymmetric particle production with $L=10\textrm{ fb}^{-1}$
at the LHC associated with a lengthy wait for a Higgs boson signal.
This does not imply that the Higgs boson could not eventually be discovered at
the LHC. For example, a full study of the process $pp\to p h_1 p$
(diffractive Higgs production) shows that $h_1\to a_1 a_1 \to
\tau^+ \tau^-\tau^+\tau^-$ could be detected with 
$L\sim 100 - 300\textrm{ fb}^{-1}$~\cite{Forshaw:2007ra}. Assuming that all trigger issues can be solved, the key to detecting this signal turns out to
be track counting, in particular keeping only events with a small number of
centrally produced tracks. The same is likely to apply to other
possible observation modes such as the $WW$ fusion mode: 
$pp\to WWX \to h_1 X\to \tau^+\tau^-\tau^+\tau^- X$. Of course, $e^+e^-\to Z h_1$ with $h_1\to a_1 a_1 \to \tau^+\tau^-\tau^+\tau^-$ would be highly visible at a linear collider.

At low $\tan \beta$ in the NMSSM, additional very natural scenarios arise
with even more exotic features.  In particular, one can have~\cite{Dermisek:2008uu}: 
(i)~strong mixing among the $h_{1,2,3}$,
leading to suppressed $WW,ZZ$ couplings for all; 
(ii)~dominant $h_2\to
h_1 h_1 $ (as well as $h_{1,2}\to a_1a_1$) decays; 
and (iii)~exotic $h^+\to
W^+a_1$ decays. The $h_1$, $h_2$, and $h^\pm$ can all have mass $\lsim
100$~GeV without conflicting with LEP and Tevatron limits.  Adding CP
violation to the model would lead to even more possibilities.

Future colliders might prove crucial to understanding the kinds of
Higgs scenarios discussed above. At a linear collider, any Higgs boson
with substantial $ZZ$ coupling will be produced via
$e^+ e^- \to Z h$ and detected as a peak in the $M_X$ spectrum of
$e^+ e^- \to ZX$ where the $Z$ decays to fully visible final states (e.g.
$e^+ e^-$, $\mu^+ \mu^- $, $jj$).  The various Higgs decay channels could then
be studied. For example, the full 
$h_1\to a_1 a_1 \to \tau^+ \tau^- \tau^+ \tau^-$ decay chain could be
reconstructed. Detection of direct $a_1$ production (e.g. via
$e^+ e^- \to \nu \bar{\nu} a_1 a_1$) would typically be quite challenging
due to the low mass of the $a_1$ and its singlet nature. Both a
$\gamma \gamma$ and a $\mu^+ \mu^-$ collider would also be excellent facilities
for studying the $h_1$. The muon collider would be particularly
interesting due to the fact that $\Gamma_{tot}^{h_1}$ is typically a
factor of 10 larger than $\Gamma_{tot}^{h_\textrm{\tiny SM}}$ due to the extra
$h_1 \to a_1 a_1$ decays. As a result, the beam energy resolution required to
scan the $h_1$ peak at the muon collider could be about a factor of 10
larger than that needed to scan the $h_\textrm{\tiny SM}$ peak. A direct scan of the $h_1$
peak would provide the most accurate measurement of
$\Gamma_{tot}^{h_1}$ and allow the highest precision measurements of
all the couplings of the $h_1$~\cite{Barger:1996jm}.

\subsubsection{Invisible Higgs decays}

As a third example of a possibly difficult scenario for Higgs searches
at the LHC, we discuss the case where a Higgs boson decays predominantly
into {\it invisible} particles. This can happen for instance in the MSSM via
the decay of a Higgs boson into a pair of the lightest neutralino, in models 
with extra dimensions via the decay of a Higgs boson into KK neutrinos, 
in models with neutrinos of a 4th generation, etc..

A priori a predominantly invisibly decaying Higgs boson does not automatically
imply that the corresponding signature will not be detectable at the
LHC. However, the Higgs search in such a scenario is expected to be more
difficult than in the SM case, which could give rise to the fact that 
no Higgs boson signal can be established in the first 10~fb$^{-1}$ at the LHC.

In order to estimate the potential for invisible Higgs searches, a model
independent variable, $\xi$, is commonly introduced to take into account
possible modifications in the Higgs production cross-section as well
as the invisible Higgs decay fraction:
\begin{equation}
\xi^2 =\frac{\sigma_\textrm{BSM}}{\sigma_\textrm{SM}} BR(H\to
\textrm{inv.})
\end{equation}

The most promising processes (see, e.g., Ref.~\cite{Eboli:2000ze}) for
the search of an invisible Higgs boson at the LHC are vector boson fusion 
$qq \to qq +
E\hspace{-.27cm} \big/ {}_\textrm{T}$ (VBF) and the production in
association with a $Z$ decaying into two leptons ($ZH$).  Both
analyses are confronted with substantial challenges arising from backgrounds
due to pile-up, beam halo, cosmic muons, and from detector effects
(instrumental noise, detector calibration). One of the most
significant experimental challenges for the VBF analysis is to record
enough events of interest while keeping the QCD background within the
allowable trigger rate.  This is particularly challenging since the VBF signature consists of two jets and large transverse missing energy which is copiously produced in a hadronic environment. 
In addition to the trigger and pileup effects
which are hard to estimate at the moment, a control of the systematic
uncertainties associated with jets and missing transverse energy will
be essential to set physics limits with the VBF channel.  Unlike the VBF
analysis, the $ZH$ analysis relies on lepton triggers which are much
cleaner and are not plagued by large QCD backgrounds. As such, the
trigger is not expected to be a concern in this case. Various
systematic uncertainties associated with the leptons, jets, and
missing transverse energy appear to be significantly less important
than for the VBF analysis.

\begin{figure}[htbp]
\centering
\includegraphics[width=0.50\textwidth]{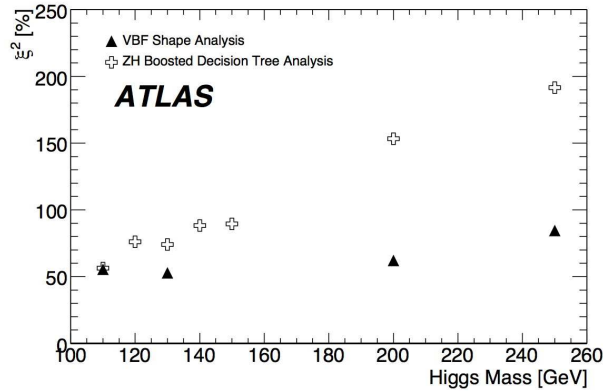}
 \caption
{Sensitivity at the 95\% C.L.\ to an invisible Higgs boson with 
ATLAS for both the VBF and $ZH$ 
channels with 30~fb$^{-1}$ of data assuming only Standard Model backgrounds.
The open crosses show the sensitivity for the $ZH$ analysis, and the solid
triangles show the sensitivity for the VBF shape analysis.
Both these results include systematic uncertainties.
If we take the uncertainty on the signal efficiency into account, the
sensitivity in $\xi^2_{95}$ could vary by up to $\pm 2.4\%$ and $\pm 10.5\%$ 
for the $ZH$ and VBF analyses, respectively
(from~Ref.~\cite{Aad:2009wy}).}
\label{fig:InvisibleHiggs}
\end{figure}

The current best estimate of the sensitivity of the ATLAS detector for
observing an invisibly decaying Higgs boson is presented in
Fig.~\ref{fig:InvisibleHiggs}.  Naively scaling the ATLAS results discussed in Ref.~\cite{Aad:2009wy}, given 
for 30~fb$^{-1}$ in Fig.~\ref{fig:InvisibleHiggs} to 
500~fb$^{-1}$ and 10~fb$^{-1}$ of 
data\footnote{This is a naive estimate only since
  systematic errors do not scale with integrated luminosity.}
yields the following results ($\xi^2_{95}$ denotes the 95\%~C.L. limit on $\xi^2$):
\begin{equation*}
\begin{array}{|c|c|c|}
\hline
m_h\ \textrm{[GeV]} & \begin{array}{c}\xi^2_{95} \textrm{ for VBF}\\ 10/30/500\ [\textrm{fb}^{-1}] \end{array} & \begin{array}{c}\xi^2_{95} \textrm{ for } ZH\\ 10/30/500\ [\textrm{fb}^{-1}] \end{array}\\
\hline
\hline
110 & .95/.55/.14 & .99/.57/.18 \\
\hline
130 & .95/.55/.14 & 1.28/.74/.23\\
\hline
200 & 1.05/.6/.15 & 2.6/1.5/.35\\
\hline
\end{array}
\end{equation*}
The results scaled to 500~fb$^{-1}$
suggest that the Higgs searches at the LHC could ultimately probe 
branching ratios at $\sim15\%$ for masses between 120--160 GeV. On the
other hand, the prospects for 10~fb$^{-1}$ are much worse, illustrating
the fact that a scenario with a sizable Higgs branching ratio into
invisible particles can imply the absence of a Higgs signal at least in
the early LHC data.

At an $e^+e^-$ LC, the detection of invisible decays of a Higgs boson
with SM-like $ZZ$ coupling that decays partly or
entirely to invisible channels would be much easier owing to the
cleaner experimental environment.
The studies presented in Refs.~\cite{Schumacher:2003ss} on the potential evidence for an invisible
decay channel of the Higgs boson at a future 350~GeV LC,
in the process $Z(\rightarrow qq)H$,
conclude that, with 500~fb$^{-1}$ of data, an invisible 
Higgs boson can be discovered down to branching ratios of $\sim2\%$ for
masses between 120--160~GeV. 

Once an invisible Higgs boson is discovered, the question of measuring its
properties remains. If the invisible channel is dominant, the mass may nevertheless be
accessible at the LHC through the measurement of the production rate.
The $ZH$ production rate, steeply falling as a function of the mass,
is sensitive to the Higgs mass. Moreover, the ratio of the production
processes can also provide an independent estimate of the Higgs mass.
Results from a parton level study~\cite{Davoudiasl:2004aj} suggest
that with 100 fb$^{-1}$ of data the mass can be determined to within
10--30 GeV.  In contrast, the mass resolution at the ILC is expected to
be significantly better using the recoil method.  Studies suggest it
should be in the tens of MeV~\cite{Djouadi:2007ik, Lohmann:2007ty}.

 Another mechanism for producing and detecting the invisible decay of
  a Higgs boson is via $pp\to php$.  The mass $m_h$ can be accurately
  determined by observing the forward going protons and measuring the
  missing mass, in close analogy to what is possible at a linear
  collider. The existence of the sharp peak in the missing mass
  spectrum allows one to determine $m_h$ with an error of $\sim \pm
  2$~GeV and dramatically reduces any background contributions. Second,
  observation of a Higgs boson in this way (i.e. via Pomeron-Pomeron fusion)
  implies that the produced boson must be neutral, colourless,
  flavourless and have natural parity, $P = (-1)^J$ , with $JP = 0^+$
  being by far the most likely. A preliminary study in Ref.~\cite{Belotsky:2004ex} 
  suggests there is reason to be optimistic
  that Higgs detection in this manner will be possible, but there are
  many issues related to triggering, backgrounds and pile-up that
  require a full study.  In the end, event rates are likely to be
  low, but if it is known from other channels that an invisibly
  decaying Higgs boson is present, even a few clean events would have a
  dramatic impact on verifying the nature of the Higgs boson and determining
  its mass.

\subsubsection{Strategies for the future}

The situation of ``no Higgs boson at the \lhcten'', even if the Higgs
mechanism is realized in nature, can result from two scenarios:
\begin{itemize}
\item[(1)]
Reduced couplings and branching ratios can lead to a 
``later Higgs discovery'' at the LHC.
\item[(2)]
The structure of Higgs couplings and decays can lead to the situation
where the Higgs boson permanently escapes detection at the LHC.
\end{itemize}
Due to the absence of a signal, it will not be possible to determine
at the \lhcten~ which scenario is realized, whereas the decision for a
future collider/experiment can depend critically on the answer to that
question. If several $\sim 2 \sigma$ effects compatible with a Higgs
boson were measured, this could speak in favor of scenario~(1).  The
absence of such effects, together with the absence of any hint for a
scenario replacing the Higgs mechanism could speak in favor of
scenario~(2). The detection of SUSY particles would clearly speak in
favor of the realization of either scenario~(1) or~(2). Any
``solution'' for scenario~(2) would also yield a solution for
scenario~(1). Consequently, in the case of a completely unclear
situation at the \lhcten~ it would be sufficient to consider
scenario~(2) and its implications.

In the following we will briefly investigate what is specific about
the two scenarios/models, and how each model can best be investigated
in the future:
\begin{itemize}
\item Scenario (1): Here, a collider would be needed that provides
  sufficiently clean Higgs production modes with a high enough 
  rate to compensate for
  reduced couplings. This could be the case of (i) possibly the sLHC with its
  very high luminosity if backgrounds and systematic uncertainties can
  be brought sufficiently well under control or (ii) 
  an $e^+e^-$ LC such as the ILC or
  CLIC running at high luminosity.  The latter will have the advantage
  of providing a clean environment allowing to detect small branching
  ratios and a detection will be possible despite other (invisible) decay modes. It has
  been shown that rare decay modes such as to $c\bar c$ and to
  $\mu^+\mu^-$ can be detected when running at high luminosity. A
  clean environment would also be helpful to detect experimentally 
  difficult decay patterns, such as the case of a Higgs boson decaying to other
  light bosons, where the latter would decay into rather soft jets
  (possibly $c\bar c$) or $\tau$ leptons.
\item Scenario (2): Here, a collider would be needed that provides a
  variety of Higgs production modes (for instance radiation off top-
  and/or bottom-quarks or other new particles, such as scalar tops in the
  case of SUSY) with a sufficiently high rate and
  controllable backgrounds and that is capable to detect any kind of
  unusual decay patterns.  If the Higgs mechanism is responsible for
  generating the masses of the weak gauge bosons $Z$ and $W^{\pm}$, one
  would expect that at least one Higgs boson should have a significant
  coupling to the weak bosons (unless the coupling to gauge bosons is
  shared among a large number of Higgs bosons, see e.g.\
  Ref.~\cite{Chang:2008cw}). The experimental capabilities for probing 
  production processes where the Higgs boson couples to gauge bosons are 
  therefore of particular importance. The experimental environment
  needed to probe this scenario could be provided by (i)
  an $e^+e^-$ LC. The particular power of the LC is its ability to look 
  for $e^+e^- \to ZH$ in the inclusive $e^+e^- \to ZX$ missing-mass, 
  $M_X$, distribution recoiling against the $Z$ boson. Even if the Higgs 
  boson decays completely invisibly or different Higgs signals overlap 
  in a complicated way, the recoil mass distribution  will reveal the 
  Higgs boson mass spectrum of the model. At a LC a Higgs boson could 
  furthermore be produced in weak-boson fusion, in association with 
  heavy fermions, $e^+e^- \to f\bar f H$, with $f\bar f = t \bar t,
  b\bar b$, or sfermions,
  $\tilde t \bar{\tilde t}, \tilde b \bar{\tilde b}$,
  in the case of SUSY. Having potentially the
  couplings to up- {\em and\/} down-type (s)fermions at hand strongly
  reduces the possibility of a complete suppression; (ii) a muon
  collider, where the Higgs boson could be produced by its couplings to
  muons directly in the $s$-channel. However, a reduction of the Higgs 
  couplings to all down-type fermions or to muons in particular 
  can be possible; 
  (iii) a high-energy $e^+e^-$ or $\gamma\gamma$ collider 
  with a large reach for heavy
  Higgs bosons that are outside the reach of the LHC.
\end{itemize}

\subsection{Higgsless/technicolor scenarios}

A general feature of Higgsless models\footnote{By a ``higgsless''
  model we mean here an effective model where no explicit fundamental
  Higgs boson is introduced.  A composite Higgs boson emerging from a strongly interacting sector can still be present in the low energy spectrum.}  is the appearance of new spin-1
states that replace the Higgs boson such that perturbative
unitarity of $W W \to W W$ scattering can be achieved 
up to a few TeV. These states
are usually the lightest non-standard particles.  While replacing the Higgs boson in
maintaining perturbative unitarity is a relatively easy goal 
in principle, the
construction of explicit Higgsless models yielding a satisfactory
description of EW precision tests is a much harder task to achieve.
In technicolor models, the SM Higgs sector is replaced
with new well defined strongly coupled four dimensional dynamics
within a renormalizable gauge theory~\footnote{ Many models of strong
  EW symmetry breaking predict the appearance of scalar resonances
  with masses comparable or even lighter than those of vector
  resonances. These scalar resonances could actually play an important
  role in the unitarization process.  }.

\subsubsection{Walking technicolor models}

Extended technicolor allows for fermion masses, but to avoid FCNCs, the scale of extended technicolor would be too high and these fermion masses would be too small unless technicolor models rely on walking dynamics, i.e.,  the idea that the
technicolor coupling remains large and nearly constant over a wide
range of energy scales. The use of fermions in higher dimensional
representations or the admixture of different matter representations
achieve near conformal dynamics for a small number of
flavors~\cite{Sannino:2004qp}.

Using novel methods to analyze the
non-perturbative dynamics of strongly coupled gauge theories, it has
been possible~\cite{Dietrich:2006cm} to find a large number of
underlying gauge theories which can be employed to break the electroweak
symmetry dynamically while alleviating problems associated with
potentially dangerous flavour changing neutral currents as well as the
tension with precision electroweak data.  The phenomenology of these
models is analyzed using the low energy effective theory developed
in Ref.~\cite{Belyaev:2008yj}. The basic ingredients are the presence of 
vector and axial-vector spin-one resonances, their coupling with the
SM fields and the presence of a scalar resonance/composite Higgs boson. To reduce the
parameter space it is convenient to make use of the modified
Weinberg sum rules (WSR) for walking
dynamics~\cite{Appelquist:1998xf} and impose, at the effective
Lagrangian level, known constraints. The LHC signatures are scanned in
terms of the mass of the axial resonance $m_A$, which gets linked with
the vector one via the modified WSRs, and the coupling $\tilde{g}$
which controls both the overall strength of the heavy spin-one and
spin-zero interactions as well as the mixing with the SM gauge
fields. The larger $\tilde{g}$ is, the less the heavy spin-one states
mix with the SM gauge bosons. Walking theories with fermions in higher 
dimensional representations have also been 
explored in Ref.~\cite{Christensen:2005cb}.

In another class of walking technicolor models (see, e.g.,
Ref.~\cite{Brooijmans:2008se}), a near conformal, $\beta(\alpha_T) \sim
0$, coupling is achieved by having a lot of technimatter. 
The Technicolor Strawman Model~\cite{Eichten:1997yq, Lane:2009ct} has often been used as a reference for Tevatron analyses and for phenomenological studies at the LHC.
As all
(electroweak charged) technimatter contributes to the weak scale, the
more matter, the lower the fundamental strong interaction scale
becomes. Consequently, these models predict light spin-one resonances,
around 700~GeV or even less. The best discovery channels for these
resonances are via their decays into pairs of SM gauge bosons. Due to
the large chiral symmetry group present in these models, there are typically
several new pseudoscalar states (technipions) as well.  

A general
classification of possible four dimensional gauge theories which can
have a near conformal (walking) behavior appeared
in Ref.~\cite{Dietrich:2006cm}. Several new models of walking technicolor
dynamics were introduced there as well. In these models there is no systematic calculation of the precision EW parameters. However, the matter content and beta function are very different from QCD, so estimates of these precision EW parameters based on rescaling the QCD values should not be trusted.

\subsubsubsection{Drell--Yan production of heavy vectors}\label{DYsection}

Heavy spin-one resonances can be produced at the LHC through the DY
processes $pp \to R_{1,2}$ where $R_{1,2}$ are the physical
eigenstates which take into account the mixing with the SM gauge
bosons. To estimate the LHC reach for DY production of the $R^0_{1,2}$
and $R^\pm_{1,2}$ resonances, the following final state lepton
processes were considered (see Ref.~\cite{Belyaev:2008yj} for
details):
 \begin{itemize}
 \item[(a)] $l^+l^-$ signature from the process $pp \to R^0_{1,2} \to
   l^+l^-$,
 \item[(b)] $3l+{\not \hskip -3.5pt E_T}$ signature from the process
   $pp \to R^\pm_{1,2} \to Z W^\pm \to 3l\nu$.
 \end{itemize}
Acceptance cuts ($|\eta^l|<2.5$ and $p_T^l>15$~GeV) on the leptons are
applied and an additional cut on the missing  transverse energy is
imposed (${\not \hskip -3.5pt E_T}>15$~GeV). For process (a), the
dilepton invariant mass distribution $M_{ll}$ is used to separate the
signal from the background, while for process (b), the analysis relies
on the transverse mass variable $M_{3l}^T$.  For $\tilde{g}=2$, clear
signals from the leptonic decays of the resonances are seen up to
masses of around 2 TeV with only 10 fb$^{-1}$, $\sqrt{s} =
10$~TeV. Both peaks from $R^0_{1,2}$ may be resolved. For larger
values of $\tilde{g}$, this signal deteriorates and will only be
observable for small vector masses. Fortunately, for large
$\tilde{g}$, the triple-vector coupling is enhanced and a signal in
the $M^T_{3l}$ distribution may be observed at large masses as
presented in Fig.~\ref{Fig:DY}. A few events could potentially be
observed with only 10~fb$^{-1}$ at $\sqrt{s} = 14$~TeV in this
channel.

\begin{figure}[htbp]
\centering
 \includegraphics[width=0.35\textwidth]{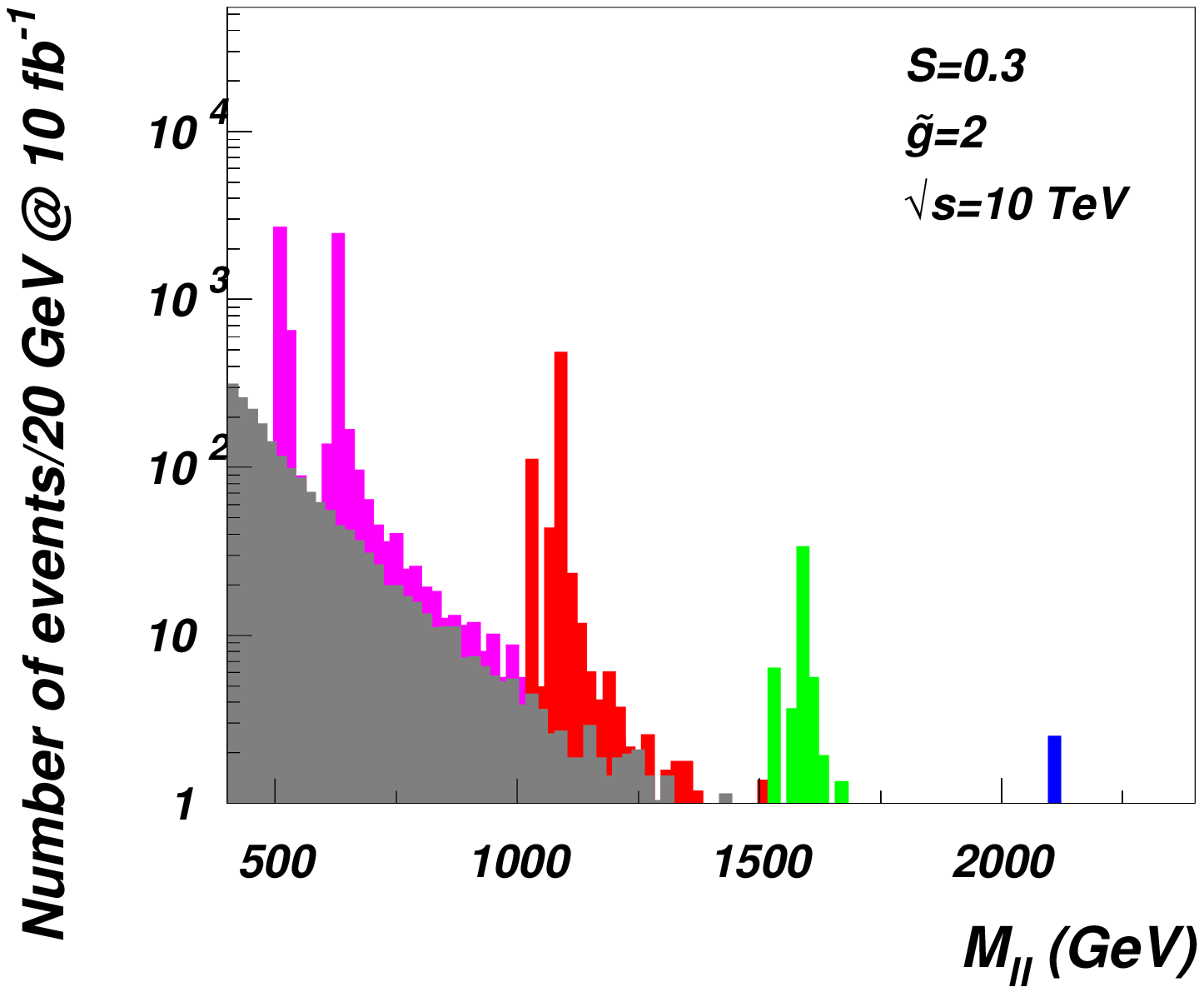}
 \hspace{.5cm}
 \includegraphics[width=0.35\textwidth]{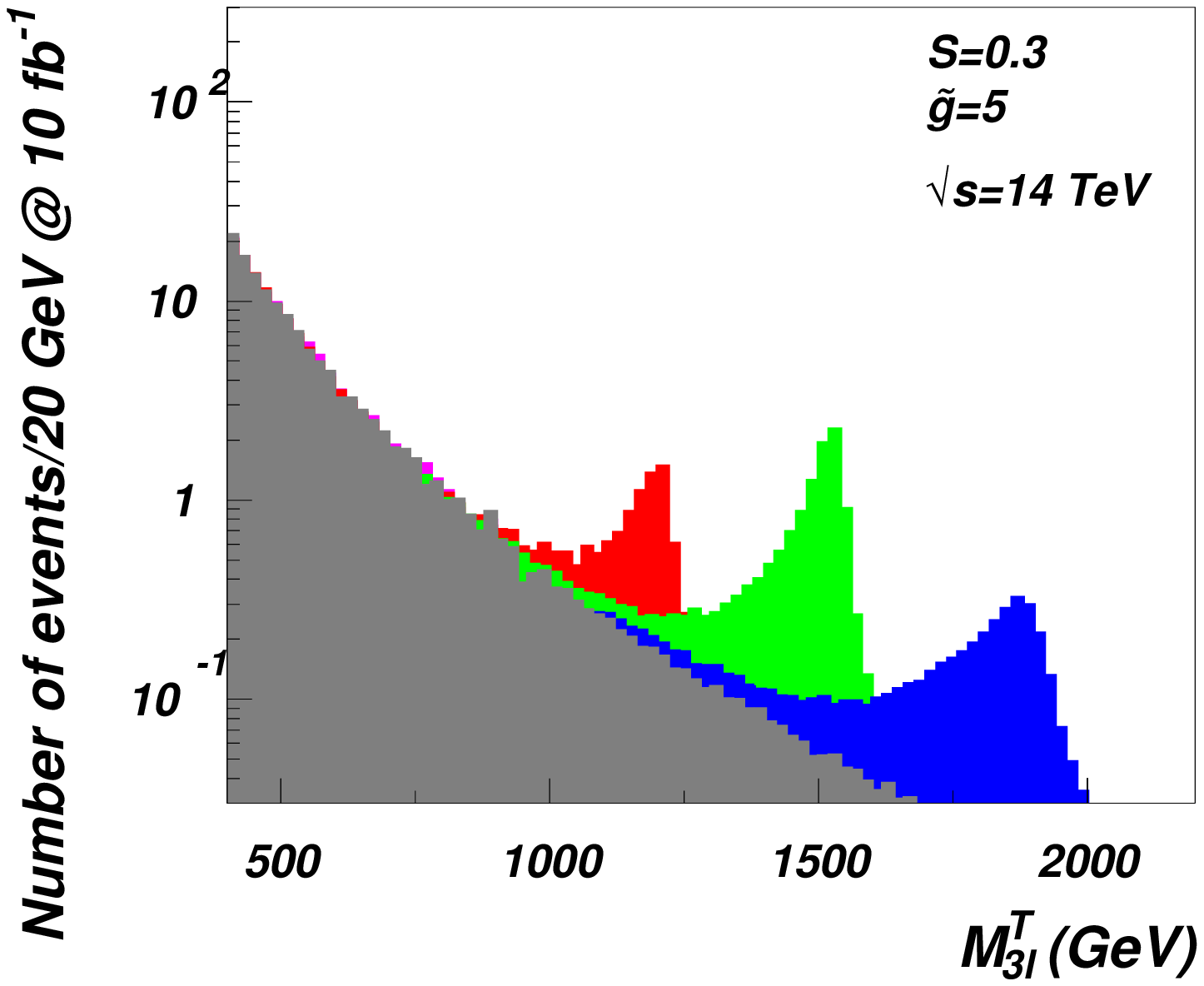}
 \caption
{Invariant and transverse mass distributions for signal and background
  processes (a) and (b) of Sect.~\ref{DYsection}. The model parameters
  considered are $\tilde{g}=2$ and $\tilde{g}=5$ in the left and
  right figures respectively (note also the different energy for the two figures) and $m_A=0.5$~TeV (purple), 1~TeV (red),  1.5~TeV (green) and 2~TeV (blue). The background appears in grey. In both cases, the $S$ parameter~\cite{Peskin:1990zt} is equal to 0.3.}
\label{Fig:DY}
\end{figure}

\subsubsubsection{Associate production of a composite Higgs boson and a SM vector boson}

When the strong dynamics produces a light scalar resonance with the
quantum numbers of the Higgs boson (composite Higgs boson), the resonant production of heavy
vectors can enhance significantly the WH and ZH production compared to
the SM. Figure~\ref{Fig:HW} presents event-rate plots for $pp \to W^\pm
H \to W^\pm ZZ \to 4l+2j$ for $\sqrt{s}=10$~TeV (resp. 14~TeV) and
10~fb$^{-1}$ (resp. 100~fb$^{-1}$) of integrated luminosity. The peaks
correspond to the contribution from a heavy axial spin-one state with
mass $m_A=0.5$ and 1~TeV. A similar analysis can be performed for the
$pp \to ZH$ channel. It is possible to discover vectors (the mostly
axial technicolor spin-one state) with masses up to 1~TeV while the
signal worsens when increasing the mass of the composite Higgs boson. With
10~fb$^{-1}$ only, the LHC could be able to observe the interplay of a
composite Higgs boson and the mostly axial vector boson only if both are
very light. Thus, the scenario of a composite Higgs boson could lead to
the observation of a Higgs-like signal at the LHC but could also give
rise to a situation where no clear signal can be established with the
first 10~fb$^{-1}$ at the LHC.

\begin{figure}[htbp]
\centering 
\includegraphics[width=0.35\textwidth]{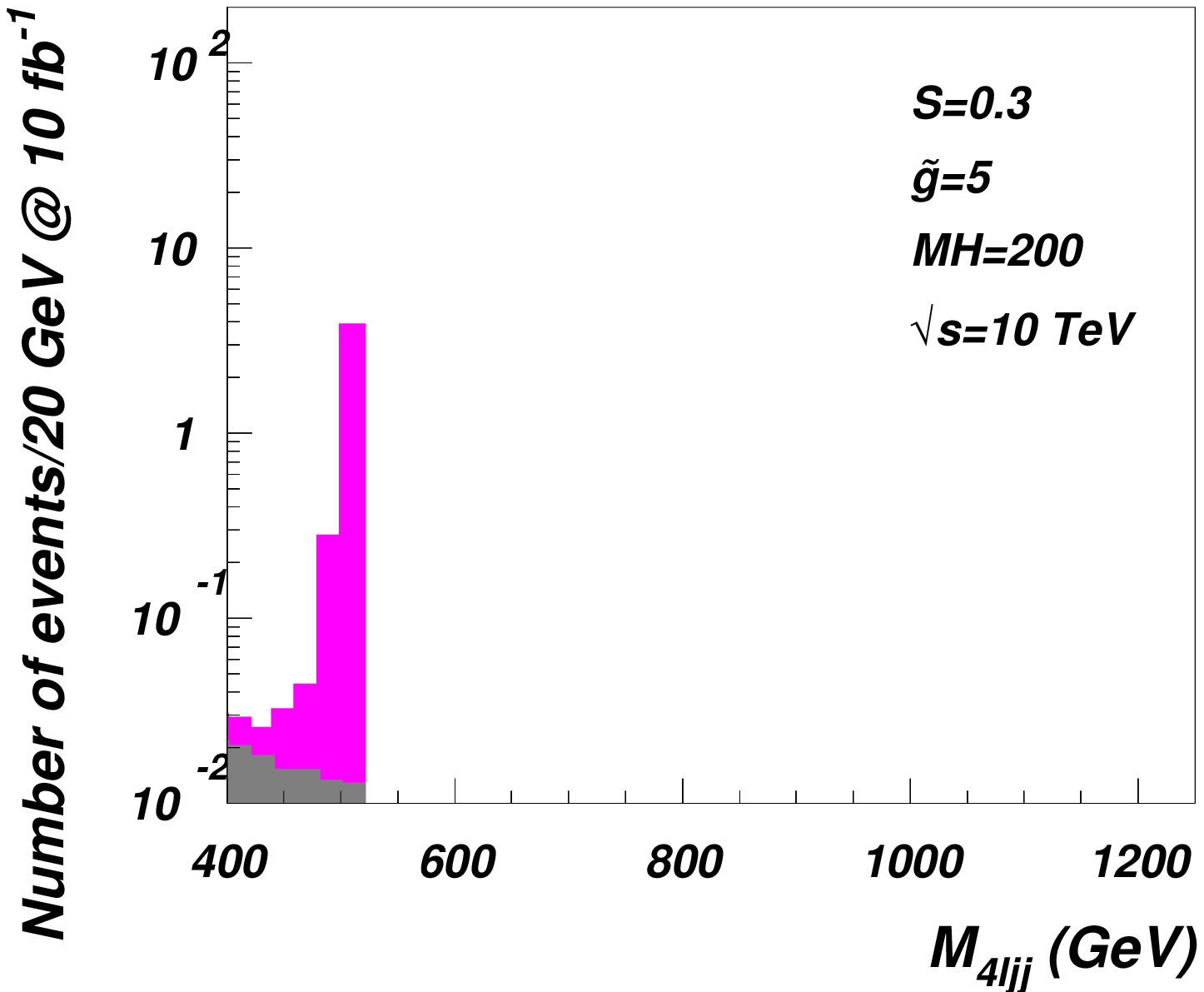}
\hspace{.5cm}
\includegraphics[width=0.35\textwidth]{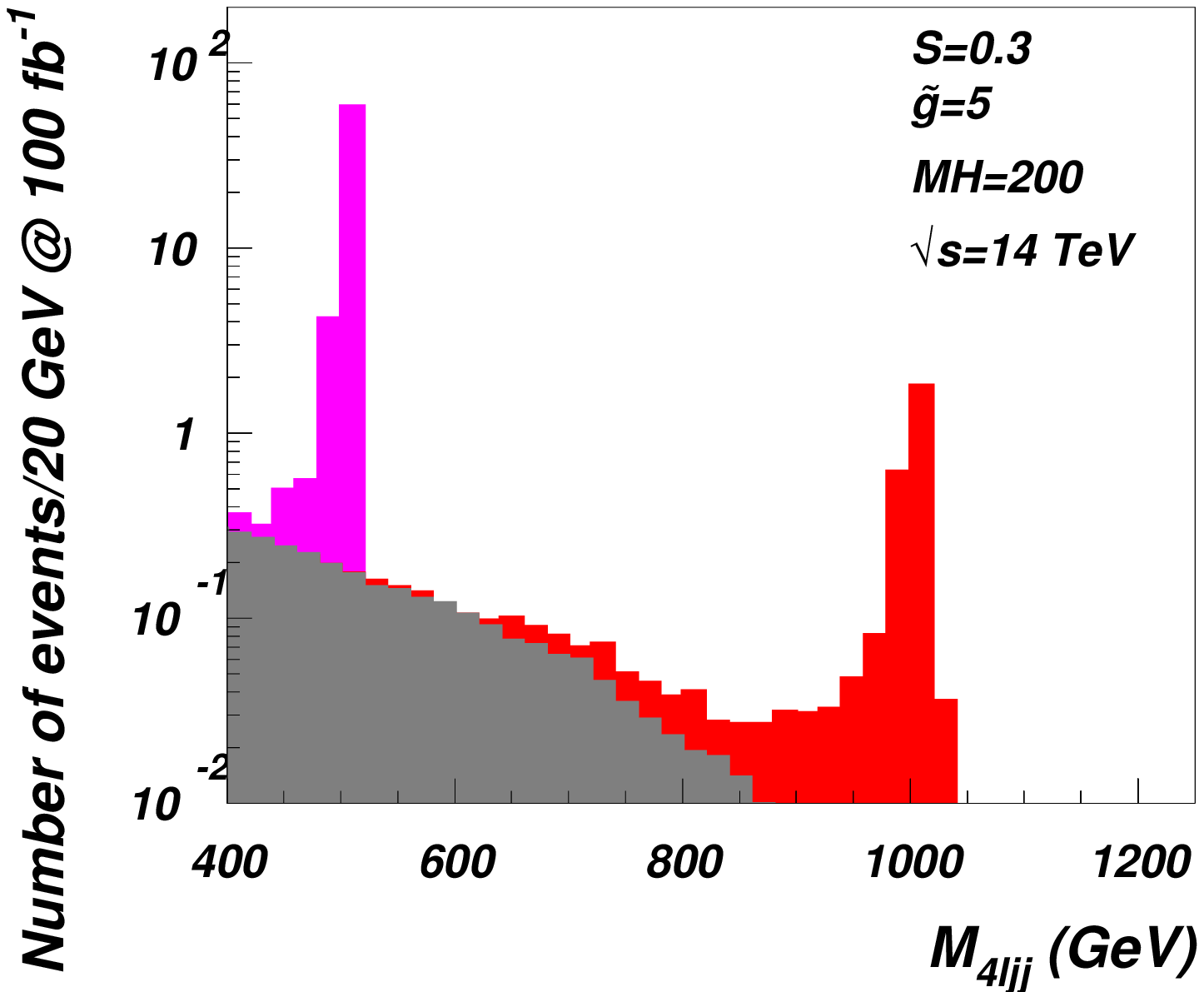}
 \caption
{$pp \to W^\pm H \to W^\pm ZZ \to 4l+2j$ with $\sqrt{s}=10$~TeV,
  10~fb$^{-1}$ and 14~TeV, 100~fb$^{-1}$ on the left and right
  respectively with $m_A=0.5$~TeV (purple), 1~TeV (red). The parameters of the model correspond to an intrinsic  $S$ parameter~\cite{Peskin:1990zt} equal to 0.3 and a heavy vector dimensionless coupling
  $\tilde{g}=5$.}
\label{Fig:HW}
\end{figure}

\subsubsection{Effective lagrangian approach to resonances}

The phenomenology of heavy vectors at high-energy colliders, as well
as their role in EW precision tests (EWPT), has been widely discussed
in the literature recently. However, most of the existing analyses are
based on specific dynamical assumptions, such as considering these
vector states as the massive gauge bosons of a hidden local
symmetry. As recently discussed in Ref.~\cite{Barbieri:2008cc}, these
assumptions may be too restrictive for generic models with strong
dynamics at the TeV scale, and going beyond these assumptions, the sole
exchange of heavy vectors can satisfy EWPT. More generally, the
construction of an appropriate effective theory including only SM
fields and these new light states is a very efficient tool to discuss
theoretical and phenomenological constraints on such states.

The effective theory proposed in Ref.~\cite{Barbieri:2008cc} is based on
the following rather general assumptions:
\begin{itemize}
\item The new strong dynamics is invariant under a global chiral
  symmetry $G=SU(2)_L \times SU(2)_R$, broken spontaneously into
  $H=SU(2)_{L+R}$ (the {\em custodial symmetry} of the SM Higgs
  potential), and under a discrete parity symmetry ($P:\ SU(2)_L
  \leftrightarrow SU(2)_R$).
\item A pair of vector ($V$) and axial-vector ($A$) states, belonging
  to the adjoint representation of $H$, are the only new {\em light}
  dynamical degrees of freedom below a cut-off scale
$\Lambda \sim (2\div3)$~TeV.
\end{itemize}
Under these assumptions the dynamics of the new spin-1 states is
controlled by three effective couplings: $G_V$, $F_V$ and $F_A$,
expected to be of $\mathcal{O}(v\approx 250~{\rm GeV})$ by na\"ive
dimensional analysis, and the masses of the two new states ($M_V$ and
$M_A$).  The coupling $G_V$ controls the effective coupling of the new
vector states to the Goldstone bosons of the theory (or the
longitudinal components of SM gauge bosons), while $F_{V}$ and $F_{A}$
control the (gauge-invariant) mixing of the new states with the
transverse components of the SM gauge bosons.

By construction, $G_V$ is rather constrained after the
unitarity condition is imposed. However, as shown in Fig.~\ref{Fig:GvMv}, the
unitarity constraint alone does not pose a significant constraint on
$M_V$. A more constrained picture is obtained if it is requires that the
sole exchange of the two spin-1 states leads to a satisfactory
description of EWPT. Under this stronger assumption it turns out that
at least the vector state must be relatively
light~\cite{Barbieri:2008cc}.

\begin{figure}[htbp]
\centering
\includegraphics[width=0.4\textwidth]{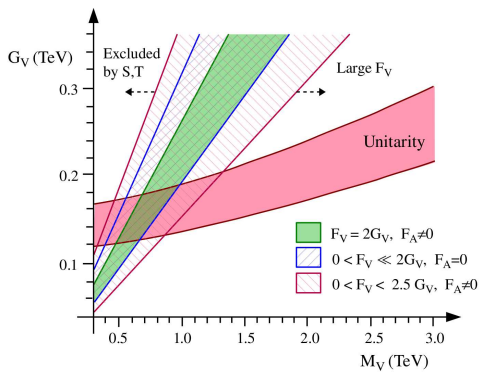}
 \caption
{Summary of unitarity and EWPT constraints (at 95\% C.L.) in the
$(M_{V},G_{V})$ plane. From Ref.~\cite{Barbieri:2008cc}.}
\label{Fig:GvMv} 
\end{figure}

The free parameters of this effective theory are crucial ingredients
for determining the possible signatures at future colliders. If only 
the unitarity constraint is used (assuming that other states play a
significant role in EWPT), the lack of information on the spectrum
does not allow one to draw firm conclusions about the detection of any of
the new states at the LHC (see also
Refs.~\cite{Foadi:2008ci,Foadi:2008xj}). In particular, the detection
of a resonance of mass above 1 TeV in $WW \to WW$ scattering will be
extremely hard, even with high luminosity.  An illustration of the
typical signal expected in this process is shown in
Fig.~\ref{Fig:WZ}. Since the appearance of these resonances in $WW \to
WW$ scattering is the most general property of Higgsless models, if at
the LHC no new light state (including the Higgs boson) is found, the natural
option to consider is a collider where $WW \to WW$ scattering can be
probed to higher energies, unless there are hints that a fundamental
Higgs boson could have been missed in the LHC searches.

If the mass of at least one of such states is relatively 
low, as suggested by the hypothesis that the sole 
exchange of such fields is relevant in EWPT, then it would 
be interesting to try to determine all the parameters
of the effective theory. In such a case, Drell-Yan 
production of the new states, and the subsequent decay 
into $\ell^+\ell^-$, $WZ$, $WW$ or states with three SM 
gauge fields ($WWW$, $WWZ$, $WZZ$)
are the most interesting final states to be studied.
For sufficiently light masses (and relatively large
couplings), some of these signals could be within 
the LHC reach (see, e.g.,~Fig.~\ref{Fig:WZ}).

\begin{figure}[htbp]
\begin{center}
\includegraphics[width=0.37\textwidth]{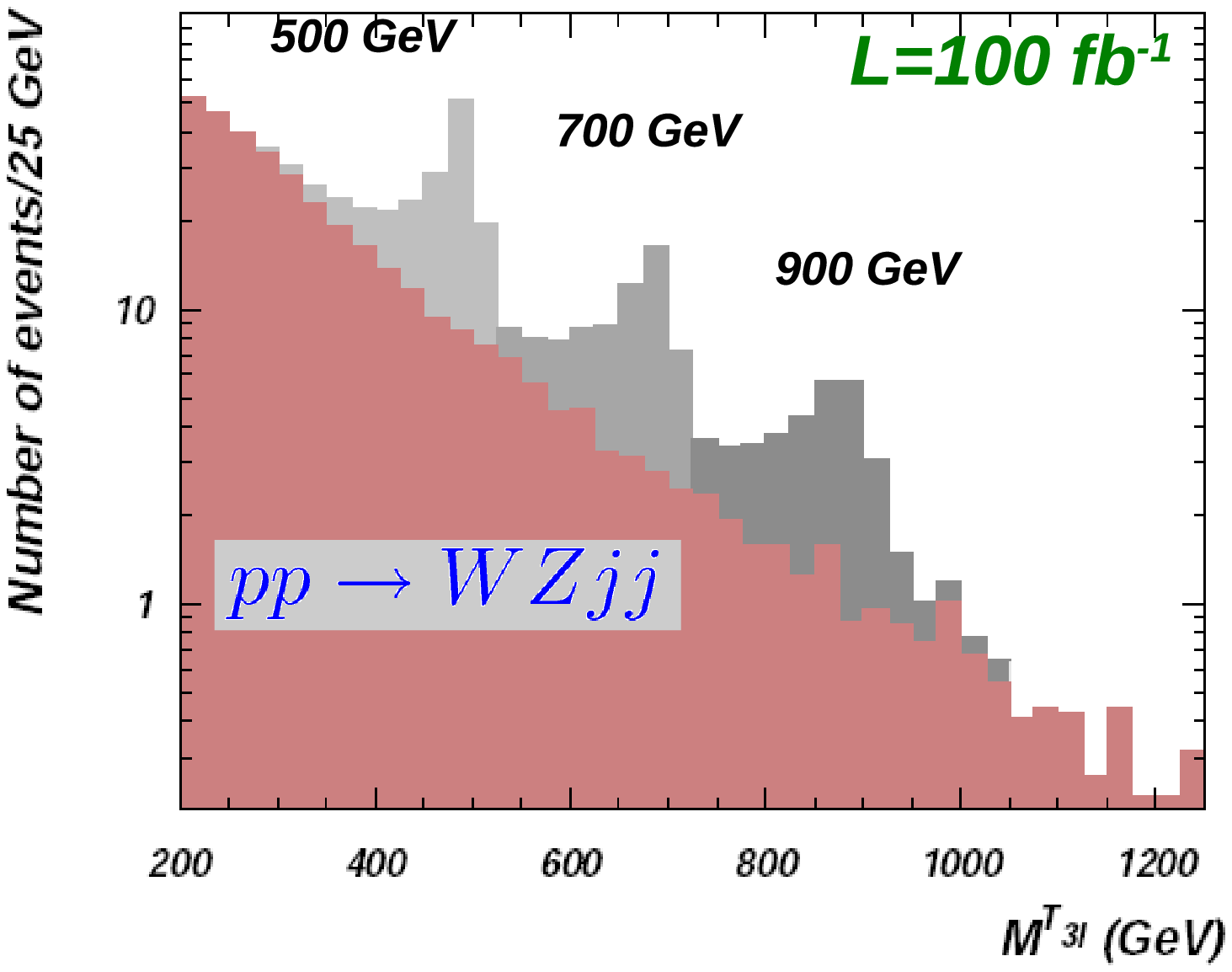}
 \hspace{1cm}
\includegraphics[width=0.45\textwidth]{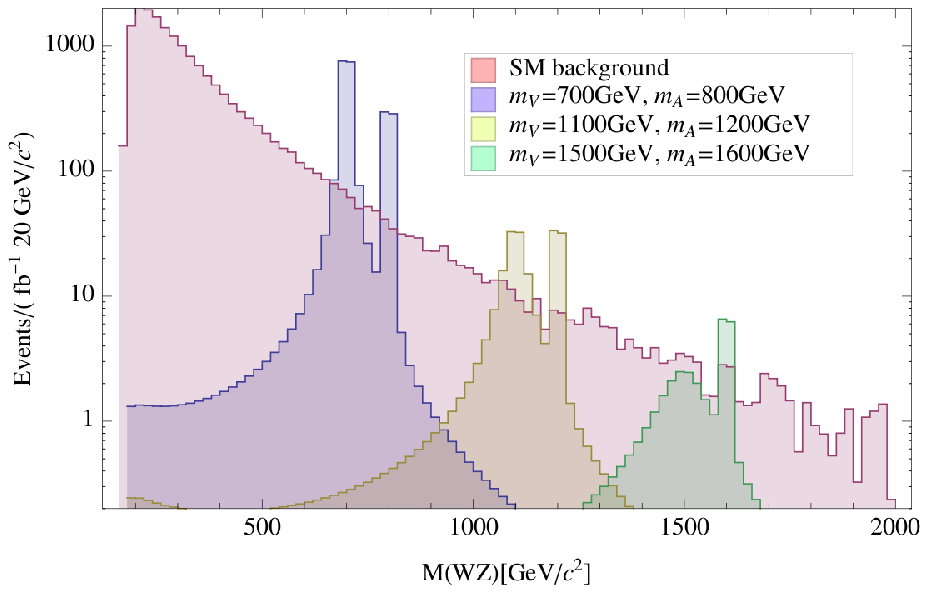} 
 \end{center}
 \caption{Left: Signal events in $pp\to WZ$+2 jets (from $WW$ fusion) 
with leptonic decays of both $W$ and $Z$ bosons and 
standard $WW$-fusion cuts on the two jets. From Ref.~\cite{Belyaev:2007ss}. 
Right: Invariant mass spectrum of $WZ$ pairs produced 
in $pp \to WZ$ at $\sqrt{s}=14$~TeV,
with contributions from  $V$ and $A$ states.
All resonance signals have been obtained 
assuming $F_A=F_V=2G_V$ (condition that maximises the 
signal). The SM background corresponds 
only to the irreducible electroweak production of $WZ$ pairs.
The plot does not include the decay branching ratios 
of $W$ and $Z$ bosons, as well as any experimental cut. 
From Ref.~\cite{Cata:2009iy}.}
\label{Fig:WZ} 
\end{figure}

\subsubsection{Extra dimensional models}

To make predictions of the LHC signatures of viable technicolor (TC)
models, we can: (i) set up 4D models with field content which we
anticipate gives walking behavior and extract whatever information we can, or
(ii) consider the subset of near-conformal TC models with a weakly
coupled (and therefore calculable) 5D description. With simplest
Anti-de Sitter setup, the agreement with EW precision constraints is
still problematic. In the absence of any symmetry to enforce a small
$S$ oblique parameter, the only option is a tuned
cancellation. Currently there are two techniques for achieving such a
cancellation: Cured Higgsless, and Holographic Technicolor. In each
case, the cancellations required for $S$ lead to distinct signatures
at the LHC.

\subsubsubsection{Cured Higgsless (CHL)}
The matching of 5D to 4D, which sets parameters involved in the EW
precision observables, is dependent on the fermion wavefunction
localization along the extra dimension. By tuning the fermions to
have very specific profiles, the EW observables can be brought back
to an acceptable range. This cancellation happens when the fermion profiles are
nearly flat. Though the couplings between the SM fermions and the KK
excitation of the $W$ and the $Z$ are small ($\sim 0.1\times\text{SM}
$), they are still important at the LHC and Drell-Yan production is
usually a dominant channel with the following
properties~\cite{Cata:2009iy,Ohl:2008ri}:
\begin{itemize}
\item Charged $W_{KK}$ appear in $WZ (3\ell + \nu)$ final states.
\item $Z_{KK}$ and its higher tiers can most easily be seen in
  dilepton channels.
\item Unlike low scale technicolor, there is no $W\gamma$ mode.
\item There are actually two $Z_{KK}$ modes -- the KK partners of the
  $Z$ and $\gamma$. These states are nearly degenerate, $\Delta M <
  \Gamma_{Z_{KK}}$, which may lead to interesting interference
  effects~\cite{Cacciapaglia:2009ic}.
\item Minimal bulk $SU(2)_L \otimes SU(2)_R$ symmetry implies there
  are no technipions in the spectrum.
\end{itemize}
Since tuning the fermion wavefunction is required to satisfy EWPT, the
observation of KK fermions and measurements of their couplings are
important. These KK fermions should have similar mass to the KK gauge
bosons, and can be either singly or doubly produced at colliders. One
interesting possibility is the production of $ q_{KK} q_{KK}$ (rather
than $q_{KK}\overline{q_{KK}}$) through $t$-channel KK boson exchange which, following the decay $q_{KK}
\rightarrow q' W^{\pm}$, can lead to final states with two same sign
leptons.

\subsubsubsection{Holographic Technicolor (HTC)}
A second option for cancelling contributions to the $S$ parameter is
to drop the assumption that all fields feel the same
metric. Specifically we can allow the vector and axial combinations of
the $SU(2)$ to feel different warp factors with the following expected
consequences\cite{Hirn:2007we,Hirn:2008tc}:
\begin{itemize}
\item by divorcing vector from axial, we can dial the individual warp
  factors such that we have nearly degenerate ($m_V \sim m_A$) or even
  inverted ($m_A < m_V$) spectra, reducing $S$ without having to
  change the fermions.
\item different warp factors can be recast as local operators and to
  achieve $S \sim 0$ the coefficients of these local operators must be
  much larger than na\"ive dimensional analysis (NDA) estimates. Thus
  the HTC scenario is also tuned~\cite{Agashe:2007mc}.
\item the signature of HTC is the presence of nearly degenerate KK
  gauge bosons, both charged and neutral. Therefore we should observe:
  (i)~{\em two} charged resonances in $WZ(3\ell + \nu)$, (ii)~two
  neutral resonances in dileptons and (iii)~a possible nonzero
  coupling between a photon, a SM $W$ and a charged resonance.
\end{itemize}
%


\subsubsection{Deconstructed/BESS models}

The discretization of the compact fifth dimension of the previous
models to a lattice generates the so-called deconstructed theories
which are chiral Lagrangians with a number of replicas of the gauge
group equal to the number of lattice sites. The delocalization of
fermions along the fifth dimension is equivalent, in the deconstructed
picture, to direct couplings between new vector bosons and SM
fermions~\cite{Casalbuoni:2005rs}. In the simplest version of this
latter class of models, corresponding to just three lattice sites and
gauge symmetry $SU(2)_L\times SU(2)\times U(1)_Y$ (the so-called BESS
model~\cite{Casalbuoni:1985kq}), the requirement of a small $S$
oblique parameter implies that the new triplet of vector bosons is
almost fermiophobic. Then the only production channels for their
search are those driven by boson-boson couplings.  However, the
minimal three-site model can be extended by inserting an additional
lattice site.  This four-site Higgsless model, based on the
$SU(2)_L\times SU(2)_1\times SU(2)_2\times U(1)_Y$ gauge symmetry,
predicts two neutral and four charged extra gauge bosons, $Z_{1,2}$
and $W^\pm_{1,2}$, and satisfies the EWPT constraints without
necessarily having fermiophobic
resonances~\cite{Accomando:2008jh}. Within this framework, the more
promising Drell-Yan processes become particularly relevant for the
extra gauge boson search at the LHC.  Clearly a future Linear Collider
operating in the TeV range has indirect sensitivity to the four-site
model and can profile the low mass $Z_1$ and $Z_2$.

\subsubsubsection{Drell--Yan production of heavy vectors at the LHC}

The four-site Higgsless model predicts six new gauge bosons $Z_{1,2}$
and $W^\pm_{1,2}$ which can be produced at the LHC through
Drell-Yan channels. There are two interesting classes of processes:
\begin{itemize}
\item[(a)] $pp\to Z_{1,2} \to l^+l^-$ characterized by two isolated charged
  leptons in the final state,
\item[(b)] $pp\to W_{1,2}^\pm \to l\nu_l$ giving rise to one isolated charged lepton
  plus missing energy.
\end{itemize}
A study for the LHC of these channels is described in Ref.~\cite{Accomando:2008jh}.
In Fig.~\ref{Fig:DY4sites}, the distributions both in the charged
and neutral Drell--Yan channels are shown. Here standard acceptance cuts ($p_T^l>20$~GeV, $|\eta_l|<2.5$) are applied.
Two particular sets of free parameters describing the
four-site Higgsless model are used with two resonances at 1~TeV and
1.25~TeV. The full Drell--Yan process, considering signal and
SM-background, is computed at the EW and QCD leading order.

 For two different sets of fermionic couplings
($b_1, b_2$) shown, for example,
the signal strength can vary significantly. With the first set, the
new gauge bosons could be discovered already at the LHC start-up, with
a minimum integrated luminosity of 1~fb$^{-1}$ while with the second set
high luminosity will be required.

\begin{figure}[htbp]
\centering 
\includegraphics[width=0.40\textwidth]{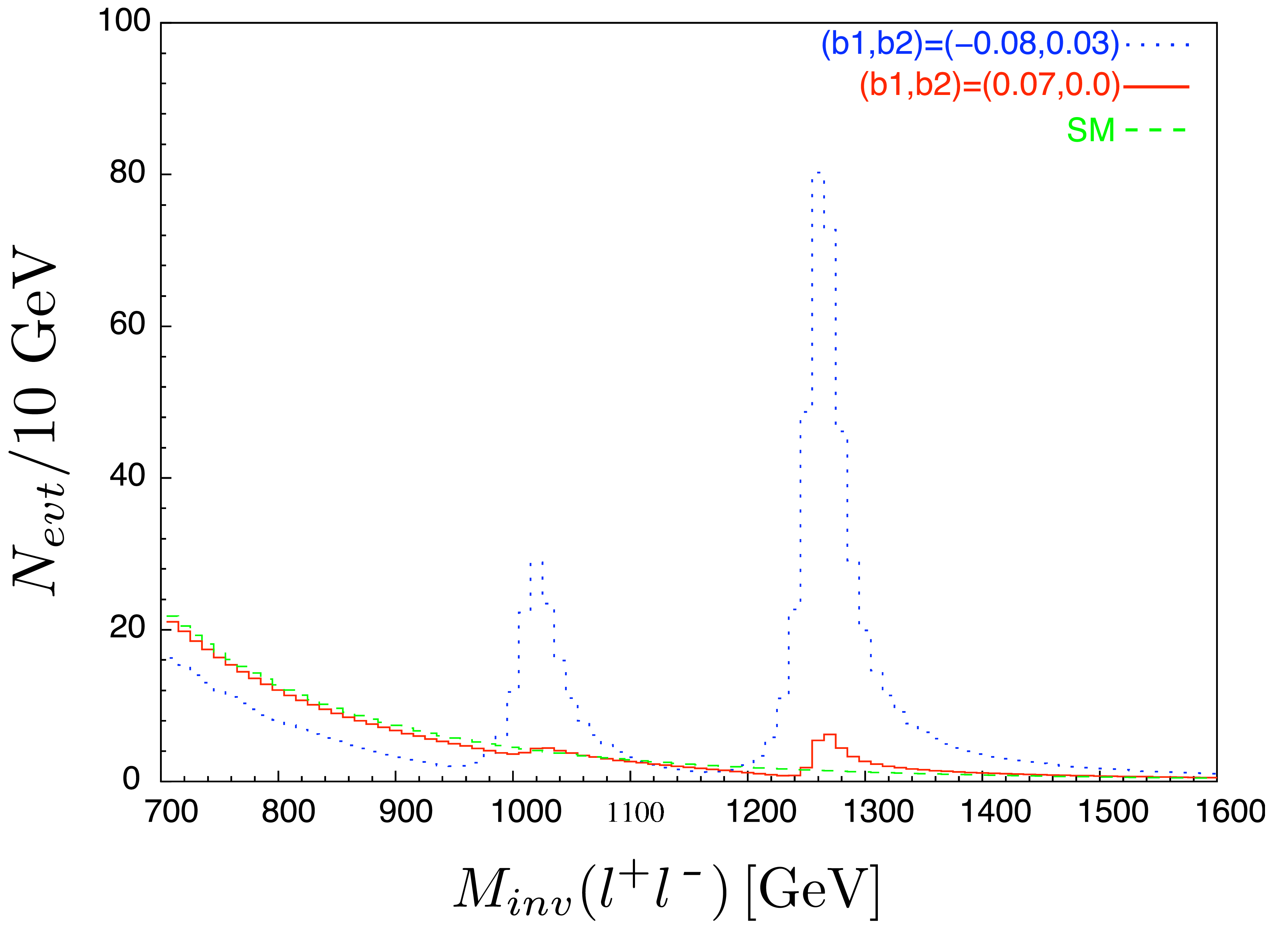}
\hspace{.5cm}
\includegraphics[width=0.40\textwidth]{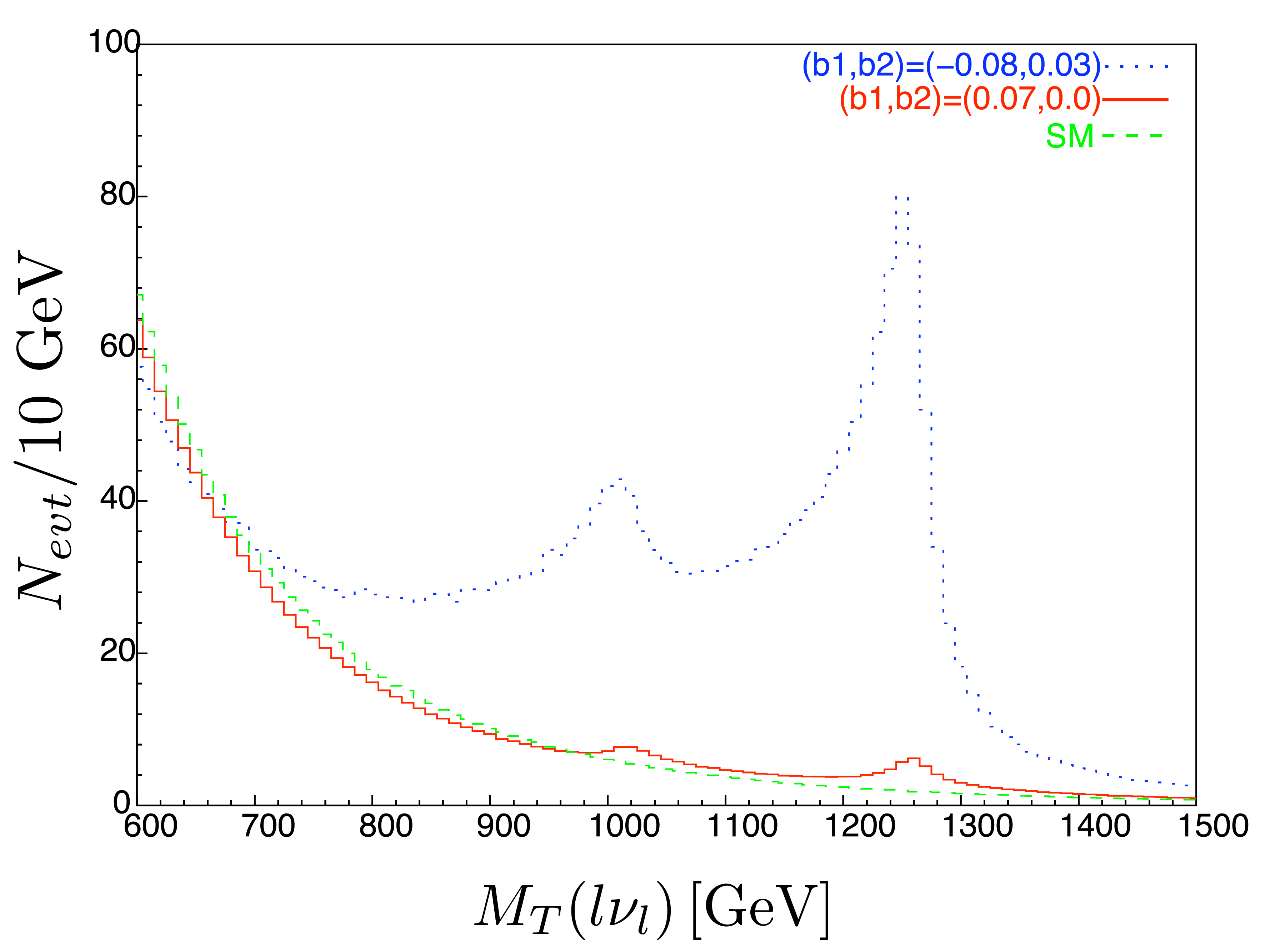}
 \caption
{Total number of events per 10 GeV versus the dilepton invariant
  mass $M_{l^+l^-}$ for the process $pp\to l^+l^-$ (left) and versus
  the lepton transverse mass, $M_T(l\nu_l)$ (right), for the process
  $p p\to l\nu_l$ with an integrated luminosity of 10~fb$^{-1}$ at the LHC with $\sqrt{s} = 14$~TeV for
  $M_{1,2}=(1000,1250)$~GeV and the two sets of parameters $(b_1,b_2)$ corresponding to different couplings to fermions. The $e,\mu$ and charge
  conjugate channels have been summed. (From Ref.~\cite{Accomando:2008jh}).}
\label{Fig:DY4sites}
\end{figure}

\subsubsubsection{Four-site model at a future linear collider}

One of the most striking manifestations of new physics at a TeV-class
$e^+e^-$ LC will come from the sudden increase of the
$e^+e^-\to f\bar f$ cross section indicating the $s$-channel
production of one or more new particles. The existence of the two
neutral gauge bosons, $Z_{1,2}$, of the four-site model and their
properties can be precisely studied, if their mass is lower than the
LC centre-of-mass energy. 
Figure~\ref{Fig:4sitesAtLC} shows two examples
of $Z_{1,2}$ scanning, namely $M_{1,2}=(680,850)$~GeV at a 1~TeV-LC and
$M_{1,2}=(1600,2000)$~GeV at a 3~TeV-LC with a luminosity of
100~fb$^{-1}$. For each scenario, two sets of fermionic couplings
inside the region allowed by the EWPT were chosen and significant signals
are observed.
In this preliminary analysis the initial state radiation is taken into
account but no beamstrahlung is included.

\begin{figure}[htbp]
\centering
\includegraphics[width=0.40\textwidth]{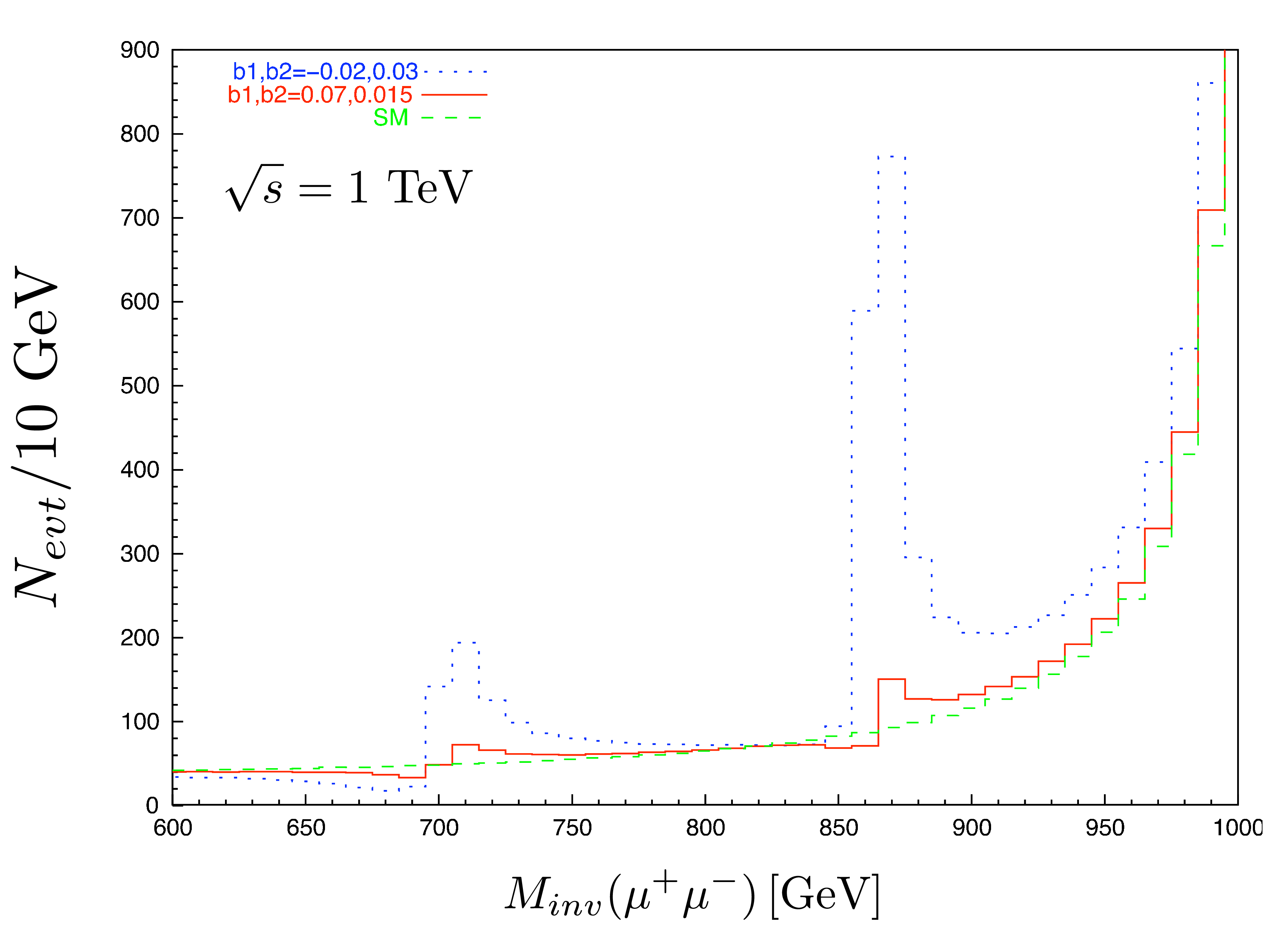}
\hspace{.5cm}
\includegraphics[width=0.40\textwidth]{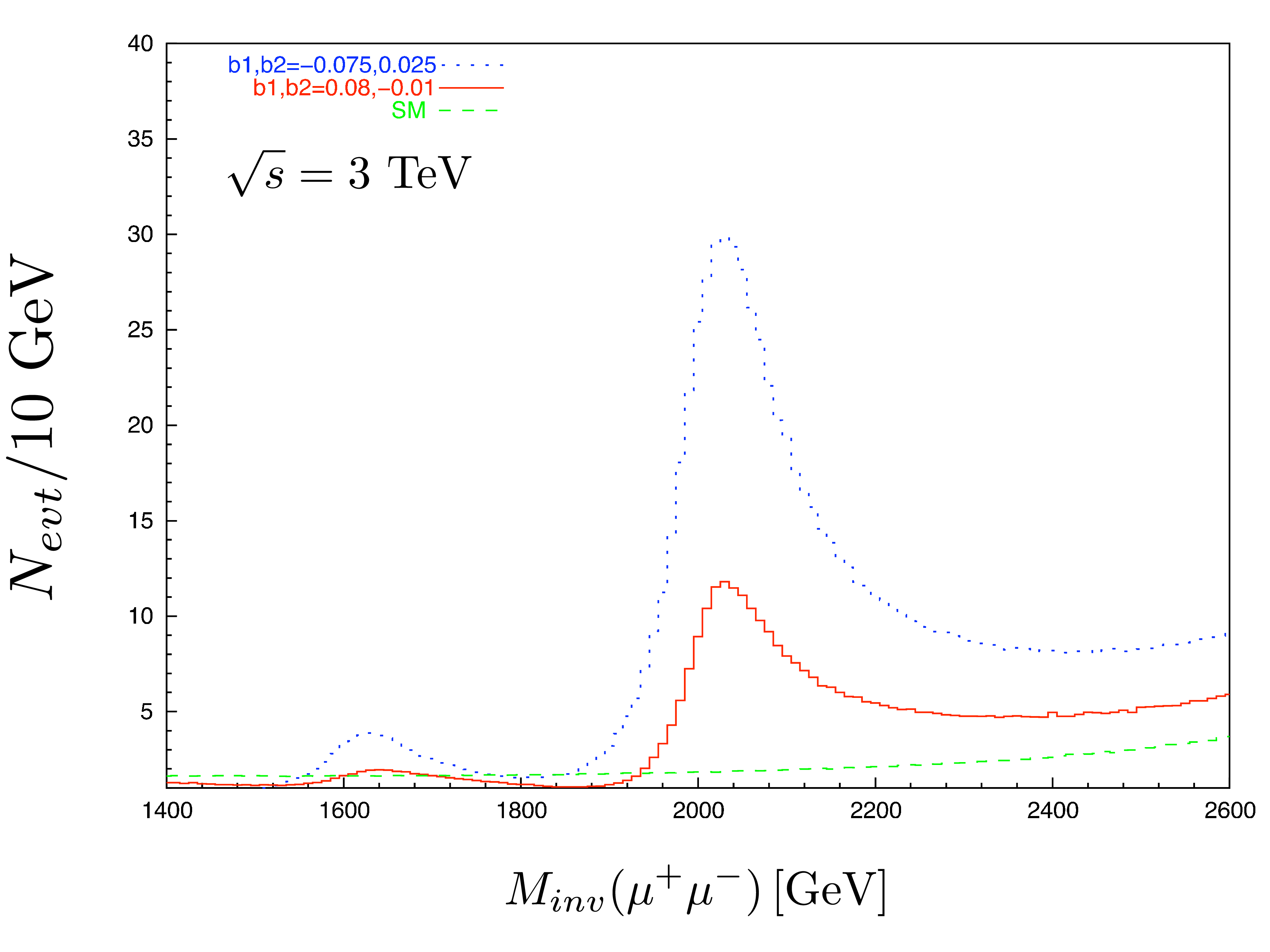}
 \caption
{Total number of events per 10 GeV versus the
dimuon invariant mass, $M_{\mu^+\mu^-}$ for the process
$e^+e^- \to \mu^+\mu^-$ at a 1~TeV (left), 3~TeV (right)
$e^+e^-$ Linear Collider with a 100~fb$^{-1}$ integrated luminosity, for
$M_{1,2}=(680,850)$~GeV (left) and  $M_{1,2}=(1600,2000)$~GeV (right)
and for two sets of fermionic couplings as quoted in the figures. Initial state raditation is included,
beamstrahlung is not taken into account. (From Ref.~\cite{Battaglia:20??}).
}
\label{Fig:4sitesAtLC}
\end{figure}

Even beyond the kinematical reach for $s$-channel production, a
TeV-LC could prove the existence of new vector resonances up to
scales of several TeV by studying the electroweak observables, e.g., 
$\sigma_{\mu\mu}$, $\sigma_{b \bar b}$,
$A_{FB}^{\mu}$, $A_{FB}^{b}$, and indirect bounds from a
1~TeV-LC can exclude a portion of the parameter space left open by
the LHC, as shown in Fig.~\ref{Fig:Visibility4sites} where the region excluded by a 1~TeV-LC is obtained by comparing the deviations of the four-site model predictions from the SM ones, with the uncertainties, assuming that they are statistically dominated. The relative statistical accuracies are rescaled from those obtained for a total luminosity of 1~ab$^{-1}$ accumulated at CLIC running at $\sqrt{s}=3$~TeV, including the effect of $\gamma \gamma \to$~hadrons background~\cite{Battaglia:2002sr}. Better sensitivity could be obtained by considering
polarized beams.

\begin{figure}[htbp]
\centering
\includegraphics[width=0.35\textwidth]{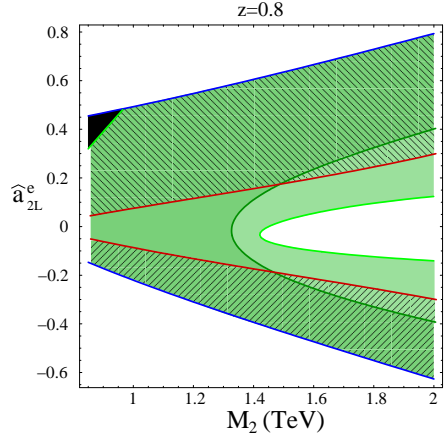}
 \caption
{3$\sigma$-exclusion plots in the plane $(\hat{a}_{2L}^e,M_2)$ for $M_1/M_2=0.8$ ($\hat{a}_{2L}^e$ is the left-handed coupling between the $Z_2$-boson and the SM electron in units of the electric charge, $M_2$ is the mass of the $Z_2$-boson). The upper and lower parts are
excluded by EWPT, the black triangle is the region excluded by the
direct search at the Tevatron for a luminosity of 4~fb$^{-1}$. The
dashed region is excluded by the Drell-Yan processes at the LHC at
L=100 fb$^{-1}$. The green (light-green) region is excluded by a
1TeV-LC by combining the measurements of $\sigma_{\mu\mu}$,
$\sigma_{b \bar b}$, $A_{FB}^{\mu}$, $A_{FB}^{b}$ for a luminosity of
10~fb$^{-1}$ (1~ab$^{-1}$).  (From Ref.~\cite{Battaglia:20??}).
}
\label{Fig:Visibility4sites}
\end{figure}

\subsubsection{WW Scattering at ATLAS/CMS}

Most of the processes described in the earlier sections rely on the
leptonic final states, which provide relatively clean samples, with
signal resonances expected to peak over backgrounds that can be estimated
from sidebands with relative ease. 

On the other hand, the LHC experiments will focus not only on the
leptonic final states, but also on hadronic decays of the vector bosons
(VB). 
For example, particularly for integrated luminosities of up to 100$\,$fb$^{-1}$, both ATLAS
and CMS analyses for VB scattering in semi-leptonic final
states, in which one of the VBs decays hadronically, appear to be
as or more promising than fully leptonic final states for observing
signals from possible new resonances (Higgs boson or other) at the 0.5--1$\,$TeV
range.

However, there are two major challenges to the reconstruction of semi-leptonic
final states. The first one is the significantly higher background,
from sources like $t\bar{t}$+jets and VB+jets. 
In the particular case of VB scattering, 
additional requirements can be put on the event topology:
the presence of two high-energy forward jets (tag jets), 
resulting from the quarks that radiated the scattering VBs, 
is one of the well-known characteristic feature of this process.
The electroweak nature of the interaction, with no color exchange in the
rapidity interval between the tag jets, further provides a handle
against backgrounds from QCD processes with central jet activity. 
Studies from both collaborations indicate that they can be used under pile-up expected
for 10$^{33}$cm$^{-2}$s$^{-1}$ luminosity$~$\cite{CMSAN2007-005,Aad:2009wy}.

The second challenge is in the reconstruction of the hadronic VBs
themselves, particularly when they are highly-boosted ($p_{T}\gtrsim250\,$GeV),
which is common in Higgless models. The decay products are often
collimated and reconstuction from a pair of jets is no longer applicable.
In such cases, it is possible to reconstruct the VB candidates as
single massive jets and perform substructure analysis to suppress
QCD backgrounds. The feasibility of such an approach was previously
shown at a hadron-level study of $WW$ scattering$~$\cite{Butterworth:2002tt}.
Recently, the ATLAS Collaboration has performed this study using both
fast and full detector simulation$~$\cite{Aad:2009wy}.

Besides complementing the inclusive searches, the VB fusion signature
can be exploited to study the VB scattering as a probe of electroweak
symmetry breaking, by investigating the $M_{VV}$ spectrum in case
of Higgs presence (where a peak is expected) and absence (where the
shape at high masses changes due to the different contributions of
longitudinal and transverse couplings of the VBs) \cite{Accomando:2005hz}.
A feasibility study in the CMS detector,
based on ${\cal O}(\alpha_{EW}^{6})+{\cal O}(\alpha_{S}^{2})$
six-fermion-final-state generator Phantom$~$\cite{Ballestrero:2007xq}, 
shows the separation potential at reconstruction level, 
after 60~fb$^{-1}$ of integrated luminosity while running at $10^{33}$~cm$^{-2}$s$^{-1}$,
with the corresponding pile-up taken into account. 
Figure~\ref{fig:CMSAN07005} shows the value of the discriminant used 
($\int_{M_{cut}}^{\infty}dM_{VV}\frac{d\sigma_{noHiggs}}{dM_{VV}}$
/ $\int_{M_{cut}}^{\infty}dM_{VV}\frac{d\sigma_{m_{H}=500\: GeV}}{dM_{VV}}$),
applied to the simulated signal samples, as a function of $M_{cut}$,
for three different final states. At a high luminosity collider, this discriminant has a potential to yield information on the unitarization mechanism.

\begin{figure}
\begin{centering}
\includegraphics[width=0.33\textwidth]{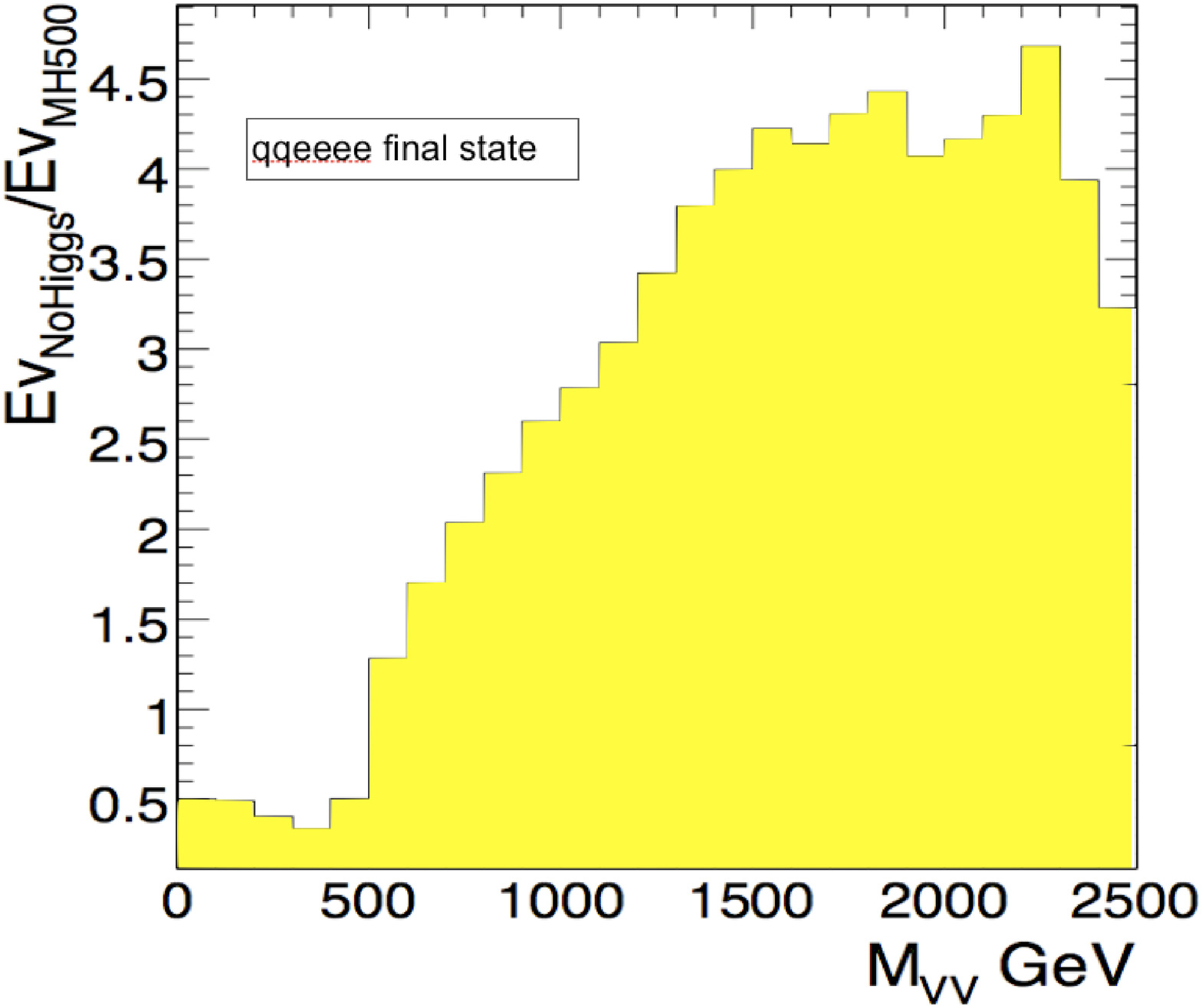}
\includegraphics[width=0.32\textwidth]{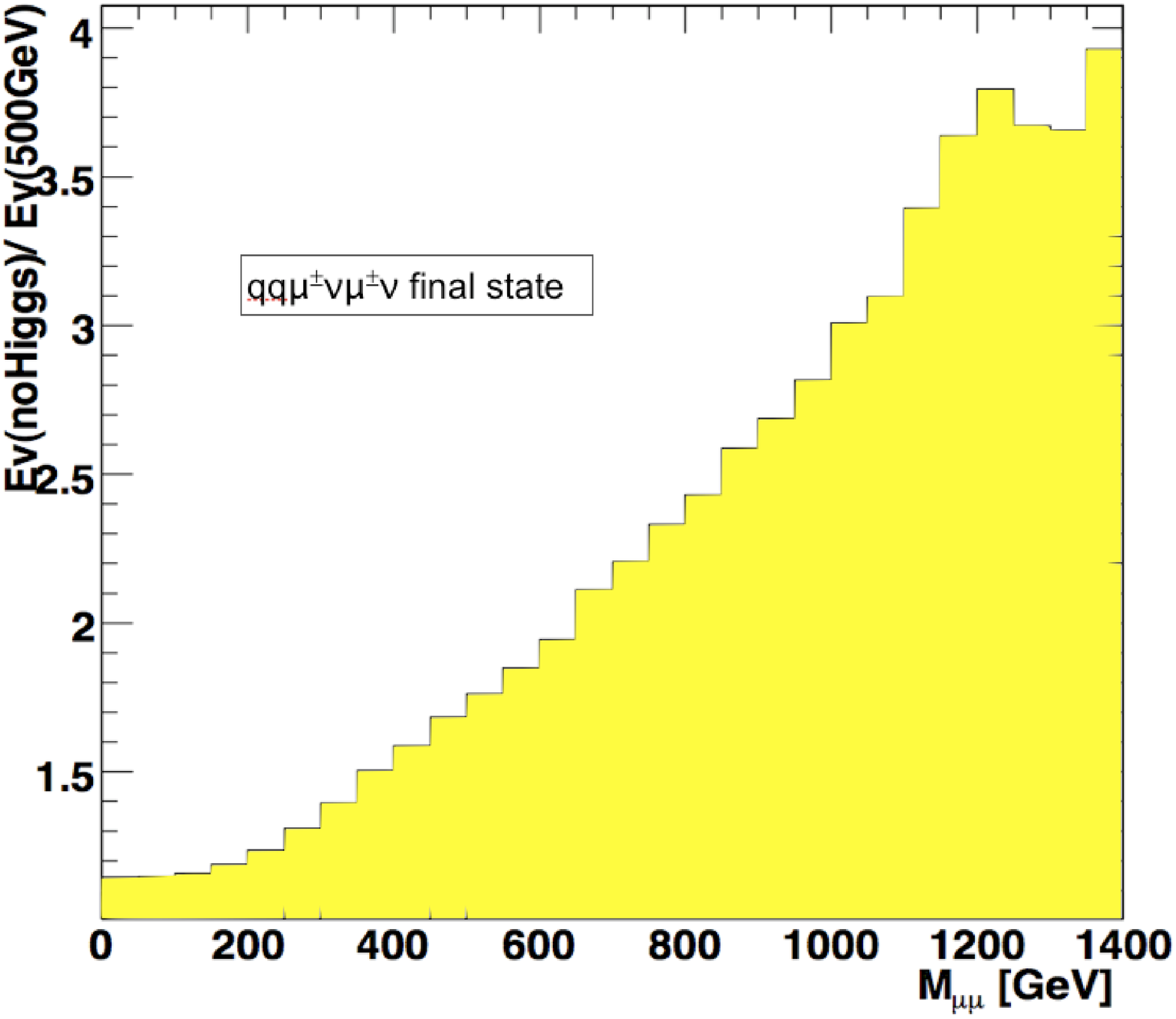}
\includegraphics[width=0.33\textwidth]{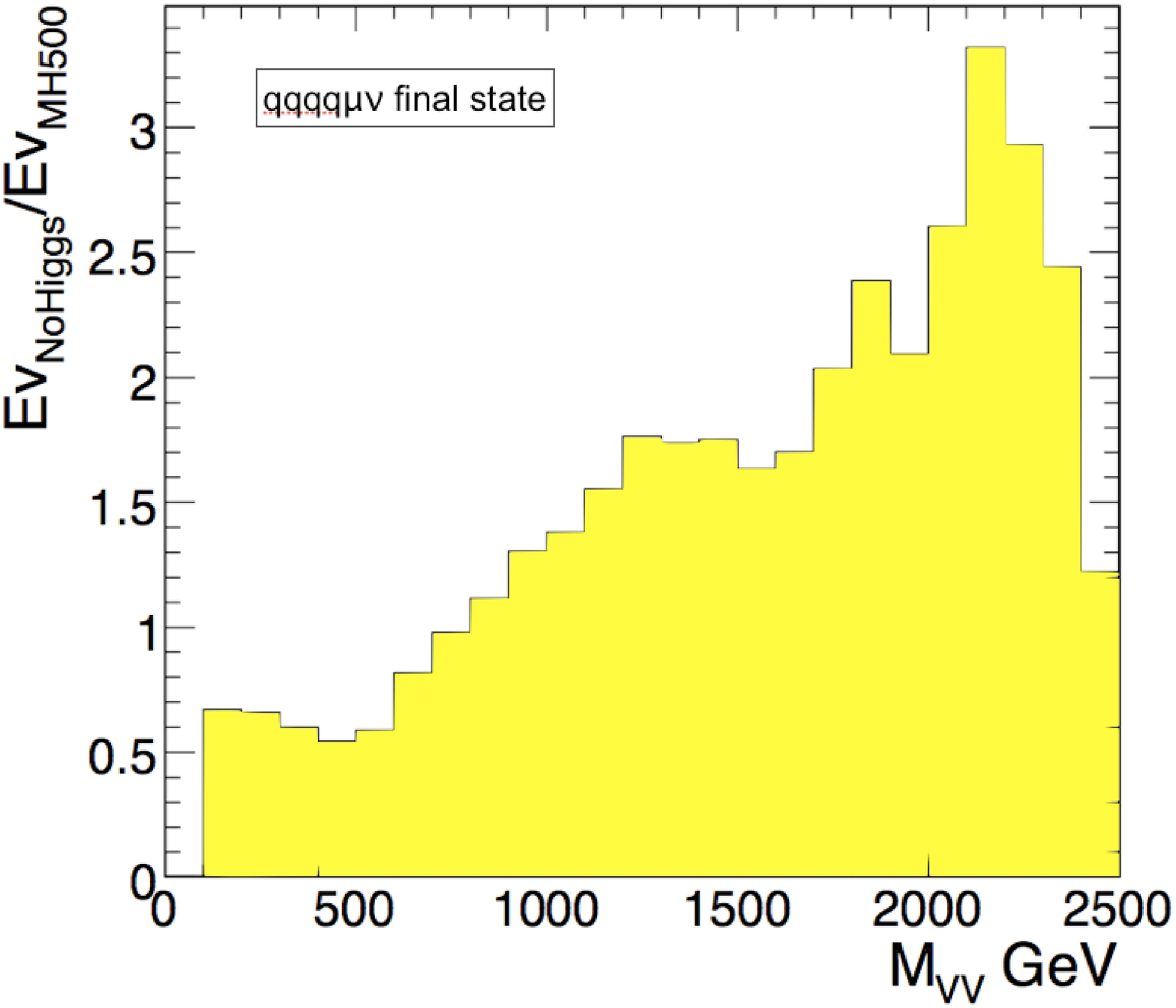}
\par\end{centering}
\caption{Higgs/no-Higgs boson signal discriminant as a function of the minimum $M_{VV}$
considered, for different final states, in the CMS detector~\cite{CMSAN2007-005}.}
\label{fig:CMSAN07005} 
\end{figure}

Likewise, recent parton-level studies show the Higgs/no Higgs separation potential in realistic conditions,
taking into account the contributions from irreducible backgrounds,
with a simple cut-based analysis developed for a higher integrated luminosity~\cite{Ballestrero:2008gf}.

Measurements of multi-boson final states
are expected to start with the extraction of Tevatron-competitive
limits for anomalous triple-gauge couplings in the first few hundred
pb$^{-1}$ of data. 
They will be followed first by searches for model-dependent
states (like techni-resonances at ${\cal O} \sim10$~fb$^{-1}$)
and later by generic searches for resonances (mass up to $\sim$1$\,$TeV
at ${\cal O}\sim50-200\,$fb$^{-1}$). 
Discoveries of much heavier resonances, 
their spin analyses and the extraction of a detailed spectrum
up to 2-3$\,$TeV will probably need few hundreds of fb$^{-1}$ and
will benefit from the sLHC and other future colliders.

\subsubsection{EW chiral lagrangian in absence of resonance}

If there exists no new particle below 2-3~TeV, the scattering of $W,
Z$ gauge bosons is well described in terms of the EW chiral
lagrangian~\cite{Appelquist:1980vg}. Assuming custodial symmetry,
there are only two coefficients contributing to the scattering
amplitudes at order $p^4$:
\begin{equation}
\mathcal{A} \left( W^a_L W^b_L \to W^c_L W^d _L \right ) = 
\mathcal{A}(s,t,u) \delta^{ab} \delta^{cd} 
+\mathcal{A}(t,s,u) \delta^{ac} \delta^{bd} 
+\mathcal{A}(u,t,s) \delta^{ad} \delta^{bc} 
\end{equation}
with~\cite{Bagger:1992vu}
\begin{eqnarray}
\mathcal{A}(s,t,u)&=& \frac{s}{v^2} 
+\frac{4}{v^4}\biggl(2\alpha_5 s^2 +\alpha_4 (t^2+u^2)
\biggr)
+\frac{1}{16 \pi^2 v^4}
\biggl\{
 -\frac{1}{12}
\biggl(3t^2+u^2-s^2
\biggr)
\log\biggl(
\frac{-t}{\mu^2}\biggr)
\nonumber \\
&&
-\frac{1}{12}
\biggl(3 u^2+t^2-s^2\biggr)\log\biggl(\frac{-u}{\mu^2}
\biggr)-\frac{s^2}{2}\log\biggl(\frac{-s}{\mu^2}\biggr)\biggr\}
\, .
\label{eq:8}
\end{eqnarray}
Figure~\ref{Fig:EWChiral} shows the region of parameter space where
perturbative unitarity breakdown is postponed from 1.2~TeV to 2~TeV
(above 2.1~TeV, there is no value of $\alpha_{4,5}$ that unitarizes all the scattering channels simultaneously). The sensitivity of the ILC to these parameters in the channels
$e^+e^- \to W^+W^-\nu\nu$ and $e^+e^- \to ZZ\nu\nu$ at $\sqrt{s} = 800$GeV
~\cite{Weiglein:2004hn} is somewhat better than what can be achieved at the LHC~\cite{Belyaev:1998ih,Eboli:2006wa}.

To avoid unitarity violation, it is common~\cite{Dobado:1999xb} to
invoke a unitarization procedure such as the Pad\'e or the K-matrix
scheme. The latter case leads to a non-resonant enhancement of the
cross section, with respect to a low-mass Higgs scenario, of
longitudinal vector boson scattering. This will be extremely difficult
to observe at the LHC. The Pad\'e unitarization scheme, which gives a
good description of meson scattering, leads to the
presence of resonances~\cite{Dobado:1999xb}. At the LHC, it will
require a few tens of fb$^{-1}$ to observe such resonances for masses
up to $\sim 1.2$~TeV~\cite{Aad:2009wy}.

\begin{figure}[htbp]
\centering
\includegraphics[width=0.60\textwidth]{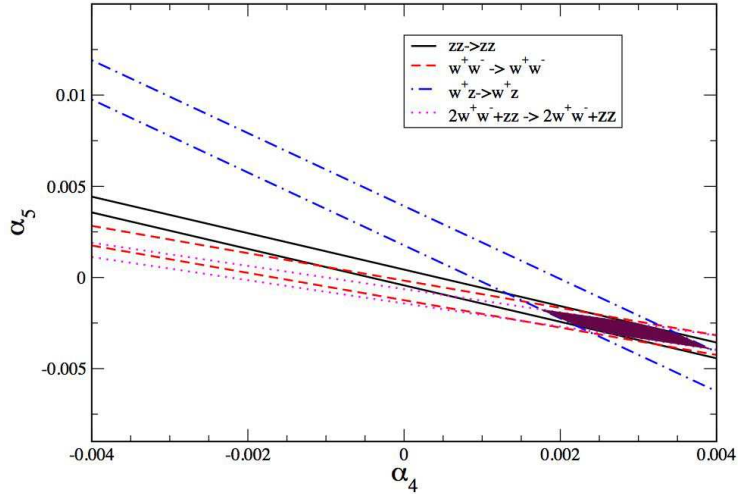}
 \caption
{Allowed values of $\alpha_{4,5}$,
see eq.~(\ref{eq:8}), for perturbative unitarity to hold at $\sqrt{s}=2$~TeV.
The  region between the red dashed lines shows the allowed values of $
\alpha_{4,5}$ to hold for the scattering process $W^+W^- \to W^+W^-$. Similarly,
black, blue, and pink lines delineate the bounds for the scattering $ZZ \to ZZ$,
$W^+ Z \to W^+ Z$ and $(2W^+W^-+ZZ)\to (2W^+W^-+ZZ)$ respectively. This covers
all the elastic scattering processes for the gauge bosons (the unitarity
condition for $W^\pm W^\pm \to W^\pm W^\pm$ is already satisfied if it holds for
$W^\pm Z \to W^\pm Z$). At $\sqrt{s}=2.1$~TeV, the different regions do not
overlap anymore.}
\label{Fig:EWChiral}
\end{figure}

\subsection{Summary of WG2}

In the case where no clear Higgs-like signal will have been established with
the first 10~fb$^{-1}$ of (understood) data at the LHC, one will be faced
with the question whether one or more Higgs bosons  exist but
have been missed in the LHC searches because of their non-standard
properties or whether there really is no fundamental Higgs boson, 
meaning that other new degrees of freedom or new dynamics beyond 
the Standard Model have to be present to achieve electroweak symmetry
breaking while maintaining unitarity at high energy.

The strategy for the future in such a scenario will clearly be
influenced by the other phenomenology observed at the LHC. If other new
physics is detected that seems to hint towards the realisation of (at
least one) fundamental Higgs state in nature, such as the production of
supersymmetric particles, and/or the gauge sector does not
show indications of strong electroweak symmetry breaking dynamics, 
then this could be a strong case for an $e^+e^-$ LC to explore the
expected mass range for the Higgs boson and to precisely determine the nature
of the other observed new physics. A particular strength of an $e^+e^-$ LC would
be to identify a Higgs boson produced in association with a $Z$~boson 
completely independently of its (possibly very unusual) decay properties 
by solely relying on the mass distribution recoiling against the $Z$
boson. If the Higgs mechanism is responsible for generating the masses 
of the weak gauge bosons, one would expect that at least one Higgs boson
should have a significant coupling to the weak bosons so that it could
be observable in this channel. An $e^+e^-$ LC would possibly provide further Higgs
production modes, for instance in association with heavy fermions or
(in the case of SUSY) their scalar superpartners, and the measurements
would allow one to determine the profile of a detected Higgs boson
with high precision. 

The sLHC could provide access to rare Higgs production and decay modes
and with a sufficient amount of accumulated luminosity could possibly
establish a Higgs signal. It could also profit from its enlarged mass
reach for heavy Higgs bosons. A muon collider, in addition to production
modes possible also at $e^+e^-$ linear colliders, could also provide Higgs
production in the s-channel. However, for motivating this option it
would certainly be helpful if the Higgs coupling to muons had already been
established independently at another collider.

If resonances or other indications of strong electroweak symmetry 
breaking dynamics are observed in gauge boson scattering, the strategy
for the future appears to be less clear. An $e^+e^-$ LC operating in the
TeV range would have good prospects to either directly produce
resonances or indirectly probe the effects of the new dynamics.

 }
\newpage
{\setcounter{equation}{0}
\setcounter{figure}{0}
\setcounter{table}{0}

\def\etmiss{{{E}_T^{\rm miss}}}
\def\SUMET{\ensuremath{\Sigma E_{T}}}
\def\slashchar#1{\setbox0=\hbox{$#1$}           
   \dimen0=\wd0                                 
   \setbox1=\hbox{/} \dimen1=\wd1               
   \ifdim\dimen0>\dimen1                        
      \rlap{\hbox to \dimen0{\hfil/\hfil}}      
      #1                                        
   \else                                        
      \rlap{\hbox to \dimen1{\hfil$#1$\hfil}}   
      /                                         
   \fi}          
\def\eslash{\slashchar{E}}

\def\eslash{\slashchar{E}}

\section{WG3: Missing Energy}
\label{wg3}
{\it 
B.~Gripaios,
F.~Moortgat,
G.~Moortgat-Pick,
G.~Polesello
(convenors)\\ 
P.~Bechtle,
K.~Desch,
B.~Foster,
V.~Morton-Thurtle,
K.~Rolbiecki,
J.~Smillie,
J.~Tattersall,
P.~Wienemann
}

\bigskip
In this section the prospects for missing energy signals from new physics are
discussed, both at the LHC and at a future linear collider. We also discuss
potential 
synergy effects between both colliders.  We
summarize discovery potential, as well as methods for measuring masses,
spins and other properties of new states. We also include searches for
Dark Matter candidates at the different colliders.

\subsection{Introduction and scenarios}
Just before the actual start of the major experiment at the
high-energy frontier, the LHC, is a particularly interesting time 
especially for new physics searches. Many physics models beyond the
the Standard Model (SM) predict new sources for missing energy,
therefore the main focus in this section is the discovery and the first
analysis of new physics sectors. One of the favoured models for
physics beyond the Standard Model is Supersymmetry (SUSY). It nicely
overcomes major shortcomings of the SM, giving rise to unification
of the coupling constants of the electroweak and strong interactions,
a natural explanation of the hierarchy problem, and a suitable dark matter
candidate. We therefore take 
SUSY as a representative new physics model, explain the respective 
physics
potential of future colliders and apply the proposed experimental methods
for analysing the properties of new particles within the SUSY context.

After giving an up-to-date discussion of the status of commisioning of
missing transverse energy in the ATLAS and CMS detectors in 
Section~\ref{misscomm}, we discuss in Section~\ref{disc} the discovery
potential. Then, in Section~\ref{mass}, we describe methods for determining
some properties of the new physics models, including accurate measurement of
masses and spins. In Section~\ref{cosmo}, we include possible
input for, as well as constraints from, dark matter experiments and
outline the precision that is required to provide reasonable
predictions.

We close the discussion in Section~\ref{linear} with the inclusion of 
foreseen next high-energy 
options, namely the linear collider, and embed a comprehensive study of
the ILC physics. In particular, we are interested in synergy
effects between lepton- and hadron-collider types and discuss which outcomes
from LHC results may have direct input on the current design efforts of the
ILC.

\subsection{Commissioning of Missing Transverse Energy in the ATLAS and CMS experiments}\label{misscomm}


Neutrinos and other hypothetical weakly interacting particles pass through 
a collider experiment without detection. However, the presence of such 
particles in a collision can be inferred from the imbalance of the event's 
total momentum. This imbalance in a plane perpendicular to the beam direction 
is called Missing Transverse Momentum. Its magnitude is referred to as 
Missing Transverse Energy (MET). MET plays a principal role in studying 
SM physics as well as in searches for physics beyond 
the SM (e.g., Lightest Supersymmetric Particles (LSPs) ). 
The traditional method for missing-transverse-momentum determination 
at hadron colliders is based on the calorimeter information.\par 
In CMS, it is calculated as the negative vector sum of the transverse energies 
deposited in the calorimeter towers (above a noise threshold). This sum is corrected for 
(i) the presence of identified muons; and (ii) the under-measurement of the 
hadronic energy in the calorimeters, as explained in \cite{METPas07}. 
First, identified muons are corrected for by replacing the minimum 
ionizing transverse energy expected in the calorimeters by the transverse 
momentum of the associated track reconstructed in the central tracker. 
Second, the transverse energies of the reconstructed jets are replaced 
by those of the jet-energy-scale corrected jets. The sequential application 
of the muon and the jet-energy-scale corrections defines the current 
standard missing transverse energy in CMS (called {\it CaloMET}).\par 
For the ATLAS experiment the baseline algorithm \cite{Aad:2009wy}
starts from the energy deposits in calorimeter cells that survive 
a noise suppression procedure. Two noise suppression methods have
been studied, one based on only using cells with energies larger than
a threshold, the other based on only using cells in 3-dimensional topological
calorimeter clusters \cite{Cojocaru:2004jk,Aad:2009wy}. 
The cells are then calibrated using global calibration weights depending on
their energy density, and the MET calculated as 
the negative vector sum of the transverse energies
deposited in the considered cells. Corrections
are applied for the muon energy and for the energy lost in the
cryostats. Only good quality muons in the muon spectrometer
with a matching track in the internal tracker  are considered,
and the muon momentum  as measured in the muon spectrometer
is taken.
The energy lost in the cryostats of the Liquid
Argon (LArg) calorimeter is estimated using the correlation of energies 
between the last layer of the LArg calorimeter and the first layer of the 
hadronic calorimeter.
In a subsequent step, the  calibration of the calorimetric term 
is refined by recalibrating cells  according
to the reconstructed high-$P_T$ object they are assigned to. 
An important performance figure is the linearity, defined by the
expression 
$$
\mathrm{Linearity} \equiv (\mathrm{MET(True)}-\mathrm{MET})/\mathrm{MET(True)}
$$
where $\mathrm{MET(True)}$ is the true value and MET is the measured value. 
The linearity for \mbox{$A\rightarrow\tau\tau$} (where $A$ denotes a CP-odd
Higgs bosons) with $M_A=800$~GeV is shown in
Fig.~\ref{fig:metl} as a function of $\mathrm{MET(True)}$ for the
different reconstruction steps described above. 
The bias of linearity at low values is due to the finite
resolution of the measurement. The reconstructed  MET~ is positive by
definition, so the linearity is negative when the true MET~ is
near to zero. For $\mathrm{MET(True)}>40$~GeV, the  
linearity is  within $2\%$.

\begin{figure}[!ht]
\begin{center}
\includegraphics[width=0.5\textwidth]{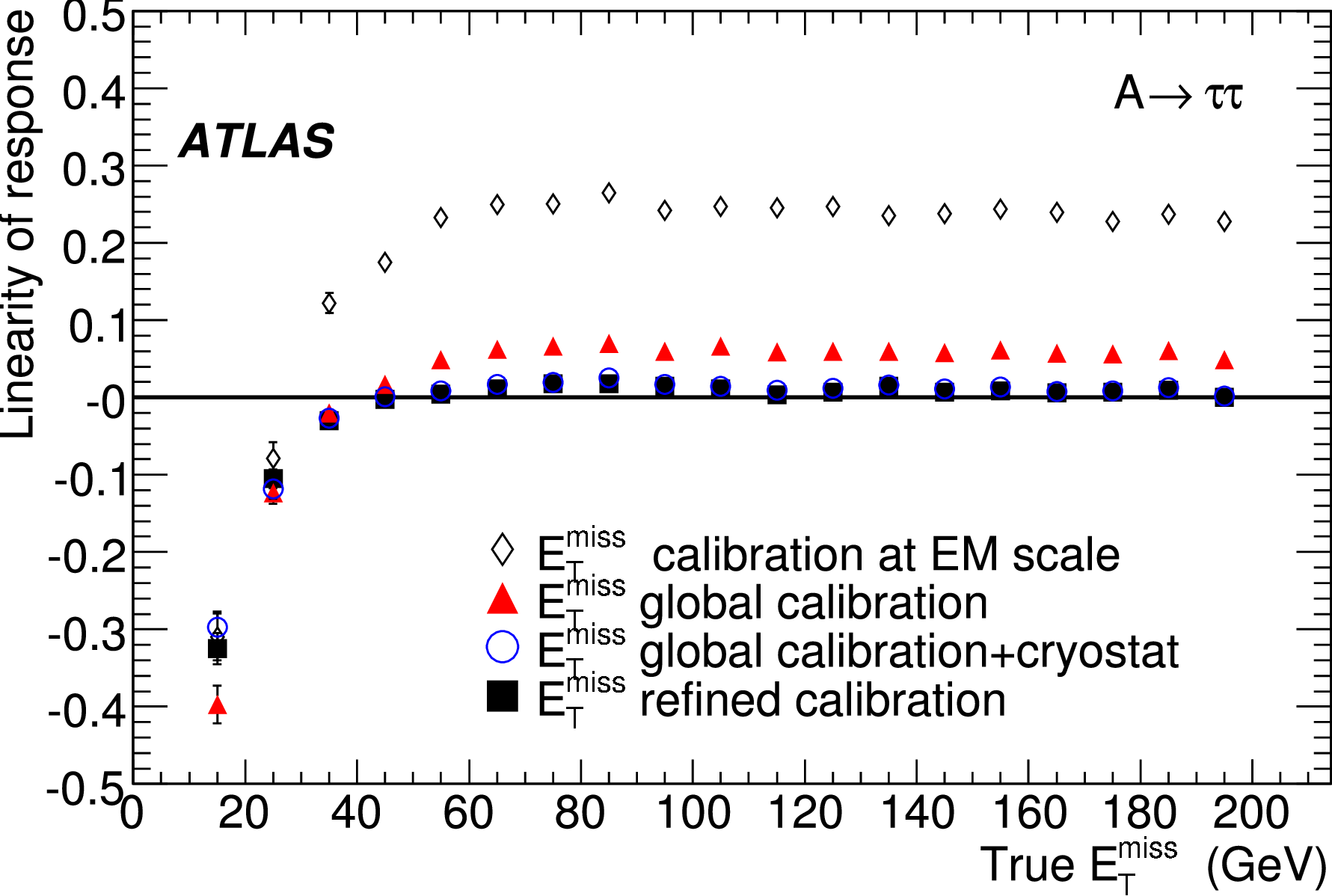}
\caption{Linearity of response for reconstructed MET as a
  function of the average true MET for \mbox{$A\rightarrow\tau\tau$} 
 events with $M_A=800$~GeV (taken from Ref.~\cite{Aad:2009wy}, see inside for details).} 
\label{fig:metl}
\end{center}
\end{figure}

The resolution of the measurement is estimated from the width of the
distribution of the difference between true and estimated
values of the MET $x$ and $y$ components in bins of the total transverse energy
deposited in the calorimeters (\SUMET).
The MET resolution, is shown in Fig.~\ref{fig:met} as a function 
of $\Sigma E_\mathrm{T}$ for different ranges of \SUMET, based 
on the study of different Monte Carlo samples. 
The dependence can be  fitted with a function $\sigma= a \cdot
\sqrt{\SUMET}$ for values of \SUMET\ between 20 and $2000$~GeV. The
parameter $a$, which quantifies the MET resolution, varies between
0.53 and 0.57. Additional details on the ATLAS MET performance can be found 
in Ref.~\cite{Aad:2009wy}

\begin{figure}[!ht]
\includegraphics[width=0.5\textwidth]{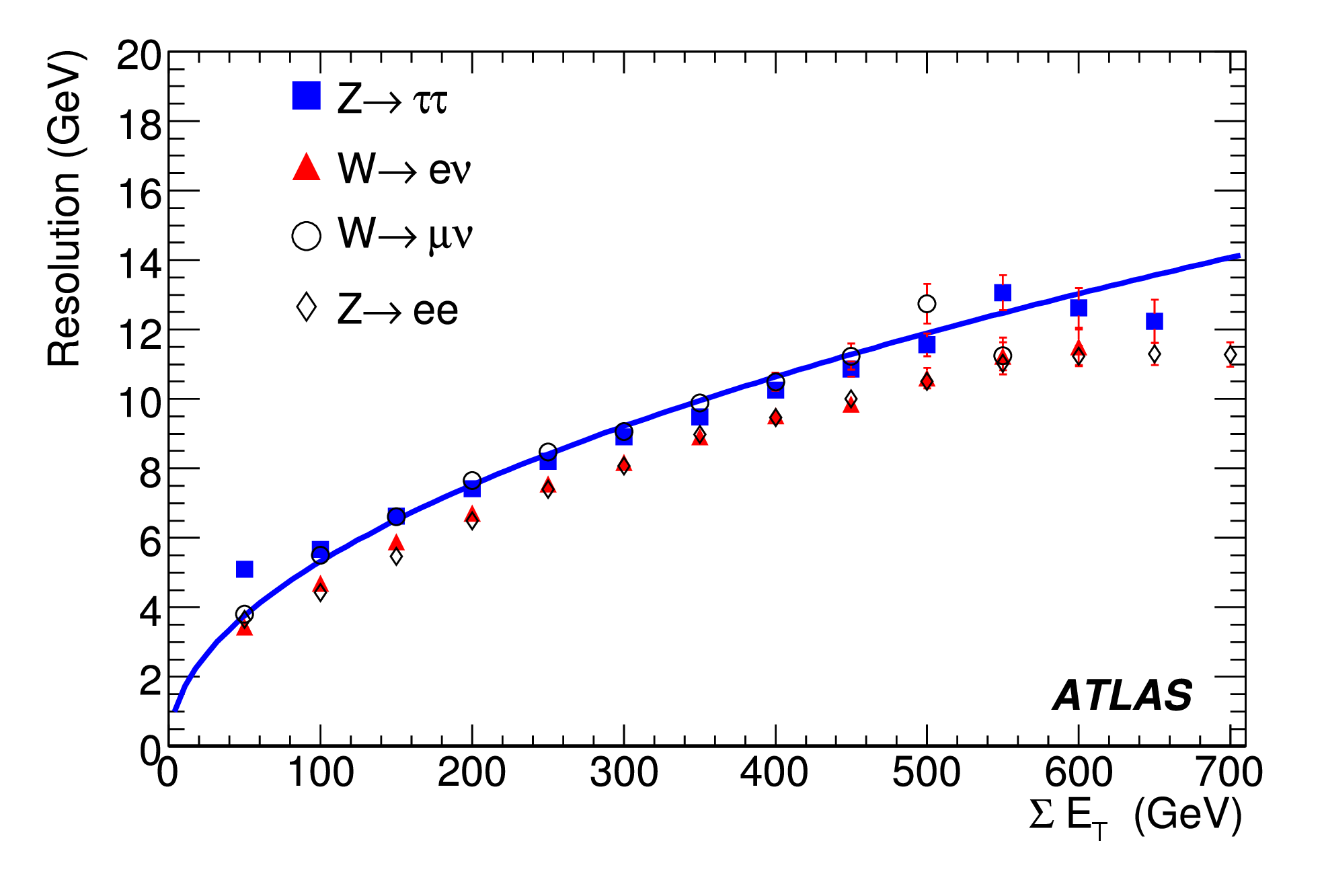}
\includegraphics[width=0.5\textwidth]{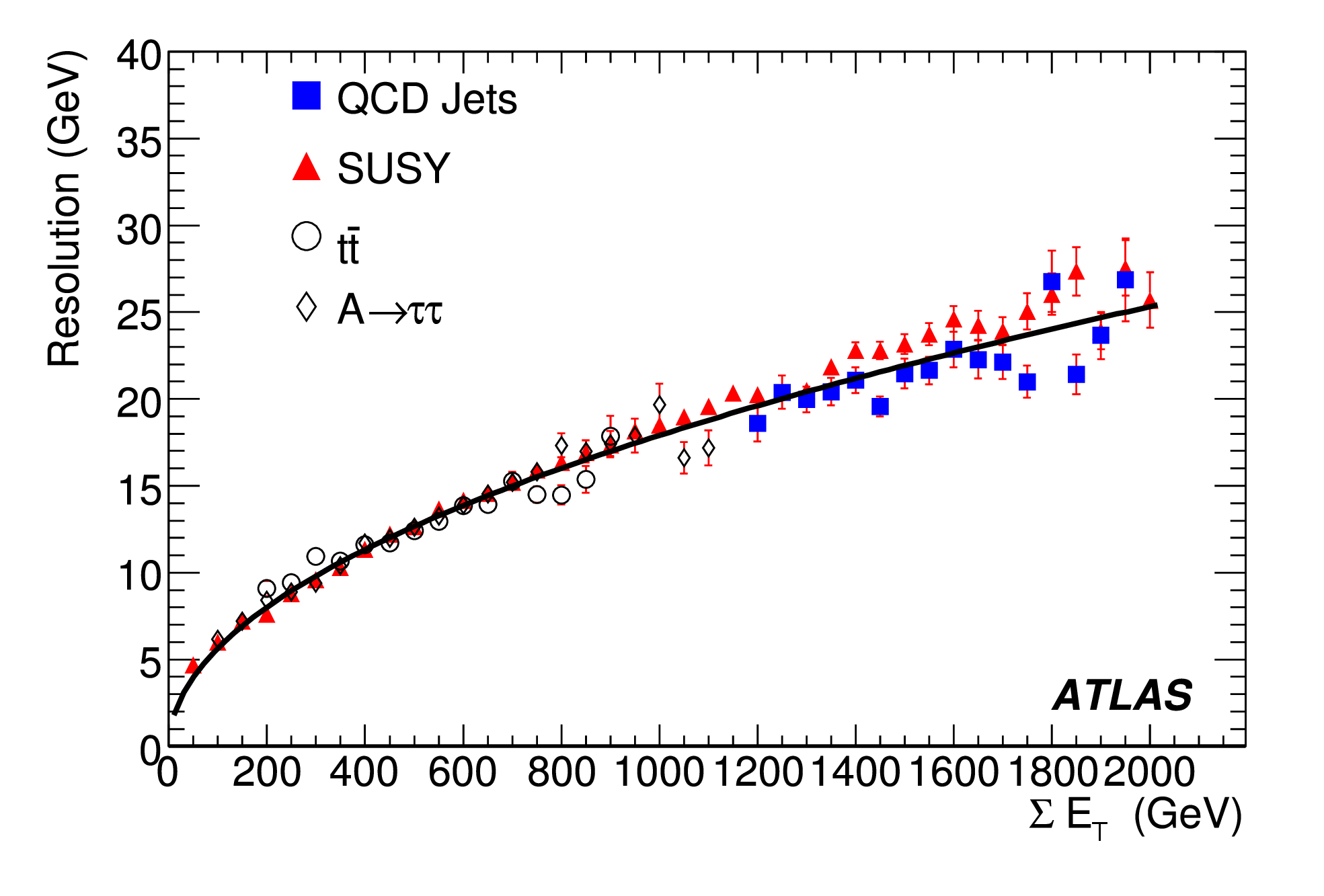}
\caption{Resolution of the two MET components
    with refined calibration as a function of the
    total transverse energy, $\Sigma E_\mathrm{T}$ for low to medium values
    (left) and for higher values (right). The curves correspond to the
    best fits of $\sigma=0.53 \sqrt{\Sigma E_\mathrm{T}}$ through the points
    from \mbox{$Z\rightarrow\tau\tau$} events (left) 
    and $\sigma=0.57 \sqrt{\Sigma E_\mathrm{T}}$
    through the points from \mbox{$A\rightarrow\tau\tau$} 
    events (right). The points from
    \mbox{$A\rightarrow\tau\tau$} events are for masses 
    $M_A$ ranging from 150 to 800~GeV
    and the points from QCD jets correspond to dijet events with
    $p_T$ between 560 and 1120~GeV.
(Both plots are taken from Ref.~\cite{Aad:2009wy}, see inside for details.)}
\label{fig:met}
\end{figure}

The resolution of the purely calorimetric MET measurement is significantly
better in ATLAS than in CMS thanks to the better energy resolution
and longitudinal energy containment of the ATLAS hadronic calorimetry.

To cope with this issue, recently, two new methods for improving the purely calorimeter-based MET 
have been developed, exploiting information from the other subdetectors in CMS.  
In the {\it Track-Corrected MET} \cite{tcMETPas09}, 
 the correction for the under-measurement of the hadronic energy in 
the calorimeters is replaced by a charged-particle-track-based 
correction: the transverse momentum of each  reconstructed charged 
particle track is added to the total missing transverse momentum, from 
which the corresponding transverse energy expected to be deposited 
in the calorimeters is subtracted. The aforementioned muon correction 
is applied in turn. The resulting track-corrected 
missing transverse energy is shown to have a slightly better MET 
resolution and a reduced MET fake rate with respect to the 
calorimeter-only MET. Fig. \ref{fig:tcmet} shows the improvement: the left plot
illustrates that in $Z \rightarrow \ell \ell$ events, the track-corrected MET algorithm reduces the number of events with MET $>$ 30 (50) GeV by a factor of 3.4 (6.8), compared to calorimetric MET 
corrected for muons only.  The reduction is a factor of 3.0 (4.3) when 
compared to calorimetric MET corrected for both muons and the jet energy scale. 
 Fig. \ref{fig:tcmet} (right) illustrates that in $t\overline{t} \rightarrow \ell \ell + X$ 
 events the accuracy of determining the MET resolution by 
track-corrected MET is improved by more than 25\% (20\%) compared to calorimetric MET corrected for muons only (muons + jet energy scale).

\begin{figure}[!ht]
\includegraphics[width=0.5\textwidth]{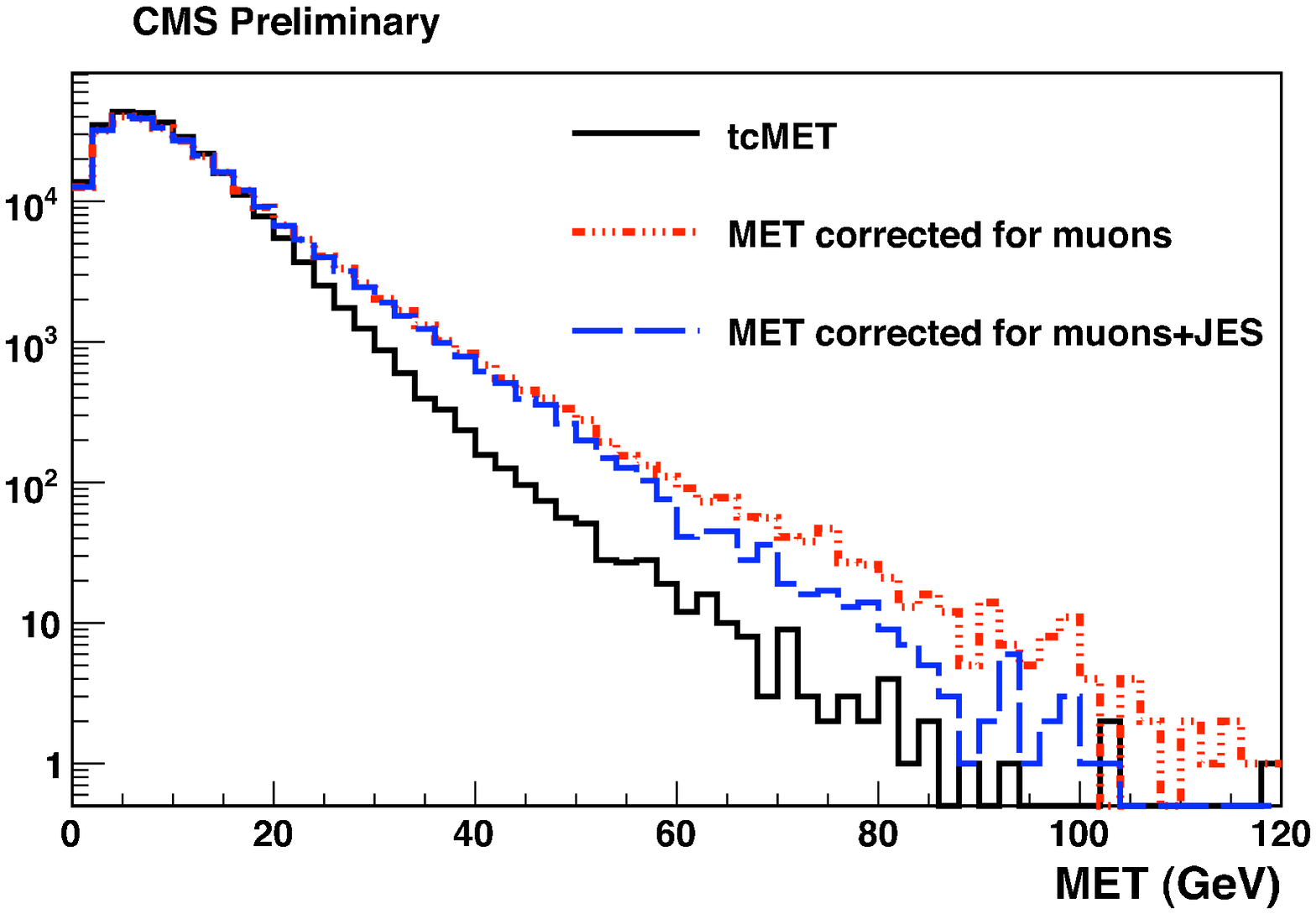} 
\includegraphics[width=0.5\textwidth]{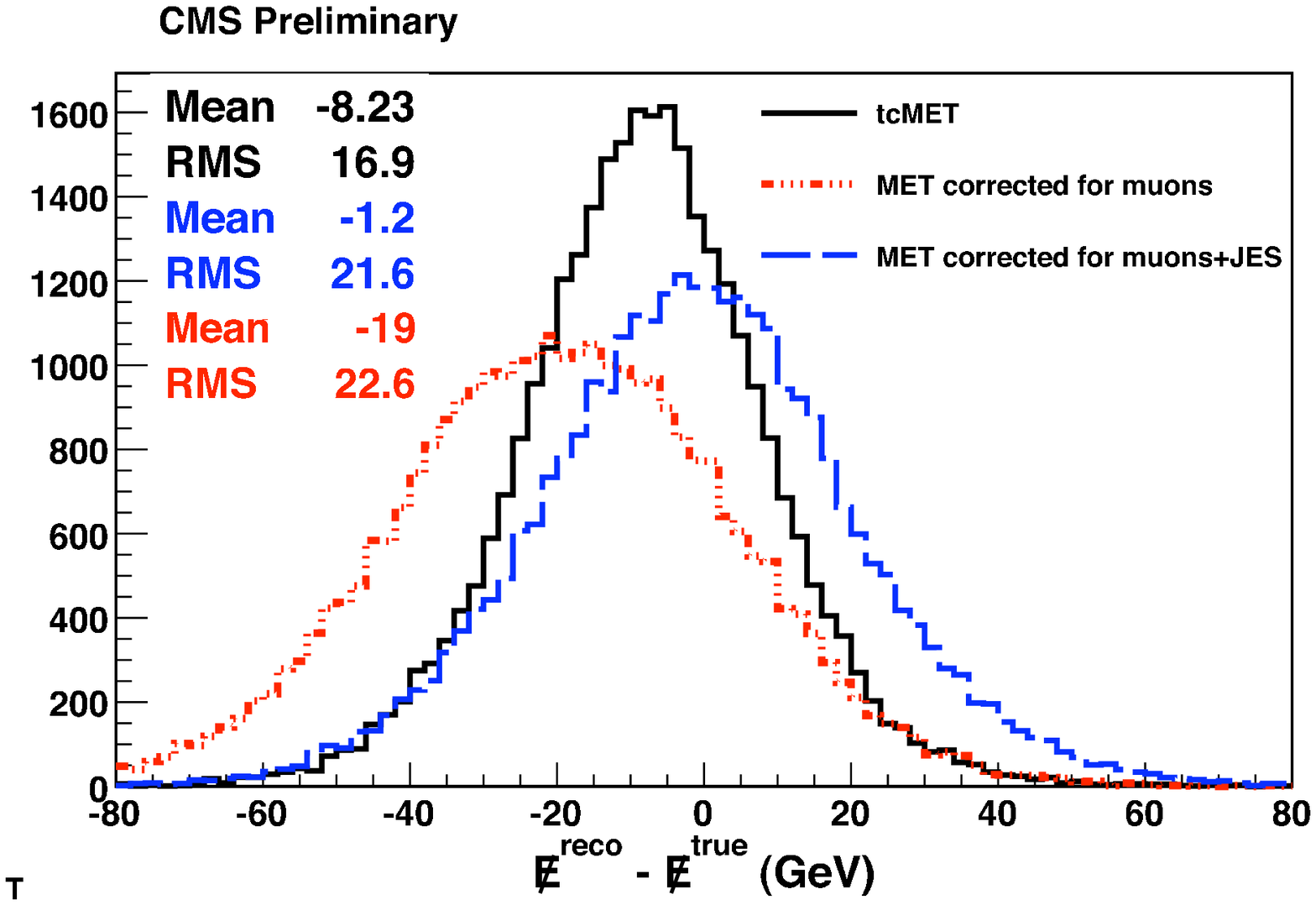}
\caption{{\em Left} Distributions for track-corrected MET (tcMET) and
  calorimeter-only MET in $Z \rightarrow \ell \ell$ events with the CMS
  detector.   {\em Right} Distributions of $\Delta \etmiss$ =
  $\eslash_{T}^{\rm reco}- \eslash_{T}^{\rm true}$ for 
  $t\overline{t} \rightarrow \ell \ell + X$ events with 
  $\eslash_{T}^{\rm true} >$ 50 GeV (taken from~\cite{tcMETPas09}) .  }
\label{fig:tcmet}
\end{figure}

A different approach is followed in the particle-flow event reconstruction: 
individual particles are reconstructed and identified by exploiting 
the characteristics of all CMS sub-detectors, 
towards efficiency and purity, and by using the built-in redundancies 
of the energy and direction measurements in the different sub-detectors, 
towards precision and accuracy. The {\it Particle-Flow MET} is determined 
as the modulus of the negative vector sum of the 
reconstructed-particle transverse momenta, with no need for posterior 
corrections. A detailed description of the algorithm, a report of its 
performance, and a number of related systematic studies can be 
found in \cite{pfMETPas09}. 
Fig. \ref{fig:pfmet} shows that the Particle Flow technique improves the MET 
resolution with almost a factor of two with respect to the calorimetric 
determination, irrespective of the true missing transverse energy.

\begin{figure}[!ht]
\begin{centering}
{\rotatebox{90}{\includegraphics[width=0.40\textwidth]{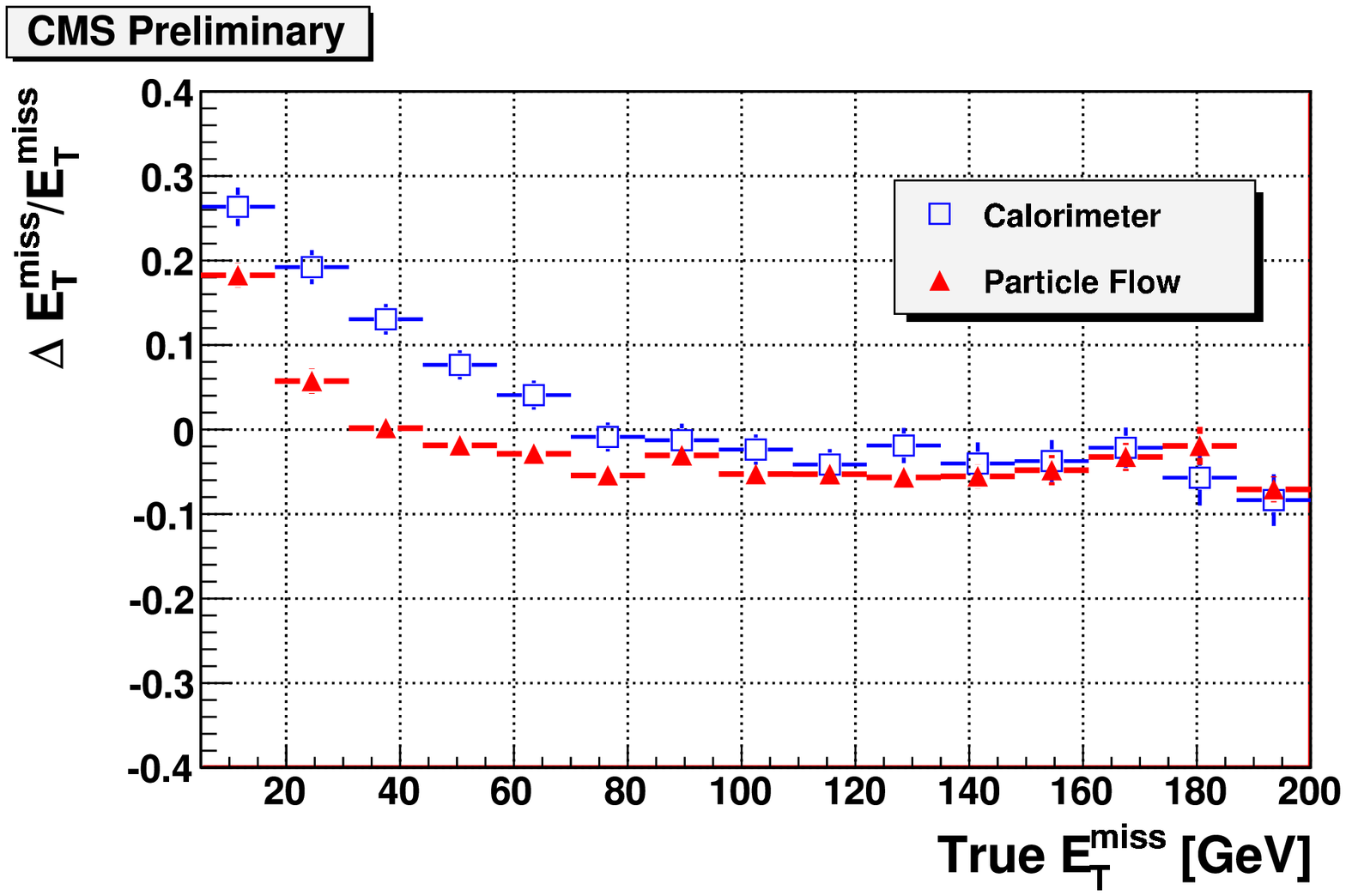}}}
\caption{ Distribution of $(E^{\rm miss}_{\rm T, reco} - E^{\rm miss}_{\rm 
T, true})/E^{\rm miss}_{\rm T, true}$ as a function 
of the $E^{\rm miss}_{\rm T, true}$, in a fully inclusive ${\rm 
t\bar{t}}$ event sample, for particle-flow reconstruction 
(solid triangles) and for calorimeter-only reconstruction (open squares) in
CMS (taken from~\cite{pfMETPas09}).
 }
\label{fig:pfmet}
\end{centering}
\end{figure}

The conceptual differences between the three approaches, from the simplest 
(CaloMET) to the most comprehensive (Particle-Flow MET), is expected to 
be a great asset when the first collision data are produced in the LHC. 
Independently of the respective performance, the largely independent 
systematic uncertainties and the undoubtedly different failure modes 
of the three methods will pave the road towards a rapid understanding 
and a robust determination of the missing transverse energy in CMS. 

The MET commissioning activities in both experiments 
can be divided in three stages: 
(i) the pre-collision phase (no beam and single circulating beam period), 
where the detector can record cosmic muons and beam halo muons; 
(ii) the 10 pb$^{-1}$ phase, where the first proton-proton collisions 
will be registered, mainly minimum bias and QCD processes;
and (iii) the 100 pb$^{-1}$ (and above) phase, after which 
sufficient statistics for physics processes such as $Z$+jets or 
$t \bar{t}$ will be collected.
In the pre-collision phase, many important instrumental procedures 
for constructing the missing transverse energy can already be commissioned. 
In particular, the handling of abnormal calorimeter cells (hot or dead) 
and the removal of detector noise can be and have been tested in 
the extensive campaigns of cosmics runs  of the two detectors in 2009 with 
complete detector configuration. 
The MET Data Quality Monitoring (DQM) system can also be commissioned during 
this period. Procedures for the identification and removal of cosmic muons 
in future collision events can be verified. During the single circulating 
beam phase (i.e. when the LHC beams are being commissioned), the filters 
for beam halo muons can be tested and improved.

In the second phase, after the collection of the first collision data, 
the large cross section of QCD di-jet events will quickly allow the 
relative calibration of the calorimeters in order to obtain a uniform 
response over $\phi$ and $\eta$. Also the absolute calibration of the 
calorimeter cells, using single isolated tracks, will be performed. 
The jet energy scale corrections, needed for the corrections of 
calorimetric MET, will be derived using photon+jet and $Z$+jet 
balance techniques.

In the third phase, when sufficient events of "standard candle" processes 
such as $W$ and  $Z$ + jets or  $t \bar{t}$ will have been recorded, 
the missing transverse energy can be validated using these control processes.
The MET scale can be
determined in-situ in \mbox{$W\rightarrow e\nu$} events 
and in \mbox{$Z\rightarrow\tau\tau$} events, where,
using the mass constraint, the MET scale can be determined
with an accuracy of 8\%.
 
The  $Z$ + jets process, where the $Z$ decays to electrons and muons, 
is ideally suited to study the MET performance since it is almost 
background-free and the two leptons can be measured with excellent precision. 
This process can also directly be used for data-driven background 
estimations for processes involving MET.
Semileptonic $t \bar{t}$ events have real  MET and
allow a test of the MET  reconstruction in the high multiplicity
environment relevant for SUSY searches.

\subsection{Discovery at LHC}\label{disc}
The discovery of R-Parity-conserving SUSY is in principle ``easy''
at the LHC. In fact, squarks and gluino are produced through 
strong interactions, yielding cross-sections at the picobarn 
level for masses of order 1~TeV. Squarks and/or gluinos will then 
decay to the LSP through, in general, complex and model-dependent decay chains.
The resulting final state will, in any case, include high $p_T$ hadronic
jets, and missing transverse momentum from the two undetected LSPs 
in the final state. In most models, the decay chains will also involve 
the presence of leptons, $b$-jets, $\tau$-jets, photons, and $Z$s.\par
There are, however, important experimental caveats. The essential 
point is that, although appropriate cuts will easily separate
a low mass SUSY signal from the Standard Model backgrounds, 
the SUSY signal has no distinctive features, such as mass peaks, 
which separate it from the background. 
We are therefore dealing with  counting experiments, where 
an accurate prediction of the backgrounds is mandatory.\par
An additional difficulty is the fact a large $\etmiss$ can be 
generated by a number of experimental effects, and a complete 
control of this variable will require long and painstaking studies.
Before a SUSY discovery can be claimed, a lot of  work
on understanding background will be needed, including both the
control of experimental effects, and the collection of the needed
control samples of Standard Model events. The time for discovery
will be driven by these considerations rather than by the
cumulated signal statistics. 
The ATLAS and CMS experiments have therefore recently 
focused their efforts on the development of methods 
based on  a combination of Monte Carlo simulation and  data-driven 
techniques for background estimation. The aim is to optimise,
for each given value of integrated statistics, the level of systematic
uncertainty on background estimate. Many different approaches to this
issue are documented in \cite{Aad:2009wy, :2008zzk}. As an example, 
for cumulated statistics of 1~fb$^{-1}$, the ATLAS collaborations quotes
an uncertainty on backgrounds from mismeasured QCD jets of 50\%,
and an uncertainty of 20\% from backgrounds with real $\etmiss$ from
neutrinos, such as $\bar{t}t$ and $W$+jets.

Based on these considerations, the ATLAS and CMS Collaborations 
have evaluated their SUSY discovery potential  for inclusive 
analyses requiring high-$p_T$ jets, $\etmiss$, and one or more 
additional leptons or other objects. The resulting reaches of the 
ATLAS and CMS experiments are given in terms of the parameters 
of a constrained SUGRA-inspired model, but the analyses are very general,
and cover most of the topologies resulting from a generic MSSM 
with equivalent squark/gluino mass scale. 
With the moderate assumed luminosity of 1~fb$^{-1}$ 
squarks and gluinos with masses of order 1.3~TeV should be discovered
for a very broad range of models. For most of the accessible parameter
space, SUSY should show up for several different signatures, thus
enhancing the robustness of the signal. 
The quoted reach  includes the systematic
uncertainties on the backgrounds quoted above. It is expected
that for lower luminosities the accuracy on background evaluation
will decrease, severely affecting the discovery reach.\par

\subsection{Measurement at LHC}\label{mass}
Although hadron colliders have an excellent reputation for discovering
new physics, the next step, that of measuring various properties
associated with the new physics, is a very challenging one. The
reasons for this are, of course, well known, and we shall not repeat
them here.

In the case of signals involving missing energy signals, the
difficulties in performing measurements are compounded by the fact
that kinematic information is lost in events: the energies and momenta
of invisible particles go unmeasured. This problem can, to a certain
extent, be offset by the fact that the total missing transverse
momentum can be inferred from the measured transverse momentum in the
event, but such a measurement in itself requires a very good global
understanding of the detector.

\subsubsection{Mass measurements}
Assuming we discover a missing energy signal at the LHC associated
with new physics, the first measurement priority will, presumably, be
to establish the mass scale or scales of the new physics. In the
absence of missing energy, such measurements are not
difficult. Imagine, for example, that some new particle undergoes a
decay into visible SM final states. To measure the mass of the new
particle, it suffices to measure the four-momenta of the final states,
to compute the invariant mass, and to look for a Breit-Wigner peak in
the invariant mass distribution. The only major requirement is the
ability to distinguish the signal amongst the background.

But for events with invisible particles, the final state cannot, in
general, be reconstructed, and more thought is needed. In recent
years, several methods to measure masses have been proposed. We
discuss several of them below. In all cases, it turns out that it is
far easier to measure mass differences between new states than to
determine the overall mass scale of new physics. Unfortunately, it is
the absolute mass scale that is of most interest to us, both in terms
of our understanding of physics and in our planning for a future
collider. It is, for example, the absolute mass scale that will be
most relevant for dark matter, and for addressing the hierarchy
problem (little or large). Similarly, it is the absolute mass scale
that determines whether or not a future collider will be able to
access the new physics.
\subsubsubsection{Invariant mass endpoints}
In cascade decays with at least two visible particles in the chain,
one can construct various invariant mass combinations.  For a particle
of mass $m_0$ decaying to a massless visible state and an intermediate
state of mass $m_1$, followed by a decay of the intermediate state
into a second massless visible particle and a final invisible state of
mass $m_2$, the endpoint of the visible invariant mass distribution is
given by $\sqrt{(m_0^2-m_1^2)(m_1^2-m_2^2)/m_1^2}$. So measurement of
this endpoint yields one combination of the masses. In a cascade decay
with $n$ visible particles, there are $n+1$ unknown masses and $2^n -
n-1$ invariant mass distributions, so it is possible to determine all
of the masses if $n\geq 3$. Assuming such cascades exist in Nature and
can be isolated in experiment, the principal difficulties are to
isolate the endpoints and also to solve the combinatoric issues of
which of the two decay chains an observed particle belongs to, and
where on the chain it belongs. For further discussion see
\cite{Paige:1996nx,Hinchliffe:1996iu,Bachacou:1999zb,Matchev:2009iw}.
\subsubsubsection{Polynomial constraints}
An alternative method \cite{Nojiri:2003tu,Kawagoe:2004rz,Cheng:2008mg,Cheng:2009fw}
for cascade decay chains is to use the observed four-momenta directly
rather than just the invariant masses. One can also combine the data
from multiple events.  If one isolates a single cascade decay with $n$
visible particles in each of $N$ events, then there are, as before,
$n+1$ unknown masses and $4N$ unknown momentum components
(corresponding to the four-momentum of the LSP in each event). But
energy conservation at each vertex implies that there are $N(n+1)$
constraints. The constrained system of polynomial equations can
therefore be solved (up to discrete ambiguities) provided $n\geq 4$.

If on the other hand, one is able to observe both cascade decay chains
in $N$ events (the chains are assumed to contain $n$ and $m$ visible
particles respectively) then the number of unknown four-momentum
components is only $6N$, because even though there are two LSPs, two
of the four-momentum components can be inferred from the missing
transverse momentum. Counting up the constraints as before, one finds
that the system of polynomial equations can be solved for $n+m \geq
5$; for identical chains with $n=m$, one has $n\geq3$, as for the
endpoint method. Even if one is in a situation with shorter cascade
decays and an underconstrained system, one can still use the
constraints from many events to pin down the masses
\cite{Cheng:2007xv}.

The advantage of methods of this type is that, unlike the endpoint
methods which employ only a subset of events (those near the endpoint)
to perform a measurement, the polynomial methods use all of the
available data. The disadvantages are that the solutions one obtains by
solving the system are often the wrong ones, or simply do not
exist, because of combinatoric ambiguities and mismeasurement
effects. Moreover, it is not clear that these methods, which combine
pool together a small number of events at a time until an exactly
constrained system is obtained, are the best way to combine data from
many events.

Is there a better way? According to the theory of statistics, this
situation of partial data from multiple events is naturally dealt with
using the likelihood function. The problem in the case at hand is that
the likelihood should be constructed from the matrix element, and
since the new physics is unknown, the matrix element is also unknown.
Short of prescribing the new physics, the best hope seems to be to
define the likelihood in an {\em ad hoc} way \cite{Allanach:2004ub,Webber:2009vm}.
\subsubsubsection{Transverse observables}
These methods are all based on method originally used to measure the
mass of the $W$-boson in its leptonic decays after its discovery at
UA1 and UA2.  This method is still used today at the tevatron
\cite{Aaltonen:2007ps}, and provides the single most precise
measurement of the $W$ mass.  In the leptonic decay of a $W$, the
neutrino is invisible in the detector. The transverse mass, defined by
\begin{gather}\label{mT}
m_T^2 \equiv m_v^2 + m_i^2 + 2 (e_v e_i - \mathbf{p}_v \cdot \mathbf{p}_i),
\end{gather}
where $\mathbf{p}$ is the momentum transverse to the beam,
$e=\sqrt{\mathbf{p}\cdot \mathbf{p} + m^2}$ denotes the transverse
energy, and $v$ and $i$ label the visible and invisible decay products
respectively, (a charged lepton and a neutrino in the case at hand),
is an observable, because $m_i$ can be neglected and because $
\mathbf{p}_i$ can be inferred from the missing transverse momentum in
the event. Moreover, $m_T$ is bounded above by $m_W$, and so its
distribution features at edge. In practice, the edge is smeared out by
resolution and finite-width effects, but provided that this can be
modelled, a good measurement of the $m_W$ can be obtained.

In the case of missing energy events associated with new physics, two
complications arise. The first is that, since the invisible particles
are typically pair produced, the individual transverse momenta of the
invisible particles cannot be inferred. This problem has been solved
by the introduction of the derived observable $m_{T2}$
\cite{Lester:1999tx,Barr:2003rg}. (Very recently, the usefulness of
$m_{T2}$ for hadron collider measurements has been confirmed in its
application to measurement of the top quark mass in the dileptonic
decay channel at the Tevatron \cite{cdfmt2}.)  The second problem is
that the masses of the new, invisible particles cannot be neglected,
so $m_T$ is no longer observable. At way out of this impasse has been
found, at least in principle, in
\cite{Cho:2007qv,Gripaios:2007is,Barr:2007hy,Cho:2007dh}: one can
compute the distribution of $m_T$ or $m_{T2}$ for some hypothetical
value of the invisible mass; a plot of the endpoints of the resulting
distributions, considered as a function of the hypothetical mass,
features a kink (is continuous, but not differentiable) exactly at the
point where the hypothetical mass equals the true invisible mass. The
co-ordinates of the kink this yield the masses of both the parent and
the invisible daughter. If one generalizes to more complex decay
topologies, one finds that all of the unknown masses can be determined
\cite{Serna:2008zk,Burns:2008va}

It was pointed out in \cite{Cheng:2008hk}, that this method is connected to the polynomial approach in the following way:
For each event, the $m_{T2}$ variable, considered as a function of the unknown LSP mass, generates a curve which delineates the boundary of the region
in mass space which is compatible with the kinematic constraints. This shows that the methods described in \cite{Cho:2007qv,Gripaios:2007is,Barr:2007hy,Cho:2007dh} are in fact the best one can hope for using kinematic information alone.

Whether or not this method can actually be successfully employed in
the real-world remains to be seen. What is clear is that this is the
only known method that works for any decay topology. In particular, it
is the only method that exists to measure the masses in two-body
decays, or decays involving off-shell intermediate states.

In another development, it has been shown in \cite{Randall:2008rw,Barr:2009wu} that observables of this type may be of some use for discovery itself, rather than for the later task of mass measurement.
Finally, we note that various other methods involving transverse
observables of one kind or another have also appeared
\cite{Lester:2007fq,Ross:2007rm,Barr:2008ba,Barr:2009mx}.
\subsubsection{Spin measurements}
Spin measurements will be crucial in determining the nature of the new
physics. In particular, various scenarios for physics beyond the SM
involve new particles which are partners of the SM particles, in that
they share the same quantum numbers as SM particles. They differ in
their spins, however. For example, in supersymmetric scenarios, the SM
partners, or rather superpartners, have spins which differ by one-half
from the corresponding SM states. In extra-dimensional scenarios, by
contrast, the spins are the same.

Just as for mass measurements, spin measurements are complicated by
the presence of missing energy in the final state. Roughly speaking,
the reason for this is that spin, being a form of angular momentum, is
a generator of rotations. Thus the physics of spins is principally
associated with the angular distribution properties of production and
decay processes. But in order to measure these angles, one needs to be
able to reconstruct some reference frame, such as the rest frame of a
decaying particle, or the centre of mass frame in a two-body
collision. But the presence of missing energy makes reconstruction
very difficult. For a recent review of spin measurement at the LHC,
see \cite{Wang:2008sw}.
\subsubsubsection{Cascade decays}
Spin measurements can still be performed, even if one is not able to
reconstruct reference frames. For example, if one looks at
Lorentz-invariant quantities, the choice of frame is irrelevant. This
observation gives rise to the first method for spin measurement
\cite{Barr:2004ze,Goto:2004cpa,Smillie:2005ar,Datta:2005zs,Athanasiou:2006ef,Burns:2008cp,Gedalia:2009ym},
which looks at the invariant mass distributions of various visible
final state particles. The invariant masses do depend on the angular
properties of decays, and do therefore contain information about the
spin, in principle.  In order that non-trivial angular correlations
exist, one requires at least that the intermediate particle in
question be polarized, and, if a fermion, that its decay be chiral
\cite{Wang:2006hk}.
\subsubsubsection{Production processes}
One can also hope to perform spin measurement via angular effects in
production processes. The main difficulty here is that, at a hadron
collider, many sub-processes, each with different angular dependence,
contribute in a given production channel.  A method for measuring the
slepton spin in dislepton production via $q\overline{q}$ production
via a $Z/\gamma$ was proposed in \cite{Barr:2005dz}.  One exploits the
fact that the pseudo-rapidity of the resulting leptons is correlated
with the $Z/\gamma$ decay angle. The viability of this method depends
strongly on the mass spectrum. Measurement of the slepton spin should
be much easier in a future lepton collider
\cite{Battaglia:2005zf,Battaglia:2005ma,Bhattacharyya:2005vm}.

A second option is to look at the azimuthal angular effects in
production \cite{Buckley:2008eb,Boudjema:2009fz}. These effects arise
whenever multiple sub-processes are able to interfere.

A third method allows one to measure the spin of pair-produced gluinos
decaying to $b\tilde{b}$ \cite{Alves:2006df}, by looking for
asymmetries in the tagged $b$ and lepton system. One may also look at
the shape of the dijet mass distribution in $q\tilde{q}$ decays
\cite{Kramer:2009kp}.
\subsubsubsection{Three-body decays}
In a situation where new particles undergo point-like three-body
decays (involving one invisible particle in the final state), two
methods have been proposed.  One is to simply look at the invariant
mass distribution of the two visible final state particles
\cite{Csaki:2007xm}. The other \cite{Cho:2008tj} uses the $m_{T2}$
variable, discussed above in relation to mass measurement, to assign
the four-momentum of the invisible particle. The distribution of
assigned momenta is peaked around the true value. Once the momentum
has been assigned, one can construct both invariant mass distributions
and the Dalitz plot, from which the spin can be inferred. It is worth
remarking that this so-called `MAOS' (for $m_{T2}$ assisted on-shell)
method, which enables one to reconstruct in a statistical fashion the
final state and, {\em ergo} a frame of reference, should be of much
wider applicability.
\subsubsubsection{Cross sections}
The last method \cite{Datta:2005vx,Kane:2008kw,Hallenbeck:2008hf} we
mention simply utilizes the fact that the production cross sections
for particles have a strong dependence on the spins of the particles
involved. Unfortunately, cross sections also depend strongly on
masses, so precise mass measurements will be necessary prerequisites
to such a method. This is exacerbated by our inability to measure or
model cross sections precisely at hadron colliders.
\subsubsection{Polarization measurements}
One idea, in a supersymmetric scenario, is to measure the polarization
of the tops arising from stop pair production followed by decays to
the top and the lightest neutralino \cite{Perelstein:2008zt}.  The top
polarization may be inferred from the distribution of the various
decays products of the top, be they hadronic or leptonic.  If
measurable, the polarization would yield information on the mixing
angle between the right- and left-handed stops. This angle is of some
interest in connection with the little hierarchy problem
\cite{Perelstein:2007st}.

For methods based on the $\tau$ polarization, see, {\em e.g.}
\cite{Godbole:2004mq,Guchait:2002xh,Choi:2006mt,Nattermann:2009gh}.

\subsubsection{Measurement of ratios of branching ratios}
The other promising observable for the determination of the stop
mixing angle $\cos\theta_{\tilde{t}}$ turns out to be the ratio of
different branching ratios in stop decays.  In \cite{Weiglein:2004hn}
it has been shown that the different branching ratios of stop decays
are sensitive to the mixing angle. In ~\cite{Rolbiecki} it is
discussed that $\tilde{t}_1$-decay into $\tilde{\chi}^{\pm}_1$ and
$\tilde{\chi}^0_{1,2}$ can be a sensitive probe of the stop mixing
angle and in some cases also of the CP-violating phase. The origin for
such sensitivity is the different coupling structure of $\tilde{t}_R$
and $\tilde{t}_L$ to gauginos and higgsinos.
Since the
measurement of the branching ratios can turn out to be very
challenging, the ratio of different branching ratios is more
promising. 
Defining
\begin{eqnarray} \label{eq:ratios}
R_1 = \frac{BR(\tilde{t}_1 \to 
\tilde{\chi}^+_1 b)}{BR(\tilde{t}_1 \to \tilde{\chi}^0_{1}t)},
\qquad R_2 = \frac
{BR(\tilde{t}_1 \to \tilde{\chi}^+_1 b)}{BR(\tilde{t}_1 \to \tilde{\chi}^0_{2}t)
}, 
\qquad R_3 = \frac{BR(\tilde{t}_1 
\to \tilde{\chi}^0_{1} t)}{BR(\tilde{t}_1 \to \tilde{\chi}^0_{2} t)}\,
\end{eqnarray}
as observables and using 1000 events of stop production, a
$\chi^2$-fit has been performed for the scenario SPS1a' to all 3
ratios. One obtains a two-fold ambiguity that needs imput from other
measurements and observables to be resolved. Assuming the correct solution can be pinned down, 
this leads to a
determination of the mixing angle and of the stop mass
\begin{eqnarray}
&&\cos\theta_{\tilde{t}}=0.56\pm 0.03,\\
&&m_{\tilde{t}_1}=366\pm 3~\mbox{\rm GeV},
\end{eqnarray}
where 1-$\sigma$ statistical error has been included so 
far~\cite{Rolbiecki}, see Fig.~\ref{fig-stop}. 
\begin{figure}[!ht]
\begin{center}\setlength{\unitlength}{1cm}
\begin{picture}(10.5,6.7)
\put(-2,0.3){\mbox{\includegraphics[width=0.45\textwidth]{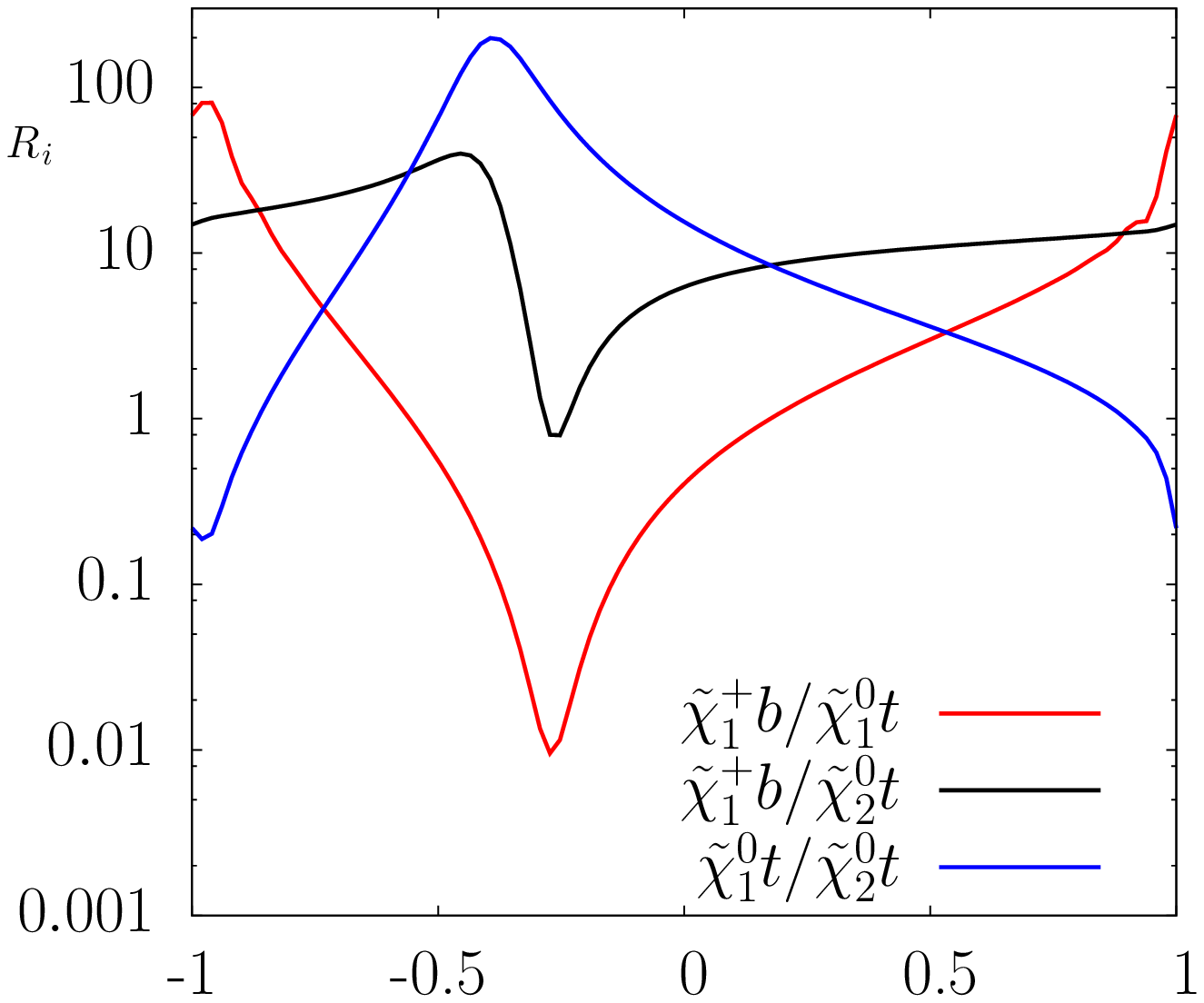}\hskip 0.7cm}}
\put(4.,.5){\small $\cos \theta_{\tilde{t}}$}
\put(6,0.3){\mbox{\includegraphics[width=0.40\textwidth]{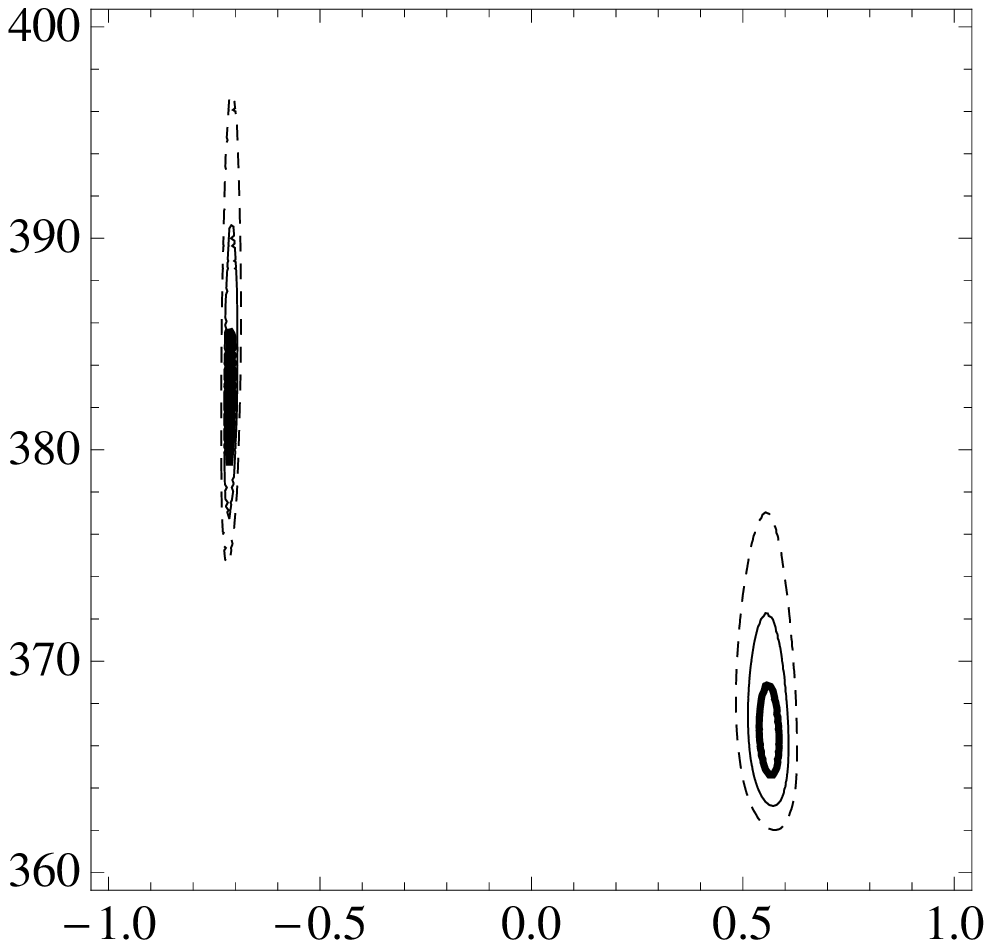}\hskip
 0.5cm}}
\put(6.,6.){\small $m_{\tilde{t}_1}$} 
\put(12.4,.5){\small $\cos \theta_{\tilde{t}}$}
\end{picture}
\caption{Ratios of different branching ratios (left panel),
  Eq.~\ref{eq:ratios},  in scenario SPS1a' as a function of the stop mixing
  angle $\cos\theta_{\tilde{t}}$. 
The $\chi^2$-test (right panel) leads to a determination of the stop mixing
angle at the 5\%-level and of the mass at the 1\%--2\%-level~\cite{Rolbiecki}
(assuming 1000~well identified events).
The bold, solid and dashed lines denote the $1\sigma$, $2\sigma$ and $3\sigma$
contours, respectively.
\label{fig-stop}  }
\end{center}
\end{figure}
Further studies are needed that will take into account pollution from
background and experimental effects.

\subsubsection{On-Shell Effective Theories}
All the methods discussed above rely on kinematic properties, and are
thus largely model-indepen\-dent. This is perhaps just as well, given
that we don't yet know what the new physics is. Nevertheless,
kinematics can only take us so far; at some point we will want to
introduce a model and compare it to the data. Indeed, the nature of a
hadron collider, with large backgrounds and slowly-varying
distributions, makes it rather likely that we will have to do this
sooner rather than later in our analysis of the data.

The question then is: what model should we choose? The traditional
paradigm in particle physics has been to specify a lagrangian, which
of course provides a complete description of the
physics. Unfortunately, the lagrangians that have been suggested for
physics beyond the SM are typically very complex, involving many new
fields and often hundreds of free parameters. Even though such a
Lagrangian may well describe the LHC data, one may wonder whether such
a complicated description is something of an overkill, given the fact
that measurements at the LHC will neither be particularly precise, nor
wide-ranging.

One recent suggestion has been to use so-called `On-shell effective
theories' \cite{ArkaniHamed:2007fw} as a more simplistic description
of new physics than a Lagrangian. To define a model of this type, one
needs only to specify the particles, as well as the principal channels
by which those particles are produced and decay. The particle masses
and the production cross sections and decay branching ratios are
included as free parameters to be fitted to the data. The initial
results seem reasonably encouraging, with indications that masses can
be determined to $O(50 GeV)$ or so. Clearly, as the number of observed
production and decay channels proliferates, the utility of such a
method diminishes rapidly, but at least in the early period of LHC
data, this approach may well turn out to be of use, at least as a
stepping-stone in guiding us towards the right Lagrangian.

\subsubsection{Numerical Tools} \label{lhc-meas}
A number of numerical analysis tools were presented during the
workshop including: Sfitter \cite{Lafaye:2004cn} and Fittino
\cite{Bechtle:2004pc} for MSSM fits; Gfitter \cite{Flacher:2008zq} for
global electroweak fits; 
MasterCode~\cite{Buchmueller:2007zk,Buchmueller:2008qe,Buchmueller:2009fn} for
fits using the combination of  
all current experimental data; fits to the constrained MSSM using Bayesian
\cite{Allanach:2005kz,Allanach:2006cc,Roszkowski:2007fd} or
frequentist
\cite{Ellis:2004tc,Ellis:2006ix,Ellis:2007fu,Heinemeyer:2008fb}
approaches, see also sect.\ref{sect-lc-cosmo}.
\subsection{The LHC--Dark Matter Connection}\label{cosmo}
The Dark Matter (DM) relic density has been measured with very high
precision by the WMAP experiment \cite{Dunkley:2008ie}.  
In order to interpret the
measurement in terms of particle physics, a large effort is being
devoted to experiments for the direct detection of a DM particle
candidate.  An alternative approach would be to produce the DM
particles with high energy accelerators. Given a DM candidate, its
relic density can be calculated from its mass and from the cross
section for its annihilation \cite{Drees:1992am,Jungman:1995df}.  
Several public programs 
\cite{Gondolo:2004sc,Belanger:2006is,Baer:2002fv} are
available to perform these calculations, and have been presented
during the workshop.  A basic result is that, in order to account for
the observed relic density, a DM candidate with a mass scale of
100~GeV and weak annihilation cross-section is needed. This is a
remarkable result, which points to the possibility of identifying the DM
candidate with the stable weak-interacting particle that terminates
the cascade decays of several new physics models.  A typical example
is the LSP in the MSSM.  If all the parameters defining the BSM model can
be measured at a combination of future high energy colliders, the
annihilation rate of the DM candidate particle can be calculated, and
the relic density predicted, based on the collider data.\par
Concentrating on SUSY MSSM, a natural question is whether the
constraints on the MSSM parameter space achieved thorough the
measurements discussed in the previous sections can significantly
constrain the relic density. This can be done either based on the
constraints on the model parameters calculated based on the numerical
tools described in section \ref{lhc-meas}, or directly based on the
measurements we expect to be able to perform at colliders for a given
benchmark point.  We find in the literature two exercises which
explore the latter approach in great detail, based on a version of the
MSSM with around fifteen parameters.  The work of \cite{Nojiri:2005ph} is focused on a very
favourable SUSY point and restricted to LHC data, whereas the work in
\cite{Baltz:2006fm} is of broader scope, addressing several different
final-state topologies and exploring both the potential of the LHC,
and the combination of the LHC and a future Linear Collider. The result is
that the LHC data alone can give a well defined prediction only for
one of the studied topologies, the case corresponding to neutralino
annihilation through exchange of a light selectron or smuon, whereas
in all other considered cases a combination of LHC and linear Collider
measurements will be needed. 
Another recent study~\cite{cosmo-lc}
shows, for instance, how accurate the gaugino component of the dark
matter candidate has to be determined in order to avoid substantial
misinterpretations, cf. Section \ref{sect-lc-cosmo}. 
Such cases of parameter points impose a big
challenge for LHC predictions and further input from a
linear collider will be required. 
This is a line of investigation 
which deserves further investigation, as its success would effectively
unify the fields of particle physics and cosmology. 
\subsection{Future Lepton Colliders}\label{linear}
\subsubsection{Introduction and characteristic features at linear 
colliders\label{sect-lc-intro}}
Due to the clean environment, unbiased measurements with high precision
are expected to be 
only achievable at a future linear collider.
Therefore linear collider physics 
will perfectly complement and extend the
physics potential of the
LHC~\cite{Weiglein:2004hn}
and a strong physics case can already  be derived from present knowledge.
The discovery
potential of a linear collider is not only defined by its energy scale
but in particular by the precision of its measurements. 

As one example for the need of precision measurement that is well-defined
already today is physics of the top quark.
Applying threshold scans for top quark mass measurement a precision of
$m_t=100$~MeV is predicted, see~\cite{talk-klaus}.
The measurement of the top mass is of utmost
importance for the electroweak precision observables and predictions of the Higgs 
sector, since $m_t$ enters up to $m_t^4$ at the quantum level. It causes intrinsic
theoretical uncertainties that will only be matched with a measured
$\Delta m_t$ at the linear collider~\cite{Heinemeyer:2003ud}.
Concerning the strong linear collider physics potential in the top and 
Higgs sector, for instance for measuring 
absolute couplings, branching ratios and widths, see~\cite{talk-klaus,talk-moenig}.
Due to the broad spectrum of different experimental analyses 
and the unprecedented potential for precision 
measurements at the linear collider, it is hard 
to imagine any LHC result that would not require an
extensive $e^+e^-$ physics programme~\cite{talk-klaus} 
(see for instance ~\cite{AguilarSaavedra:2001rg} and references therein).

Characteristics of a linear collider are the precisely defined and
known centre-of-mass energy, the clean environment,
the possibility to avoid any hardware trigger, and
the ability to fully reconstruct both leptonic and hadronic final states.
The additional features of providing a
precise tunable energy together with polarized beams open a broad
spectrum of available different analyses and prepare the LC ideally for the
`unexpected'.  Threshold scans allow most precise mass measurements
and open up also experimental possibilties for probing
the existence of CP--violation,
for instance, in the chargino sector of SUSY.  
Another spectacular experimental tool is the availability of
the polarization of both beams which has a rich field of physics 
applications~\cite{MoortgatPick:2005cw}. It 
offers, for instance, unique access to the chiral structure of all kind of
interactions and strengthens strongly
the potential to perform model-independent analyses.

\subsubsubsection{Parameter determination in the SUSY sector\label{sect-lc-cosmo}}
Any model-independent determination of parameters in new physics
models as, for instance, in SUSY, is expected to be achievable
only in the clean linear collider enviroment. Applying the full
programme of polarized beams in their different
configurations, threshold scans and continuum measurements
and in combination with LHC results on the coloured spectrum it
may even be possible to extrapolate the high-energy behaviour of new
physics and, for instance, perform tests of the SUSY breaking
mechanisms in supersymmetric and string models~\cite{Blair:2005ui}. 
The programs 
Fittino~\cite{Bechtle:2004pc} and 
Sfitter~\cite{Lafaye:2004cn}, see sect.~\ref{lhc-meas},
determine SUSY parameters in a multi-parameter fit for several models, including 
extended SUSY models. The strong impact of linear collider results when combined with results 
from the LHC at an integrated luminosity of 300~fb$^{-1}$ for the determination of the 
parameters in the MSSM with 18 parameters can impressively be seen in 
Fig.~\ref{fig-bechtle},~\cite{Bechtle:2009ty}.
\begin{figure}[!ht]
\begin{center}
\includegraphics[width=0.45\textwidth]{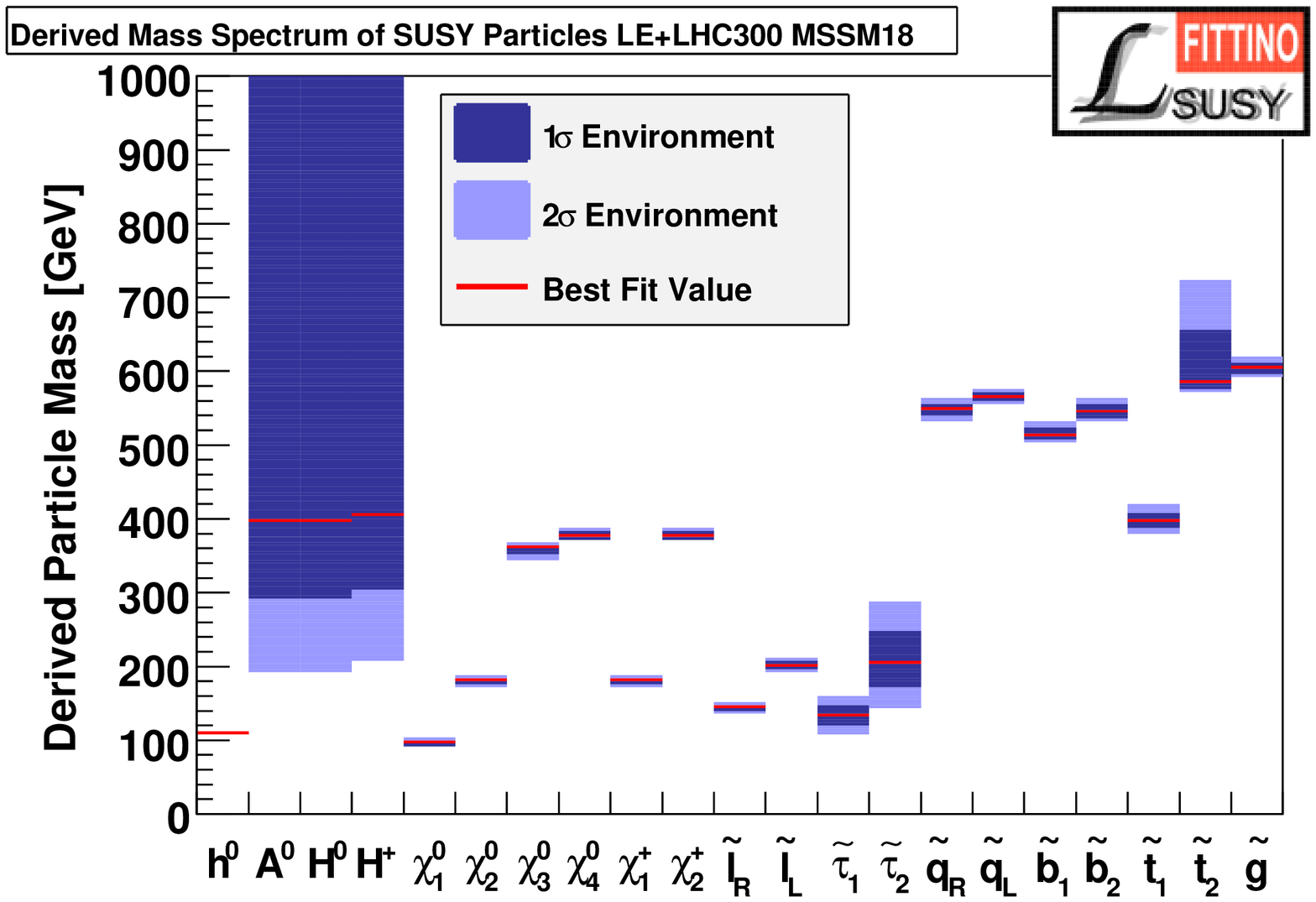} \hskip 0.5cm
\includegraphics[width=0.45\textwidth]{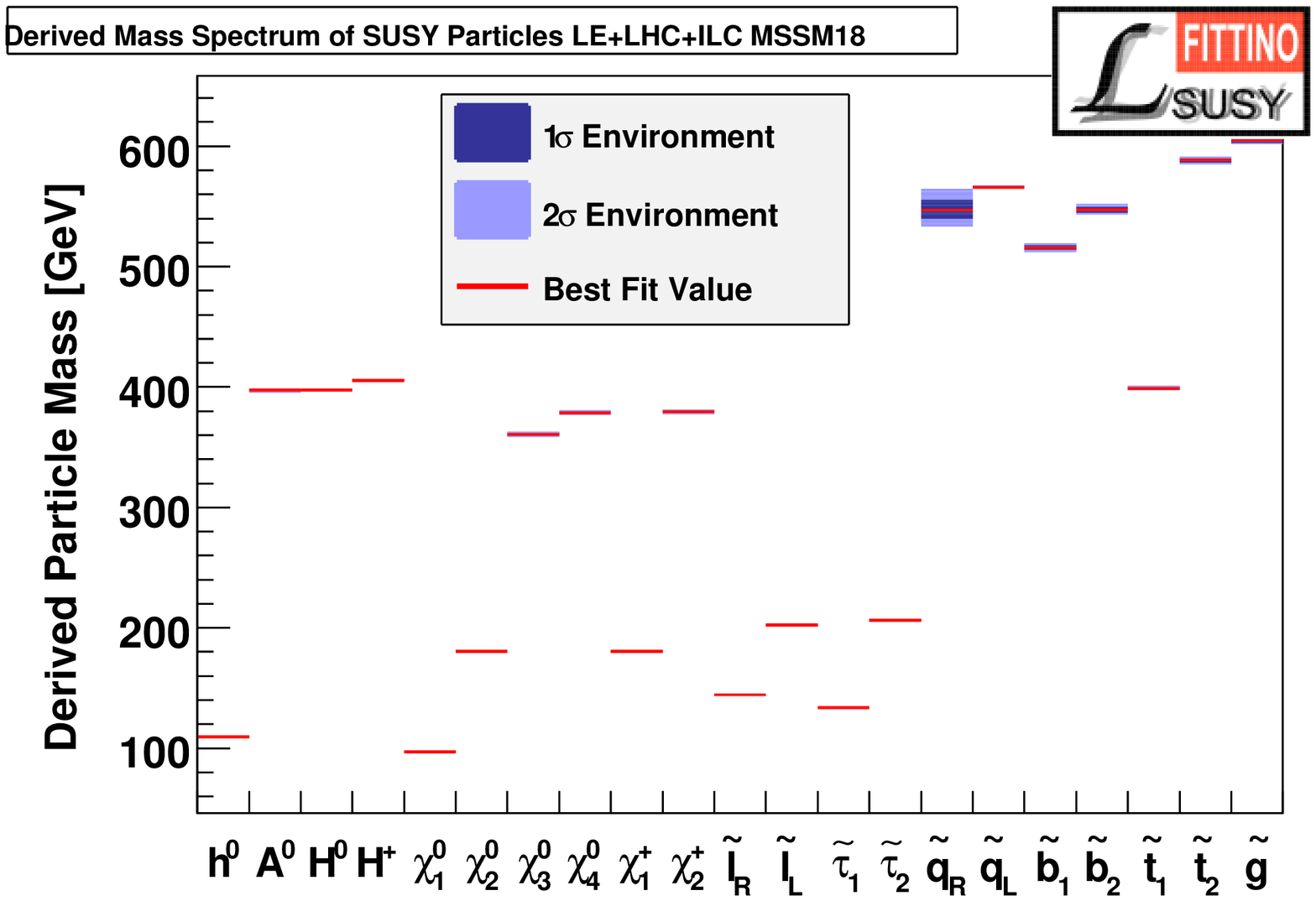}
\caption{Left panel: 
SUSY mass spectrum derived from low-energy observables
and expected LHC measurements at 
${\cal L}^{\mathrm{int}}=300\,\mathrm{fb}^{-1}$ for the MSSM18 model.
The uncertainty ranges represent model dependent uncertainties
of the sparticle masses and not direct mass measurements~\cite{Bechtle:2009ty}. 
Right panel: Derived mass distributions of the SUSY particles in the MSSM18
model, using low energy observables, expected 
results from LHC with ${\cal L}^{\mathrm{int}}=300\,\mathrm{fb}^{-1}$ and
expected results from ILC~\cite{Bechtle:2009ty}.
\label{fig-bechtle}  }
\end{center}
\end{figure}

\subsubsubsection*{\it Dark matter predictions at required accuracy}
Accurate parameter determination is also crucial for testing the dark matter properties of
the new particles. 
With entering the
cosmology in an era of precision measurements, it might be possible to match
results from missing energy signals with precise dark matter predictions.
However, high precision in the model-independent determination
of the dark matter gaugino eigenstate is necessary to avoid severe misinterpretations in parts 
of the SUSY parameter space where, for instance,  the LSP changes abruptly its interaction 
character~\cite{cosmo-lc}. Such abrupt changes can happen due to the swapping of the mass 
eigenstates.
As shown in Fig.~\ref{fig-victoria} (left panel) 
for a general MSSM model without unfication 
assumptions, a high precision on the measurement of $M_2$, only achievable at the linear collider, is required.
Such a precision is crucial in order to
reliably test whether the measured SUSY
scenario leads to a contribution of $\tilde{\chi}^0_1$ 
that is consistent with the WMAP bounds or not, see
Fig.\ref{fig-victoria}(right panel)~\cite{cosmo-lc}.  
\begin{figure}
\begin{center}
\includegraphics[width=0.45\textwidth]{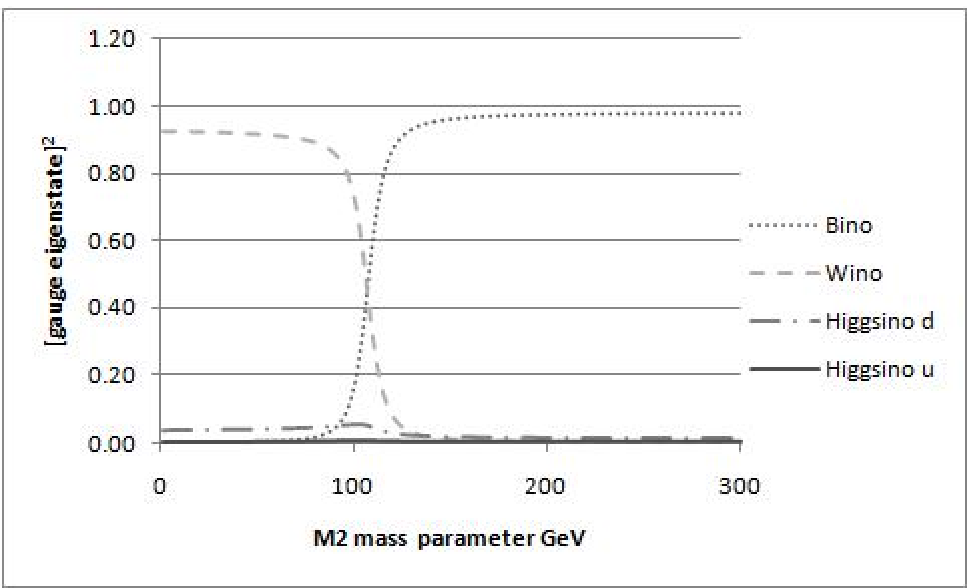} \hskip 0.5cm
\includegraphics[width=0.51\textwidth]{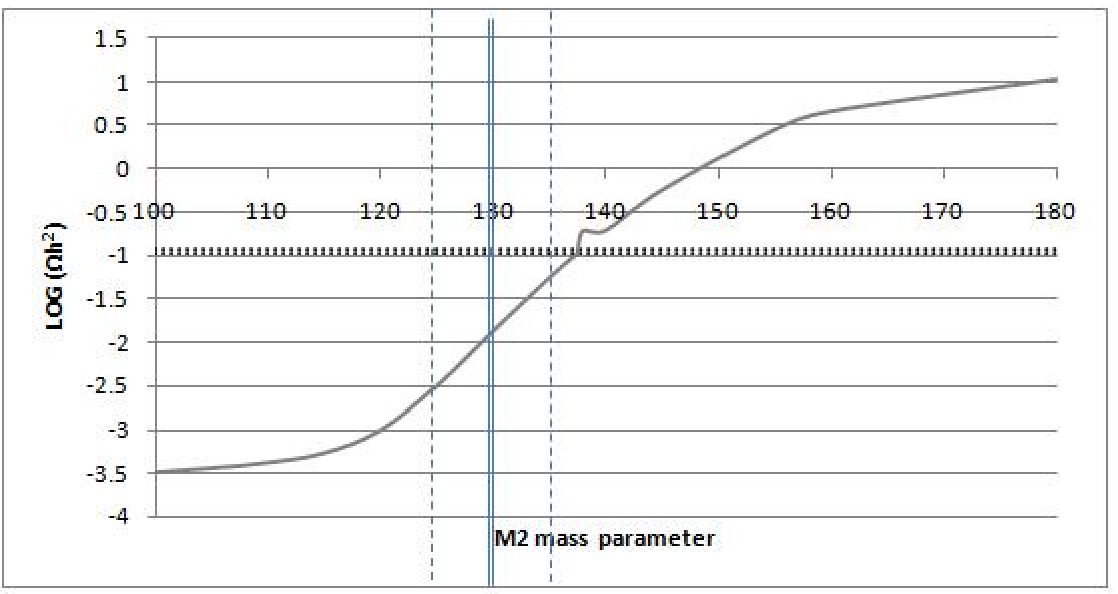}
\caption{Left panel: 
The gaugino and higgsino components of the LSP in SPS1a' scenario where only the $M_2$ 
parameter is varied. Around $M_2~\sim 104$~GeV the mixing character of the LSP changes abruptly 
due to the swapping of the mass eigenstates.
Right panel: the corresponding change in the dark matter contribution of the LSP for this scenario 
with variation of $M_2$~\cite{cosmo-lc}. 
\label{fig-victoria}  }
\end{center}
\end{figure}
For some further
highlights of expected physics results at the linear collider, 
see~\cite{talk-klaus,AguilarSaavedra:2001rg} and references therein.

\subsubsection{Technical design status and technology options}
Two accelerator technologies are discussed for a linear collider.
\begin{itemize}
\item[a)] In the energy range of
$\sqrt{s}=0.5$--1~TeV the superconducting technology, as implemented in the
International Linear Collider (ILC), is
the mature 
concept~\cite{Augustin:2004nq} to provide the expected unique
scientific opportunity and enter a new precision frontier. 
The Reference Design Report (RDR)
has been finished in 2007~\cite{Brau:2007sg}.
No technical obstacles are
predicted for the ILC design, but careful studies of 
possible cost saving changes for the current design have to be done
with regard to their
possible impact on the physics potential of the machine. 
The Technical Design Phase (TDP) of the ILC is under the responsibility of the
`Global Design Effort' (GDE). 

Starting the industrial engineering phase, the optimization of, for instance, the 
cavities shapes are under study. Higher gradients up to 59 MV/m have already
been 
achieved in single cells with so-called 're-entrant' cavities, developed
by Cornell and KEK. Concerning the industrial cavity production, an average
of 36 MV/m in nine-cell cavities has been achieved~\cite{talk-brian}.

\item[b)] Further exploitation of new physics scenarios as well as the
more precise determination of the electroweak symmetry breaking
mechanism may require a linear collider design with a higher cms
energy in the multi-TeV range as foreseen for the CLIC
design~\cite{talk-battaglia} (see also Sect.~\ref{wg1}).  

High mass particles in the range beyond 1~TeV, e.g. heavy squarks in SUSY,
can be studied at CLIC, should they be discovered at the LHC or inferred at the ILC. It is
also expected to
determine the triple Higgs couplings that are important for the
verification of the Higgs mechanism about a factor 2 more precisely at
higher cms energy than at 500~GeV if a similar clean experimental
environment is achievable at the multi-TeV option.

For achieving higher energies of $\sqrt{s}$ in the multi-TeV range a
normalconducting two-beam acceleration concept is discussed, the Compact
Linear Collider (CLIC). A conceptual design report is foreseen for 2010, where
the key 
feasibility issues of the CLIC technology are foreseen to be demonstrated as
well as the preliminary performance and a first cost estimation.\\[-.7cm]
\end{itemize}
A fruitful ILC/CLIC collaboration has been started to address common
 R\&D issues in the civil engineering \& conventional facilities,
beam delivery system, beam dynamics and detectors, more details
 see~\cite{talk-delahaye}. 
Comparing the potential of the two linear collider technologies several
technical issues have to be taken into account that may have 
 impact on the physics
potential.  Many precision measurements at a linear collider depend
crucially on machine parameter more than on the achievable detector
precision.  For instance, the average energy loss i.e. beamstrahlung,
has impact on the precision achievable via threshold scans and (polarized) cross
sections. Beamstrahlung is predicted to be 2.4\% at ILC with
$\sqrt{s}=500$~GeV (ILC500), 7\% at CLIC with $\sqrt{s}=500$~GeV
(CLIC500) and 29\% at CLIC technology with $\sqrt{s}=3000$ (CLIC3000).
A formidable experimental challenge arises from the short (0.5~ns) bunch
spacing at CLIC. Severe impact on the achievable
precision due to pile-up of soft hadronic interactions can arise unless
unprecedented time-stamping capability both for charged
and neutral particles can be implemented into the CLIC detectors.Detailed
simulations will be needed for achieving 
conclusive results concerning the physics potential of the different designs.
Therefore a staged approach between the different design may be
beneficial~\cite{talk-klaus}. 

\subsubsection{Impact of early LHC results on the LC   \label{sect-lc-syn}}

\noindent Important information from early LHC results could infer
technical but also theoretical requirements for linear collider
physics. Sensitive science areas in this context are, for instance:\\[-2em]
\begin{enumerate}
\item defining further specific LC detector capabilities,
\item providing a way out of worst case scenarios in the interpretation of LHC data 
\item staging the required energy steps at the LC, optimizing 
the running scenarios and outlining linear and hadron collider upgrade
options.\\[-1cm]
\end{enumerate}

\subsubsubsection{Specific detector capabilities \label{sect72}}
Results from LHC data could have an impact on the required ILC 
detector capabilities. For instance, if one finds hints for the existence of new  
CP-violating phenomena as expected 
in supersymmetry, high resolution in the capabilities
of $b$- and $c$-charge tagging may be required for the LC detectors in order to 
optimize respectively the physics potential of the ILC.

Since some of the CP-violating phases are strongly experimentally
constrained by the electric dipole moments of the electron, neutron and
some atoms, only small phases are expected in the
SUSY sector, although large phases can still be accomodated in the free
parameter space of the MSSM,  
see~\cite{talk-sabine,Kraml:2007pr,Hesselbach:2007dq,Kittel:2009fg} and references therein.
Hints for unique signs of
CP-violation may be difficult to detect in the harsh environment of
the LHC. It has been studied 
in~\cite{Langacker:2007ur,Ellis:2008hq,Deppisch:2009nj,MoortgatPick:2009jy} how to detect
the effects of SUSY CP-phases in squark decays into neutralinos at the LHC. 

In~\cite{Ellis:2008hq} 
CP-odd sensitive observables, i.e. asymmetries composed by 
triple product correlations, have been analyzed in detail 
for the process $gg\to\tilde{t}_1\tilde{t}_1$, 
$\tilde{t}_1\to t \tilde{\chi}^0_2$, 
$\tilde{\chi}^0_2\to\tilde{\chi}^0_1\ell^+\ell^-$ and $t\to Wb$ 
at parton and at 
hadronic level. Although this process offers several asymmetries that show 
different sensitivity to CP-phases, so that effects of $\Phi_{M_1}$ and 
$\Phi_{A_t}$ might be distinguishable, the statistics are quite low.
Even in the case of large asymmetries up to O(10\%) at the parton level,
they are significantly reduced by about a factor 4 due to dilution at the
hadronic level. 
Therefore high luminosity is desired, see Fig.~\ref{fig-stop-asy}.
\begin{figure}
\begin{center}\setlength{\unitlength}{1cm}
\begin{picture}(10.5,5.7)
\put(-2,4.8){\mbox{\includegraphics[width=0.3\textwidth,angle=270]{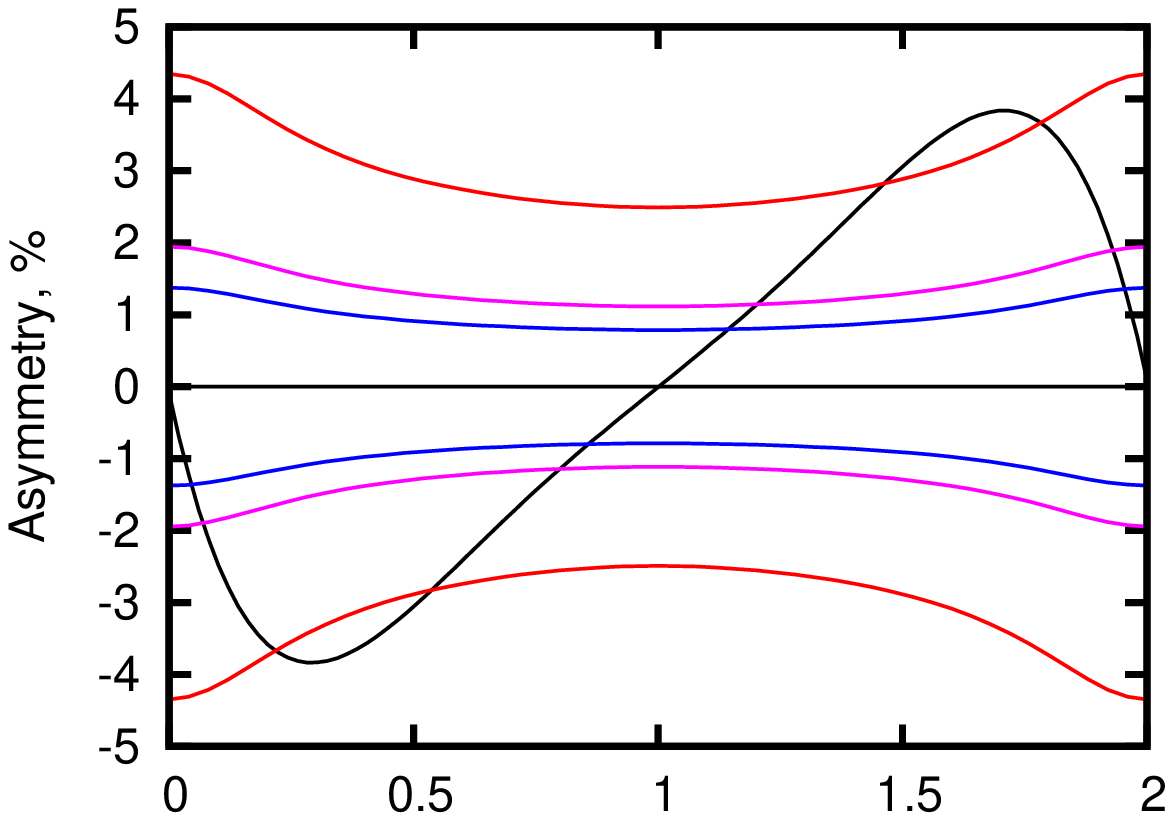} \hskip 0.5cm}}
\put(6,4.8){\mbox{\includegraphics[width=0.3\textwidth,angle=270]{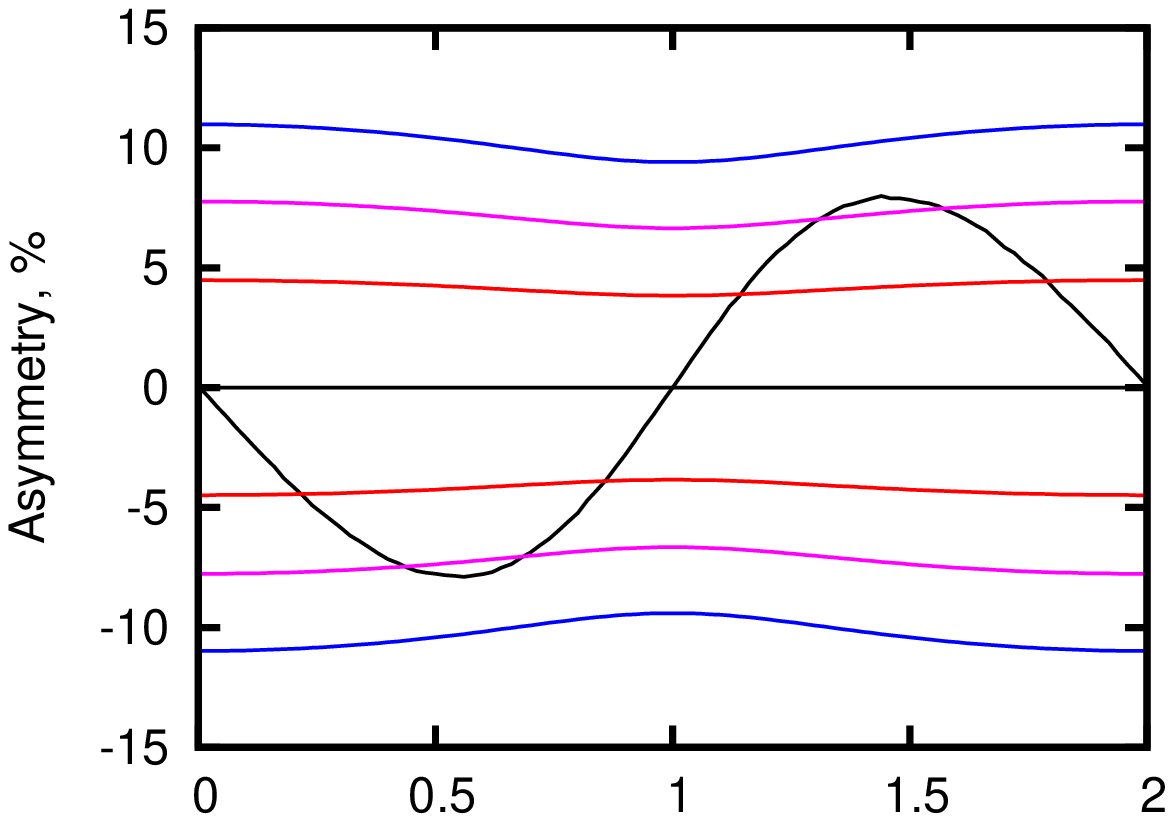}}}
\put(3.,-0.2){$\Phi_{M_1}/\pi$}
\put(.1,4.8){${\cal A}_{CP}\{\vec{p}_t\cdot(\vec{p}_{\ell^+}\times\vec{p}_{\ell^-})\}$}
\put(2.5,4){\small 1$\sigma$-level}
\put(11.,-0.2){$\Phi_{M_1}/\pi$}
\put(8.2,4.8){${\cal A}_{CP}\{\vec{p}_q\cdot(\vec{p}_{\ell^+}\times\vec{p}_{\ell^-})\}$}
\put(10.4,4.){3$\sigma$-level}
\end{picture}
\caption{Left panel: CP-odd asymmetry 
sensitive to $\Phi_{M_1}$ in stop decays at the LHC
 with parton density functions 
included in the production process 
at the 
$1\sigma$-level  with the luminosities, $\mathcal{L}$=(100 fb$^{-1}$, 500 fb$^{-1}$, 
1 ab$^{-1}$)~\cite{Ellis:2008hq}. Right panel:
CP-odd asymmetry  in squark decays
with parton density functions included in the production process
but after momentum 
reconstruction has been performed. The selection cuts have been applied and
the momenta  
of the final state particles have been smeared to replicate
 the LHC detectors effects. 
The coloured lines show the size of the asymmetry needed for a $3\sigma$ observation at the given
 luminosity, $\mathcal{L}$=(50 fb$^{-1}$, 100 fb$^{-1}$, 
300 fb$^{-1}$)~\cite{MoortgatPick:2009jy}. 
\label{fig-stop-asy}  }
\end{center}
\end{figure}
Studying CP-effects in the process,
$qg\to\tilde{q}_L\tilde{g}$, $\tilde{q}_L\to q \tilde{\chi}^0_2$,
$\tilde{\chi}^0_2\to\tilde{\chi}^0_1\ell^+\ell^-$ leads to a
significantly higher statistics and although the dilution reduces the
asymmetry again by about a factor 3-4 when going from the partonic to
the hadronic level, a large range of $\Phi_{M_1}$ may lead to
observable asymmetries even at the $3\sigma$-level, in particular after momentum 
reconstruction of the LSP has been done, see
Fig.~\ref{fig-stop-asy} (right panel). Some selection cuts and smearing effects at the
LHC have already been included~\cite{MoortgatPick:2009jy}.

Another example where LHC results may have an impact on the optimization of the LC detector 
are physics scenarios that require to measure precisely the
polarization of $\tau$'s~\cite{Nojiri:1994it,Bechtle:2009em} in order to, for
instance, 
decompose the caracter of the lightest SUSY particle and provide model 
distinction of different physics scenarios~\cite{talk-roy}, aim to determine 
the SUSY $A_{\tau}$ parameter~\cite{Boos:2003vf} 
or to test $CP$ quantum numbers in the Higgs 
sector~\cite{Desch:2003rw,Desch:2003mw}.  

Particularly challenging scenarios of new physics consist of new
particles that are almost mass-degenerate. In case that LHC results point to
SUSY scenarios where a close mass degeneracies of the light neutral
and charged gauginos, $\tilde{\chi}^{\pm}_1$, $\tilde{\chi}^0_2$ is expected, 
one needs an excellent jet energy resolution in the LC
detector in order to resolve whether the final di-jet pair originates from
a $W^{\pm}$ or a $Z^0$ decay~\cite{Suehara:2009bj} and the LC detector might be optimized 
correspondingly.

\subsubsubsection{Worst case scenarios from early LHC results\label{sect73}}
Although hints from electroweak precision observables point
to a rather light scale of new physics as, for instance, 
in supersymmetry~\cite{Buchmueller:2009fn, Bechtle:2009ty}, 
it may happen that nothing new or only a light SM-like Higgs
has been detected within early LHC data. In such a worst case scenarios
what may be the conclusions for collider physics?

If really no hints for new physics can be seen at early LHC data it
is mandatory to exploit as soon as possible the top quark with highest
priority. This particle plays a key role in
the understanding of the electroweak breaking mechanism since it enters via
loops effects at the quantum level. The first energy stage at a
linear collider may therefore not be $\sqrt{s}=500$~GeV but the $m_t$
threshold of about $\sqrt{s}=350$~GeV. Such a reduction for the required
LC energy in its first stage has, of course, a strong impact on cost issues of the
LC. Another high priority physics in such a worst case scenario will be the
exploitation of electroweak physics at the Z-pole. Only high
luminosity at the Z-pole is required to achieve an unprecedented
precision in the measurements of $\sin^2\theta_{\rm eff}$. Even only small
traces of new physics contributions in the electroweak sector can be detected 
via this observable. One should remember that although a deviation of the
measured $\sin^2\theta_{\rm eff}$ from its SM prediction does not
specify the new physics model, but it would, however, 
give evidence for physics beyond the Standard Model.
Such an information may be crucial for 
continuing collider data analysis at the highest precision level.  

In case the early LHC data detect only a SM-like Higgs signal but nothing
else (see also Sect.~\ref{wg1} for a discussion), 
the foreseen first energy stage of a LC has to be
questioned. Running at the threshold for $Z^0+$ Higgs production would probably offer cost savings
as well as provide a tremendous benefit for physics. In such a 
scenario, it is also advantageous to perform a high luminosity run at
the Z-pole instead of aiming at higher energies in the first stage of the LC.
For such worst case scenarios, studies of $\sin^2\theta_{\rm eff}$ at the
quantum level have been made. These studies clearly demonstrate that a high
precision measurements of $\sin^2\theta_{\rm eff}$ could point to the
existence of supersymmetry, even if the coloured SUSY particles were in
the multi-TeV range and not detectable at the early 
LHC~\cite{Heinemeyer:2007bw,MoortgatPick:2005cw}, see 
Fig.~\ref{fig-arne}. Furthermore, such high precision measurements would indicate 
the new physics scale and point to the required energy stage of the LC.
\begin{figure}[!ht]
\begin{center}
\includegraphics[width=0.65\textwidth]{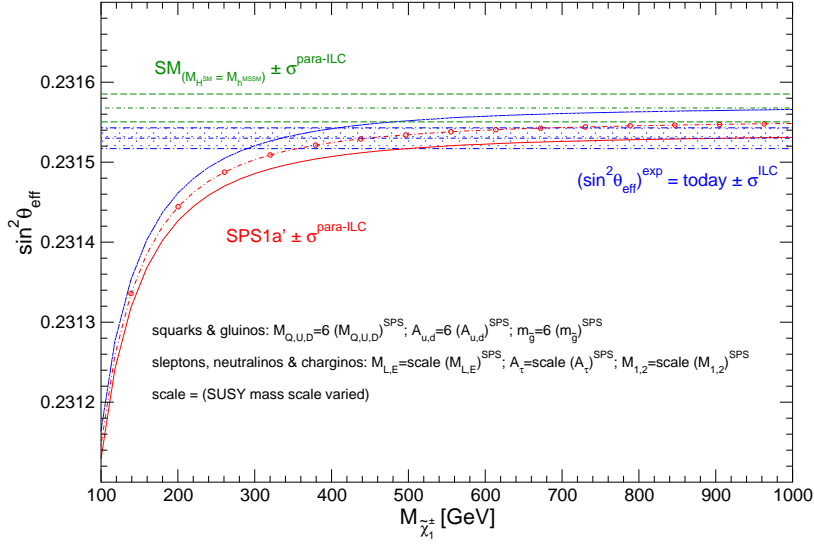} \hskip 0.5cm
\caption{Theoretical prediction for $\sin^2\theta_{\rm eff}$ in SM and MSSM compared to
future ILC precision~\cite{Heinemeyer:2007bw,MoortgatPick:2005cw}.
The SUSY scales are varied with a common scalefactor, 
but squark and gluino masses are fixed to be heavy,
i.e.\ not observable at the LHC.
The anticipated parametric uncertainty of ILC is indicated as part of the
theory predictions. The plot shows the sensitivity to contributions of 
$m_{\tilde\chi^\pm_1}> \sqrt{s}/2$ in such worst case scenarios.
\label{fig-arne}  }
\end{center}
\end{figure}

\subsubsubsection{Impact on running scenarios and upgrade options}
In case new physics is detected in early LHC data, this important
information can be used to optimize the different LC running scenarios and future collider
options. Given that a new physics scenario with many new particles 
has been detected, the technical potential of a linear collider to
perform threshold scans can be optimized. The most precise mass measurements as well as 
hints for CP-violation as, for instance, in SUSY
are available by measuring the corresponding threshold behaviour.
However, each threshold scan uses a specific
amount of the total integrated luminosity and therefore not an arbitrary number of scans
 can be done. Therefore optimization of
threshold scans via defining the suitable energy steps for the scans
as well as selecting the most crucial scans to 
derive the maximal information on the new physics sector are the key issues of 
the LHC input in such cases.

Another important LC sector where immediate input from LHC data is
important is narrowing the choices for the different upgrade options of a
linear collider. 
In some cases, the most efficient way may be to go
straight to the $\gamma\gamma$ option but --as pointed out in the
paragraphs above-- there may also be cases where 
the GigaZ option is the best solution. 
Synergy between early LHC data and a LC with a first energy stage of
    $\sqrt{s}=500$~GeV may also be crucial to predict the needed energy
    scale for energy upgrades of the LC. For instance, even in SUSY
    scenarios where only light gauginos/higgsinos may be accessible but
    the sfermion sector is within the multi-TeV range, the combined
    interpretation of early LHC results together with the precision
    measurements at the 500~GeV~LC allow a model-independent determination of
    the fundamental MSSM parameters and enable a rather accurate prediction of
    the masses of squarks and sleptons in the multi-TeV
    range~\cite{Desch:2006xp}. The required scale for a possible multi-TeV LC
    option 
    can therefore successfully be envisaged, based on early LHC and LC
    results.
Another example where LHC data may influence
the LC running scenarios concerns the suppression of background processes of new physics.
For instance in SUSY, the most severe background
processes are expected to be supersymmetric background
processes themselves. Therefore a crucial information from LHC data
could provide substantial information on
choosing the optimal configuration of $e^+e^-$ beam polarization in order
to reduce the background most sufficiently.

There exist also scenarios in supersymmetry where only the $e\gamma$ and
$e^-e^-$ options of a LC would lead to a significant improvement in SUSY analyses.
For instance, in cases with heavy selectrons, where at least a single
production may kinematically be accessible within these collider options.  
Weighting these options with
regard to a total life time of a future LC, however, it is mandatory to fold in
all possible information coming from early LHC data in order to 
maximally exploit the full potential of a LC with all its variable running
options. Many of these collider options also have impact on costs,
 on collider as well as detector R\&D issues. The early LHC input will
therefore be crucial also from the economical point of view. However, many
detailed studies are still missing in this context and should be addressed in the near future.

\subsection{Summary and conclusions of WG3}
\label{sect:sumwg3}

The energy imbalance in a plane perpendicular to the beam direction is
called Missing Transverse Energy (MET). MET plays a crucial role for
studying Standard Model physics as well as for
detecting physics beyond the SM, for instance, for detecting the stable 
lightest supersymmetric particles in SUSY models with R-parity conservation. 

Concerning the experimental performance, it is expected that the
purely calorimetric MET measurement in ATLAS is more precise
than in CMS due to the better energy resolution and the higher
longitudinal energy containment in the ATLAS hadronic calorimeter. 
In CMS, two methods that take into account other subdetectors have been developed to improve
the purely calorimeter-based
MET: the {\it Track-Corrected MET} and the {\it Particle-Flow MET}. It
is expected that these methods improve the MET resolution up to a
factor of about two. In particular the conceptual differences between
the three approaches will guide and lead to a rather quick
experimental determination of the MET.  The $Z +$~jets process, where the $Z$ decays 
to electrons and muons, is  ideally suited to study the MET performance since it is almost 
background-free and the two leptons  can be measured with excellent precision. This process can also directly be used for data-driven background estimations for processes involving MET.
The $Z\to\tau^+\tau^-$ process
can be used to determine the MET scale to about 8\% accuracy.  A
global good understanding of the detector, however, will be mandatory to infer
on event-by-event basis the total missing transverse momentum
accurately from the measured transverse momentum.

Although it is expected that even for a moderate assumed luminosity of
1 fb$^{-1}$ squarks and gluinos with masses of order 1.3~TeV should be
discovered for a very broad range of R-parity conserving SUSY models,
one should be aware of a few experimental caveats: a SUSY signal has
no distinctive features, such as mass peaks which would separate it from
possible SUSY background processes. Only SM backgrounds may be easily
separated via appropriate cuts.  Therefore one has to deal with
counting experiments where an accurate prediction of the possible
background processes may be mandatory for a correct interpretation.
Excellent prospects exist for mass measurements, in particular in measuring mass differences.
 Several methods have
been proposed to reconstruct the masses of particles in events
including invisible particles. The absolute mass scale that is of most interest 
for planning future experiments, however,
is particularly challenging to determine in the experiments.

Invariant mass endpoints or polynomial constraints of the
observed four-momenta lead to a good determination of masses in
sufficiently long cascade decays.  The only known method working for
any decay topology is the recently renewed method of measuring the
transverse mass $m_T$. This method has originally been performed to
measure $m_W$ precisely. However, in cases with missing energy
complications arise. For instance, due to the expected non-negligible
mass of new invisible particles, the transverse mass is no longer
observable.  This problem may be overcome by calculating the
distributions for some hypothetical values.  It is expected that kinks
are featured exactly at the point where the hypothetical mass equals
the true invisible mass. Generalization to more complex topologies might lead 
to a determination of all unknown masses. Whether ot not this method can actually
successfully be employed in experiments remains to be seen.

Another important topic concerns the spin determination of new particles. Reconstructing 
different reference frames seems mandatory to reveal the underlying
spin information. Invariant mass distributions
of various visible final state particles and the determination of angular
effects in the production processes give complementary information on the spin property.
Exploiting further the spin property, as for instance, via analyzing
the polarization of top quarks
might be important for determining the mixing properties of the SUSY partners, i.e. the 
stop mixing angle. In this context a rather new set of observables turns out to be 
promising for the determination of the stop mixing angle: measuring ratios of different
branching ratios in stop decays: accuracies at percent level in the determination of the stop mass and 
mixing angle might be achievable. Further studies of this promising set of observables
including precise simulations of 
background processes and detector effects are desirable. 
 
Signals with missing energy consist, for instance, of the lightest
stable SUSY particle (LSP). The LSP is a promising cold dark matter
candidate.  Entering a new precision era with the results from WMAP,
precise predictions of the respective relic density and its dark
matter contributions in new physics models are required.  Only for a
rather restricted number of topologies, LHC data alone can provide a
sufficiently accurate prediction of the DM candidate. It is expected
that precision results from a linear collider in combination with LHC
results are needed to finally determine the question of dark matter and
achieve consistency with the current experimental precision bounds.

Results from a linear collider will also be mandatory to reveal the
underlying physics, to determine the underlying SUSY parameters in a
model-independent way and to determine the properties of the new
particles. These results have been obtained by several physics analyses and are also confirmed
by multi-parameter fits implemented in several numerical codes.
 
Two technologies for a linear collider are under
discussion: the ILC and the CLIC concept.  The ILC, with a first
energy stage of $\sqrt{s}=500$~GeV is already on a mature design
stage, feasible and under further responsibility of the global GDE.  The
feasibility of the CLIC concept has still to be demonstrated in the future, but has
the potential to be applicable up to the multi-TeV range.  Results
from early LHC data may be important for specifying the detector
requirements of the LC, for instance in cases where hints for
CP-violation may be found in new physics. 
Early LHC results may also
be decisive for defining the required energy stages of the later phase of the LC. 
Even in worst case scenarios, 
i.e. observing nothing or 
only a SM-like Higgs,    precision measurements at the LC at a first energy stage  
of the top quark threshold or even only 
at the Z-pole are scientifically well motivated and have a large potential for revealing effects of new physics.
Since the LC has its great potential in discoveries via precision measurements,
future design considerations of a LC may therefore
take into account a possible technical impact of design issues
on the subsequent precision potential. 
A reliable and economic prediction of the 
required high energy scale of a future TeV machine may only be achievable 
if LHC data are interpreted in combination with precision results from a LC in its first energy 
stage of $\sqrt{s}=500$~GeV.
Therefore a staged 
approach of a future LC, providing precision physics from the Z-pole up to the new physics scale, 
seems to be reasonable and highly desirable.

 }
\newpage
{\setcounter{equation}{0}
\setcounter{figure}{0}
\setcounter{table}{0}

\def\ie{{\it i.e.}}
\def\eg{{\it e.g.}}
\def\etc{{\it etc}}
\def\etal{{\it et al.}}
\def\ibid{{\it ibid}.}
\def\lsim{\mathrel{\mathpalette\atversim<}}
\def\gsim{\mathrel{\mathpalette\atversim>}}
\def\to{\rightarrow}
\def\lsim{\raise0.3ex\hbox{$\;<$\kern-0.75em\raise-1.1ex\hbox{$\sim\;$}}}
\def\gsim{\raise0.3ex\hbox{$\;>$\kern-0.75em\raise-1.1ex\hbox{$\sim\;$}}}
\allowdisplaybreaks

\def\llpp{\ell^+\ell^+}
\def\llmm{\ell^-\ell^-}
\def\llpm{\ell^\pm\ell^\pm}
\def\llij{\ell^+_i \ell^+_j}
\def\HH{H^{++}H^{--}}
\def\WW{W^{\pm}W^{\pm}}
\def\yll{Y_{\ell\ell}}
\def\br{{\rm BR}}
\def\vd{v^{}_\Delta}
\def\lsim{\mathrel{\raise.3ex\hbox{$<$\kern-.75em\lower1ex\hbox{$\sim$}}}}
\def\gsim{\mathrel{\raise.3ex\hbox{$>$\kern-.75em\lower1ex\hbox{$\sim$}}}}
\def\etmiss{E\!\!\!\!\slash_{T}}
\def\ptmiss{p\!\!\!\slash_{T}}
\def\na{{\chi^0_1}}
\def\nb{{\chi^0_2}}
\def\ni{{\chi^0_i}}
\def\ca{{\chi^\pm_1}}
\def\cb{{\chi^\pm_2}}
\def\ci{{\chi^\pm_i}}
\def\s{\tilde}
\def\nn{\noindent}
\def\non{\nonumber}
\def\ie{{\it i.e.}}
\def\eg{{\it e.g.}}
\def\etc{{\it etc}}
\def\etal{{\it et al.}}
\def\ibid{{\it ibid}.}
\def\tev{\,{\rm TeV}}
\def\gev{\,{\rm GeV}}
\def\mev{\,{\rm MeV}}
\def\ev{\,{\rm eV}}
\def\to{\rightarrow}
\def\slash{\not\!}
\def\beq{\begin{equation}}
\def\eeq{\end{equation}}
\def\be{\begin{equation}}
\def\ee{\end{equation}}
\def\bea{\begin{eqnarray}}
\def\eea{\end{eqnarray}}
\def\lsp{\chi_1^0}
\def\dm{\Delta m}

\section{WG4: Other new physics signatures}
{\it 
A.~De\,Roeck,
T.~Han,
J.L.~Hewett,
S.~Riemann
(convenors)\\ 
G.~Azuelos, 
M.~Carena, 
K.F.~Chen, 
H.~Dreiner, 
A.~Giammanco, 
S.~Gopalakrishna,
W.S.~Hou, 
G.~Isidori,
J.~Kalinowski,
E.~Kou, 
D.~Milstead,
T.G.~Rizzo,
S.~Sultansoy,
B.~Webber
}

\bigskip
In this section we examine the ability of the Tevatron, LHC, and future
facilities to discover and interpret 
new phenomena that does not involve a missing energy signature.  We discuss
the production  
of a new Z boson as well as other states that lead to a leptonic resonance, the fourth family,
testing seasaw mechanisms at colliders, exotic signatures of new physics, 
black hole production, and the impact of high precision flavor physics.


\newcommand{\lhcten}{LHC$_{10}$}
\newcommand{\lhchl}{LHC$_{300}$}

\subsection{Introduction and scenarios}

A strategy of planning for the future is to consider a wide variety of
possible new phenomena and discern which types of facilities would be best to
first discover new physics and then to elucidate its properties.  This
approach will hopefully prepare us for a broad set of possible signatures and
for the surprises that Nature undoubtedly has in store, whatever they may be. 

The goal of the investigations in this section was to explore signatures for
new physics that do not involve missing energy or are not related to the
production of a Higgs-like boson.  These include: 
\begin{itemize} 
\item Leptonic and other resonances
\item Multi gauge boson resonances
\item Leptoquark type signatures
\item Fourth generation and exotic quark production
\item TeV scale gravity signatures
\item New signatures such as heavy stable charged particles.
\item New physics related to flavor physics
\end{itemize}
These scenarios were covered in review talks or in discussion sessions during the workshop; a number of 
these topics are discussed in this report. A benchmark signature for new physics is a leptonic resonance
from, {\it e.g.}, a new gauge boson ($Z'$) and is discussed here in detail.  This signal
is illustrative of what can be learned at the LHC, but also of what will remain elusive, and how we can 
advance our knowledge with future machines. Another detailed study was performed for the presence
of a 4th generation
of quarks and leptons. Here, the LHC may discover or exclude this scenario in the early stages of
operation, with implications for a future collider.  We next report on the possibility of distinguishing at
colliders the various approaches of implementing the seasaw mechanism in the neutrino sector.  We then discuss
the ability of LHC detectors, with implications for future experiments, to observe exotic signatures 
of new physics such as stable charged particles, stopped particles, and non-pointing photons.  The production
of micro black holes is a possibility in theories with visible extra spatial dimensions at the Terascale.
The signatures for this reaction are striking, although there are numerous associated theoretical
uncertainties which could be pinned down by experimental measurements.  Lastly, we discuss the implications
of ultra-high precision measurements in the heavy quark sector and the role such measurements play
in the discovery and elucidation of new interactions.

\subsection{\boldmath{$Z'$} production at future colliders}
\label{sec:tom}

\bigskip
Once the LHC turns on, one of the cleanest potential signals for new
physics beyond the Standard  
Model (SM) will be a $Z'$-like resonance in the dilepton/Drell-Yan channel. Such an object 
is predicted to exist in an ever-widening set of new physics 
scenarios~\cite{rev1,rev2,rev3,rev4,rev5,rev6,rev7,rev8}. 
In some cases the $Z'$-like object 
is also accompanied by an analogous $W'$-like state whose presence will help us to identify the nature of 
the underlying physics. Present constraints from the Tevatron, employing the standard-candle Sequential 
Standard Model (SSM) 
scenario wherein the $Z'$ and $W'$ are just heavier versions of the usual SM states, imply that the masses 
of such particles typically lie above $\sim 1$ TeV. If their couplings happen to be somewhat weaker than  
the typical electroweak strength, far lighter $Z'/W'$-like states may exist which could have 
been missed at the Tevatron.  
The LHC, even running at 10 TeV and with an initial integrated luminosity of 
$100-200$ pb$^{-1}$ has a chance to make a $Z'$ or $W'$ discovery as can be seen in Fig.\ref{fig1}.

\begin{figure}
\centering
\includegraphics[width=7cm,height=8cm,angle=90]{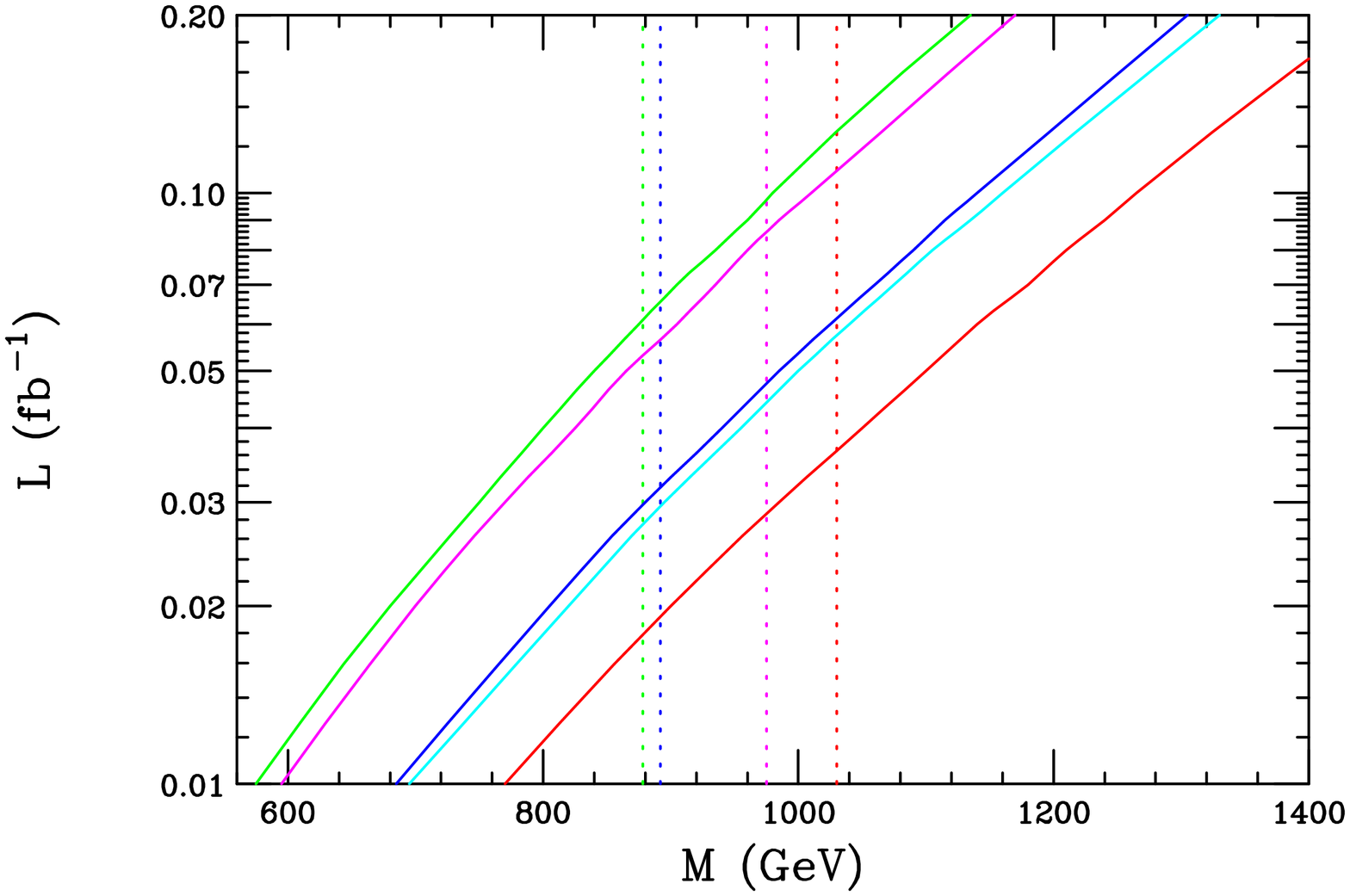}
\includegraphics[width=7cm,height=8cm,angle=90]{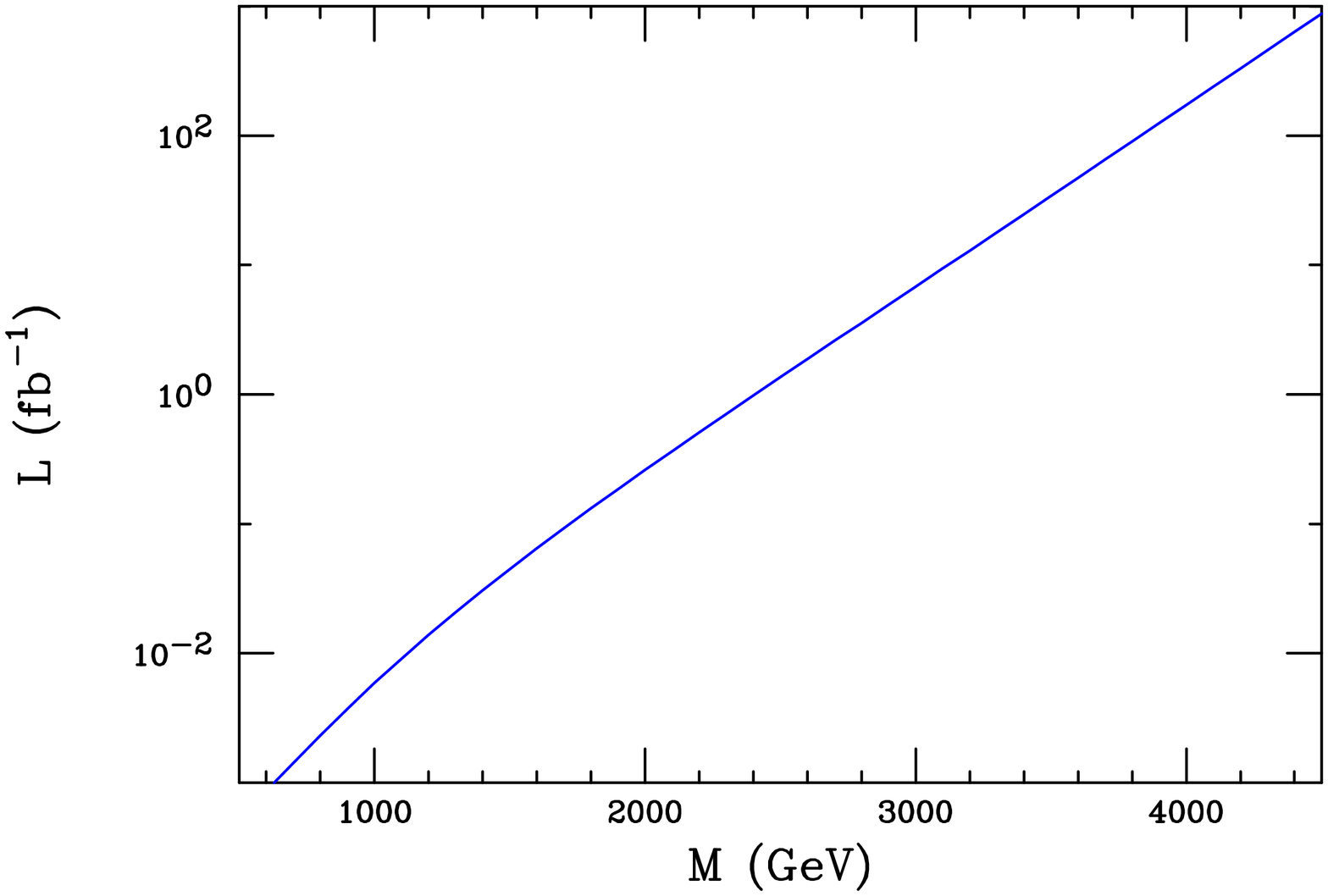}
\caption{$5\sigma$ discovery reaches for (left) $Z'$ in the $\psi$(green), $\chi$(cyan), $\eta$(magenta), 
Left-Right Model(blue) and SSM(red) cases and for (right) $W'$  (SSM case)at the 10 TeV LHC as a function of the 
integrated luminosity. The vertical dotted lines are the present Tevatron bounds for the corresponding 
color-coded models. }
\label{fig1}
\end{figure}

Resonances that are very weakly coupled to the SM fields are present in many models; in many cases their SM couplings are 
generated only via mixing with one or more of the SM gauge bosons. 
In this case, the state will be rather narrow and may be hard to find due to issues of mass resolution. The LHC, 
however, can go fairly deep into the small coupling parameter space provided sufficient luminosity is available. For 
example, a 1 TeV SSM-like $Z'$ with a coupling $\sim 1/20$ of the usual SM strength should be easily visible above the 
SM background at the 14 TeV LHC with a luminosity of 100 fb$^{-1}$.

Once a $Z'$ state is discovered we will want to discern its properties. Our goal will be to try to identify 
the underlying theoretical structure from which it arose. In many models the new resonance is accompanied by 
a number of other new particles. Here we will try to address what we can learn from the new resonance itself. 

The first thing to determine  
is its lineshape, \ie, its mass and width and whether or not it is a (single) Breit-Wigner (BW) resonance. Unparticle 
resonances provide a good example of a non-Breit-Wigner lineshape that we may hope to distinguish from something 
more conventional{\cite {Rizzo:2008vr}}. The shape of this un-resonance is controlled by the unparticle mass and coupling 
strength as well as the effective anomalous dimension.  The unusual lineshape of this un-resonance
can be seen in the left panel 
of Fig.\ref{fig2} for various values of the parameters. 
Detector resolution can be of significant importance in performing the detailed 
measurements necessary to identify this non-BW structure. 
For some values of the parameters it is 
clear that the non-Breit-Wigner shape will be apparent although a detailed study has yet to be performed to 
determine the parameter ranges for which this differentiation can be performed. 

Another possibility to consider is that there are 2 or more (almost)degenerate resonances 
which may interfere with each other, as well as the SM exchanges, thus distorting the expected line-shape. 
This can happen, {\it e.g.}, in the case of string resonances or in extra-dimensional models where KK excitations of 
both the $\gamma$ and $Z$ can appear. This scenario also needs further study. 

Next, we will want to determine the particle's spin;  as is well-known, 
this can be acheived by examining the angular distribution of 
the final state leptons.  While only $\sim 10$ clean 
events will be necessary to discover a $Z'$-like state, hundreds of events will be needed to perform a 
measurement of the angular distribution. This implies that the `reach' for spin determinations is significantly 
lower than for discovery and likely to be less than $\sim 3$ TeV for the SSM $Z'$ with design machine
parameters.  In this case, if the LHC finds  evidence for a Z'-like object in the data, an
LHC luminosity upgrade will be extremely beneficial for a more detailed understanding of this new object.

\begin{figure}[htbp]
\centering
\includegraphics[width=6.7 cm,height=7.7 cm, angle=90]{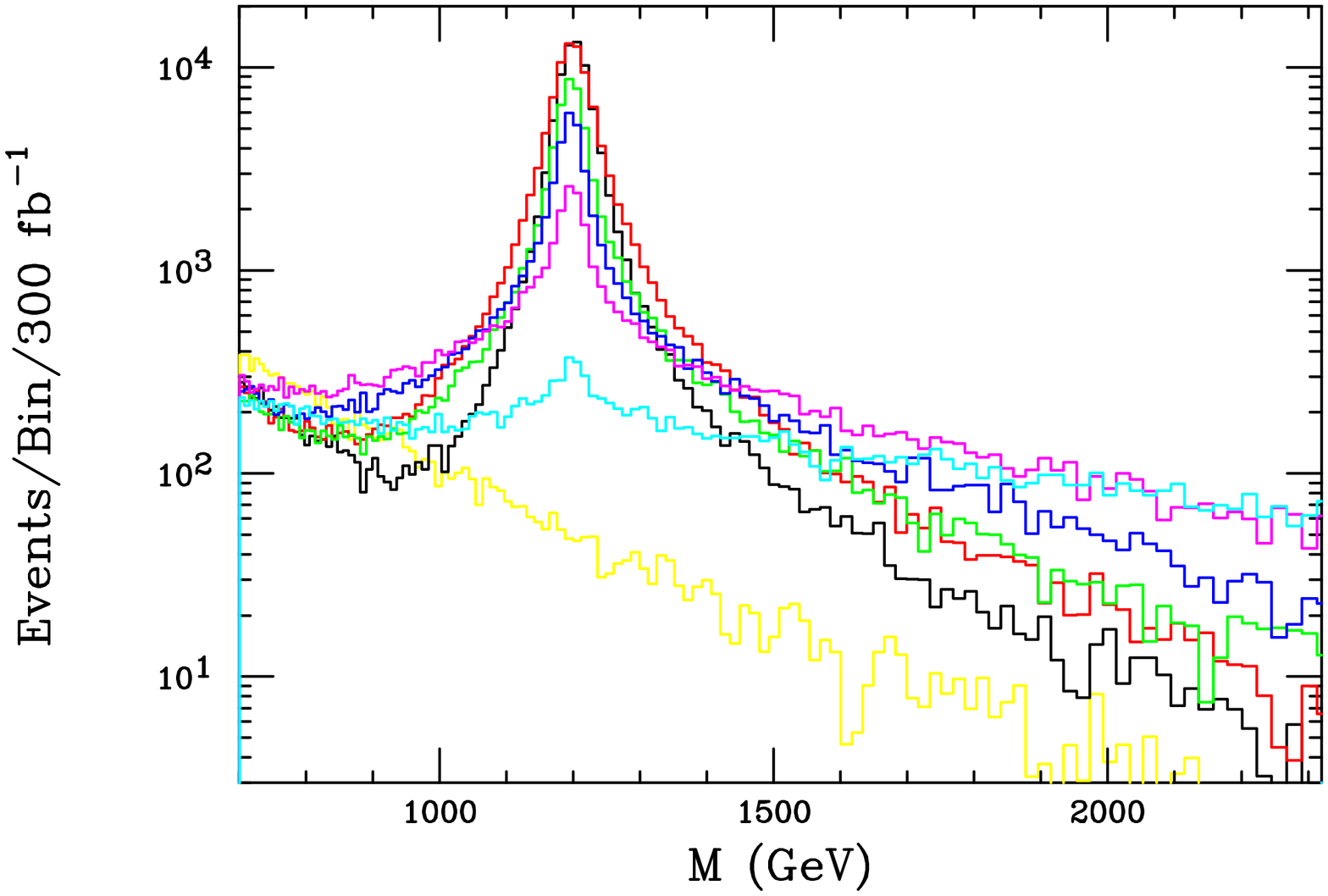}
\hspace{.5cm}
\includegraphics[width=6.7 cm,height=7.7 cm, angle=90]{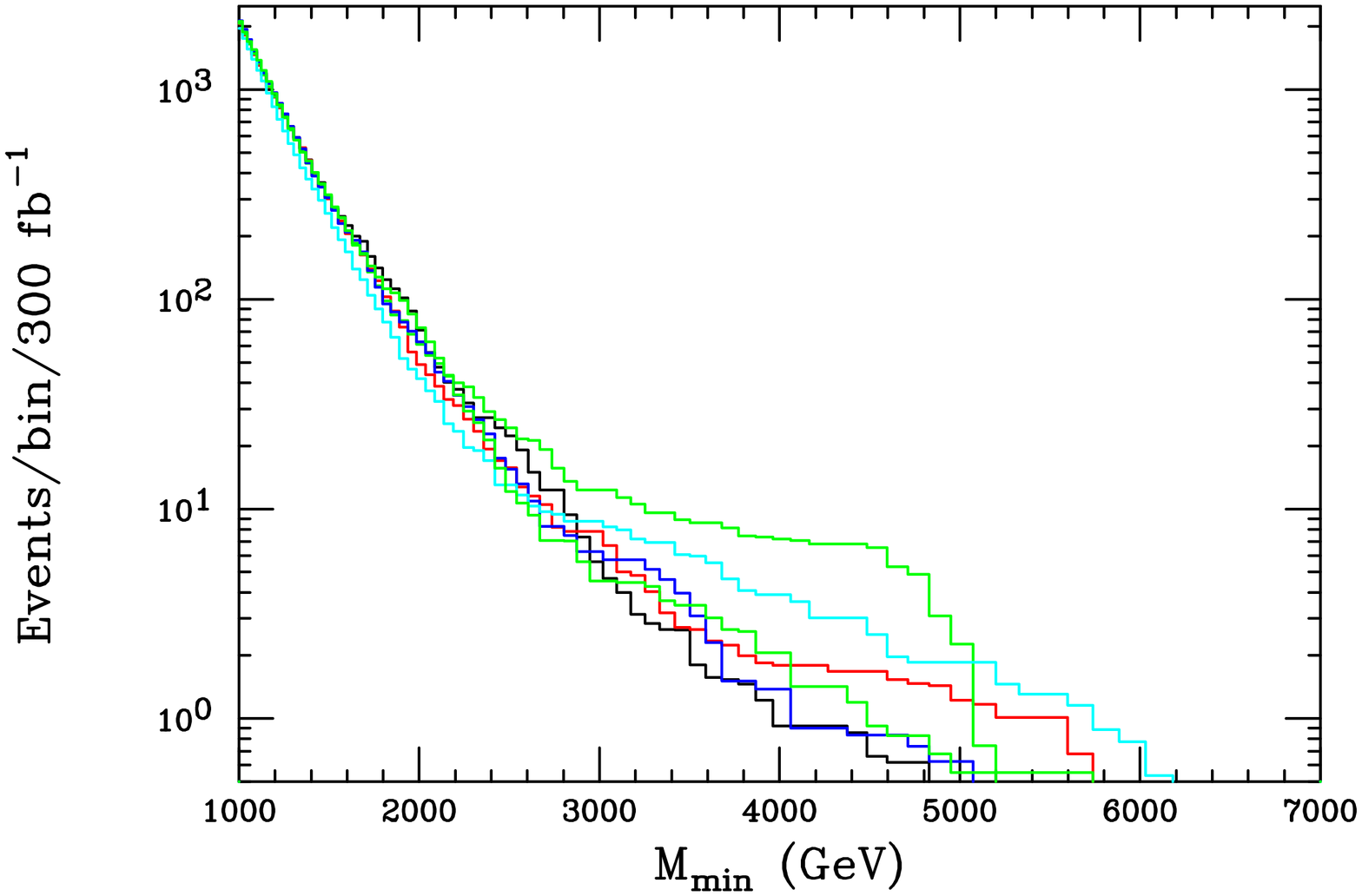}
\caption{(Left) Unparticle lineshapes (colored) in comparison to a SSM $Z'$ (black) with a mass of 1.2 TeV at 
the 14 TeV LHC; a $1\%$ mass resolution has been assumed. The yellow histogram corresponds to the SM expectation~\cite{rev2}. 
(Right) Dilepton event rate expected above a minimum invariant mass for the SM(black), the SSM with a 6 TeV 
$Z'$(red), 
a photon/$Z$ gauge KK state with a mass of 6 TeV(cyan), a Randall-Sundrum graviton with a 5 TeV mass(blue) and 
$k/M_{pl}=0.04$ and a 5 TeV (the $5\sigma$ discovery reach in this case) $R$-parity violating sneutrino(green) with 
electromagnetic strength couplings to both quarks and leptons. An additional (lower) green histogram is also 
present for 
the spin-0 $R$-parity violating sneutrino case assuming a resonance mass of 6 TeV. Detector smearing has been 
included and the reader should remember Poisson statistics.
}
\label{fig2}
\end{figure}

After the lineshape and particle spin are determined we will want to know the couplings of this new state to 
the fields of the SM. (Recall that if it decays to $\gamma\gamma$ it cannot be spin-1.) This subject has been 
widely discussed in the literature{\cite {rev1,rev2,rev3,rev4,rev5,rev6,rev7,rev8,Petriello:2008zr}} so we will be brief here. In the simplest case 
where the couplings are generation-independent and isospin invariant, there are 5 independent parameters to 
determine, one corresponding to each of the basic SM fields. Traditionally, one combines measurements of the 
resonance production cross section and width, the forward-backward asymmetry, $A_{FB}$, of the dilepton pair 
both on and off the resonance as well as the dilepton rapidity distribution to restrict the various couplings. 
Two important observations are: (a) this requires a rather large amount of integrated luminosity, 
$\sim 100$ fb$^{-1}$, even for a relatively light 1.2 TeV $Z'$ and (b) there are not enough observables in 
this list to make a unique determination of the 5 coupling parameters. 
There are, however, further
measurements~\cite{rev1,rev2,rev3,rev4,rev5,rev6,rev7,rev8} that may be
helpful when high luminosities are available:  
($i$) associated $Z'V$ production with $V=\gamma,Z,W^\pm$ 
($ii$) rare $Z'$ decay branching fractions to $f\bar fV$, ($V=Z,W^\pm, f=\ell\nu)$, ($iii$) polarization of 
$\tau$'s from $Z'$ decay, ($iv$) $Z'\to W^+W^-,Zh, b\bar b$ and $t\bar t$. Most of 
these have not yet been studied in any detail for the LHC. 

Clearly, a future linear collider will be the ideal machine to study the properties of a Z'-like object, in particular
to accurately measure its mass, width, couplings and spin properties.  This is particularly true if the new object is within the collider's energy range and can be produced directly. In addition a high energy 
electron-proton collider, such as proposed in the LHeC study~\cite{LHeC}, with polarized lepton beams, 
can yield additional information for the determination 
of the couplings~\cite{Rizzo:2008fq}.

In addition, one could ask whether a new $Z'$-like state may be discovered {\it indirectly} 
at the LHC if its mass is too large to produce an observable resonance{\cite {Rizzo:2009pu}}. As is well known, 
high energy $e^+e^-$ colliders can make precise measurements of the various $e^+e^- \to f\bar f$ processes and 
look for deviations due to the exchange of high mass states. Furthermore, it is possible that $e^+e^-\to f\bar f$ 
measurements made at different $\sqrt s$ values can be also be combined to determine the mass of the new 
resonance itself, provided the mass is not too far above the range of $\sqrt s$ values at which the data were 
taken{\cite {Rizzo:1996rx}}. At the LHC we have access to the entire Drell-Yan dilepton mass distribution; can we use it to 
see heavier states indirectly and determine their masses? Note that here we are not looking for contact interaction 
effects, but are trying to determine the mass of an essentially invisible object. 
In Ref.{\cite {Rizzo:2009pu}} an analysis 
was performed to address this issue. Specifically, it sought indirect evidence for states with masses 1 TeV beyond 
the 14 TeV $5\sigma$ discovery reach assuming an integrated luminosity of 300 fb$^{-1}$. States with spin-0,1 and 2 
were considered. Based on this study it appears that, except for the very special case of degenerate $Z/\gamma$ KK 
resonances (where there is strong destructive interference with the SM exchanges far below the resonance), the 
answer to this question appears to be `no'.  Results of this preliminary analysis are shown in the right
panel of Fig.~\ref{fig2}.
Evidently, the rapid fall-off of the parton 
densities at the required high-x values prevent us from gathering sufficient statistics to perform the same 
procedure as in the $e^+e^-$ case. If we see any deviation in the spectrum we will already see the resonance itself.

If a $W'$ accompanies the $Z'$ and decays into $\ell \nu$, the most important thing to determine is the helicity 
of its couplings to the SM fermions as these are usually chiral. This simple measurement broadly splits all possible 
models into 2 classes depending upon whether these couplings are LH or RH. This measurement cannot be performed 
on the `peak' of the transverse mass distribution since there is no sensitivity `on-resonance' as 
only pure $W'$ exchange is being sampled. However, there is a significant effect 
in the transverse mass region below the peak as in the case of RH couplings there is no interference with the SM $W$ 
contribution while there is for LH couplings. This is shown explicitly in Fig.~\ref{fig4} from Ref.{\cite {Rizzo:2007xs}}. 

\begin{figure}[htbp]
\centering
\includegraphics[width=6.7 cm,height=7.7 cm, angle=90]{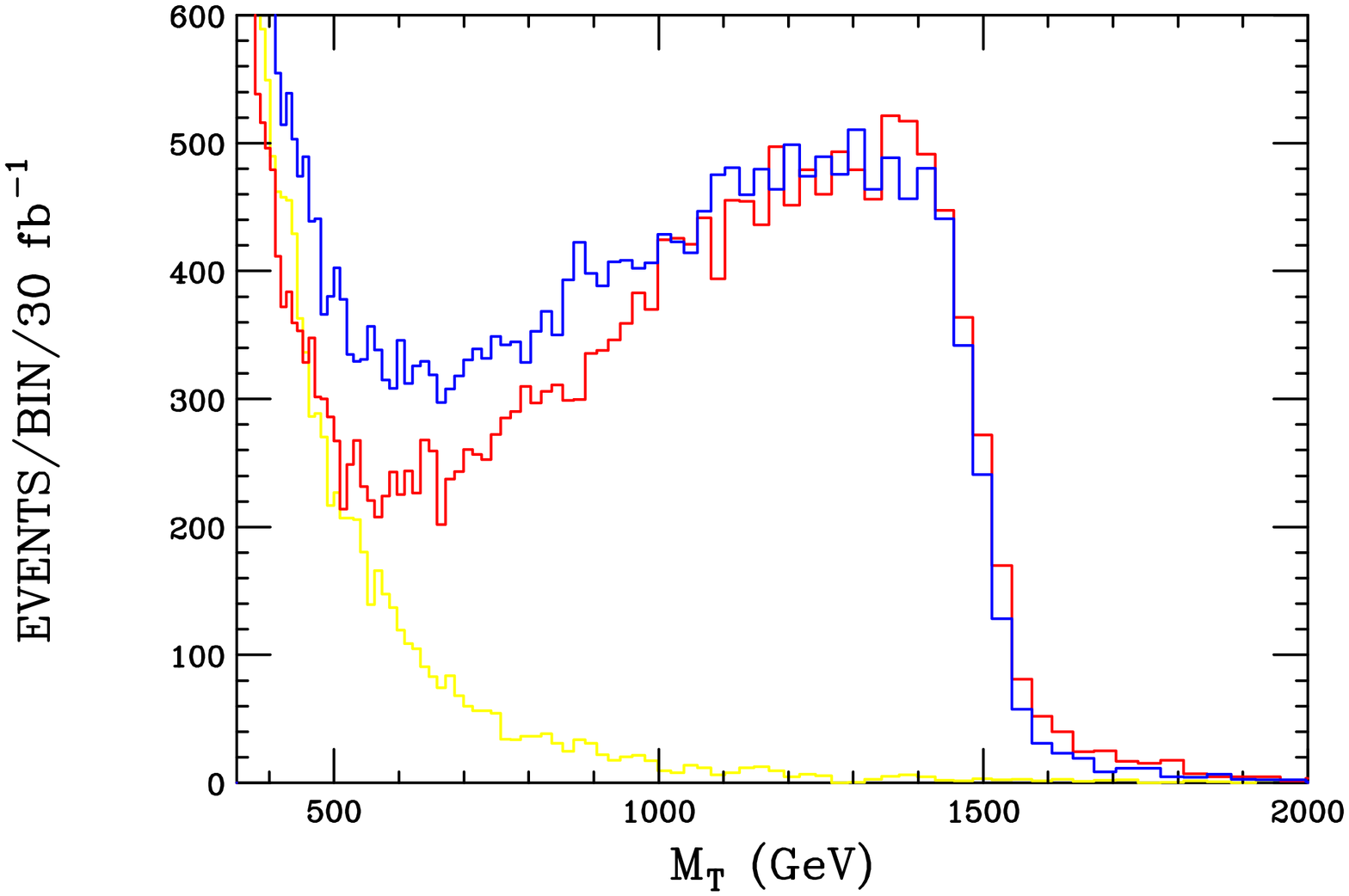}
\hspace{.5cm}
\includegraphics[width=6.7 cm,height=7.7 cm, angle=90]{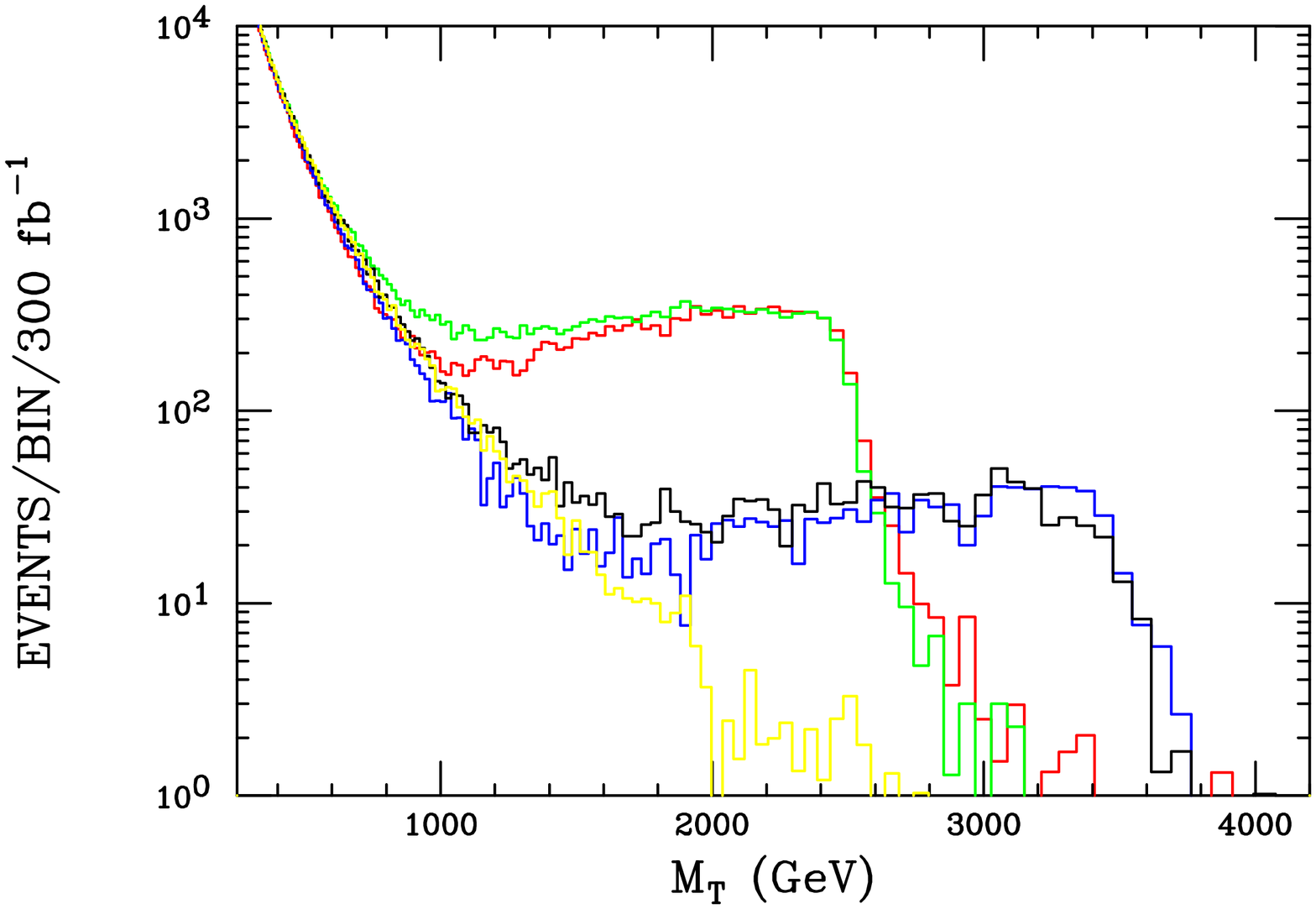}
\caption{$W'$ transverse mass distributions for (Left) $M_{W'}=1.5$ TeV and for (Right) $M_{W'}=$2.5 and 3.5 TeV~\cite{rev8}. 
The upper(lower) histogram in each case corresponds to right(left)-handed couplings to the SM fields. A $2\%$ 
$M_T$ smearing has been included in these results. }
\label{fig4}
\end{figure}

\subsection{4th Generation of Fermions}
\label{sec:4th}

\bigskip
The addition of a fourth generation of chiral fermions (4SM) to the SM has long 
been investigated as one of the simplest
extensions of the SM.  At the advent of the LHC start-up, it is pertinent
to reconsider the physics potential of the 4SM. The Tevatron 
has excluded the $t^{\prime}$ quark of mass less than around 300 GeV. This limit can be 
further pushed up to around 500 GeV during the first year of LHC data taking. In this report, we review the 
impacts of the forthcoming LHC data on the theory as well as on the prospects of a future Linear Collider
experiment. 

\subsubsection{4th generation: is it suggested by the present data? }  

CDF sets a 311 GeV mass limit~\cite{cdf} at 95\% CL for a $t^{\prime}$ decaying into a light quark plus a 
$W$ with a reported  excess in a mass region centered at 450 GeV. 
At present there is an ongoing search for a $t^{\prime}$ quark at D\O which may help clarify this situation.  
The $b^{\prime}$ quark has been searched for in the $bZ$~\cite{Affolder:1999bs} channel. This decay   
occurs via a loop and may compete with the tree-level process $b^{\prime}\to tW$, with 
the latter mode being more difficult to reconstruct (due to combinatorials with up to 10 jets).

In the 4SM, the $3\times 3$ CKM matrix is only part of the full matrix, thus the unitarity of this 
$3\times 3$ portion could be broken. It is often considered that the $3\times 3$ CKM matrix is unitary at 
high precision, however, the 3rd row is not determined precisely by tree-level processes.
Constraints on the $4\times 4$ 
CKM matrix may be obtained by computing the $t^{\prime}/b^{\prime}$ contributions to FCNC and electroweak
observables.  Such studies have been recently updated \cite{Alwall:2006bx,Bobrowski:2009ng,Chanowitz:2009mz}.
In addition, recent Tevatron data for the CP asymmetry in $B_s$ mixing shows a 
2-3$\sigma$ deviation from the 3SM \cite{Bona:2008jn}, which can be explained by the two new 
CP violating phases in the 4SM.

In addition, it has recently been emphasized that the electroweak oblique parameters {\it do not exclude} the 
4SM \cite{Alwall:2006bx,Bobrowski:2009ng,Chanowitz:2009mz,Kribs:2007nz}. It is usually considered that 
the large contribution from the 4th generation quarks  to the $S$ parameter  
creates a tension with data. However, if there is an extra contribution to the $T$ parameter 
(e.g. from non-degeneracy of $t^{\prime}$ and $b^{\prime}$ or the 4th generation leptons), one can easily 
accomodate the 4SM within the experimental allowed ranges.

\subsubsection{Impact of a 4th generation quark around \boldmath{$500$} GeV}

Here, we consider the scenario where the 4th generation quark has a mass around 500 GeV. We focus 
on this mass range not only because it is the early LHC early discovery reach but also because this 
corresponds to the perturbative unitarity limit for the 4th generation \cite{Chanowitz:1978uj} 
and thus it provides an important theoretical benchmark.   A mass scale heavier than this would imply:
i) we are 
entering the strongly coupled regime and thus we would expect some consequences such as a condensate of the 
heavy fermions, and ii)  new physics would need to be introduced in order to satisfy the unitarity limit.

Such a heavy $t^{\prime}$ impacts the Higgs mass limits, in particular the bounds from stability/triviality. 
This occurs due to the $t^{\prime}$ Yukawa coupling contribution 
to the loop correction to the Higgs quartic coupling, $\lambda$.
The solution to the renormalization group equation (RGE) for this quartic coupling 
gives the Higgs mass in terms of the cut-off scale $\Lambda$. 
When  $\lambda$ becomes negative, the Higgs potential no longer has a stable minima. This 
limit leads to a lower bound for the Higgs mass (the so-called stability limit). For the 3SM case, the 
obtained Higgs mass limit is $M_H> 130 (70)$ GeV for $\Lambda\simeq M_{\rm Planck} (1\ {\rm TeV})$
\cite{Casas:1994qy}. The RGE shows that the large 4th generation Yukawa couplings
increase this lower bound. On the other hand, as the Higgs mass increases, there is a point where the 
quark Yukawa term exactly cancels the other contributions. 
This leads to an upper bound on the Higgs mass (the so-called triviality 
limit). The RGE shows that this fixed point yields
the upper bound of $m_H=500-800 (200)$ GeV for $\Lambda=1 (M_{\rm Planck})$ 
TeV~\cite{Hambye:1996wb} in the 3SM, while the larger value of the fourth generation Yukawas dramatically lowers 
this upper bound. 
In fact, it has been shown in~\cite{Kribs:2007nz} that even for $m_{t^{\prime}}=m_{b^{\prime}}=260$ GeV, 
the Higgs mass is constrained to the range $200<M_H<470$ GeV for $\Lambda=1$ 
TeV and most importantly, the famous chimney ({\it i.e.} for $130<M_H<200$, the 3SM is valid up to the Planck 
scale) is closed out. In summary,  i) the Higgs mass lower limit (the stability bound) increases as the 
4th generation quark masses increase, and ii) the 4SM ceases to be valid for $\Lambda\gsim 10^{\rm a\ few}$ TeV if a 
4th generation quark heavier than the current Tevatron limit is discovered.
It has been emphasized that this situation does not change even if we extend 
the Higgs sector~\cite{Gunion:1994zm}. 

It has also been shown \cite{Fok:2008yg} 
that electroweak baryogenesis occurs at sufficient rates to produce the observed
matter anti-matter asymmetry in the case where the 4SM is further extended by SUSY (4MSSM), provided
that the 4th generation is fairly heavy and its scalar partners have similar mass.
We discuss a potential discovery of such scenario at the LHC and future colliders below.

\subsubsection{What do we expect during the early years of LHC?}

At the LHC, the cross section for heavy quark production is large and allows us to reach the 
unitarity mass limit with a few 
100 $pb^{-1}$. Detection of the $t^{\prime}$ seems to be easy. The dominant channel $t^{\prime}\to bW$ has been 
studied in detail using the same strategy as for observing the top quark. The $b^{\prime}$ 
search is more involved {as the decay $b^{\prime}\to tW\to bWW$ leads to a soft $W$'s}. 
The recent study estimates that the $t^{\prime}$ quark with mass  $m_{t^{\prime}}=450$ GeV can be discovered 
at 5$\sigma$ level with the first year of data taking (100 pb$^{-1}$)~\cite{Ozcan:2008yp}. In this workshop, 
CMS also reported that the $b^{\prime}$ with mass  $m_{b^{\prime}}\leq 300$ GeV can be discovered at the
7.5$\sigma$ level with the same amount of data. On the contrary, leptons are produced by the Drell Yan process 
and requires $\gsim$10 fb$^{-1}$ for a meaningful search.

It is well-known that the addition of two heavy quarks increases the coupling of the Higgs to two gluons 
by a factor of  9  for $m_H$=125 GeV and of 5  for $m_H$=500 GeV~\cite{Arik:2007vi}.  Therefore, naively 
speaking, to get a 5$\sigma$ discovery, one needs a factor of $81$ less luminosity than for the 3SM at
14 TeV. Thus, the first year of data would allow for a Higgs discovery in the $\sim$350-500 GeV region. 
It is interesting to note that given the factor of 9 enhancement for the gluon-gluon process, one can already say 
that the mass region 135-200 GeV is excluded by Tevatron in the 4SM. The search
for a light Higgs remains difficult at the LHC in spite of the enhancement factor for the cross section since 
$Br(H\to \gamma\gamma)$ is reduced by a factor of $\sim$10. 
The consequences are that the most efficient 
channel will be $H\to ZZ^*$. A detailed analysis is needed to provide a more quantitative answer but 
following ~\cite{Kribs:2007nz}, one expects that a few fb$^{-1}$ would be needed.

Within the SUSY+baryogenesis scenario, squarks and heavy quarks are being produced with similar rates 
(except for the $\beta$ factor for scalar production) which may result in a rich but confusing situation. For instance 
$b^{\prime}\to tW$ final states may overlap with a squark $b^{\prime}$ giving a top quark and a chargino.

\subsubsection{What is the role of the future linear collider?}

The mass limit for heavy quarks set by unitarity calls for a LC reaching a centre of mass 
energy of $\sim$1 TeV. Very precise masses and couplings of the new heavy fermions can be measured using 
polarized beams and constrained reconstruction. 
Branching fractions for Higgs decays into the 4th leptons, if kinematically allowed, would be 
well measured at a LC. One could also observe stable or long lived neutrinos through neutrino counting. 

Within the 4MSSM there could be a Higgs lighter than 135 GeV giving access to the fermionic decay modes with 
incomparable accuracies. Also the precise measurement of the branching ratio of $H\to gg$ would be crucial 
to determine the Yukawa couplings of the heavy quarks. A consideration of the successful baryogenesis 
through 4MSSM requires mass degeneracy of the heavy quarks and squarks. In such a situation, the LC becomes 
most powerful: threshold scans would allow to disentangle the various degenerate quark and squark states.

\subsection{Testing Seesaw mechanisms at the LHC}
\label{sec:seesaw}

\bigskip
The neutrino mass and flavor oscillations are arguably the first indication for physics beyond the 
Standard Model (SM). The smallness of the neutrino mass is attributed to a large energy scale of
new physics $\Lambda$, via the ``Seesaw" relation $m_\nu \sim y^2 v^2/\Lambda$
where $y$ is a Yukawa coupling. If $\Lambda$ is near the Terascale, one would have the hope
to test the seesaw mechanism by searching for lepton-number violating signals.

\subsubsection{Type I Seesaw \cite{TypeIa,TypeIb,TypeIc,TypeId,TypeIe}: Heavy Majorana Neutrinos}

Searching for heavy Majorana neutrinos at hadron colliders have long been
considered by many authors \cite{goran}, however the interest for the LHC
has been lately renewed \cite{Han:2006ip,more,Franceschini:2008pz}.
We calculate the exact process, but
it turns out to be an excellent approximation to parameterize the
cross section as
\beq 
\label{eq:bare} 
\sigma(pp\rightarrow \ell_1^\pm\ \ell_2^\pm\ W^\mp)\approx
\left( 2 - \delta _{\ell_1 \ell_2 }\right) \sigma(pp\rightarrow
\ell_1^\pm N_4)Br(N_4 \rightarrow \ell_2^\pm W^\mp) \equiv 
\left( 2 - \delta _{\ell_1 \ell_2 }\right)\ S_{\ell_1 \ell_2}\ \sigma_0(N_4), 
\eeq
where $\sigma_0(N_4)$, called the ``bare cross section",  is only
dependent on the mass of the heavy neutrino and is independent of all
the mixing parameters when the heavy neutrino decay width is
narrow. We calculate  the exact cross section for the dilepton production
and use the definition Eq.~(\ref{eq:bare}) to find the bare cross
sections $\sigma_0(N_4)$.  These are shown in
Fig.~\ref{Fig:KvsMass} at the Tevatron and LHC energies versus
the mass of the heavy Majorana neutrino, where we have calculated the
cross sections at both 10 and 14 TeV c.m.~energies. The production rate
is increased at the higher energy by a factor of 1.5, 2.0, 2.5 for
$m_4=100,\ 550$ and 1000 GeV, respectively. 

\begin{figure}[tb]
\center
\includegraphics[width=10truecm,clip=true]{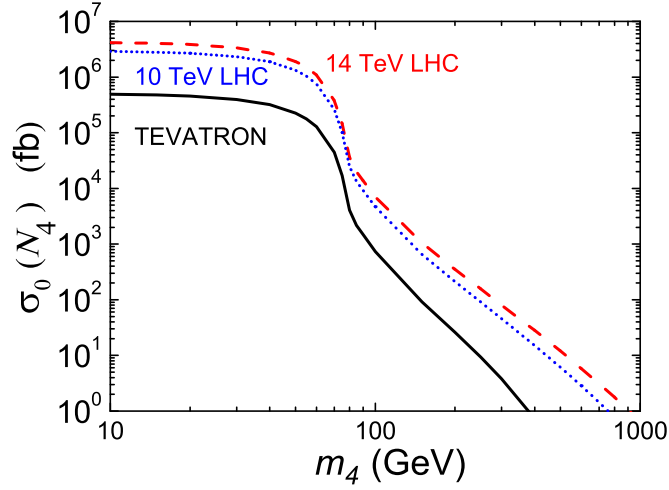}
\caption{The bare cross section $\sigma_0(N_4)$ versus mass of
heavy Majorana neutrino $m_4$ for the Tevatron ($p\bar p$ at 1.96
TeV, solid curve) and the LHC ($pp$ at 10 and 14 TeV, dotted and
dashed curves, respectively)~\cite{Atre:2009rg}. } 
\label{Fig:KvsMass}
\end{figure}

The flavor information of the final state leptons is parameterized
by \beq
 S_{\ell_1\ell_2}=\frac{ \left|V_{\ell_1 4} V_{\ell_2 4}
 \right| ^{2}}{\sum_{\ell=e}^{\tau} \left|V_{\ell 4}\right| ^{2}}.
\eeq 
In general the two final state charged leptons can be of any
flavor combination, namely,
\beq e^\pm e^\pm,\ \  e^\pm \mu^\pm,\ \  e^\pm \tau^\pm,\ \
\mu^\pm \mu^\pm ,\ \  \mu^\pm \tau^\pm \quad  {\rm and}\quad
\tau^\pm \tau^\pm. \eeq
The constraint from $0 \nu \beta \beta$  is very strong and makes it difficult to observe
like-sign di-electrons $e^\pm e^\pm$. The events with $\tau$
leptons will be challenging to reconstruct experimentally. We will
thus concentrate on clean dilepton channels of $\mu^\pm\mu^\pm$
and  $\mu^\pm e^\pm$.

In Fig.~\ref{Fig:DetectLimit}(a) and Fig.~\ref{Fig:DetectLimit}(b),
we summarize the sensitivity  for $S_{\mu\mu}$ and $S_{e \mu}$
versus $m_4$, respectively. The solid (dashed) curves correspond
to $2 \sigma$ ($5 \sigma$) limits on $S_{\ell \ell'}$ with the
exclusion of the Higgs decay channel. The dotted (dash dotted)
curves are similar but with the inclusion of the Higgs decay
channel for $m_H = 120$ GeV. The horizontal dotted line
corresponds to constraints on $|V_{\mu4}|^2 < 6 \times 10^{-3}$
from precision EW measurements. In
Fig.~\ref{Fig:DetectLimit}(b) the dashed line at the bottom
corresponds to the limit from $0\nu\beta\beta$.

\begin{figure}[tb]
\includegraphics[width=7.8truecm,clip=true]{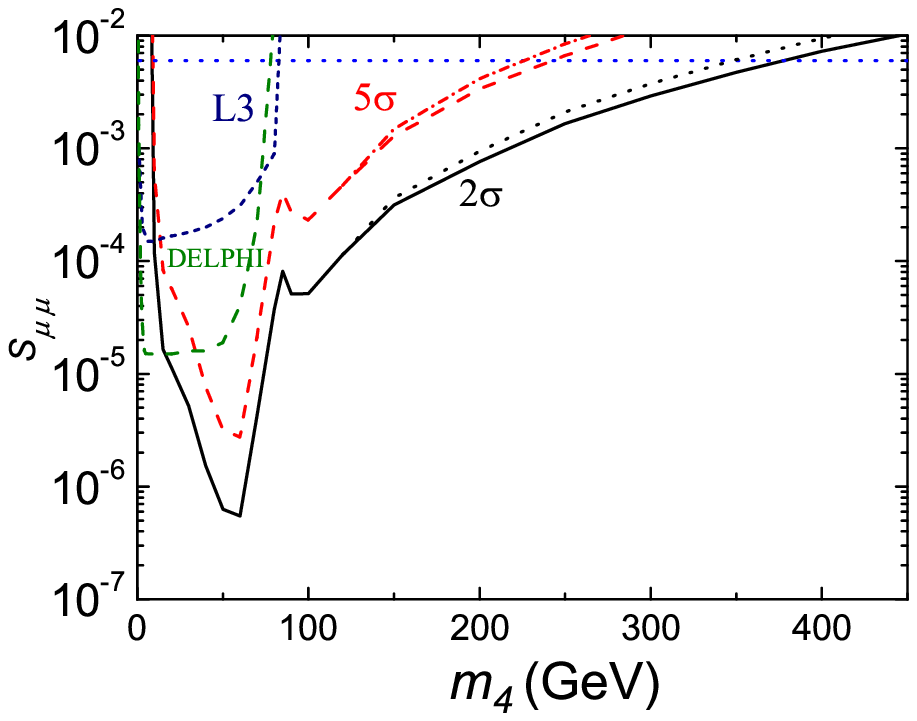}
\includegraphics[width=7.8truecm,clip=true]{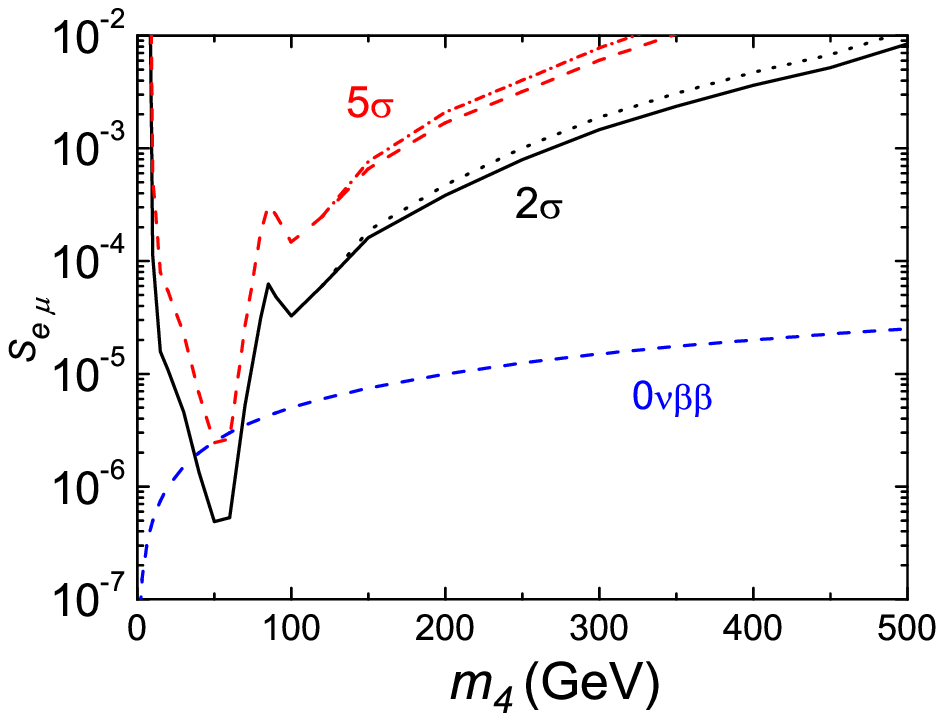}
 \caption{ (a) Left: $2\sigma$ and  $5\sigma$ sensitivity for
  $S_{\mu\mu}$
  versus $m_4$  at the LHC with 100 fb$^{-1}$ integrated luminosity; (b)
  right: same as (a) but for $S_{e\mu}$ (both plots taken from
  Ref.~\cite{Atre:2009rg}). The solid and dashed (dotted and 
  dash dotted) curves correspond to limits with the exclusion (inclusion) of
  the Higgs decay channel for $m_H = 120$ GeV. The horizontal dotted line
  corresponds to the constraint on 
$S_{\mu\mu}\simeq  |V_{\mu4}|^2 < 6 \times 10^{-3}$ from precision
EW measurements. }
\label{Fig:DetectLimit}
\end{figure}

 We find that, at the
Tevatron with $8\ \mbox{fb}^{-1}$ integrated luminosity,
 there could be $2\sigma$ ($5\sigma$)
 sensitivity for resonant production of a Majorana neutrino in the $\mu^\pm \mu^\pm$
 modes in the mass range of $\sim 10 - 180\ \mbox{\gev} \ (10 - 120\ \mbox{\gev})$. This reach
can be extended  to $\sim 10 - 375\ \mbox{\gev}\ (10 - 250\
\mbox{\gev})$ at the LHC of 14 TeV with $100\ \mbox{fb}^{-1}$.
The production cross section at the LHC of 10 TeV is also presented for comparison.
We study the $\mu^\pm e^\pm$ modes as well and find that the signal
could be large enough even taking into account the current bound from
neutrinoless double-beta decay.
However, it is believed
that any signal of $N$ would indicate a more subtle mechanism beyond
the simple Type I seesaw due to the otherwise naturally small mixing
$V_{N\ell}^2 \sim m_\nu/M_N$ between $N$ and the SM leptons.

\subsubsection{Type II Seesaw \cite{TypeIIa,TypeIIb,TypeIIc,TypeIId,TypeIIe}: Doubly Charged Higgs Bosons}

Several earlier studies of certain aspects of the Type II seesaw model
at the LHC exist \cite{Chun}.
We find that in the optimistic scenarios, by
identifying the flavor structure of the lepton number violating
decays of the charged Higgs bosons at the LHC, one can establish the neutrino mass
pattern of the Normal Hierarchy, Inverted Hierarchy or Quasi-Degenerate.
We emphasize the crucial role of the singly charged
Higgs boson decays. The associated pair production
of $H^{\pm\pm} H^{\mp}$ is essential to test the triplet
nature of the Higgs field. The observation of either
$H^+ \to \tau^+ \bar \nu$ or $H^+ \to e^+\bar \nu$ will be
particularly robust for the test since they are independent
of the unknown Majorana phases. Combining with the doubly charged
Higgs decay, for instance $H^{++} \to e^+ \mu^+, e^+\tau^+,\mu^+\tau^+$,
one will even be able to probe the Majorana phases.

The production cross sections for all three channels are shown in
Fig.~\ref{total}(a) for the LHC at 10 TeV, and in Fig.~\ref{total}(b) at 14 TeV.
With negligible SM backgrounds,
the only limitation would be the event rate, that determines
the statistical error for the BR measurements, {\it i. e.}, a relative error
$1/\sqrt N$ if Gaussian statistics is applicable.
%
\begin{figure}[tb]
\includegraphics[scale=1,width=8cm]{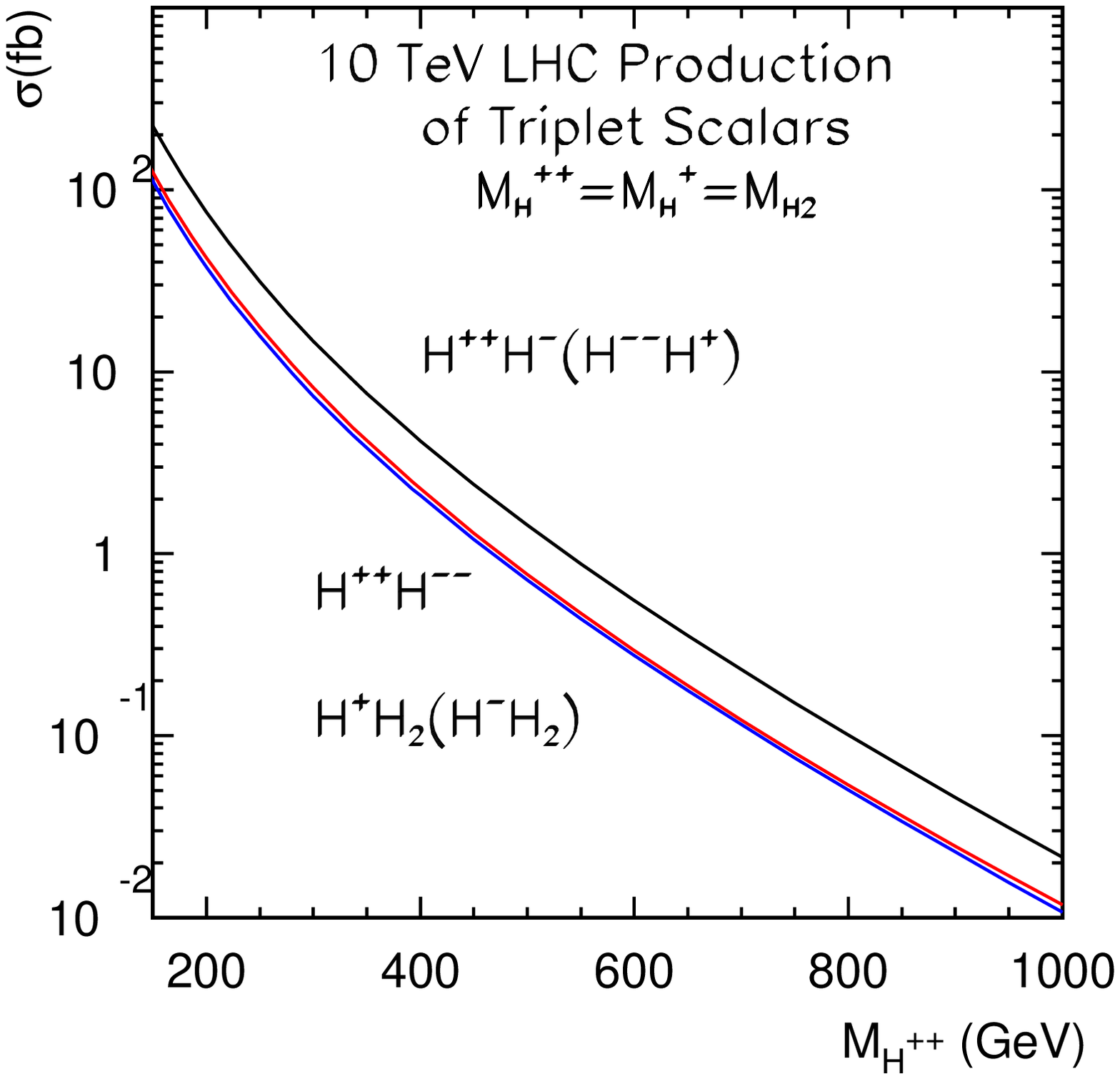}
\includegraphics[scale=1,width=8cm]{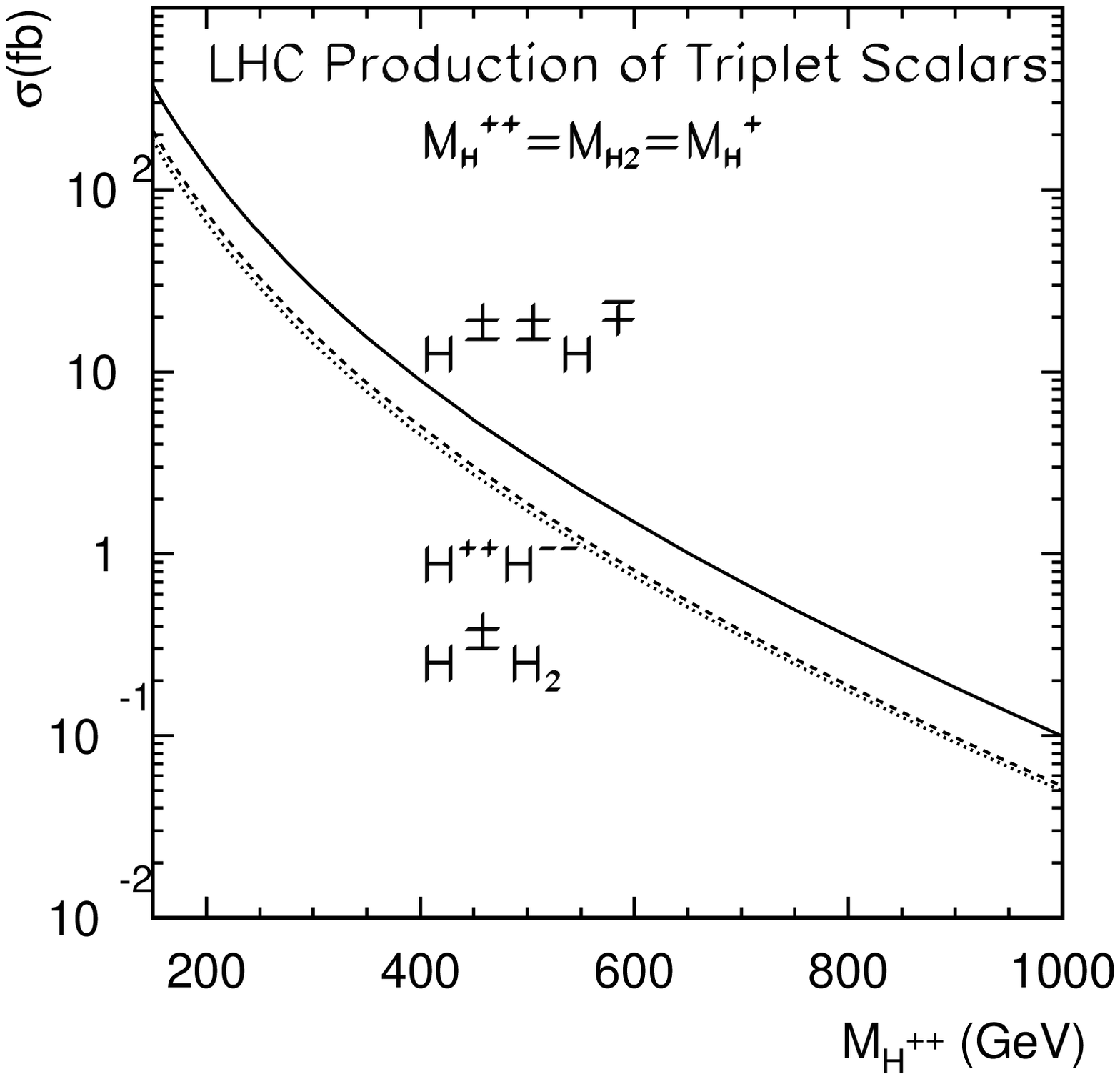}
\caption{Total production cross section at the LHC versus the heavy Higgs mass
for (a) at 10 TeV, (b)  at 14 TeV~\cite{Perez:2008ha}. }
\label{total}
\end{figure}

\subsubsection{Type III Seesaw \cite{TypeIII}: Heavy Leptons}

In Fig.~\ref{fig:total1}, we present the total production cross
sections for the leading electroweak processes versus the heavy lepton mass  $M_T$
at the LHC ($pp$ at $\sqrt s =14$ TeV and 10 TeV). 
To view the generic feature, we have pulled out the effective couplings $\lambda^2$
 in the plots, which is normalized to unity for the pair production, 
 and to the Yukawa coupling squared for the single production. 

\begin{figure}[tb]
\begin{center}
\scalebox{0.6}{\includegraphics[angle=0]{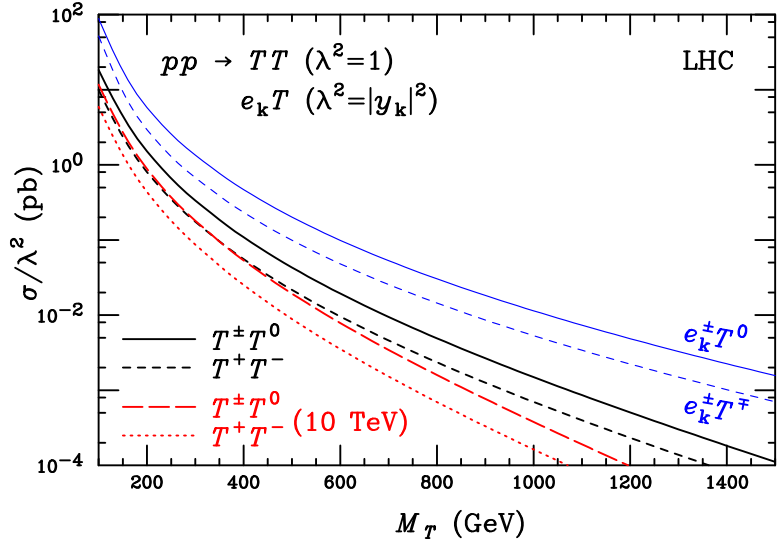}}
\caption{Cross sections of single and pair productions of
 $T^\pm/T^0$  as a function of its mass
 at the LHC (14 TeV and 10 TeV). 
 The scaling constant $\lambda^2$ is 1 for $TT$, and $|y_k|^2$ for 
 $e_k T$~\cite{Arhrib:2009mz}. }
\label{fig:total1} 
\end{center}
\end{figure}

The smoking gun is the production of lepton number violating same-sign dileptons 
plus four jets without significant missing energy. 
Our analysis shows that via the unique channel, 
\beq
  T^0 T^\pm \to (\ell^\pm W^\mp) (\ell^\pm Z/h), 
  \ {\rm or }\ \ell^\pm \ell^\pm W^\mp Z/h
\eeq
the heavy lepton can be searched
for up to a mass of 200 GeV at the Tevatron with 8 fb$^{-1}$, and  up to 450 (700) GeV at the LHC of 14 TeV 
C.M.~energy with 10 (100) fb$^{-1}$. 
The signal rate at the 10 TeV LHC is reduced by a factor
of $60\% - 35\%$ for a mass of 200$-$700 GeV.

In conclusion, if the scale for the neutrino mass generation is near the Terascale, it is possible 
to test the seesaw mechanism by searching for lepton-number violating signals at the LHC. 
The signatures are rather unique and clean. 
Higher integrated luminosity and energy would be beneficial for  extending the search.  A future
$e^+e^-$ linear collider would also be a sensitive probe of the Seesaw mechanism.

\subsection{New signatures and  implications for  detectors on new colliders}
\label{sec:det-implication}

\bigskip
Recently suggested theoretical scenarios predict the  possibility of exotic signatures at the LHC, such as heavy stable charged particles, particles that may stop in the detector, non-pointing photons, monopoles etc.
Present experiments such as ATLAS and CMS are often not specifically designed for these 
type of signatures and thus
it has been an interesting exercise over the past few years to evaluate and design triggers and analysis methods 
to tackle the search for these new physics scenarios. Overall, and sometimes surprisingly, the 
detectors can handle these new physics signatures in general very well. Both ATLAS and CMS have been studying these 
new physics scenarios and a  few examples are given in this section.

In some of these scenarios, heavy stable charged particles can be produced and a fraction 
of these can, via their energy loss in ionization and hadronic interactions, stop
in the detector, sit there for a while (seconds, hours, days) and
then decay. It is a challenge for the experiments to be ready for these
signatures, in particular  to trigger on these events. So far the
experiments are found to be up to this challenge. CMS made a study for stopped R-hadrons driven by long  lived
gluinos. A good fraction, as much as a third of the produced R-hadrons can stop in the dense structure 
of CMS. After some time the R-hadron finally decays
in a cluster of jets. Experimentally this will be most easily observable when such a decay happens 
during abort accelerator gaps or other empty bunch crossings, or when there is no beam in the machine all together.
Backgrounds will mainly come from cosmics or noise in the detector, and these have 
already been studied with the CMS detector while waiting for LHC collisions. An example of a simulated
signal event 
is shown in Fig.~\ref{gluinos}(left). CMS has designed a trigger that
will detect energy clusters in the calorimeter during no-collision time. The sensitivity is shown in 
Fig.~\ref{gluinos}(right). Already with a few weeks of good luminosity (10$^{32}$cm$^{-2}$s$^{-1}$) 5$\sigma$
significance discoveries can be made for gluinos with mass of 300 GeV.

Other signatures include those where the particles pass through the detector, but since they move with 
a speed which is only a fraction of the speed of light, these will arrive late in  the outer detectors. 
If the velocity $\beta $ of the particle is larger than 0.6 then a good fraction can be 
detected and reconstructed in CMS and ATLAS and the 
time delay can be measured, as shown in Fig.~\ref{hcsp} (left).  Also the energy loss by ionization in the tracker should be unusually high.
Some of these signatures could well be visible with early data. 
For example Fig.~\ref{hcsp}(right) shows the luminosity needed for observing 3 events
(in the anticipated absense of background) for different scenarios of new physics that lead to
heavy stable charged particles. This includes KK taus, gluinos, stable stops
and GMSB staus. The prospects are excellent!
Now let's see what Nature really has in
store for us...

\begin{figure}[tb]
\begin{center}
\includegraphics[scale=1,width=6cm]{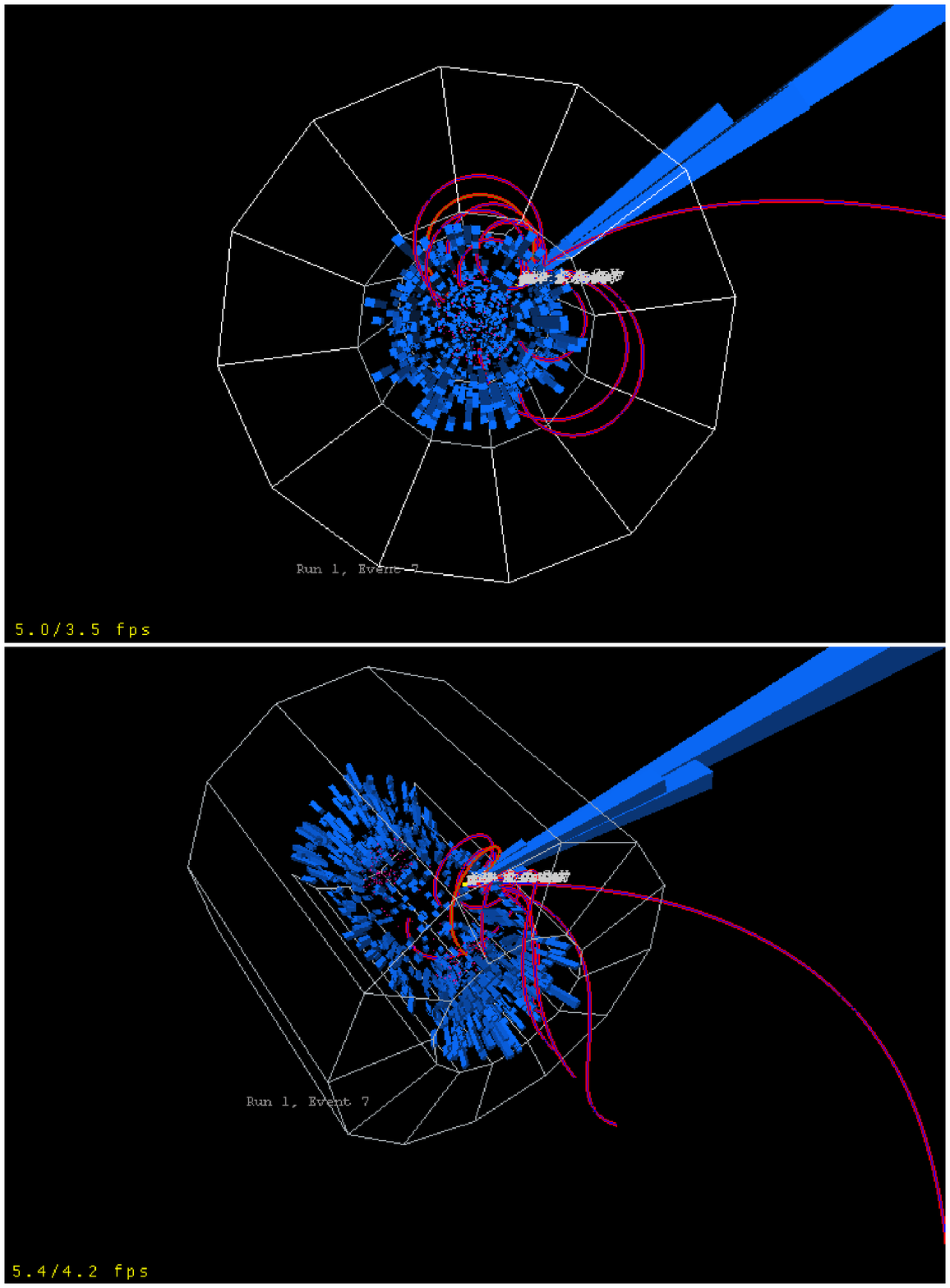}
\hspace{.5cm}
\includegraphics[scale=1,width=8cm]{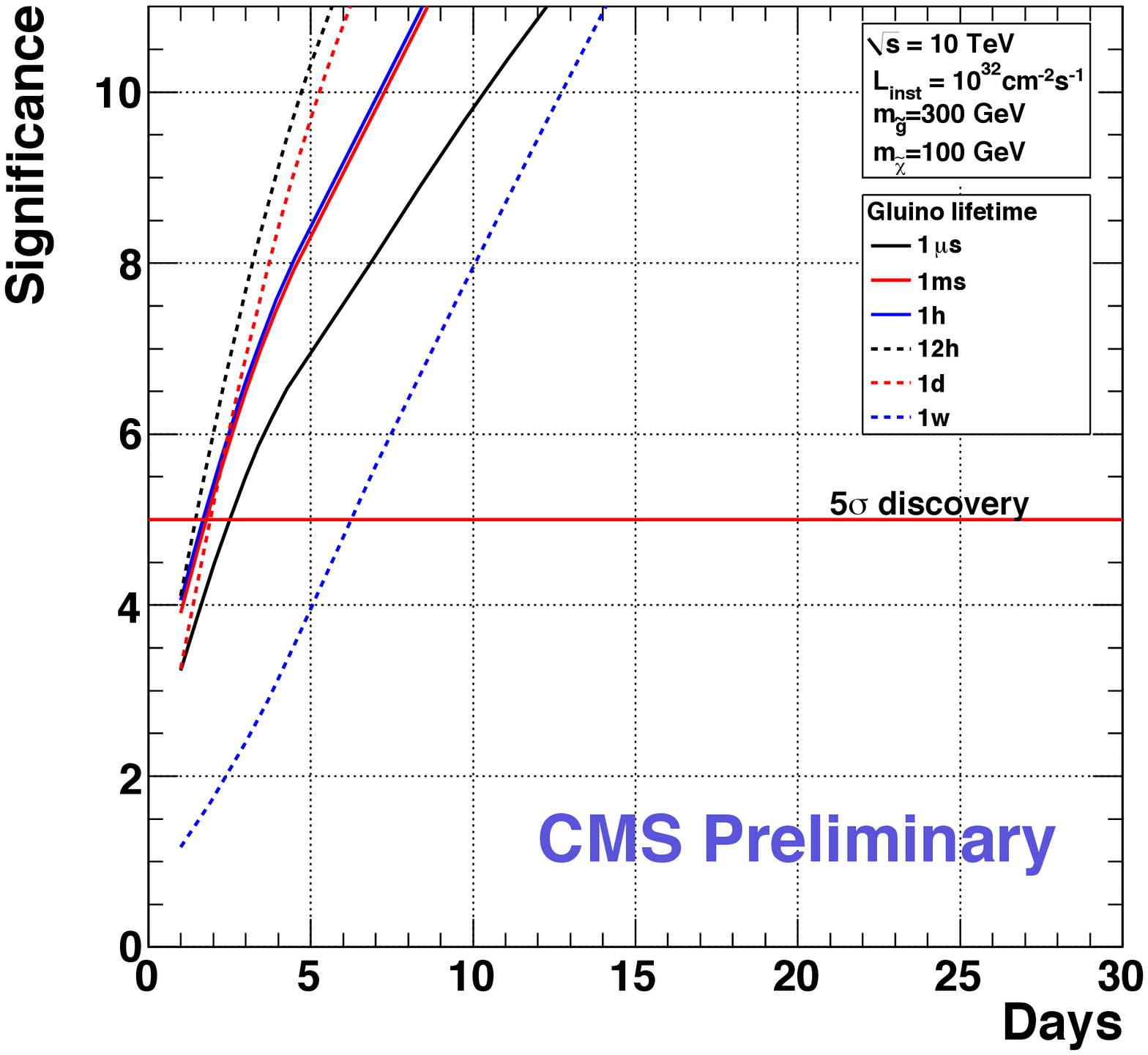}
\end{center}
\caption{ (left) Simulated signal
 events in the detector, in the search for stopped gluinos; (right) significance that can be 
obtained with a luminosity of 10$^{32}$cm$^{-2}$s$^{-1}$ for different gluino
lifetimes (both plots taken from Ref.~\cite{EXO09}).}
\label{gluinos}
\end{figure}

Other new signatures include monopoles, events with many displaced vertices (such as from Hidden Valley
models), non-pointing photons (from {\it e.g.}, GMSB models).  ATLAS has demonstrated that the trigger 
for {\it e.g.}, hidden valley models is under control for a good part of the expected phase space.

When such new particles are discovered, this will have a huge impact on the design of the detectors 
at future colliders or even the LHC detector upgrade. Precise time-of flight will become much more important, preferably
at the 100 ps seconds or better. For the LHC itself it may be that one needs to keep data for more than one
bunch crossing, ie longer than 25 ns, in order to accept and measure particles with low $\beta$
values. If particles will move slower than roughly $\beta < 0.6$ then they will
reach the muon systems in the experiments at a time compatible with an interaction of the next bunch crossing.
Good timing in the electromagnetic  calorimeter in particular would allow also to detect cleanly non-pointing 
photons: the path crossed to reach a calorimeter cell will be larger when they come from a decay, compared 
to when these photons arrive in a straight line from the interaction point.
Clearly any information from the LHC on the masses of the new particles will be important to define the next 
colliders minimum energy threshold.

\begin{figure}[tb!]
\begin{minipage}[h]{8cm}
\includegraphics[height=10cm]{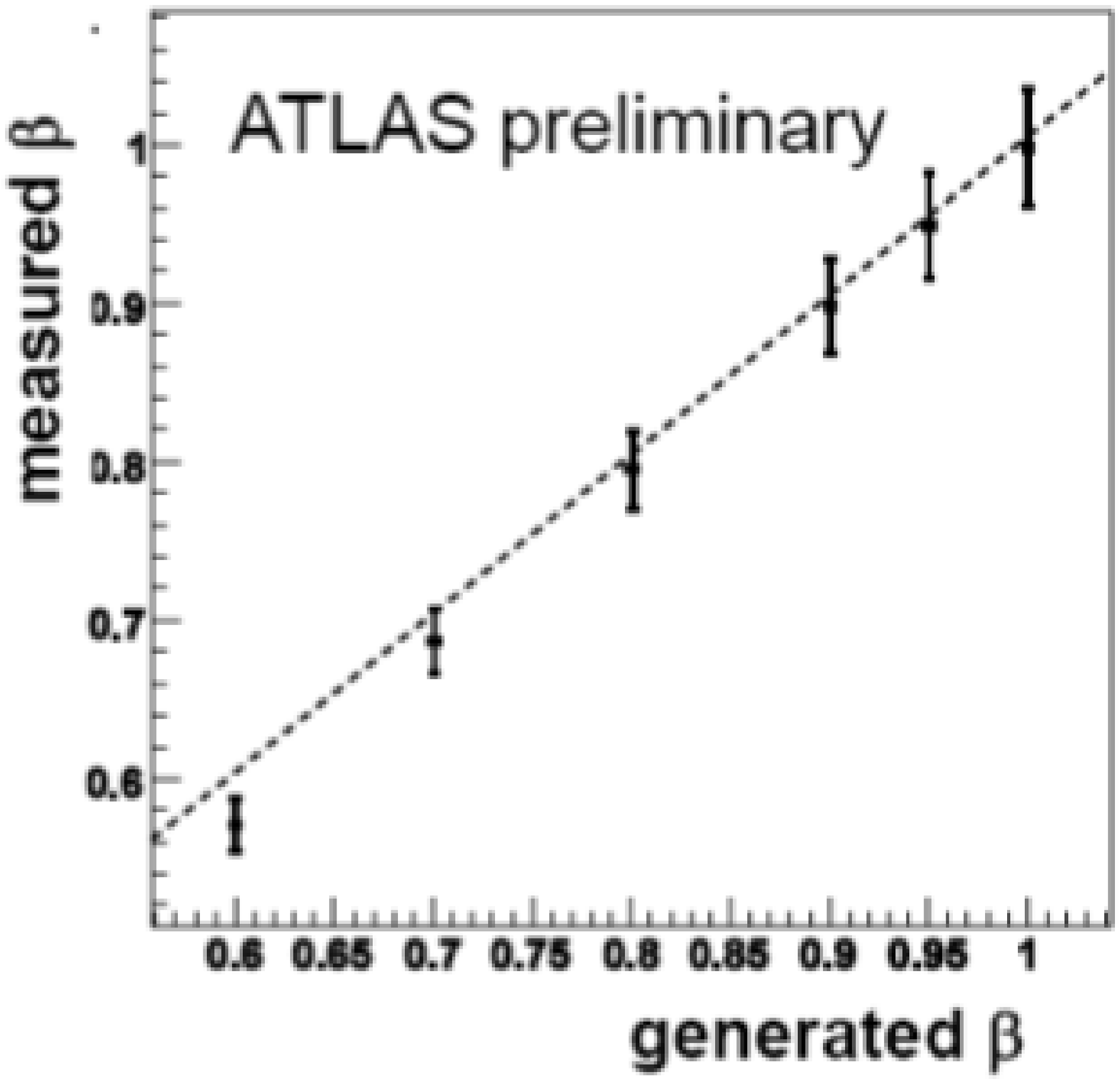}
\end{minipage}
\begin{minipage}[h]{8cm}
\includegraphics[height=6cm]{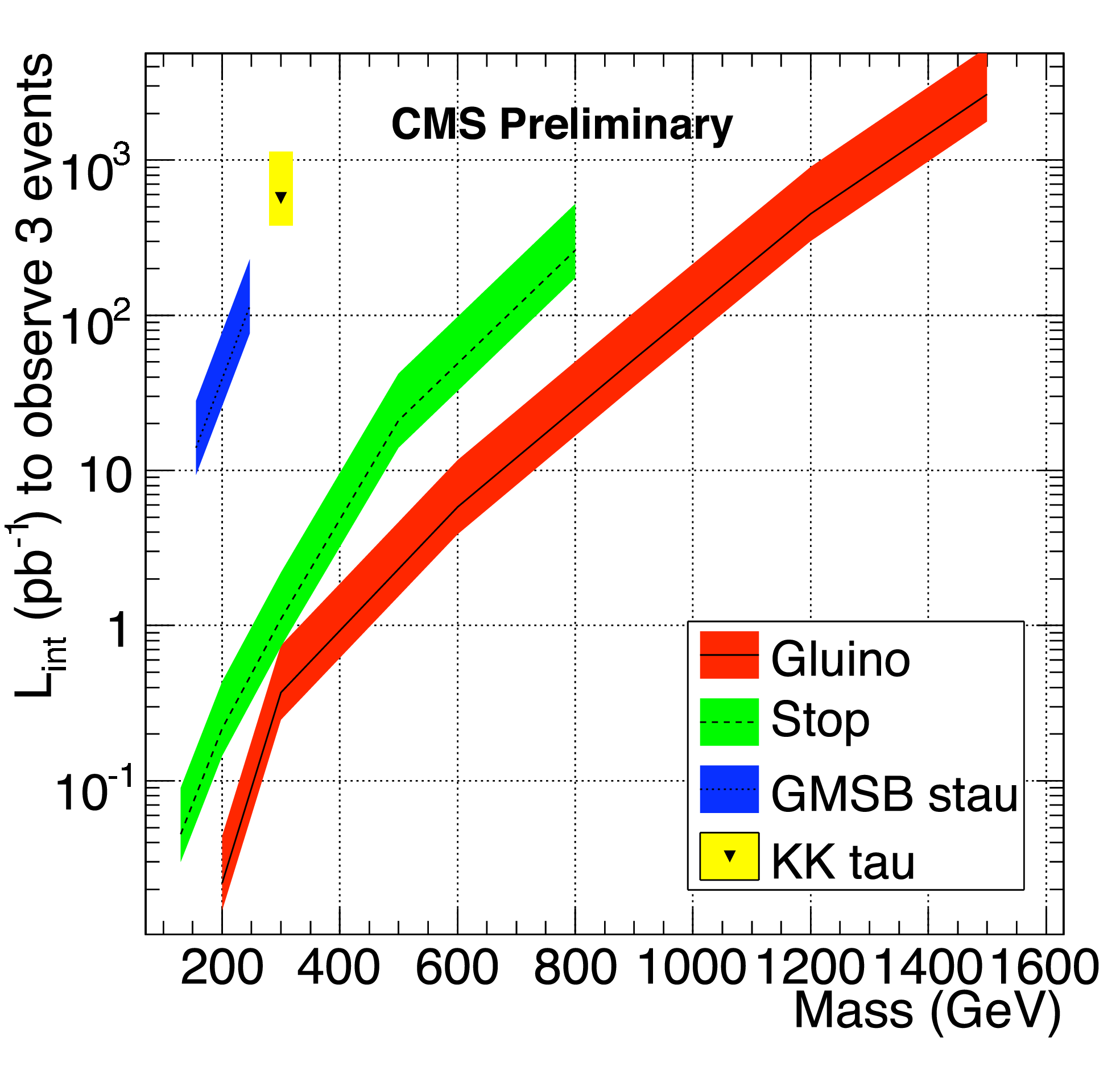}
\end{minipage}
\vspace{-3em}
  \caption{(left) The timing of a heavy charged
particle as measured in ATLAS compared with the true timing~\cite{Aad:2009wy}; 
(right) Luminosity needed for a discovery requiring 3 events 
(for no background) for various heavy stable charged particles in
CMS~\cite{EXO08}.} 
  \label{hcsp}
\end{figure}

\subsection{Black Holes}
\label{sec:black}

\newcommand{\CH}{C{\small HARYBDIS2}}
\newcommand{\BM}{B{\small LACK}M{\small AX}}

\bigskip
In scenarios with large or warped extra dimensions, the higher-dimensional
Planck mass $M_D$ could be as low as the TeV scale.  For three or more extra
dimensions, this is not excluded by astrophysical observations. Then gravity
would be a much stronger force at short distances and black holes could be
formed in multi-TeV particle collisions.  For a recent review and references,
see~\cite{Kanti:2008eq}. 

At energies sufficiently far above $M_D$, it should be possible to treat black hole formation in particle 
collisions using general relativity extended to higher dimensions.  Even in this classical approximation 
the dynamics of such a highly nonlinear process is not yet fully worked out.  For a recent study in four 
dimensions, see~\cite{Shibata:2008rq}.  For extra dimensions we have only lower limits on the impact 
parameter for formation and on the mass  of the resulting black hole~\cite{Yoshino:2005hi}.  
The collision energy that is not trapped in the black hole is emitted in gravitational radiation, which escapes 
detection.  However, studies at zero impact parameter suggest that the losses of energy and angular momentum in 
this so-called balding process do not saturate the limits derived by Yoshino and Rychkov in \cite{Yoshino:2005hi}.

The simulation program \CH~\cite{Frost:2009cf} includes a model for the balding process that satisfies the Yoshino-Rychkov bounds while being consistent with the results of other approaches at zero impact parameter.  Typically $\sim 80$\% of the collision energy and angular momentum is trapped.   The model is activated by setting the parameter {\tt MJLOST=.TRUE.}.  The production cross section for different values of the Planck scale {\tt MPL}$=M_D$ and {\tt MJLOST} options is shown in Fig.~\ref{fig:BHcs}. 

\begin{figure}
  \centering
   \includegraphics[height=7 cm]{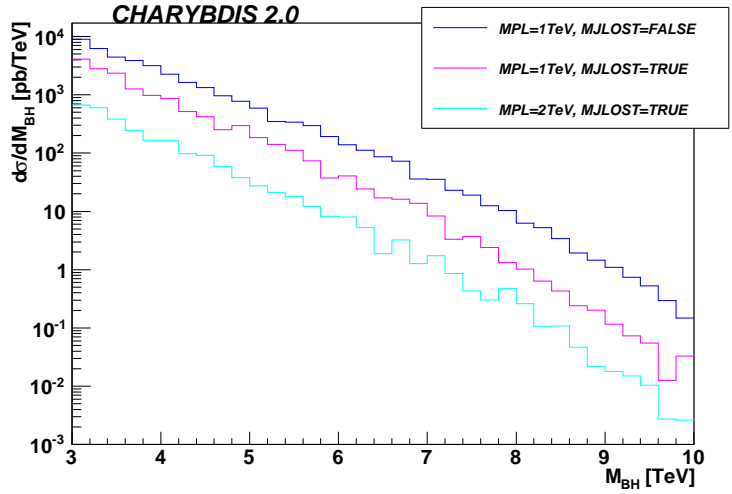}
\caption{Black hole cross section at LHC for $n=4$ extra
   dimensions~\cite{Frost:2009cf}.} 
\label{fig:BHcs}
\end{figure}

After formation, the black hole decays rapidly by Hawking evaporation.  In \CH\ and also in the program \BM~\cite{Dai:2007ki,Dai:2009by}, but not in earlier programs (including earlier versions of C{\small HARYBDIS}), angular momentum is taken into account.  This affects the spectra and angular distributions of the emitted particles, and the relative abundances of different particle species.  All Standard Model particles are assumed to be emitted ``on the brane'', i.e.\ they do not propagate into the extra dimensions. Their differential fluxes are given by
\begin{equation}
\frac{d^4N_\lambda}{d\cos\theta\,d\phi\,d\omega\,dt} =\frac 1{4\pi}
\sum_{jm}\frac{T_{jm}}{e^{\frac{\omega-m\Omega}{T}}\pm 1}|_\lambda S_{jm}(\theta,\phi)|^2
\end{equation}
where $\lambda$, $\omega$, $j$ and $m$ are the helicity, the energy and the total and azimuthal angular momentum quantum numbers of the emission,  $T$ is the Hawking temperature, $\Omega$ is the horizon angular velocity, $T_{jm}$ is the coefficient for transmission from the horizon to infinity (the ``greybody factor'' modifying the purely thermal spectrum), and $_\lambda S_{jm}$ is a (generalized) spheroidal harmonic function.
The dependence of the thermal factor on $\omega-m\Omega$, which is just the energy in a frame co-rotating with the horizon, favour emissions with high values of $m$, which help the black hole to shed its angular momentum.  The same dependence skews the spectrum toward higher energies, relative to the non-spinning case.

The spheroidal harmonic angular dependence leads to a distribution of higher-energy emissions concentrated around the equatorial plane of the black hole.  Since the original angular momentum vector is approximately perpendicular to the beam directions, this implies a somewhat broadened rapidity distribution.  Emission of particles with non-zero spin and appropriate helicity along the polar axes also enables the black hole to lose angular momentum and becomes more favourable at lower energies.  Thus for example neutrinos and antineutrinos are preferentially emitted in the southern and northern hemispheres respectively.  Similarly the decay angular distributions of emitted W bosons are strongly correlated with their polarization and hence also with the orientation of the black hole.

Owing to the high number of colour degrees of freedom, the Hawking emission is dominated by quarks and gluons.  The polar emission mechanism mentioned above enhances the flux of vectors relative to fermions, and of fermions relative to scalars (including longitudinal vector bosons), compared to the fluxes from a non-rotating black hole.  However, the shift in the spectra to higher energies means that the overall multiplicity of emitted particles is reduced.

One missing component of the existing simulations is Hawking emission of gravitons.  This is because gravitons are emitted into the higher-dimensional bulk and the corresponding greybody factors and angular distribution functions for rotating black holes are unknown.\footnote{Bulk graviton emission from non-rotating black holes is included in \BM.}  For numbers of dimensions that are not too large, the relatively low number of graviton degrees of freedom probably make this a small effect, comparable with the uncertainties in the amount of gravitational radiation in the formation process.

As the Hawking radiation carries off energy and angular momentum, the black hole becomes lighter and loses its spin.  On the average these processes occur in parallel, rather than as distinct spin-down and static evaporation phases.  As the mass decreases the Hawking temperature rises, until the mass and/or temperature approach the Planck scale. At this stage the process leaves the classical realm and a quantum theory of gravity would be required to follow it further.  The simulation programs include a variety of models for this Planck phase of black hole decay, ranging from a stable exotic remnant particle to a string-inspired option in which
particles ``boil off'' at a fixed limiting temperature.  See refs.~\cite{Frost:2009cf,Dai:2007ki,Dai:2009by} for details.

In conclusion, the production of higher-dimensional black holes in particle collisions remains a possibility
 worth exploring.  The formation process is still not well understood and there are uncertainties in
the fractions of the collision energy and angular momentum that would be trapped in the black hole.  Once
the initial conditions are established, the main phase of decay via Hawking radiation is under better control.
The only missing component is bulk graviton emission.  The terminal Planck-scale phase of decay is not
understood but a variety of models are available and its contribution to the final state is probably not large.

The effects of black hole rotation during the Hawking emission phase are substantial: the spectra, angular
distributions and relative abundances of particles are all affected.  This will complicate the extraction of
the fundamental parameters, i.e.\ the number of extra dimensions $n$ and the Planck scale $M_D$. 
More sophisticated analyses than those formulated for the non-rotating case are required and are
currently under study.

Clearly if the LHC enters the energy region where  extra dimensions can be probed  and  
 micro-black-holes can be produced, the data will give us insights on the energy scales involved. Any  
 future colliders will then no doubt be around or above the energy threshold for these phenomena, expected to be
 well over the TeV range. It would be useful to be able to span a range up to  five to ten times the fundamental
 Planck scale, to study quantum gravity effects and reach a region where the black hole dynamics is expected to be described by general relativity. While multi-TeV lepton colliders can perhaps offer precision measurements in the  domain just beyond the Planck scale, an energy upgrade of the LHC (DLHC) or even a Very Large Hadron
 collider may be required for a complete mapping of this new regime.
 In some cases it is also possible that the cross sections of the black hole production will be low 
 (eg if apparent horizon effects are important) so that  the LHC will only see some evidence of black hole
 production but does not allow for a detailed study. Then the LHC luminosity upgrade may be essential.

\subsection{Flavour Physics}
\label{Sect:flavour}

\bigskip
In the last few years there has been great experimental progress 
in quark and lepton flavour physics. In the quark sector, 
the validity of the Standard Model has been strongly 
reinforced by a series of challenging 
tests~\cite{Bona:2007vi,Charles:2004jd}.
All the SM parameters controlling quark-flavour dynamics 
(quark masses and CKM angles)
have been determined with good accuracy.
More important, the measurements of several suppressed 
observables, such as $\Delta M_{B_d}$, $\Delta M_{B_s}$, 
${\mathcal A}^{\rm CP}(B\to K \Psi)$,
${\mathcal B}(B\to X_s \gamma)$, $\epsilon_K$, 
do not show significant deviations from the SM.
The situation is somehow similar to the flavour-conserving 
electroweak precision observables (EWPO) after LEP: 
the SM works very well and genuine one-loop electroweak effects 
have been tested with relative accuracy in the $10\%$--$30\%$ range.
Similarly to the EWPO case, also in the quark flavour sector 
New Physics (NP) effects 
can only appear as a small correction to the leading SM contribution.

If NP respects the SM gauge symmetry, as we expect from general 
arguments, the corrections to low-energy 
flavour-violating amplitudes in the quark sector 
can be written in the following general form
\begin{equation}
\mathcal{A}(q_i \to q_j + X) = \mathcal{A}_{0} \left[  \frac{c_{\rm SM}}{M_W^2} 
+ \frac{c_{\rm NP}}{\Lambda^2} \right]~,
\label{eq:fl1}
\end{equation} 
where $\Lambda$ is the energy scale of the new degrees of freedom.
This structure is very general: the coefficients 
$c_{\rm SM(NP)}$ may include appropriate CKM factors 
and eventually a $\sim 1/(16\pi^2)$ suppression 
if the amplitude is loop-mediated. Given our ignorance 
about $c_{\rm NP}$, the values of the scale $\Lambda$ probed by present 
experiments vary over a wide range. However, the general result 
in Eq.~(\ref{eq:fl1}) allow us to predict how these bounds will 
improve with future experiments: the sensitivity on $\Lambda$
scale as $N^{1/4}$, where $N$ is the number of events 
used to measure the observable. This implies that 
is not easy to increase substantially the energy 
reach with indirect NP searches only. 

On the other hand, if $\Lambda$ is just above the electroweak 
scale (and the LHC will soon provide a clear evidence of some
new states), then we are already learning a lot 
about the couplings of these new degrees of freedom
from flavour observables. Indeed to keep $\Lambda$ 
close to the electroweak scale we need some alignment 
between the SM and NP flavour structures. Natural 
possibilities are the Minimal Flavour Violation (MFV)
hypothesis~\cite{D'Ambrosio:2002ex}, which could easily be 
implemented in supersymmetric models 
or the so-called RS-GIM protection, which is at work in models 
with warped extra dimensions~\cite{Agashe:2004cp}.
In all these cases improving the existing constraints 
on low-energy flavour-violating observables is necessary 
to improve our knowledge about  
some of the fundamental couplings of the NP model. 

An example of the interest of improving measurements 
in the quark flavour sector, even in a framework where the 
impact of flavour observables is minimal, such as the 
constrained MSSM, is shown in  Fig.~\ref{fig:drill}.
Here we illustrate the present impact of various 
low-energy measurements
in constraining the parameter-space of the model, and the 
possible future impact assuming a reduction of the present 
errors~\cite{Buchmueller:2008qe}.
Even if the flavour structure of the CMSSM is completely 
specified, helicity-suppressed observables such as 
${\mathcal B}(B\to \tau \nu)$, ${\mathcal B}(B\to X_s \gamma)$, 
and ${\mathcal B}(B_{s}\to \mu^+ \mu^-)$
(not explicitly shown in Fig.~\ref{fig:drill} because 
of the present weak bound),
are very useful in constraining the model because 
of their large sensitivity to $\tan\beta=v_d/v_u$. 
Since $\tan\beta$ cannot be determined with high 
accuracy from high-$p_t$ physics, these 
low-energy observables will remain key measurements 
also in the LHC era.
 
\begin{figure}[t]
\begin{center}
\includegraphics[width=13 cm]{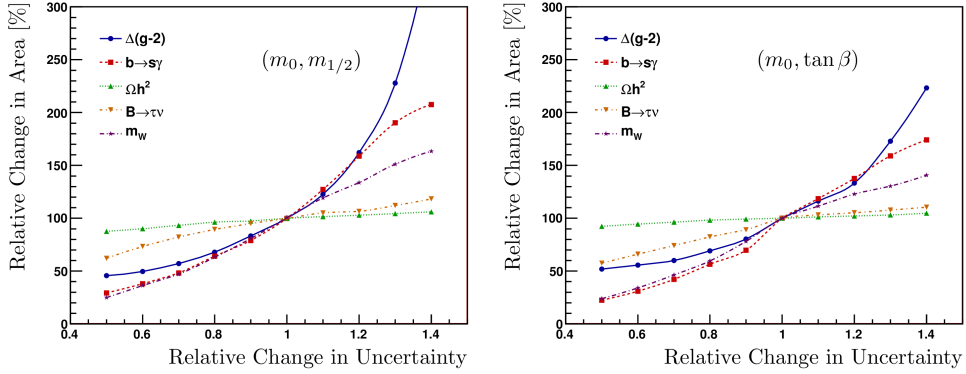}
\end{center}
\caption {Relative sizes of the 95\%~C.L.\ areas in the $(m_0, m_{1/2}$)
  plane (left) and in the $(m_0, \tan\beta)$ plane (right) as a function
  of the hypothetical errors of various low-energy observables
  (plus $M_W$ and $\Omega_{\rm CDM}h^2$)~\cite{Buchmueller:2008qe}.
  The error
  scaling is relative to the current combined theory and experimental
  error. 
}
\label{fig:drill}
\end{figure}

While helicity suppressed $B$-physics observables are
very interesting in the CMSSM, $K$ and $D$ decays are 
more interesting in different NP frameworks.
Altogether the set of low-energy observables to be measured 
with higher precision in the quark sector is quite limited. 
In several cases we are already dominated by irreducible 
theoretical uncertainties: the theoretical error 
on $c_{\rm SM}$ in Eq.~(\ref{eq:fl1}) prevents the 
observation of possible NP effects. 
However, there are a few windows for very interesting 
dedicated new experiments. A notable example are
the ultra-rare $K\to \pi\nu\bar\nu$ decays. 
Here the irreducible theoretical errors are very 
small~\cite{Buras:2005gr}
and these decays modes are quite interesting in the MSSM 
with non-MFV sources of flavour symmetry 
breaking~\cite{Isidori:2006qy}, and in various 
models with extra dimensions~\cite{Blanke:2008yr}.

Compared to the quark sector, the situation of 
flavour physics in the lepton sector is more uncertain 
but also more exciting. The discovery of 
neutrino oscillations has clearly revealed new 
flavour structures beside the three 
SM Yukawa couplings. We have not yet enough information
to unambiguously determine how the SM Lagrangian 
should be modified in order to describe the 
phenomenon of neutrino oscillations. However, 
natural explanations point toward the 
existence of new degrees of freedom 
with explicit breaking of lepton number
at very high energy scales,
in agreement with the expectations of 
Grand Unified Theories (GUT). 
In several realistic supersymmetric 
frameworks, 
the new sources of lepton-flavour violation (LFV)
should give rise to visible               
effects also in the charged-lepton sector. 
For this reason improved searches of LFV 
processes such as  $\mu\to e\gamma$, 
$\tau \to \mu\gamma$, or $\mu\to e$ conversion in nuclei,  
are particularly interesting. A significant step forward 
in this field is expected soon by the MEG experiment, 
which should reach a sensitivity on $\mathcal{B}(\mu\to e\gamma)$ 
around $10^{-13}$, two orders of magnitude below the present one, 
covering a significant parameter region 
of realistic supersymmetric models~\cite{Hisano:2009ae}. 
However, we stress that also 
for LFV processes the general decomposition in Eq.~(\ref{eq:fl1})
is valid: in this case  $c_{\rm SM}=0$, hence we are never limited by 
irreducible theoretical errors, but the mild sensitivity on 
$\Lambda$ implied by the $N^{1/4}$ scaling is still valid. 

\subsection{Summary and conclusions of WG4}

A broad spectrum of signatures for new phenomena have been extensively studied for the LHC.
Many of the new physics signals discussed here are accessible with low luminosity
at the LHC, extending the search reach of the Tevatron.  In some cases, {\it e.g.}, 
the 4 generation SM, early operations at the LHC can either discover or exclude
the model.  In other cases, the increasing  luminosity (and energy) of the LHC with time  
will extend the reach and/or will  allow for measurements of the properties of the new
states.  Some scenarios, such as the  Little Higgs Model, may require several tens of fb$^{-1}$
of data. For $Z'$-like objects, as discussed in detail in this report,  discovery
depends on the mass and the couplings: early observation is possible but 
in other cases 100 fb$^{-1}$ at 14 TeV may be barely enough to claim evidence.
In addition, new physics could well be observed  with unusual signatures. The LHC experiments have
prepared for that
to the best of their abilities and a signal could be observed early on.

Once a discovery has been made at the LHC, it will be imperative to determine the underlying theory
which gives rise to the new phenomena.  This has been the hallmark of the physics case for a
high energy $e^+e^-$ linear collider \cite{AguilarSaavedra:2001rg,Weiglein:2004hn,ilc}.  The sLHC luminosity upgrade has the capability
to add crucial information on the properties of new physics (in addition to increasing the
search sensitivity), although full studies have yet to
be performed.  A linear collider, with its clean environment,
known initial state, and polarized beams, is unparalleled in terms of its abilities to conduct
ultra-precise measurements of new (as well as SM) phenomena, as long as the new physics scale
is within reach of the machine.  The physics case has yet to be established for the LHeC and
the muon collider.  In the former case, it is clear from HERA data that much ground can be
covered.  In the latter case, a background saturated environment, challenging
vertex measurements and lack of polarized beams, as well as a significant loss
of forward 
coverage due to shielding will make precision measurements challenging.

A roadmap of sensitivities of various colliders for new physics scenarios is a useful
guide to determine a machine's ability to probe new interactions.
To that end, the sensitivities of various future facilities for discovering some scenarios
are reproduced in 
Table~\ref{new_physics} (from \cite{DeRoeck:2001nz}).  While 
lepton colliders allow for much more precise and complete measurements, as stated above,
their effectiveness depends on the scale of the new physics.  However, for the scenarios listed
in this Table, the ILC sensitivity essentially matches that of the LHC, while CLIC matches
that of the VLHC. 
Equally important is to understand the consequences if NO new signal is observed at the LHC,
and how to tackle that scenario with future machines, although all evidence points to new
physics at the Terascale.

In conclusion, we look forward to exciting times ahead with spectacular discoveries at the LHC,
and to these discoveries pointing the way forward to the next machine.

\begin{table}
\begin{center}
\begin{tabular}{|c|c|c|c|c|c|c|}
\hline 
Process    & LHC   & sLHC & DLHC & VLHC & ILC & CLIC \\
 & 14 TeV & 14 TeV & 28 TeV & 200 TeV& 0.8 TeV & 5 TeV \\
 & 100 fb$^{-1}$ & 1000 fb$^{-1}$ & 100 fb$^{-1}$ & 100 fb$^{-1}$ & 500 fb$^{-1}$ & 1000 fb$^{-1}$ \\
 \hline
 $Z'$  & 5 & 6 & 8 & 35 & 8 & 30 \\
 ED\;($\delta = 2$)  & 9 & 12 & 15 & 65 & 5-8.5 & 30-55\\
 excited quarks & 6.5 & 7.5 & 9.5 & 75 & 0.8 & 5 \\
 $\Lambda_{\rm compositness}$  & 30 & 40 & 40 & 100 & 100 & 400\\
 \hline
 \end{tabular}
 \caption{Illustrative reach in TeV for the different new physics scenarios. The Z' and EDs for the LCs are
 indirect reach limits from precision measurements.}
 \label{new_physics}
 \end{center}
 \end{table}


 }

\newpage

\renewcommand{\appendixname}{Appendix}

\addcontentsline{toc}{section}{Appendix: Future Colliders Overview Series}
\begin{appendix}
\section*{Future Colliders Overview Series}

The primary theme of the Theory Institute was to organise our thoughts around
four broadly-based physics signature categories, and then to ask how various
colliders can uncover and study the corresponding physics theories. The talks,
discussions and write-ups of the four Working Groups are the products of that
effort. Special attention was given to the near-term capabilities of the LHC,
as  the start-up of that collider is currently approaching. Results from the
Tevatron were also integrated in the studies, along with some anticipations of
gains one might obtain from the projected LHC luminosity upgrade (sLHC), an
LHC-electron collider (LHeC), a future high-energy $e^+e^-$ collider (ILC
and/or CLIC), and a muon collider.   

In addition to this effort placed within the Working Group structure, we also
commissioned presentations that were focused exclusively on future
technologies. We wanted summaries of the physics case for each possible future
facility and a summary of its technology status. We were fortunate to have ten
leaders of their respective colliders agree to participate in this ``Future
Colliders Overview Series." 

Aurelio Juste's presentation was on the latest developments at the Tevatron
and the expectations for Tevatron machine running, studying the Standard
Model, and discovering new physics in the next few years. In particular,
slides 48-55 have a discussion of current Higgs boson limits and future search
prospects at the Tevatron.  Lyn Evans gave a comprehensive talk on the LHC
accelerator status, with comments about the path to upgrading to a
higher-luminosity machine (the sLHC). Michelangelo Mangano gave the physics
case for the sLHC, and summarised some of the attending challenges as well.
Max Klein and Emmanuelle Perez spoke about the technology challenges and
physics opportunities for turning the LHC machine into an electron-proton
collider (the LHeC).  

We had four talks devoted to future $e^+e^-$ colliders.  Brian Foster and
Klaus Desch discussed the technology and physics case for the ILC, which  is a
mature design for sub-TeV $e^+e^-$ collisions with potential for expanding to
larger energies depending on site and design specifics.   Jean-Pierre Delahaye
and Marco Battaglia gave parallel talks for CLIC, which is in its R\&D phase,
with the goal of providing a design option that enables colliding $e^+e^-$
beams at energies of several TeV in the center of mass.  

Finally, the status of the muon collider was discussed by Robert Palmer, who
explained the advantages and limitations of  pursuing muon collider
technologies. The ``easy parts" and the ``hard parts" of reaching a final
design for a collider were carefully detailed. An R\&D programme was proposed
that would be necessary to determine the feasibility of a muon collider. 

Below we list all ten talks in the series in alphabetical order of the
speakers names. The slides for each talk can be accessed directly by the URLs
given. In addition to the full URL, we have provided for each talk a shortened
URL in parentheses from the http://is.gd/ service for the reader's
convenience. 

\newpage
\bigskip
\noindent
{\bf Talks delivered in the Future Colliders Overview Series}
\bigskip

Marco Battaglia, {\it Physics case for CLIC}, 19 Feb 2009 (is.gd/3gdkx) \\
\small{\tt http://indico.cern.ch/contributionDisplay.py?contribId=11\&confId=40437}
\bigskip

Jean-Pierre Delahaye, {\it Technology path to CLIC}, 19 Feb 2009  (is.gd/3gdsG)\\
\small{\tt http://indico.cern.ch/contributionDisplay.py?contribId=13\&confId=40437}
\bigskip

Klaus Desch, {\it Physics case for the ILC}, 17 Feb 2009 (is.gd/3geif)   \\
\small{\tt http://indico.cern.ch/contributionDisplay.py?contribId=1\&confId=40437}
\bigskip

Lyn Evans, {\it LHC accelerator status and upgrade plans}, 10 Feb 2009 (is.gd/3gelI) \\
\small{\tt http://indico.cern.ch/contributionDisplay.py?contribId=0\&confId=40437}
\bigskip

Brian Foster, {\it Technology progress report of the ILC}, 17 Feb 2009 (is.gd/3genR)  \\
\small{\tt http://indico.cern.ch/contributionDisplay.py?contribId=3\&confId=40437}
\bigskip

Aurelio Juste, {\it Recent results and prospects from the Tevatron}, 26 Feb 2009 (is.gd/3gcij) \\
\small{\tt http://indico.cern.ch/contributionDisplay.py?contribId=19\&confId=40437}
\bigskip

Max Klein, {\it Towards a LHeC at the LHC}, 18 Feb 2009  (is.gd/3gerX)\\
\small{\tt http://indico.cern.ch/contributionDisplay.py?contribId=9\&confId=40437}
\bigskip

Michelangelo Mangano, {\it Physics opportunities with the sLHC}, 20 Feb 2009 (is.gd/3gevW) \\
\small{\tt http://indico.cern.ch/contributionDisplay.py?contribId=5\&confId=40437}
\bigskip

Robert Palmer, {\it Muon collider technology status}, 24 Feb 2009 (is.gd/3gey0)  \\
\small{\tt http://indico.cern.ch/contributionDisplay.py?contribId=14\&confId=40437}
\bigskip

Emmanuelle Perez, {\it Physics opportunities with the LHeC}, 18 Feb 2009 (is.gd/3geAB) \\
\small{\tt http://indico.cern.ch/contributionDisplay.py?contribId=7\&confId=40437}
\bigskip

\end{appendix}


\end{document}